\documentclass[12pt,twoside,openany,letterpaper]{book}

\usepackage[utf8]{inputenc}
\usepackage{graphicx}
\usepackage[english]{babel}

\usepackage{cite}
\usepackage{amsmath}
\usepackage{amssymb}

\usepackage{relsize}
\usepackage{graphics}
\usepackage[export]{adjustbox}
\usepackage{subfig}

\usepackage{pifont}

\usepackage{tensor}
\usepackage{mathrsfs}
\usepackage{relsize}
\usepackage{amsfonts}
\usepackage[makeroom]{cancel}

\usepackage[onehalfspacing]{setspace}

\usepackage{bbm}

\usepackage{geometry}
\usepackage{parskip}

\usepackage[nottoc]{tocbibind}

\usepackage{color}
\definecolor{mygray}{rgb}{0.5,0.5,0.5}
\usepackage{listings}
\usepackage{fancyvrb}
\usepackage{fancyhdr}
\usepackage{etoolbox}

\fancypagestyle{chapter_new}{%
	\fancyhead[RE]{\thepage}%
	\fancyhead[LE]{}%
	\fancyhead[LO]{}%
	\fancyhead[RO]{\thepage}%
}
\patchcmd{\chapter}{plain}{chapter_new}{}{}

\fancypagestyle{plain}{%
	\fancyhead[RE]{\thepage}%
	\fancyhead[LE]{}%
	\fancyhead[LO]{}%
	\fancyhead[RO]{\thepage}% % except the center
	
	}

\newcommand{\dd}{\mbox{d}}

\newcommand{\ind}[1]{{\mbox{\scriptsize #1}}}

\newcommand{\ts}[1]{{\boldsymbol{#1}}}         % equivalent, just shorter
\newcommand{\lap}{\triangle}

\newcommand{\phisq}{\langle\varphi^2(x)\rangle_\text{ren}}

\usepackage[colorlinks=true,citecolor=green,linkcolor=red,filecolor=cyan,urlcolor=magenta,backref=page]{hyperref}
\renewcommand*{\backref}[1]{}
\renewcommand*{\backrefalt}[4]{%
    \ifcase #1 (Not cited.)%
    \or        (Cited on page~#2.)%
    \else      (Cited on pages~#2.)%
    \fi}

\usepackage{makeidx}
\makeindex

\allowdisplaybreaks

\usepackage{caption}
\makeatletter
\renewcommand\listoftables{%
        \@starttoc{lot}%
}
\renewcommand\listoffigures{%
        \@starttoc{lof}%
}
\makeatother

\begin{document}

%\maketitle
%\begin{titlepage}
\thispagestyle{empty}

{\raggedleft
	{\huge Effects of Non-locality}\\[0.2\baselineskip]
	{\huge in Gravity and Quantum Theory}\\[3\baselineskip]
}

\vspace{30pt}
\begin{center}
\includegraphics[height=.3\textwidth]{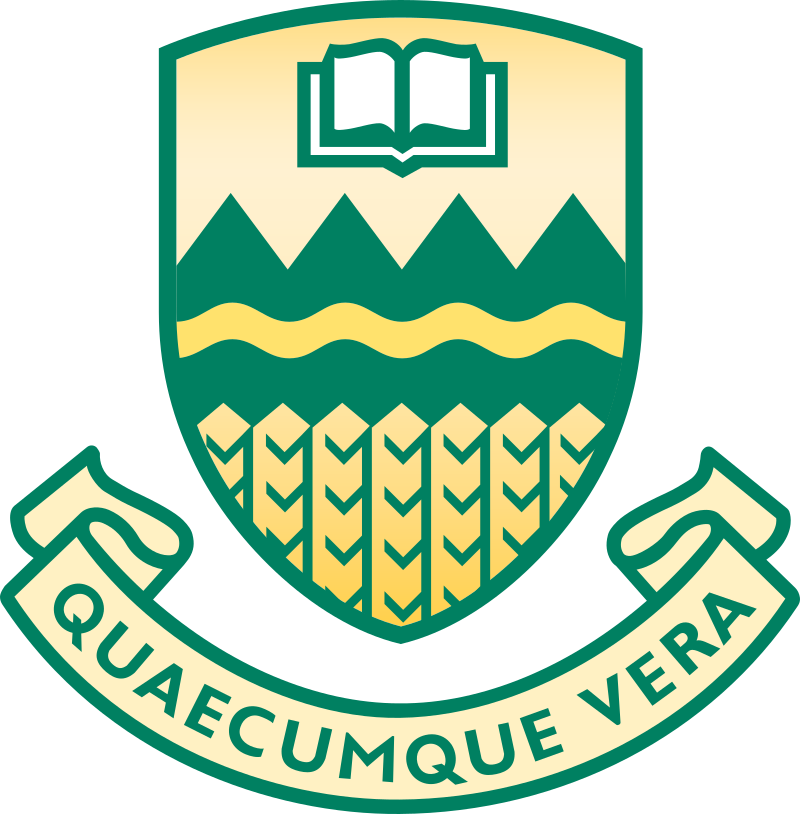}
\end{center}

\vspace{2cm}

\begin{center}
A thesis submitted to the Faculty of Graduate Studies and Research\\
in partial fulfillment of the requirements for the degree of Doctor of Philosophy

\vspace{1.5cm}

%{\Large\color{magenta} --- PRELIMINARY VERSION ---}
{\Large\color{magenta} --- arXiv version ---}

\vfill

University of Alberta\\
Department of Physics\\
Edmonton, Canada\\
\today

\end{center}

\vfill

%\noindent
%\parbox{.5\textwidth}{\hfill}
%\parbox{.5\textwidth}{\hfill \emph{Supervisor}}
\noindent
\parbox{.5\textwidth}{\emph{Author }\hfill}
\parbox{.5\textwidth}{\hfill \emph{Supervisor}}
\noindent
\parbox{.5\textwidth}{Jens Boos\hfill}
\parbox{.5\textwidth}{\hfill Prof.\ Valeri P.\ Frolov}

%\end{titlepage}

%\thispagestyle{empty}
%\cleardoublepage
%\frontmatter

\chapter*{Abstract}
\pagenumbering{roman}
\setcounter{page}{2}
\thispagestyle{chapter_new}
\addcontentsline{toc}{section}{Abstract}
\markright{Abstract}
\doublespacing

Spacetime---the union of space and time---is both the actor and the stage during physical processes in our fascinating Universe. In Lorentz invariant local theories, the existence of a maximum signalling speed (the ``speed of light'') determines a notion of causality in spacetime, distinguishing the past from the future, and the cause from the effect.

This thesis is dedicated to the study of \emph{deviations} from locality. Focussing on a particular class of \emph{non-local} theories that is both Lorentz invariant and free of ghosts, we aim to understand the effects of such non-local physics in both gravity and quantum theory. Non-local ghost-free theories are accompanied by a parameter $\ell$ of dimension length that parametrizes the scale of non-locality, and for that reason we strive to express all effects of non-locality in terms of this symbol. In the limiting case of $\ell=0$ one recovers the local theory, and the effects of non-locality vanish.

In order to address these questions we develop the notion of non-local Green functions, study their causal properties, and demonstrate that non-locality leads to a violation of causality on small scales but may be recovered at macroscopic distances much larger than the scale of non-locality. In particular, we utilize non-local Green functions to construct the stationary gravitational field of point particles and extended bodies in the weak-field limit of non-local gravity and demonstrate explicitly that non-locality \emph{regularizes gravitational singularities} at the linear level. Boosting these solutions to the speed of light in a suitable limit, we obtain a class of geometries corresponding to non-local regular ultrarelativistic objects.

In the context of quantum mechanics we demonstrate that non-locality affects the scattering coefficients of a scalar field in the presence of a $\delta$-shaped potential: for a critical frequency, the potential becomes completely opaque and reflects 100\% of the incoming wave of that frequency. In the realm of non-local quantum field theory we first illustrate that non-locality smoothes the vacuum polarization and thermal fluctuations in the vicinity of a $\delta$-shaped potential and then prove the fluctuation-dissipation theorem in this particular case.

Turning towards quantum field theory in curved spacetime, we construct a non-local ghost-free generalization of the Polyakov effective action and evaluate the resulting quantum average of the energy-momentum tensor in the background of a two-dimensional black hole. While non-locality does not affect the asymptotic flux of Hawking radiation in this model, the entropy of the black hole is sensitive to the presence of non-locality.

The results presented in this thesis establish several effects of a Lorentz invariant, ghost-free non-locality in the areas of both gravitational and quantum physics.

keywords: non-locality, ghost-free, Green functions

\onehalfspacing

\chapter*{Preface}
\thispagestyle{chapter_new}
\addcontentsline{toc}{section}{Preface}
\markright{Preface}

\subsubsection*{Statement of Originality}
I hereby declare that this Ph.D.\ thesis reflects original research; all results have been obtained using none other than the stated references. All references have been properly cited.

\vspace{30pt}

\begin{flushright}
  \underline{\hspace{7cm}}\\[0.2cm]
  Jens Boos, Edmonton, September 2020
\end{flushright}

\subsubsection*{Relevant partial publications/papers}

This thesis is based on several research papers \cite{Boos:2018bxf,Boos:2018bhd,Boos:2018kir,Boos:2019fbu,Boos:2019zml,Boos:2019vcz,Boos:2020kgj,Boos:2020ccj}. In particular, Refs.~\cite{Boos:2018bxf,Boos:2018kir,Boos:2019fbu,Boos:2019zml,Boos:2019vcz} are the product of close collaboration with Prof.~Valeri Frolov and Dr.~Andrei Zelnikov, and the author was significantly involved in all steps of the research. Ref.~\cite{Boos:2020ccj} is the result of collaborative work with Prof.~Valeri Frolov and Jose Pinedo Soto, where the author was again significantly involved in all steps of the research. Refs.~\cite{Boos:2018bhd,Boos:2020kgj} are single-authored papers.

Chapters \ref{ch:ch2}--\ref{ch:ch7} are based on a subset of the previously mentioned publications/papers, and we highlight the relevant papers at the beginning of each chapter. Chapters \ref{ch:ch1} and \ref{ch:ch8} are of more general nature since they provide an introduction/conclusion/outlook based on the results presented in this thesis. Appendices \ref{app:calc} and \ref{app:2d} consist of supplemental material for the relevant chapters, and they are to be considered part of their corresponding citing chapters. The material presented in Secs.~\ref{sec:ch3:rotating-p-branes}, \ref{sec:ch4:gyratonic-branes}, and \ref{sec:ch5:qnm} contains unpublished results that may be part of future publications.

\begin{itemize}

  \item[1.]
	J.~Boos, V.~P.~Frolov and A.~Zelnikov,
	``Gravitational field of static $p$-branes in linearized ghost-free gravity,''
	Phys.\ Rev.\ D {\bf 97} (2018) no.~8, 084021;
	\href{http://arxiv.org/abs/1802.09573}{arXiv:1802.09573 [gr-qc]}.

  \item[2.]
	J.~Boos,
	``Gravitational Friedel oscillations in higher-derivative and infinite-derivative gravity?,''
	Int.\ J.\ Mod.\ Phys.\ D {\bf 27} (2018) no.~14, 1847022;
	\href{http://arxiv.org/abs/1804.00225}{arXiv:1804.00225 [gr-qc]};
	Honorable Mention in the Gravity Research Foundation Essay Competition 2018.

  \item[3.]
	J.~Boos, V.~P.~Frolov and A.~Zelnikov,
	``Quantum scattering on a delta potential in ghost-free theory,''
	Phys.\ Lett.\ B {\bf 782} (2018), 688;
	\href{http://arxiv.org/abs/1805.01875}{arXiv:1805.01875 [hep-th]}.
	
  \item[4.]
	J.~Boos, V.~P.~Frolov and A.~Zelnikov,
	``Probing the vacuum fluctuations in scalar ghost-free theories,''
	Phys.\ Rev.\ D \textbf{99} (2019) no.~7, 076014;
	\href{http://arxiv.org/abs/arXiv:1901.07096}{arXiv:1901.07096 [hep-th]}.

  \item[5.]
	J.~Boos, V.~P.~Frolov and A.~Zelnikov,
	``On thermal field fluctuations in ghost-free theories,''
	Phys.\ Lett.\ B \textbf{793} (2019), 290;
	\href{https://arxiv.org/abs/1904.07917}{arXiv:1904.07917 [hep-th]}.
	
  \item[6.]
	J.~Boos, V.~P.~Frolov and A.~Zelnikov,
	``Ghost-free modification of the Polyakov action and Hawking radiation,''
	Phys.\ Rev.\ D \textbf{100} (2019) no.~10, 104008;
	\href{http://arxiv.org/abs/1909.01494}{arXiv:1909.01494 [hep-th]}.

  \item[7.]
	J.~Boos,
	``Angle deficit \& non-local gravitoelectromagnetism around a slowly spinning cosmic string,''
	\href{http://arxiv.org/abs/2003.13847}{arXiv:2003.13847 [gr-qc]}, invited for publication in Int.\ J.\ Mod.\ Phys.\ D;
	Honorable Mention in the Gravity Research Foundation Essay Competition 2020.

  \item[8.]
	J.~Boos, J.~P.~Soto and V.~P.~Frolov,
	``Ultrarelativistic spinning objects in non-local ghost-free gravity,''
    Phys.\ Rev.\ D \textbf{101} (2020) no.~12, 124065;
	\href{http://arxiv.org/abs/2004.07420}{arXiv:2004.07420 [gr-qc]}.

\end{itemize}

\subsubsection*{Unrelated partial publications}

The author has also been part of additional research projects delineated below. Refs.~\cite{Boos:2017pyd,Boos:2017qbx} are the result of close collaboration with Prof.~Valeri Frolov, where the author was significantly involved in all stages of the research project. In Ref.~\cite{Itin:2018dru} the author contributed a small part that corresponds to Ch.~3 and Appendix~A in the published paper and was not involved with other parts of the research.

\begin{itemize}

  \item[9.]
	J.~Boos and V.~P.~Frolov,
	``Stationary black holes with stringy hair,''
	Phys.\ Rev.\ D {\bf 97}, no.~2, 024024 (2018);
	\href{http://arxiv.org/abs/1711.06357}{arXiv:1711.06357 [gr-qc]}.

  \item[10.]
	J.~Boos and V.~P.~Frolov,
	``Principal Killing strings in higher-dimensional Kerr--NUT--(A)dS spacetimes,''
	Phys.\ Rev.\ D {\bf 97}, no.~8, 084015 (2018);
	\href{http://arxiv.org/abs/1801.00122}{arXiv:1801.00122 [gr-qc]}.

  \item[11.]
	Y.~Itin, Y.~N.~Obukhov, J.~Boos and F.~W.~Hehl,
	``Premetric teleparallel theory of gravity and its local and linear constitutive law,''
	Eur.\ Phys.\ J.\ C {\bf 78}, no.~11, 907 (2018);
	\href{http://arxiv.org/abs/1808.08048}{arXiv:1808.08048 [gr-qc]}.

\end{itemize}

\chapter*{}
\thispagestyle{chapter_new}
\addcontentsline{toc}{section}{Dedication}
%\markright{Dedication}
\textit{F\"ur meine Eltern, Regina und Hans,\\ohne die ich nicht auf diese Reise h\"atte aufbrechen k\"onnen.}

\vspace{20pt}

To my parents, Regina and Hans,\\without whom I could not have embarked upon this journey.

\chapter*{Acknowledgements}
\thispagestyle{chapter_new}
\addcontentsline{toc}{section}{Acknowledgements}
\markright{Acknowledgements}
First and foremost, I would like to express my sincere gratitude to my advisor Prof.~Valeri P.~Frolov for his guidance, patience, and support over the last four years. I also benefited greatly from countless discussions with Dr.~Andrei Zelnikov. I am grateful to Prof.~Don Page for hosting my first visit to the University of Alberta in March 2016, for making me feel welcome in Edmonton from the first minute, and for his many thoughtful remarks on a preliminary version of this thesis.

I would like to thank the members of my doctoral examination committee, Prof.~Don Page, Prof.~Dmitri Pogosyan, and Prof.~Richard Sydora, as well as my external examiner Prof.~Anupam Mazumdar for their time.

Outside of Edmonton I want to thank Prof.~David Kubiznak for his support, and Prof.~Friedrich W.~Hehl for more than seven years of mentorship and encouragement.

Over the last years I deeply appreciated the friendship and thoughtful discussions on physics with Dr.~Kento Osuga and Dr.~Yasaman K.~Yazdi, as well as the many interesting exchanges during our weekly Student Meetings with Ameir Bin Akber Ali, Logan Cooke, Joel Hutchinson, Youssef Kora, Sepideh Mirabi, Sneh Modi, Jose Pinedo Soto, Mason Protter, David Purschke, Md Samiur Rahman Mir, Ahmed Rayyan, Sina Safarabadi Farahani, Pramodh Senarath Yapa, Benjamin Smith, Pourya Vakilipourtakalou, and Hennadii Yerzhakov. I am also indebted to Prof.~Frank Marsiglio and Prof.~Craig Heinke for their enthusiastic support of our weekly meetings, as well as the Graduate Physics Student Association for logistical support and caffeinated sustenance.

I would like to express my gratitude to Sarah MacKinnon and Prof.~Richard Marchand for their invaluable administrative support during the first few months; Alexis Brown and Jia Amy Wang for their patient help with financial aspects; Suzette Chan, Sandra Hamilton, Daniel Lamden, Tara Mish, and Carolyn Steinborn for help with all kinds of organizational questions, and Kailey Robertson for keeping me on course during the final months of my Ph.D.

During my doctoral studies I have been a grateful recipient of a Vanier Canada Graduate Scholarship administered by the Natural Sciences and Engineering Research Council of Canada, as well as the Dean's Excellence Recruitment Scholarship Award, a Doctoral Recruitment Scholarship, the President's Doctoral Prize of Distinction, the Golden Bell Jar Graduate Scholarship in Physics, the Andrew Stewart Memorial Graduate Prize, and a Graduate Student Travel Award, all administered by the University of Alberta. 

Last but certainly not least I would like to thank my wonderful family, in Germany and Canada alike, for all their support and encouragement. I could not have done this without you.

\chapter*{Acknowledgement of\\traditional territory}
\thispagestyle{chapter_new}
\addcontentsline{toc}{section}{Acknowledgement of traditional territory}
\markright{Acknowledgement of traditional territory}
The University of Alberta and the present author respectfully acknowledge that we are located on Treaty 6 territory, a traditional gathering place for diverse Indigenous peoples including the Cree, Blackfoot, Metis, Nakota Sioux, Iroquois, Dene, Ojibway/Saulteaux/Anishinaabe, Inuit, and many others whose histories, languages, and cultures continue to influence our vibrant community.

\pagebreak

%\chapter*{Conventions and notation}
%\addcontentsline{toc}{chapter}{Conventions and notation}
%For a more detailed exposition of our conventions in exterior calculus see appendix \ref{ch:appendix_exterior_calculus}.

{\hypersetup{linkcolor=black}
%%\listoffigures
\begingroup
\let\clearpage\relax
\tableofcontents
\endgroup
}

\chapter*{List of figures}
\thispagestyle{chapter_new}
\addcontentsline{toc}{section}{List of figures}
\markright{List of figures}
{\hypersetup{linkcolor=black}
\listoffigures
}

\chapter*{List of symbols}
\thispagestyle{chapter_new}
\phantomsection\label{ListOfSymbols}
\addcontentsline{toc}{section}{List of symbols}
\markright{List of symbols}

This is a list of the most frequent symbols used in this thesis.

\vspace{30pt}

\bgroup
\def\arraystretch{1.1}
\begin{tabular}{l|l}
$D$ & dimension of spacetime \\
$d = D-1$ & dimension of space \\
$\mu,\nu\,\rho,\sigma,\dots$ & letters from the middle of the Greek alphabet denote spacetime indices \\
$\alpha,\beta,\gamma,\delta,\dots$ & letters from the beginning of the Greek alphabet denote spatial indices \\
$\eta{}_{\mu\nu}$ & Minkowski metric with ``mostly plus'' signature $(-1,1,\dots,1)$ \\
$g{}_{\mu\nu}$ & spacetime metric with ``mostly plus'' signature $(-1,1,\dots,1)$ \\
$h{}_{\mu\nu}$ & gravitational perturbation with ``mostly plus'' signature $(-1,1,\dots,1)$ \\
$(\mu\nu)$ & symmetrization of indices, $T{}_{(\mu\nu)} = \tfrac12 (T{}_{\mu\nu} + T{}_{\nu\mu})$ \\
$[\mu\nu]$ & antisymmetrization of indices, $T{}_{[\mu\nu]} = \tfrac12 (T{}_{\mu\nu} - T{}_{\nu\mu})$ \\
$\overset{*}{=}$ & equation only valid under additional conditions \\
$\ell$ & scale of non-locality \\
$G_d(r)$ & free, local and static Green function in $d$ spatial dimensions \\
$\mathcal{G}_d(r)$ & free, non-local and static Green function in $d$ spatial dimensions \\
$G_\omega(r)$ & free local Green function in the temporal Fourier domain \\
$\mathcal{G}_\omega(r)$ & free non-local Green function in the temporal Fourier domain \\
$\Delta\mathcal{G}_\omega(r)$ & non-local modification $\mathcal{G}_\omega(r)-G_\omega(r)$ in the temporal Fourier domain \\
$\kappa = 8\pi G$ & Einstein's gravitational constant, where $G$ is Newton's gravitational constant \\
$\mathrm{GF_N}$ & ghost-free theory with form factor $\exp[(-\ell^2\Box)^N]$ where $N=1,2,\dots$
\end{tabular}
\egroup

%\mainmatter

%%%%%%%%%%%%%%%%%%%%%%%%%%%%%%%%%%%%%%%%%%%%%%%%%%%%%%%%%%%%%%%%%%%%%%%%%%%%%%%%%%%%%%%%%%%%%%%%%%%
%
% Chapter: Introduction
%
\chapter{Introduction}
\pagenumbering{arabic}
\label{ch:ch1}

If there is any principle that has guided humanity on its journey to understand the Universe, it is probably that of cause and effect. The question of \emph{why} permeates the sciences like ripples spread in a pond or gravitational waves emanate space and time itself.

In physics, the principle of cause and effect is called \emph{causality}, and theories that incorporate this principle are called causal. Causality can be seen as a \emph{physical requirement} imposed on theories, and hence---as the cause precedes the effect---causality is deeply interwoven with our definition of the direction of time, and our understanding of the phrases ``before'' and ``after.''

A related notion is that of \emph{locality.} Unlike causality, locality may be viewed as a \emph{technical requirement} to be imposed on a theory. An example of a local theory is wave mechanics: the propagation of a wave in a medium, from the ripples in a pond to gravitational waves in our Universe, depends solely on the local properties of that medium. In \emph{non-local} theories this may not be the case.

Einstein's principle of relativity \cite{Einstein:1905,Einstein:1956} provides a quantitative tool for relating the concepts of causality and locality. If events are spacelike separated in spacetime, there cannot be any causal interaction between them. If they are null or timelike separated, however, then a causal interaction may take place. While it is possible to extend the principle of relativity to incorporate gravity \cite{Einstein:1915,Hawking:1973,MTW:1974,Poisson:2004,Frolov:2011} with great experimental success \cite{Will:2006}, quantum field theories in accordance with the principle of relativity have also been remarkably successful \cite{BjorkenDrell,Weinberg,PeskinSchroeder,Ryder,Zee}. It is highly desirable to construct a theory of quantum gravity that unites these theories in a suitable way \cite{Kiefer:2012}, wherein it is conceivable that the notions of causality and locality are drastically different and only attain their experimentally established roles when either gravity or quantum physics can be neglected. 

Hence, in this thesis we will study the effects of non-locality in both gravity and quantum theory.

\section{Why non-locality?}

Besides being of general interest for quantum theories of gravity---as argued above---non-locality also offers interesting avenues for solutions of various long-standing problems in theoretical physics. Here we would like to address a few of them.

\subsection{Regularization and singularity resolution}
Consider the Poisson equation of Newtonian gravity for a point particle,
\begin{align}
\lap\phi(r) = 4\pi G m\delta(\ts{r}) \quad \Rightarrow \quad \phi(r) = \frac{-Gm}{r} \, .
\end{align}
The resulting gravitational potential is singular at $r=0$, leading to an infinite gravitational force. It is possible to regularize the potential by introducing a heavy mass scale $M$ via
\begin{align}
\lap\left( 1 - \frac{\lap}{M^2} \right)\phi_\text{reg}(r) = 4\pi G m\delta(\ts{r}) \quad \Rightarrow \quad \phi_\text{reg}(r) = \frac{-Gm}{r}\left( 1-e^{-Mr} \right) \, .
\end{align}
The resulting gravitational field $\phi_\text{reg}(r)$ is finite at $r=0$, and if $Mr \gg 1$ one recovers the Newtonian result $\phi(r)$. In the limiting case of $M\rightarrow\infty$ the modification disappears and and we recover the original Newtonian theory. In this particular example one may show that $\phi'_\text{reg}(0)\not=0$, which is somewhat unphysical, but this can be ameliorated by introducing additional mass scales $M_i$ with $i=1,\dots,N$. The Green functions of such theories have the following property:
\begin{align}
G(x',x) \sim \frac{1}{\lap\prod\limits_{i=1}^N\left(1 - \frac{\lap}{M_i^2} \right)} = \frac{1}{\lap} + \sum\limits_{i=1}^N \frac{c_i}{\lap-M_i^2} \, , \quad 1 + \sum\limits_{i=1}^N c_i = 0 \, .
\end{align}
It is important to note that the latter property implies that at least one $c_i$ is negative. At the quantum level such a negative sign can lead to problems, and propagators that appear with the wrong overall sign are commonly referred to as ``ghosts.'' 

This is where the idea of non-locality comes in: if one considers a modification that involves an infinite amount of derivatives one also obtains a regularized potential:
\begin{align}
\lap e^{-\lap/M^2}\phi_\text{GF}(r) = 4\pi G m\delta(\ts{r}) \quad \Rightarrow \quad \phi_\text{GF}(r) = \frac{-Gm}{r}\text{erf}\left(\frac{Mr}{2}\right) \, . 
\end{align}
This potential is finite at $r=0$, and for $Mr\gg 1$ one again recovers $\phi(r)$. The notable difference to the previous case lies in the fact that the propagator of such a theory is \emph{ghost-free} at tree level. Classically, this statement is related to the fact that $e^{-\lap/M^2}$ is never zero and hence its inverse is a regular function without any poles. Because this modification involves infinitely many derivatives the theory becomes non-local on a scale $\ell \sim M^{-1}$. To see why an infinite amount of derivatives corresponds to non-locality recall that the first and second derivative of a discrete function $f_i$ on a lattice with spacing $a$ at a lattice site $n$ are given by
\begin{align}
\left. \frac{\dd f}{\dd x}\right|_{i=n} \equiv \frac{f_{n+1}-f_n}{a} \, , \qquad \left.\frac{\dd^2 f}{\dd x^2}\right|_{i=n} \equiv \frac{\dd}{\dd x} \frac{\dd f}{\dd x} = \frac{f_{n+2} - 2f_{n+1} + f_n}{a^2} \, .
\end{align}
That is, a derivative of higher order requires information on the function in a larger domain. While this provides a somewhat intuitive argument for non-locality we shall make the statement more precise in the following sections.

It is well known that gravitational singularities plague Einstein's theory of gravity as they are prominently featured in the interior of many exact black hole solutions, in cosmological scenarios, and under fairly general assumptions \cite{Droste:1916,Schwarzschild:1916,Kerr:1963ud,Plebanski:1976gy,Hawking:1973}. While it is commonly believed that a consistent theory of quantum gravity should resolve these singularities \cite{Kiefer:2012}, perhaps it is also possible to construct a non-local ghost-free gravitational theory that admits singularity-free gravitational fields for black holes at the non-linear level.

\subsection{Black hole information loss problem}
Next to their singularities, black holes remain somewhat enigmatic when considering their interaction with quantum fields. An initially pure quantum state, after the black hole has fully evaporated, has been argued to evolve into a mixed state. This process would then violate unitarity and has hence been dubbed the black hole information loss paradox \cite{Hawking:1976,Page:1993up,Hawking:2005kf,Unruh:2017uaw}.

It has recently been pointed out that non-local interactions might play a role in a possible resolution of the paradox \cite{Giddings:2006vu,Giddings:2006sj}, with concrete proposals involving non-local information transfer \cite{Giddings:2012gc}, for example involving non-local qubits \cite{Osuga:2016htn}. While we will not address the black hole information loss paradox any further in this thesis it nevertheless provides yet another reason to study concrete models involving non-local physics.

\subsection{Non-locality in quantum effective actions}

Last, we would like to point out that non-locality is not an entirely new ingredient for quantum field theory and has been studied for a long time in the context of effective actions. The effective action, if evaluated to a certain level of approximation, gives rise to field equations that already include the quantum corrections up to that order, and these effective field equations can be viewed as field equations for the quantum average.

Typical expressions appearing in an effective action of a second-order theory include formal expressions like $\log(\Box)$, which are non-local objects since they are defined via a power series; for more information see also Barvinsky \textit{et al.} \cite{Barvinsky:1994cg}.

\section{Historical aspects}

At this point we would like to give some historical context to the developments of non-local physics in the context of high energy physics, since many of the key concepts presented in this thesis have been known for more than 70 years.

We find the first mention of exponential form factors in the work of Wataghin \cite{Wataghin:1934ann} in 1934, who introduces a factor $\exp[-(-\omega^2+\vec{p}^{\,2})/\Pi^2]$ in a differential cross section and stipulates a value of $\Pi \approx 10^8\text{eV} \,\hat{=}\, 10^{-15}\text{m}$ in order to solve ``divergence difficulties'' in Lorentz invariant quantum theory, even though the phrase ``non-locality'' is not used in this work.

Thirteen years later, Yukawa \cite{Yukawa:1947,Yukawa:1948,Yukawa:1949a,Yukawa:1949b,Yukawa:1950a,Yukawa:1950b}, in a series of six papers from 1947--1950, introduces the notion of non-localizable fields that are not mere functions of space and time, but depend on other coordinates as well. These coordinates are introduced by a modification of the canonical commutation relations and may be interpreted as giving elementary particles of zero radius a finite radius instead.

In their seminal work around the same time, Pauli and Villars \cite{Pauli:1949zm}, striving for regularization methods in the emerging field of quantum field theory, consider higher-order differential equations that offer an improved short-distance behavior, albeit at the cost of introducing new propagating degrees of freedom. In what is perhaps the most complete study of regularization methods up to this date, Pais and Uhlenbeck \cite{Pais:1950za} offer a concise treatment of both higher-order Pauli--Vilars-type regularization methods as well as exponential form factors, identical to those treated in this thesis. In fact, many key results on non-local Green functions were already derived in Ref.~\cite{Pais:1950za} and have hence been ``rediscovered'' by many authors, including the present one.

Around the same time from 1950--1952, Bloch, Kristensen, and M{\o}ller \cite{Bloch:1950,Bloch:1952,Kristensen:1952} focused considerable efforts on regularizing meson theory, that is, the interaction of spinors with massive scalar fields. Since the electroweak interaction would not be hypothesized for another ten years, around this time non-renormalizable interactions such as the the four-fermion contact interaction had to be regularized somehow, and mesons, as hypothesized intermediate particles, were a promising candidate. There was some intermediate success, expressing the coupled spinor-scalar field equations as integro-differential equations, but there remained a point of controversy around conserved energy integrals. This point was addressed by Pauli in 1953 \cite{Pauli:1953}, who proved that there is a well-defined Hamiltonian structure in non-local field theories, but could only construct it explicitly within first order perturbation theory.

Fourteen years later, Efimov's work \cite{Efimov:1967,Efimov:1968,Efimov:1972,Efimov:1972ac,Efimov:1974,Efimov:1977} shed additional light on non-local quantum field theory: the role of non-local form factors as UV cutoffs in loop integrals, the perturbative unitarity of the S-matrix, theory and applications of integral kernel representations, non-locality in Abelian gauge theories, as well as its observational signatures, for example, the electron self-energy.

However, these studies, well into the 1970s, merely postulated the existence of non-local form factors. While they were appealing in their short-scale regulatory faculties without spoiling causality on a macroscopic level, they yet had to be derived from first principles.

The advent of bosonic string theory around that time should change that. It was soon realized that vertices of the form $e^\Box$ appear in string field theory \cite{Witten:1986} and survive the effective field theory limit \cite{Frampton:1988}; see also the related discussions on the role of non-locality in string theory in Refs.~\cite{Eliezer:1989cr,Kostelecky:1990mi,Harms:1993bt,Tseytlin:1995uq,Siegel:2003vt}.

After that, non-local theories have been studied in the context of superrenormalizable theories \cite{Tomboulis:1997gg} and have hence found fruitful applications in high energy phenomenology, cosmology, and black hole physics: It was realized by Biswas \textit{et al.} that non-local form factors allow for bouncing universes in cosmology \cite{Biswas:2005qr,Biswas:2010zk}, Koshelev studied non-local cosmology based on $p$-adic string field theory \cite{Koshelev:2007fi}, and Modesto \textit{et al.} considered the regulatory effect of these factors in the context of black holes \cite{Modesto:2010uh}. Those studies were then extended to more general gravitational settings in the seminal papers by Modesto \cite{Modesto:2011kw} and Biswas, Gerwick, Koivisto, and Mazumdar \cite{Biswas:2011ar}, pioneering the notion of non-local ``infinite-derivative'' gravity.

This concludes our brief historical overview and we will address the more recent results in Sec.~\ref{sec:ch1:recent}.

\section{Non-local form factors and kernel representations}
\label{sec:ch1:kernel}
Let us consider a non-local action of the form
\begin{align}
\label{eq:ch1:action-gf1}
S[\varphi] = \int\dd^D x \left[ \frac12 \varphi e^{-\ell^2\Box} \Box\varphi - J(x) \varphi \right] \, ,
\end{align}
where $\varphi$ is a non-local field and $J$ is an external current. The object $e^{-\ell^2\Box}$ is a simple example of a \emph{non-local form factor}\index{form factor}, where $\ell$ denotes the \emph{scale of non-locality}. In this thesis we will consider form factors that can be written as functions of the $\Box$-operator $f(\Box)$, to be understood as a formal power series and hence contain infinitely many derivatives. For non-local ghost-free form factors we also demand that they are non-zero everywhere and satisfy $f(0) = 1$. This latter property guarantees for the above action that in the limiting case of $\ell=0$ one recovers a standard second-order action, and as we will see in Ch.~\ref{ch:ch2} it also guarantees the same far-distance asymptotics of non-local fields as encountered in local theories. If not specified otherwise, all form factors discussed in this thesis are assumed to be non-local ghost-free form factors, and a particular class is given by
\begin{align}
f(\Box) = e^{(-\ell^2\Box)^N} \, , \quad N=1,2,\dots \, .
\end{align}
We shall refer to these theories as $\mathrm{GF_N}$ theories. In this introductory section, for simplicity, we will work with $N=1$, but our considerations hold for all $N$.

Turning our attention back to \eqref{eq:ch1:action-gf1}, introducing a field redefinition $\phi = e^{-\ell^2\Box/2}\varphi$ yields
\begin{align}
\label{eq:ch1:action-smeared}
\tilde{S}[\phi] = \int\dd^D x \left\{ \frac12\phi \Box\phi - [e^{\ell^2\Box/2} J(x)] \phi \right\} \, ,
\end{align}
where we have integrated by parts assuming the field $\varphi$ and the external current $J$ decrease sufficiently fast. Comparing the two actions $S[\varphi]$ and $\tilde{S}[\phi]$, the effect of the infinite-derivative non-locality has been shifted from the kinetic term to the interaction term. Utilizing an integral kernel $K(x-y)$ defined via
\begin{align}
e^{\ell^2\Box/2}\delta(x-y) = K(x-y) \, ,
\end{align}
we can rewrite the interaction term as follows:
\begin{align}
\label{eq:ch1:action-int}
\tilde{S}_\text{int}[\phi] = - \int\dd^D x \int\dd^D y \phi(x) K(x-y) J(y) \, .
\end{align}
In order to develop an intuition for the object $K(x-y)$ it helps to focus our attention temporarily to a purely spatial setting when $\Box\rightarrow\lap$. Then the integral kernel $K$ can be determined analytically in the present case and it is nothing but the $(D-1)$-dimensional heat kernel,
\begin{align}
K(x-y) = \frac{1}{(\sqrt{2}\pi\ell)^{(D-1)/2}} \exp\left[-\frac{(x-y)^2}{\ell^2}\right] \, .
\end{align}
This implies that the interaction of the field $\phi$ with the external current $J$ is non-local in a neighbourhood of $\mathcal{O}(\ell)$, and since the typical range of this effective non-local interaction is determined by the parameter $\ell$, it is often called the ``scale of non-locality.''\index{non-locality!scale of} In the limiting case of $\ell\rightarrow 0$ the effect of non-locality vanishes, and $\phi=\varphi$, and in this limit the heat kernel becomes the $(D-1)$-dimensional delta function,
\begin{align}
\lim\limits_{\ell\rightarrow 0} K(x-y) = \delta(x-y) \, .
\end{align}
So far, we have employed real-space representations of all objects. Sometimes, however, it is convenient to work with the Fourier representation of non-local form factors in order to construct Green functions. In the present case, there are two options:
\begin{itemize}
\item[(i)] Work with the action \eqref{eq:ch1:action-smeared} with the manifestly non-local interaction terms of the form \eqref{eq:ch1:action-int}. This has the benefit that the kinetic term can be canonically normalized, and shifts the non-locality into the spatial properties of the integral kernel $K(x-y)$.
\item[(ii)] Work with the action \eqref{eq:ch1:action-gf1}, which has a non-standard kinetic term but otherwise local interactions. Then, it is possible to study the modification of the Green function for the infinite-derivative form factor in Fourier space, while the vertices themselves retain their original form. In this approach, non-locality is described by the properties of the non-standard Green function, which in turn may be studied in more detail in Fourier space.
\end{itemize}
While there are reasons for working with either method, in this thesis we will follow the second avenue. The reason lies in the fact the study of causal properties is simplified when one considers only local interactions with a modified Green function, and we will discuss this in substantial detail in Ch.~\ref{ch:ch2}. However, method (i) has one decisive advantage: It proves that in the absence of interactions there is no effect of non-locality since it can be absorbed by a field redefinition without leaving any trace. We will revisit this issue in the following section at the level of the corresponding field equations.

\section{On-shell vs.\ off-shell properties}
\label{sec:ch1:on-shell-off-shell}
Let us now focus on the non-local field equations in more detail. Specifically, let us consider an inhomogeneous non-local field equation for the field $\varphi$,
\begin{align}
e^{-\ell^2\Box} \Box \varphi = J \, ,
\end{align}
where $J$ is an external current. We may invert this relation and write the equivalent expression
\begin{align}
\Box \varphi = J_\text{eff} \, , \quad J_\text{eff} =  e^{+\ell^2\Box} J \, , 
\end{align}
and inverting the ghost-free form factor is always possible since, by construction, it is nowhere zero. Based on our previous considerations we can interpret this relation as the original, local equation where the current $J$ has been replaced by an effective current $J_\text{eff}$ which is a smeared version of the original current $J$.

As is well known, the d'Alembert operator $\Box$ is hyperbolic on Lorentzian manifolds such as Minkowski spacetime or the curved spacetimes of General Relativity. This means that even on non-compact manifolds there exist normalizable zero mode\index{zero mode} solutions to this operator satisfying $\Box\chi=0$. In other words: zero mode solutions $\chi$ are eigenfunctions of the d'Alembert operator with vanishing eigenvalue. Let us consider now such a homogeneous equation,
\begin{align}
e^{-\ell^2\Box} \Box \varphi = 0 \, .
\end{align}
Since the exponential form factor never vanishes, the only solution to this equation is $\varphi=\chi$ where $\Box\chi=0$ is a zero mode:\index{zero mode} homogeneous solutions are insensitive to the presence of non-locality.

The previous considerations hold true for \emph{ghost-free form factors} $f(\Box)$, and all of the form factors considered in this thesis fall into that class. A simple example where this does \emph{not} work is given by higher-derivative differential equations of the form
\begin{align}
(\Box-m^2) \Box \varphi = 0 \, .
\end{align}
The additional operator $\Box-m^2$ admits new zero mode solutions, and hence $\varphi\not=\chi$. But this operator is not ghost-free since $(\Box-m^2)\chi=0$ has non-trivial solutions. This is a distinctive difference between ghost-free theories presented in this thesis and higher-derivative theories.

Quantities that solve free field equations are often referred to as \emph{on-shell},\index{on-shell} and quantities that are not solutions to free field equations, conversely, are called \emph{off-shell}. An interesting case arises when a ghost-free factor $f(\Box)$ acts on an on-shell quantity $\chi$, since then
\begin{align}
f(\Box)\chi = f(0)\chi = \chi \, ,
\end{align}
and the effect of non-locality disappears.

An example for off-shell quantities are the manifestly off-shell loop momenta one encounters in perturbative quantum field theory. Yet another case for off-shell quantities are renormalized quantities, since subtraction schemes often take formally infinite on-shell quantities and transform them into finite off-shell expressions. An example of \emph{classical} off-shellness is an interaction term, as we have seen above.

These considerations show that ghost-free non-locality can be difficult to observe experimentally, as off-shell quantities are required to create measurable, non-zero effects, and it is helpful to keep this in mind for the remainder of this thesis. We will revisit this point extensively in Chs.~\ref{ch:ch6}--\ref{ch:ch7} in the context of homogeneous and inhomogeneous Green functions in the presence of non-locality.

\section{Remarks on non-locality and the variational principle}
\label{sec:ch1:action}
In the above we discussed some properties of non-local field equations. As is well known, in local field theory it is often possible to \emph{derive} field equations from a variational principle involving an action functional. In this section we would therefore like to address the role of actions within non-local field theories, and to that end we shall consider two principles:
\begin{itemize}
\item[1.] An action can be viewed as a formal integral that can be utilized to derive the field equations. We shall call this the ``local variational principle.''
\item[2.] Within variational calculus, the surface terms obtained during the first variation of the action provide insights on boundary conditions required for a self-consistent variational problem. We will refer to this as the ``global, self-consistent variational principle.''
\end{itemize}
In what follows, we will briefly compare and contrast the local and global, self-consistent variational principle for a scalar field in the absence of non-locality. In a second step we shall demonstrate that the local variational principle readily extends to the class of non-local ghost-free theories discussed in this thesis, whereas the role of a global, self-consistent variational problem in the presence of non-locality remains somewhat unclear and deserves more future attention (compare also the remarks on the ghost-free initial value problem in Sec.~\ref{sec:ch1:initial-value}).

Let us consider a local, second-order action for a free scalar field $\varphi$ in $D$-dimensional Minkowski spacetime $\mathcal{M}$ for which the action and its first variation take the form
\begin{align}
S_1[\varphi] &= -\frac12 \int\limits_\mathcal{M} \dd^D x (\partial{}_\mu\varphi)(\partial{}^\mu \varphi) \, , \\
\delta S_1[\varphi] &= \int\limits_\mathcal{M}\dd^D x \delta\varphi \Box \varphi - \int\limits_{\partial\mathcal{M}} \dd\Sigma^{D-1}_\mu \delta\varphi \partial{}^\mu \varphi \, .
\end{align}
The variational principle demands that $\delta S=0$, which means that all terms on the right-hand side need to vanish. The first term yields the expected field equation $\Box\varphi=0$. If we assume that the value of $\varphi$ is fixed on the boundary $\partial\mathcal{M}$, the variation $\delta\varphi$ has to vanish there such that $\delta\varphi|_{\partial\mathcal{M}} = 0$, rendering the second term zero, and the variational problem is self-consistent.

In order to contrast these findings, let us consider instead the action
\begin{align}
\label{eq:ch1:action-2}
S_2[\varphi] &= \frac12 \int\limits_\mathcal{M}\dd^D x \, \varphi \Box \varphi \, , \\
\delta S_2[\varphi] &= \int\limits_\mathcal{M}\dd^D x \delta\varphi\Box\varphi + \frac12\int\limits_{\partial\mathcal{M}} \dd\Sigma^{D-1}_\mu \left( \varphi \partial{}^\mu \delta\varphi - \delta\varphi \partial{}^\mu \varphi \right) \, ,
\end{align}
where the two actions $S_1[\varphi]$ and $S_2[\varphi]$ are related by a surface term,
\begin{align}
S_1[\varphi] = S_2[\varphi] - \frac12 \int\limits_{\partial\mathcal{M}} \dd\Sigma^{D-1}_\mu \varphi \partial{}^\mu \varphi \, .
\end{align}
The field equations are again $\Box\varphi=0$, but the variational problem for $S_2[\varphi]$ is not self-consistent. To see this, note that the boundary term includes both $\delta\varphi$ and $\partial{}^\mu\delta\varphi = \delta\partial{}^\mu \varphi$. Therefore, if the boundary terms are to vanish identically as is required for a self-consistent variational principle, one needs to fix both the value of $\varphi$ and its normal derivative $n{}_\mu\partial{}^\mu\varphi$ at the boundary and thus $\delta\varphi|_{\partial\mathcal{M}} = 0$ \emph{and} $\delta (n{}_\mu \partial{}^\mu\varphi)|_{\partial\mathcal{M}} = 0$. Then, however, the equations of motion $\Box\varphi=0$ only allow for the trivial solution.

Hilbert and his student Courant (see more details in Courant \cite{Courant:1943}) developed the notion of a global self-consistent variational problem, and the above considerations show the following: Partial differential equations, when supplemented with somewhat ad hoc but seemingly appropriate boundary conditions, can offer much insight into the dynamics of a physical problem. However, it is not always obvious how to select the appropriate boundary conditions in a specific case. The self-consistent variational problem solves this by demanding that the surface terms---obtained during the first variation of the action---vanish, and within that context these boundary conditions are referred to as ``natural boundary conditions.'' Therefore, as Courant puts it: the appropriate boundary conditions for a differential equation are obtained as the natural boundary conditions of the associated self-consistent variational problem \cite{Courant:1943}; see also the book \cite{Frolov:2011}.

For these reasons Eq.~\eqref{eq:ch1:action-2} is an inconsistent action from the global, self-consistent variational principle. However, one may formally consider only variations $\delta\varphi$ of a special kind: namely, those that decrease in a way that both the variation $\delta\varphi$ as well as all its derivatives vanish identically outside a compact subregion $\mathcal{C}\subset\mathcal{M}$. Then the first variation of the action vanishes, albeit only for this special class of variations. Moreover, one has to stipulate the appropriate boundary conditions for the differential equation at hand.

Do these concepts extend to non-local theories? Let us consider a non-local version of the action $S_2[\varphi]$ that we shall employ frequently in the rest of this thesis. For brevity we will omit a potential external current since we are interested in the derivative structure of the kinetic term. The free action can be written as
\begin{align}
\label{eq:ch1:action-2-gf}
S^\text{non-local}_2[\varphi] = \frac12 \int\limits_\mathcal{M}\dd^D x \varphi \Box e^{-\ell^2\Box} \varphi \, .
\end{align}
Clearly in the limit $\ell\rightarrow 0$ we recover Eq.~\eqref{eq:ch1:action-2}, and for this reason we already know that this action is not self-consistent in the sense of Hilbert and Courant. Performing a somewhat cavalier variation one still obtains the field equations $e^{-\Box\ell^2}\Box\varphi=0$, but the precise form of the boundary terms is rather complicated: they arise from expressing the exponential function as a power series, and integrating each individual finite-derivative term by parts. Schematically one obtains
\begin{align}
\delta S^\text{non-local}_2[\varphi] = \int\limits_\mathcal{M}\dd^D x \delta\varphi \Box e^{-\ell^2\Box} \varphi + \text{surface terms} \, .
\end{align}
The surface terms contain both the variation $\delta\varphi$, the field $\varphi$, and derivatives thereof. Utilizing the local variational principle for a restricted class of variations one readily obtains the field equations. 

In the present thesis we explicitly demonstrate in all relevant examples how to choose appropriate boundary conditions for various non-local ghost-free field equations. For this reason the local variational principle is sufficient. It would however be interesting to study a possible non-local self-consistent variational principle and we shall leave this open question for a future study.

\section{Initial value problem}
\label{sec:ch1:initial-value}
With the structure of non-local linear field equations known there is one immediate concern: since the equations contain infinitely many derivatives they might also require an infinite amount of initial data, and this is directly connected with the considerations on appropriate boundary conditions presented in the previous section. It is well known that any second-order linear differential equation requires both the value of the field $\varphi$ as well as its derivative $\partial_t\varphi$ on a $t=\text{const.}$ spacelike hypersurface in order to be deterministic. Is an infinite set of initial data required for infinite-derivative field equations?

Barnaby and Kamran \cite{Barnaby:2007ve} have shown that this is not always the case. They prove that each pole in the propagator contributes two pieces of initial data. This reproduces the case of Sturm--Liouville theory, where the differential operator is of second order, and as such its associated polynomial in Fourier space has two complex roots, corresponding to the two pieces of initial data required. In the case of \emph{ghost-free} infinite-derivative field theory, the number of poles \emph{coincides} with that of the local theory, as we indicated already above. For this reason the amount of initial data remains the same.

However, there is a connected problem as to \emph{where} this initial data needs to be specified. Clearly, in the presence of non-locality, the notion of a precise $t=\text{const.}$ hypersurface is no longer meaningful. Barnaby \cite{Barnaby:2010kx} has shown that the notion of an infinitely thin Cauchy surface has to be replaced by a smeared out Cauchy surface of finite thickness of $\mathcal{O}(\ell)$, which is the scale of non-locality. Hence, a small perturbation induced by a source located before an initial Cauchy surface will have a non-vanishing influence on the future, even if one keeps the data fixed on that Cauchy surface. This contradicts strict causality but, as we will see in Sec.~\ref{sec:ch2:asymptotic-causality}, may still be compatible with asymptotic causality or macrocausality.

%Additional insight might be gained from the following consideration: Suppose that there are no sources in spacetime, and hence the non-local field equations are homogeneous. Let us call this scenario ``globally source-free.'' We already know from Sec.~\ref{sec:ch1:on-shell-off-shell} that the effect of non-locality disappears in that case and the theory becomes local, and for this reason the initial value problem reduces to the local one as well, and it is sufficient to prescribe $\varphi$ and $\dot{\varphi}$ on a Cauchy slice $\Sigma_t$ in order to obtain deterministic equations. Imagine now that an infinitesimal source is turned on, $J(\ts{x})\sim \delta(t-t_0)\delta(x-x_0)$, at only one point in time and at a fixed spatial location $(t_0,x_0)$. Then, the field equations are no longer globally source-free, and we shall instead refer to them as ``source-free in a domain'' for the times $t>t_0$. However, it is no longer sufficient to simply prescribe $\varphi$ and $\dot{\varphi}$ on $\Sigma_t$ since the future evolution will depend on the presence of the sources in the past.

Additional insight might be gained from the following consideration: Suppose we are interested in the dynamics of a scalar field, subject to the local field equation
\begin{align}
\Box\varphi = j \, .
\end{align}
Let us consider a Cauchy surface $t=t_1 = \text{const.}$, and call it $\Sigma_{t_1}$. Specify both $\varphi$ and $\dot{\varphi}$ there, and let us assume that the current $j$ has the form $j \sim \delta(t-t_0)$ with $t_0<t_1$. This is a well-defined initial value problem for the future dynamics of $\varphi$ in the domain $t \ge t_1$, and because $j=0$ in this domain the dynamics are described by $\Box\varphi=0$. In other words: the presence of the current at $t=t_0$ is irrelevant.

\begin{figure}
    \centering
    \includegraphics[width=0.95\textwidth]{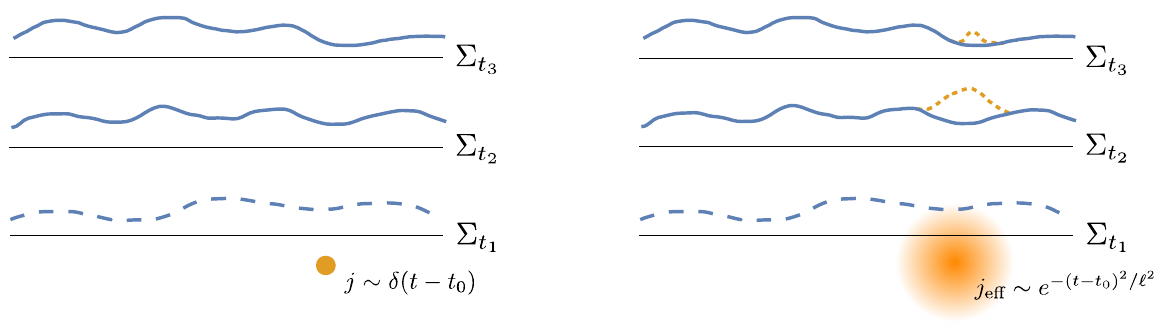}\vspace{5pt}
    \caption[Local and non-local Cauchy problem.]{In the local case (left), the presence of a $\delta$-shaped current $j$ at an earlier time $t_0$ does not influence the future evolution of a field $\varphi$, provided $\varphi$ and $\dot{\varphi}$ are specified on the Cauchy slice $\Sigma_{t_1}$. In the non-local case (right) this is no longer true: specifying the same initial values $\varphi$ and $\dot{\varphi}$ on $\Sigma_{t_1}$, the future evolution of the scalar field $\varphi$ will differ from the local case because the current $j$ is smeared out due to the presence of non-locality, visualized as $j_\text{eff}$ in the image above. The difference in the future evolution of the field is highlighted by the dotted lines. As one tracks the future evolution of $\varphi$ in over later times $t_2$ and $t_3$, the influence of the effective current decreases over time due to the notion of asymptotic causality in ghost-free theories.}
    \label{fig:ch1:cauchy}
\end{figure}

Let us now switch on non-locality and track the deviations from the local theory, while keeping the identical form of the current $j$. The field equations are
\begin{align}
e^{-\ell^2\Box}\Box\varphi = j \, .
\end{align}
We again specify the values of $\varphi$ and $\dot{\varphi}$ on $\Sigma_{t_1}$, identical to those of the local case. However, the future dynamics are now susceptible to the presence of the current $j$ in the past. This can be seen from an equivalent representation of the non-local field equations,
\begin{align}
\Box\varphi = j_\text{eff} \, , \quad j_\text{eff} = e^{\ell^2\Box}j \sim e^{-(t-t_0)^2/\ell^2} \, ,
\end{align}
where the form of the effective smeared current can be derived in a similar fashion to Sec.~\ref{sec:ch1:on-shell-off-shell}. Clearly this current is non-zero in the domain $t \ge t_1$ and its ``tail'' hence affects the future dynamics of $\varphi$. In ghost-free theories this smeared current can always be found since the infinite-derivative expressions can be inverted. Moreover, as we will show in more detail later, these smeared sources decrease on both purely temporal and purely spatial distances scales that are much larger than the scale of non-locality. We visualize this in Fig.~\ref{fig:ch1:cauchy}.

A helpful way to distinguish these situations is the following: if the equations are globally source-free for all times $t$, then the initial value problem coincides with that of the local theory. If the equations are only source-free in a certain domain ($t>t_0$ in the above example), then the presence of non-locality has observable consequences for the future evolution of the field $\varphi$.

In the present thesis we will study non-locality as modelled by ghost-free infinite-derivative theories in a variety of situations and construct explicit solutions for non-local physical fields. We will primarily make use of the notion of non-local Green functions, and our findings are consistent with the mathematical considerations presented by Barnaby and Kamran. For a study of initial conditions specifically in the context of non-local ghost-free gravity we refer to Calcagni \textit{et al.} \cite{Calcagni:2018lyd} as well as Giaccari and Modesto \cite{Giaccari:2018nzr}.

\section{Recent work}
\label{sec:ch1:recent}

In the recent years, non-local physics as mediated by infinite-derivative form factors has received considerable attention. In this section we would like to attempt to summarize the main developments while categorizing them into the three groups (i) quantum field theory, (ii) applications in cosmology, and (iii) applications in black hole physics and gravitational physics. We hope that this summary is interesting and useful for readers who consider researching non-locality themselves.

\subsection{Quantum field theory}

There is a growing amount of literature on calculational techniques in non-local quantum field theory involving loop structure and scattering amplitudes \cite{Talaganis:2014ida,Talaganis:2016ovm}, perturbative unitarity \cite{Carone:2016eyp,Briscese:2018oyx}, reflection positivity \cite{Christodoulou:2018jbn}, Green function methods \cite{Boos:2019fbu,Boos:2019zml}, and non-local effective actions \cite{Teixeira:2020kew}; see also the review \cite{Buoninfante:2018mre}. For discussions of non-locality in string theory see \cite{Calcagni:2013eua,Calcagni:2014vxa}. The presence of (complex) ghosts has been discussed in \cite{Shapiro:2015uxa,Modesto:2015ozb,Asorey:2018wot,Salles:2018ccb}; for stability analyses we refer to \cite{Mazumdar:2018xjz,Briscese:2018bny}. The concept of non-locality has also been extended to non-local supergravity \cite{Giaccari:2016kzy} and notions of quantum gravity \cite{Modesto:2017sdr}, and more general form factors have recently been considered in \cite{Buoninfante:2020ctr}. Applications to the Higgs mechanism are presented in \cite{Ghoshal:2017egr,Hashi:2018kag}. While it has been argued that for scattering processes with a large number of external legs the presence of non-locality leads to a macroscopically accessible non-local region \cite{Buoninfante:2018gce}, others argue that violations of macrocausality are unattainable in the laboratory \cite{Briscese:2019twl}.

In the context of quantum mechanics, monochromatic radiation \cite{Frolov:2016xhq} as well as wave packets \cite{Buoninfante:2017kgj,Buoninfante:2017rbw} have been studied, including superradiance effects \cite{Frolov:2018bak} and potential observational signatures in the connection with transmission and reflection coefficients \cite{Boos:2018kir,Buoninfante:2019teo}.

\subsection{Cosmology}

%Non-local inflationary models have been developed \cite{Biswas:2012bp,Koshelev:2016xqb,Koshelev:2017tvv,SravanKumar:2018dlo}, and non-locality has also been studied in the context of massive gravity \cite{Modesto:2013jea,Mazumdar:2018xjz}. The resolution of cosmological singularities has been discussed in \cite{Biswas:2005qr,Biswas:2010zk,Kumar:2020xsl} and has recently been extended to anisotropic scenarios \cite{Koshelev:2018rau}.

Non-local inflationary models have been developed \cite{Biswas:2012bp,Craps:2014wga,Koshelev:2016vhi,Koshelev:2016xqb,Koshelev:2017ebj,Koshelev:2017tvv,SravanKumar:2018dlo,Dimitrijevic:2019pct,Dimitrijevic:2020dzo}, with observational signatures studied in \cite{Koshelev:2020foq,Koshelev:2020xby,Koshelev:2020fok}. Non-locality has also been studied in the context of massive gravity \cite{Modesto:2013jea,Mazumdar:2018xjz}. The resolution of cosmological singularities has been discussed in \cite{Biswas:2005qr,Biswas:2010zk,Kumar:2020xsl} and has recently been extended to anisotropic scenarios \cite{Koshelev:2018rau}.

\subsection{Black hole physics and gravitational physics}

A significant amount of work has been devoted to the study of non-local gravity. With the general structure of the non-local gravitational theory known \cite{Biswas:2013cha,Biswas:2013kla,Conroy:2014eja}, a large amount of exact regular solutions in the weak-field regime has been constructed \cite{Modesto:2014eta,Li:2015bqa,Edholm:2016hbt,Giacchini:2016xns,Buoninfante:2016iuf,Buoninfante:2018xiw,Kilicarslan:2018yxd,Giacchini:2018wlf,Kumar:2018jjb,Buoninfante:2019pzr}, including rotating solutions \cite{Cornell:2017irh,Buoninfante:2018xif}, electrically charged \cite{Buoninfante:2018stt} and conformally flat solutions \cite{Buoninfante:2018rlq}, relativistic objects \cite{Kilicarslan:2019njc,Boos:2020ccj}, as well as extended sources \cite{Boos:2018bxf,Boos:2020kgj,Kolar:2020bpo}.

At the present stage there appears to be some confusion on the fate of the Schwarzschild singularity: whereas a stability analysis indicates that non-locality might not be enough to remove the singularity \cite{Calcagni:2017sov}, a study of the Ricci-flat field equations shows that the $1/r$-behavior does not present an exact solution \cite{Koshelev:2018hpt}. For more non-perturbative results see \cite{Calcagni:2018gke}. In the context of black hole thermodynamics it has been shown that non-locality allows for a sensible notion of Wald entropy \cite{Conroy:2015wfa}, and for an exact result of the entropy correction due to non-locality we refer to \cite{Boos:2019vcz}. Steps towards an exact Ricci flat solution have been taken in \cite{Buoninfante:2018xif}.

Phenomenologically, the effects of non-locality have been studied in the context of black hole formation from gravitational collapse \cite{Frolov:2015bia,Frolov:2015bta,Frolov:2015usa,Bambi:2016uda}, on the defocusing of null geodesics \cite{Conroy:2016sac} and the fate of the compacts objects without event horizons \cite{Koshelev:2017bxd}, as well as effects on the bending of light \cite{Giacchini:2018twk,Buoninfante:2020}, the memory effect \cite{Kilicarslan:2018unm}, and spatial oscillations of the effective energy density in the vicinity of point particles \cite{Boos:2018bhd}. A model of non-local stars has been proposed in Ref.~\cite{Buoninfante:2019swn}.

The stability of black holes in non-local gravity has been studied in \cite{Calcagni:2018pro,Briscese:2019rii}, and a diffusion method for non-linear non-local equations as encountered in non-local gravity has been presented in \cite{Calcagni:2018fid}. More recently, non-local gravity has been extended to include non-vanishing torsion \cite{delaCruz-Dombriz:2018aal}\index{torsion!non-vanishing} and has been extended to Galilean theories \cite{Buoninfante:2018lnh}. A non-local theory with string-inspired worldline inversion symmetry is presented in \cite{Abel:2019zou}, and a non-local Hamiltonian framework for the scalar sector of non-local gravity has been developed in \cite{Kolar:2020ezu}.

\subsection{Related approaches}
Let us also mention two related approaches. First, in the context of non-commutative geometry one may obtain non-local form factors \cite{Spallucci:2005bm,Spallucci:2006zj,Nicolini:2008aj} with regular spatial propagators of the form
\begin{align}
G(r) = -\frac{1}{4\pi^2} \frac{1}{r^2+\ell_0^2} \, ,
\end{align}
where $\ell_0$ is an intrinsic length scale of the theory, compare also \cite{Nicolini:2019irw}.

In a different approach developed by Mashhoon and Hehl \cite{Mashhoon:2006mj,Mashhoon:2008vr,Hehl:2008eu,Hehl:2009es,Blome:2010xn,Mashhoon:2019jkq}, non-locality enters via a non-local constitutive law. In  electrodynamics, for example, the constitutive law represents the relation between the Faraday field strength 2-form $F{}_{\mu\nu}$ (corresponding to the $\vec{E}$ and $\vec{B}$ field) and the excitation $H{}_{\mu\nu}$ (which encodes the matter response in the $\vec{D}$ and $\vec{H}$ fields). The constitutive tensor $\chi{}_{\mu\nu}{}^{\rho\sigma}$ relates these two quantities, and it if is assumed to be non-local,
\begin{align}
H{}_{\mu\nu}(\ts{x}) = \frac12 \int \dd^D y \chi{}_{\mu\nu}{}^{\rho\sigma}(\ts{x}-\ts{y}) F{}_{\rho\sigma}(\ts{y}) \, ,
\end{align}
the resulting field equations are non-local, too. At the linear level in vacuum, when it is possible to write the constitutive tensor in terms of an integral kernel and the totally antisymmetric symbol, $\chi{}_{\mu\nu}{}^{\rho\sigma} \sim \epsilon{}_{\mu\nu}{}^{\rho\sigma}K(\ts{x}-\ts{y})$, it is possible to relate this approach to the infinite-derivative formulation of non-local physics, and it would be interesting to explore this in more detail.

\section{Overview of thesis}
This thesis is the product of roughly three years worth of research on different aspects of non-local physics. Since a large portion is related to the properties of non-local Green functions, we devoted Ch.~\ref{ch:ch2} exclusively to that topic. Therein we compare non-local Green functions to their local counterparts and find their similarities and differences. Moreover, we devote some space to the study of static Green functions and find their explicit expressions since they will be used quite frequently in the remainder of the thesis.

In Ch.~\ref{ch:ch3} we turn towards the weak-field limit of non-local ghost-free gravity and construct various static and stationary solutions by employing the previously derived Green functions, and in Ch.~\ref{ch:ch4} we obtain the metric of ultrarelativistic objects by boosting stationary metrics to the speed of light in a suitable limit.

Changing gears towards quantum physics, we determine the scattering coefficients on a $\delta$-potential in non-local ghost-free quantum mechanics in Ch.~\ref{ch:ch5}, before considering the vacuum polarization and thermal fluctuations around that $\delta$-potential in Ch.~\ref{ch:ch6} within non-local ghost-free quantum field theory.

In Ch.~\ref{ch:ch7} we combine the topics of gravity and quantum physics: studying a ghost-free modification of the Polyakov effective action, we prove that non-locality has no impact on the flux of Hawking radiation measured at spatial infinity but does affect the entropy of a two-dimensional dilaton black hole.

\begin{figure}
    \centering
    \includegraphics[width=0.95\textwidth]{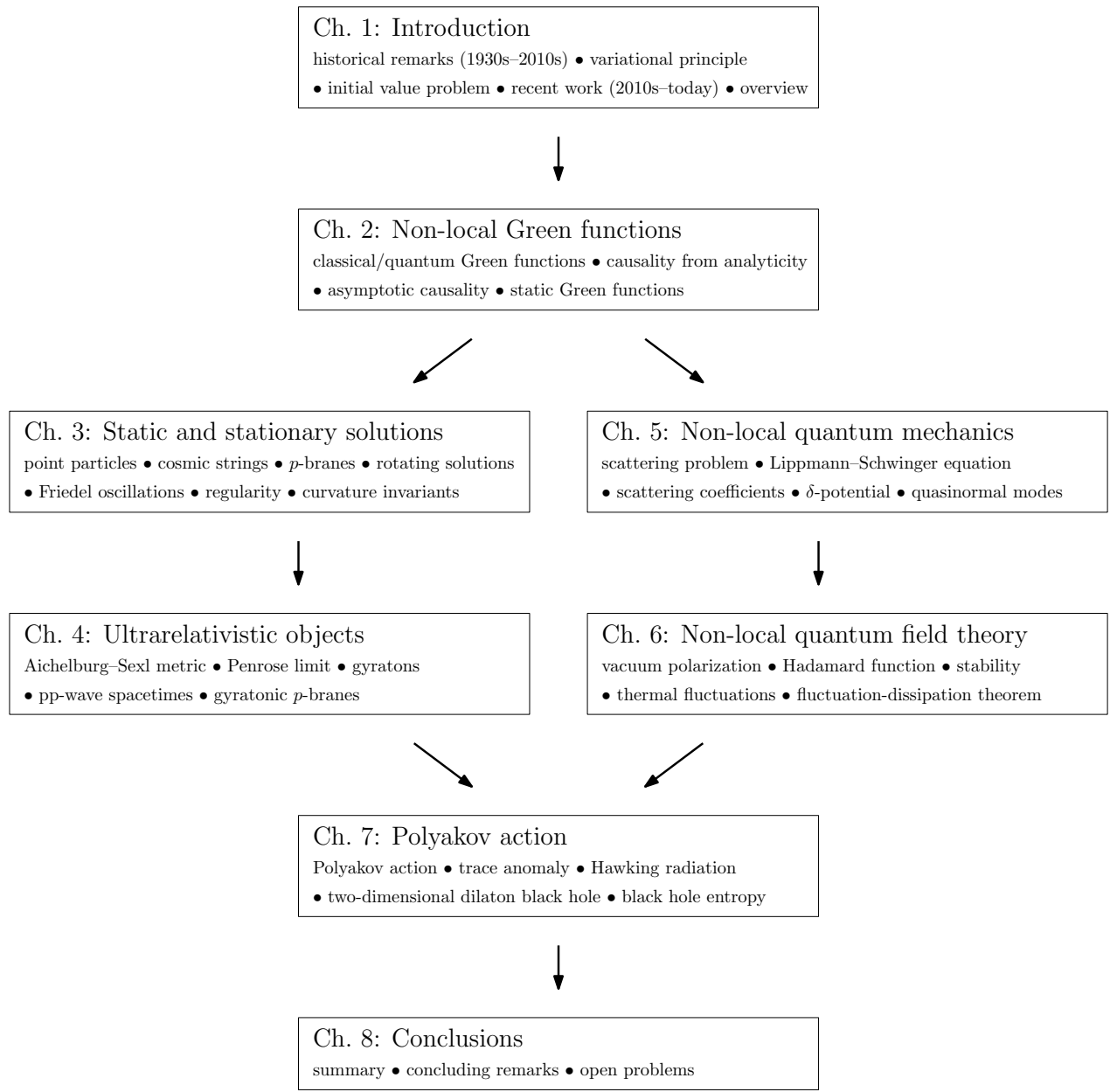}\vspace{50pt}
    \caption[Structure of this doctoral thesis.]{Graphical representation of the structure of this doctoral thesis. Ch.~7 requires knowledge of all preceding chapters, whereas Chs.~3--4 and Chs.~5--6 can be studied independently.}
    \label{fig:ch1:overview}
\end{figure}

Chapter \ref{ch:ch8} is devoted to a brief summary of our key findings and contains a list of open problems. See also Fig.~\ref{fig:ch1:overview} for a graphical representation of the logical structure of this thesis.

%%%%%%%%%%%%%%%%%%%%%%%%%%%%%%%%%%%%%%%%%%%%%%%%%%%%%%%%%%%%%%%%%%%%%%%%%%%%%%%%%%%%%%%%%%%%%%%%%%%
%
% Chapter: Green functions in non-local theories
%
\chapter{Green functions in non-local theories}
\label{ch:ch2}
\textit{This chapter serves as a prerequisite for the rest of this thesis. We develop the notion of Green functions in the presence of non-locality and derive explicit expressions for static Green functions that will be used throughout the remainder of this thesis. It is based on Refs.~\cite{Boos:2018bxf,Boos:2018kir,Boos:2019fbu,Boos:2019zml,Boos:2020ccj}.}

\section{Introduction}

Green functions are useful for solving linear differential equations exactly. Moreover, they may be employed in perturbative techniques as well. Developed by British mathematician George Green \cite{Green:1828}, see more about the history in Refs.~\cite{Cannell:1993,Cannell:1999}, they are ubiquitous in contemporary physics and are sometimes also referred to as ``Green's functions'' or ``propagators,'' and in the present work we shall use the expression ``Green functions'' and abstain from using the possessive. In the context of quantum field theory, see Ch.~\ref{ch:ch6}, we may also employ the term ``propagator.''

In the scope of the present work, it is useful to extend the theory of Green functions from local theories with a finite number of derivatives to non-local theories with an infinite number of derivatives. There are a number of interesting observations to be made, see also the previous discussion in Sec.~\ref{sec:ch1:initial-value}, and in this chapter we would like to dedicate some space to these insights. They form the mathematical foundation of a major portion of this thesis and may provide useful also beyond the scope of this thesis and hence are presented independently from a concrete application, which will follow in ample detail in Chs.~\ref{ch:ch3}--\ref{ch:ch7}.

\section{Green functions in classical field theory}
Consider a linear differential equation of the form
\begin{align}
\label{eq:ch2:eom}
\mathcal{D}\varphi(X) = \sigma(X) \, ,
\end{align}
where $X$ denotes spacetime coordinates, $\varphi(X)$ is the physical field under consideration, $\mathcal{D}$ is a differential operator, and $\sigma(X)$ denotes a source term that may be linear in $\varphi(X)$ as well. A Green function\index{Green function} $G(X',X)$ is defined as a solution of the following equation:
\begin{align}
\label{eq:ch2:g-inhomogeneous}
\mathcal{D} G(X',X) = -\delta^{(D)}(X'-X)
\end{align}
Then, a solution to \eqref{eq:ch2:eom} can be written as
\begin{align}
\varphi(X) = \varphi_0(X) + \int \dd^D Y G(X,Y) \sigma(Y) \, ,
\end{align}
where $\varphi_0(X)$ is a solution of the homogeneous equation whose precise form is fixed by the appropriate boundary conditions.

We emphasize that there are typically many Green functions for a given differential operator $\mathcal{D}$, corresponding to different boundary conditions. Some care has to be administered to find the right Green function for a physical problem under consideration, and we will address this in more detail in Ch.~\ref{ch:ch3}.

\section{Green functions in quantum field theory}
\label{sec:ch2:green-functions-quantum-theory}

As is well known, in quantum field theory the notion of classical fields is abandoned in favor of operator-valued fields or ``field operators.'' These operators are defined on a Fock space, which in turn is the direct sum of many-particle Hilbert spaces, and for that reason this procedure is also sometimes referred to as ``second quantization.'' In this quantum framework there exists a similar notion of Green functions which we would like to address next.

Let us consider a quantum field operator $\hat{\Phi}(t,\ts{x})$ on $D$-dimensional Minkowski spacetime with coordinates $(t,\ts{x})$ such that $\dd s^2 = -\dd t^2 + \dd\ts{x}^2$. Denoting the quantum expectation value as $\langle\dots\rangle$, which is to be performed in the vacuum state, and writing the commutator as $[\hat{A},\hat{B}]=\hat{A}\hat{B}-\hat{B}\hat{A}$, one may define the following Green functions \cite{DeWitt:1965}:\index{Green function!retarded}
\index{Green function!advanced}
\index{Green function!Feynman}
\begin{alignat}{5}
\text{retarded:} \quad &&
G^\text{R}(\ts{x'},\ts{x}) &\equiv i \theta(t'-t) \langle [\hat{\Phi}(\ts{x'}), \hat{\Phi}(\ts{x})] \rangle \, , \\
\text{advanced:} \quad &&
G^\text{A}(\ts{x'},\ts{x}) &\equiv -i \theta(t-t') \langle [\hat{\Phi}(\ts{x'}), \hat{\Phi}(\ts{x})] \rangle \, , \\
\text{Feynman:} \quad &&
G^\text{F}(\ts{x'},\ts{x}) &\equiv i \theta(t'-t) \langle \hat{\Phi}(\ts{x'}) \hat{\Phi}(\ts{x}) \rangle + i \theta(t-t') \langle \hat{\Phi}(\ts{x}) \hat{\Phi}(\ts{x'}) \rangle \, .
\end{alignat}
In the context of quantum field theory these objects are also frequently referred to as ``propagators.''\index{propagator} As expectation values of field operators they are seemingly unrelated to classical Green functions, but imposing the field equations for the field operator $\hat{\Phi}$ one may show that they satisfy
\begin{align}
\mathcal{D}G^\bullet(\ts{x'},\ts{x}) = -\delta(\ts{x}-\ts{x}') \quad \text{for} \quad \bullet = \text{R}, \text{A}, \text{F} \, .
\end{align}
For this reason it makes sense to refer to propagators as ``Green functions,'' and use the same symbol to denote them in both classical and quantum theory. In the context of quantum theory it is also possible to define different expectation values of the field operator via
\begin{alignat}{5}
\text{Hadamard:} \quad &&
G^\text{(1)}(\ts{x'},\ts{x}) &\equiv \langle \hat{\Phi}(\ts{x'}) \hat{\Phi}(\ts{x}) + \hat{\Phi}(\ts{x}) \hat{\Phi}(\ts{x'}) \rangle \, , \\
\text{Wightman:} \quad &&
G^+(\ts{x'},\ts{x}) &\equiv \langle \hat{\Phi}(\ts{x'}) \hat{\Phi}(\ts{x}) \rangle \, , \\
&& G^-(\ts{x'},\ts{x}) &\equiv \langle \hat{\Phi}(\ts{x}) \hat{\Phi}(\ts{x'}) \rangle \, .
\end{alignat}
These objects are called Hadamard and Wightman functions\index{Green function!Hadamard}\index{Green function!Wightman}, and are sometimes also called Hadamard Green function and Wightman Green function. Imposing the field equation for $\hat{\Phi}$ one may show that 
\begin{align}
\mathcal{D}G^\bullet(\ts{x'},\ts{x}) = 0 \quad \text{for} \quad \bullet = (1), +, - \, .
\end{align}
For this reason we also refer to the retarded, advanced, and Feynman Green function as ``inhomogeneous Green functions'' and to the Hadamard\index{Green function!Hadamard} and Wightman\index{Green function!Wightman} functions as the ``homogeneous Green functions.'' By construction these Green functions are related via
\begin{align}
G{}^\text{F}(\ts{x'},\ts{x}) &= \frac12 \left[ G{}^\text{R}(\ts{x'},\ts{x}) + G{}^\text{A}(\ts{x'},\ts{x}) + i\,G{}^{(1)}(\ts{x'},\ts{x}) \right] \, , \\
G{}^{(1)}(\ts{x'},\ts{x}) &= G^+(\ts{x'},\ts{x}) + G^-(\ts{x'},\ts{x}) \, .
\end{align}
This follows by the definition as well as from the identity $\theta(t) + \theta(-t) = 1$. 

All of the above relations have a well-defined meaning within local quantum field theory. How do these concepts generalize to the non-local case? This is not always clear because the presence of non-locality---modelled by infinite-derivative form factors---affects the construction of a Hilbert space, which poses some difficulty for defining a suitable averaging process to obtain the above expectation values. As we will see in Ch.~\ref{ch:ch6}, however, it is sometimes possible to proceed formally and obtain physically well-behaved non-local Green functions similarly to the local case.

Since the local Green functions solve the inhomogeneous (or homogeneous) field equations, we may simply \emph{define} non-local Green functions as the corresponding solutions of the non-local equations. To that end, let us assume that we can write the differential operator $\mathcal{D}$ as the concatenation
\begin{align}
\mathcal{D} = \mathcal{D}_\text{non-local} \circ \mathcal{D}_\text{local} \, .
\end{align}
Here, $\mathcal{D}_\text{non-local}$ is a non-local operator that, in the context of the present work, is related to the scale of non-locality and contains infinitely many derivatives, whereas $\mathcal{D}_\text{local}$ is a local operator that contains a finite amount of derivatives. One possible example could be
\begin{align}
\mathcal{D} = e^{-\ell^2\Box}\Box \, , \quad \mathcal{D}_\text{non-local} = e^{-\ell^2\Box} \, , \quad \mathcal{D}_\text{local} = \Box \, .
\end{align}
As already indicated in Sec.~\ref{sec:ch1:on-shell-off-shell}, ghost-free form factors have an empty kernel. In other words: the ghost-free operators considered in this work do not allow for non-trivial homogeneous solutions. This is important because it implies that the homogeneous Green functions \emph{coincide} for local and ghost-free theories since
\begin{align}
\label{eq:ch2:g-homogeneous}
\mathcal{D}G(X',X) = 0 \quad \Leftrightarrow \quad \mathcal{D}_\text{local} G(X',X) = 0 \, .
\end{align}
For ghost-free theories the inverse operator $\mathcal{D}^{-1}_\text{non-local}$ always exists and can be applied to generate the equivalence in Eq.~\eqref{eq:ch2:g-homogeneous}. We will examine this property in more detail in the form of the Hadamard function\index{Green function!Hadamard} in Ch.~\ref{ch:ch6}.

There is another point to be addressed in the next section: In local quantum field theory, the causal properties of the various Green functions can be related to analytic properties of their Fourier transform components in the complex energy plane. Since non-local modification terms, as they are studied in this thesis, do not possess any new poles (thus justifying their name ``ghost-free'') it seems plausible to us that a definition of non-local Green functions purely in terms of integration contours in the complex plane may still be valid in non-local theories of the above class.

\section{Causality from analyticity: local case}
Let us consider the local case first. In particular, we will focus on the inhomogeneous Green functions of the d'Alembert operator in flat, two-dimensional spacetime that is spanned by the coordinates $t$ and $x$. Let us call solution of
\begin{align}
\Box G(t'-t,x'-x) = -\delta(t'-t)\delta(x'-x)
\end{align}
a Green function. We are not primarily interested in the exact form of the Green functions, but rather we would like to focus on the following question: how do the causal properties of the Green function arise?

Due to the translational symmetry in both the temporal and spatial directions of Minkowski space let us introduce the momentum vector $k{}^\mu=(\omega,k)$ and consider the Fourier representation of the Green function. Inserting $\Box = -\partial_t^2 + \partial_x^2$ yields
\begin{align}
\label{eq:ch2:green-function-fourier-1}
G(t'-t,x'-x) = \int\limits_{-\infty}^\infty \frac{\dd\omega}{2\pi} e^{-i\omega(t'-t)} \int\limits_{-\infty}^\infty \frac{\dd k}{2\pi} e^{+ik(x'-x)} G_{\omega,k} \, , \quad G_{\omega,k} = \frac{-1}{\omega^2-k^2} \, .
\end{align}
In order to evaluate integrals of the above type, it is helpful to think of $\omega$ as a complex variable instead, such that the above integration over $\omega$ corresponds to a line integral in the complex plane along the real axis. The function $G_{\omega,k}$ has a single pole at $\omega=\pm k$, and for that reason an additional prescription is required to evaluate the integral. In what follows we will briefly recap how certain prescriptions affect the causal properties of the resulting Green function. While we will focus on the two-dimensional case, it is clear that these steps can be repeated for higher dimensions and other differential operators in a similar fashion, mutatis mutandis.

\subsection{Retarded Green function}
\begin{figure}
    \centering
    \includegraphics[width=0.47\textwidth]{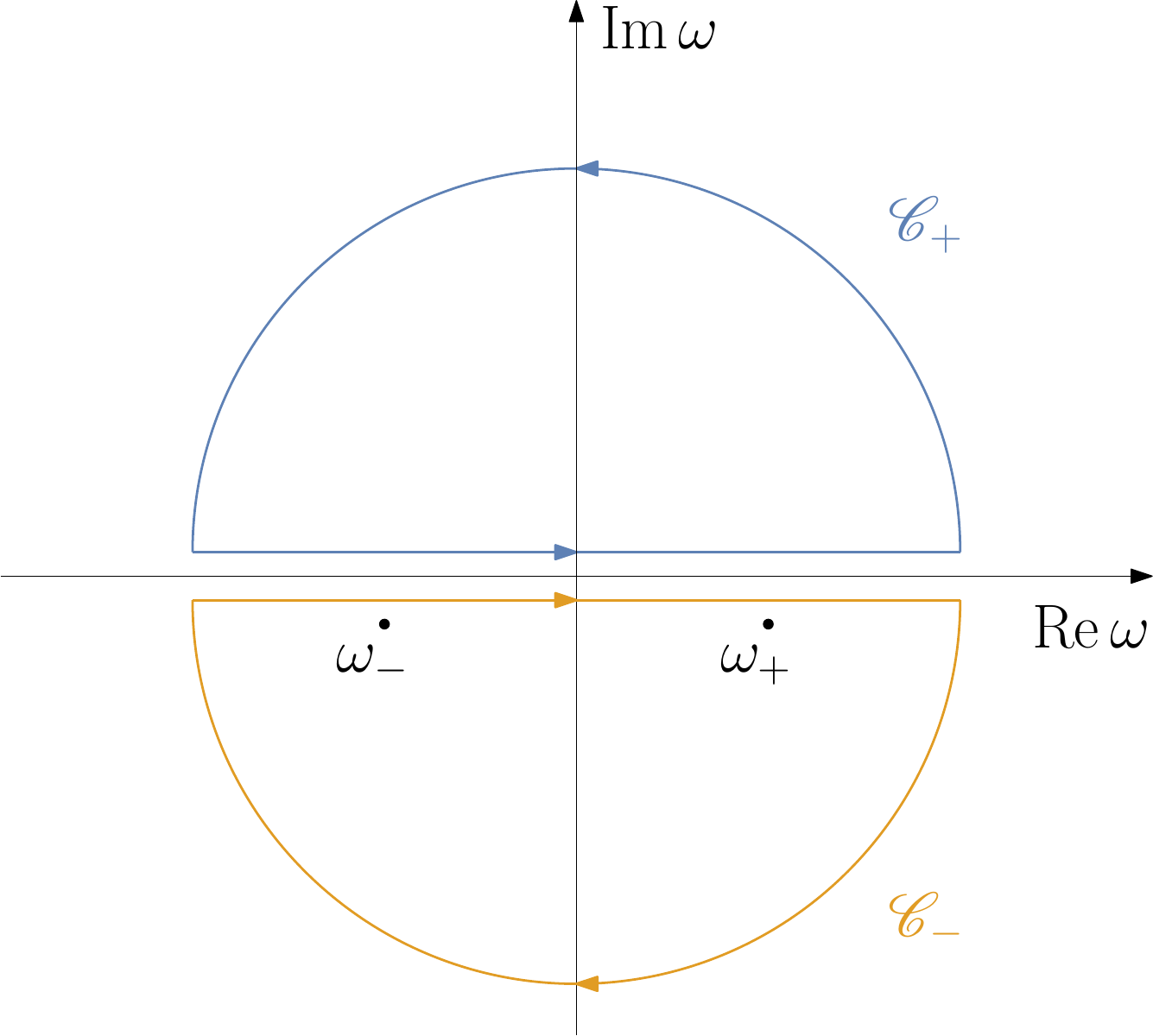}
    \caption[Integration contour for the retarded Green function.]{This choice of integration contours for the retarded Green function guarantees that the retarded Green function $G^\text{R}(t'-t,x'-x)$ vanishes if $t'<t$, which implements causality.}
    \label{fig:ch2:contour-retarded}
\end{figure}

The following choice of prescription, which is sometimes also referred to as $i\epsilon$-prescription, leads to the retarded Green function\index{Green function!retarded} (which we hence decorate with the superscript ``R''):
\begin{align}
G^{\text{R}}_{\omega,k} = \frac{-1}{(\omega+i\epsilon)^2-k^2} = \frac{-1}{(\omega+i\epsilon+k)(\omega+i\epsilon-k)}
\end{align}
Let us prove this. The poles are located in the complex $\omega$-plane at the values
\begin{align}
\omega_+ = k - i\epsilon \, , \quad \omega_- = -k - i\epsilon \, .
\end{align}
Assigning the $x$-axis the real part of $\omega$ and the $y$-axis to the imaginary part of $\omega$, because $\epsilon>0$ one sees that $\omega_\pm$ both lie below the $x$-axis. In order to evaluate the frequency integral in Eqs.~\eqref{eq:ch2:green-function-fourier-1} we realize that the integrand vanishes for large imaginary values of the frequency:
\begin{align}
\lim\limits_{\omega\rightarrow -i\infty} \frac{e^{-i\omega(t'-t)}}{\omega^2-k^2} = 0 \quad \text{if~~} t'>t \, , \qquad \lim\limits_{\omega\rightarrow +i\infty} \frac{e^{-i\omega(t'-t)}}{\omega^2-k^2} = 0 \quad \text{if~~} t > t' \, .
\end{align}
We may think now think of the $\omega$-integral in Eq.~\eqref{eq:ch2:green-function-fourier-1} as a line integral in the complex $\omega$-plane. Provided $t'>t$ we can close this line integral in the lower half-plane (and call the resulting contour $\mathscr{C}_-$), and if $t>t'$ instead we can close the line integral in the upper half-plane and call the contour $\mathscr{C}_+$. This is useful because now we can apply the residue theorem and relate the value of the integral to the poles enclosed by the contour $\mathscr{C}_\pm$; see a diagram in Fig.~\ref{fig:ch2:contour-retarded}.

For this reason one finds
\begin{align}
G^\text{R}(t'-t,x'-x) = 0 \quad \text{if~~} t>t' \, .
\end{align}
If $t'>t$, on the other hand, the retarded Green function does not vanish, and one may calculate
\begin{align}
G^\text{R}_k(t'-t) &= \lim\limits_{\epsilon\rightarrow 0} \oint\limits_{\mathscr{C}_-} \frac{\dd\omega}{2\pi} e^{-i\omega(t'-t)} \frac{-1}{(\omega+i\epsilon+k)(\omega+i\epsilon-k)} = \frac{\sin k(t'-t)}{k} \, , \\
G^\text{R}(t'-t,x'-x) &= \int\limits_{-\infty}^\infty \frac{\dd k}{2\pi} e^{ik(x'-x)} G^\text{R}_k(t'-t) = \frac12 \theta\left[ (t'-t)^2 - (x'-x)^2 \right] \quad \text{if~} t' > t \, .
\end{align}
For details of this calculation we refer to appendix \ref{app:2d-retarded}. The result can be collected as
\begin{align}
G^\text{R}(t'-t,x'-x) = \frac12 \theta(t'-t) \theta\left[ (t'-t)^2 - (x'-x)^2 \right] \, .
\end{align}
It proves two essential properties of the retarded Green function:
\begin{itemize}
\item[(i)] $G^\text{R}(t'-t,x'-x) = 0$ if $t'<t$.
\item[(ii)] $G^\text{R}(t'-t,x'-x) \not=0$ only inside the future-directed light cone.
\end{itemize}
Higher-dimensional calculations can be performed in a similar fashion, and one may also include a mass parameter $m$. We refer to the textbooks of quantum field theory \cite{BjorkenDrell,Weinberg,PeskinSchroeder,Ryder,Zee,DeWitt:1965} for more details.

The main conclusion of these considerations is that the $i\epsilon$-prescription in the momentum-space representation of the Green function gives rise to causality. The prescription, in turn, is dictated by the analytic properties of the function $G_{\omega,k}$ in the complex $\omega$-plane, and for this reason one may say that \emph{causality stems from analyticity.}

\subsection{Advanced Green function}
It is clear that a similar prescription to the retarded Green function gives rise to the advanced Green function, corresponding to the time-reversed scenario. The regularization is given by
\begin{align}
\label{eq:ch2:epsilon-advanced}
G^{\text{A}}_{\omega,k} = \frac{-1}{(\omega-i\epsilon)^2-k^2} = \frac{-1}{(\omega-i\epsilon+k)(\omega-i\epsilon-k)} \, ,
\end{align}
and we use the superscript ``A'' to denote the advanced Green function. From Eq.~\eqref{eq:ch2:epsilon-advanced}  it is clear that the change in sign of $\epsilon$ can be recast into a sign flip of $\omega$, and by means of \eqref{eq:ch2:green-function-fourier-1} that corresponds to an exchange of $t$ and $t'$. For that reason the advanced Green function, in two dimensions, takes the form
\begin{align}
G^\text{A}(t'-t,x'-x) = -\frac12 \theta(t-t') \theta\left[ (t'-t)^2 - (x'-x)^2 \right] \, .
\end{align}
This, in turn, proves two essential properties of the advanced Green function:
\begin{itemize}
\item[(i)] $G^\text{A}(t'-t,x'-x) = 0$ if $t<t'$.
\item[(ii)] $G^\text{A}(t'-t,x'-x) \not=0$ only inside the past-directed light cone.
\end{itemize}
One may also verify the above calculations by making use of the residue theorem. In this case the poles are located above the $x$-axis on the complex $\omega$-plane, and hence the contours give contributions in the opposite cases (with an overall negative sign since the contour now closes in the counter-clockwise, mathematically positive direction).

\subsection{Feynman Green function}
For completeness we would also like to mention the Feynman Green function, arising from the following $i\epsilon$-prescription:
\begin{align}
\label{eq:ch2:epsilon-feynman}
G^{\text{F}}_{\omega,k} = \frac{-1}{\omega^2-k^2 + i\epsilon} \approx \frac{-1}{(\omega-k+i\epsilon)(\omega+k-i\epsilon)} \, ,
\end{align}
where ``F'' stands for Feynman. There is also the reversed version, called anti-Feynman Green function, which arises from $\epsilon\rightarrow-\epsilon$ in the same sense as the advanced Green function arises from the retarded Green function under a similar substitution,
\begin{align}
\label{eq:ch2:epsilon-anti-feynman}
G^{\bar{\text{F}}}_{\omega,k} = \frac{-1}{\omega^2-k^2 - i\epsilon} \approx \frac{-1}{(\omega-k-i\epsilon)(\omega+k+i\epsilon)} \, ,
\end{align}
where now we decorate it with the superscript ``$\bar{\text{F}}$.'' Comparing the Feynman Green functions to the retarded (or advanced) Green functions, it becomes clear that they feature poles in the complex $\omega$ plane both above and below the $x$-axis. With everything else unchanged, however, the integrals of the type \eqref{eq:ch2:green-function-fourier-1} can still be solved using contour integration, and from that aspect alone it is clear that the Feynman Green function will be non-zero both if $t'>t$ and $t>t'$.

For this reason the Feynman Green function is not useful for classical physics but rather plays an important role in quantum field theory, which we will discuss in Ch.~\ref{ch:ch6} in more detail.

\subsection{Physical interpretations}
\label{sec:ch2:physical-interpretation}
\begin{figure}
    \centering
    \includegraphics[width=0.47\textwidth]{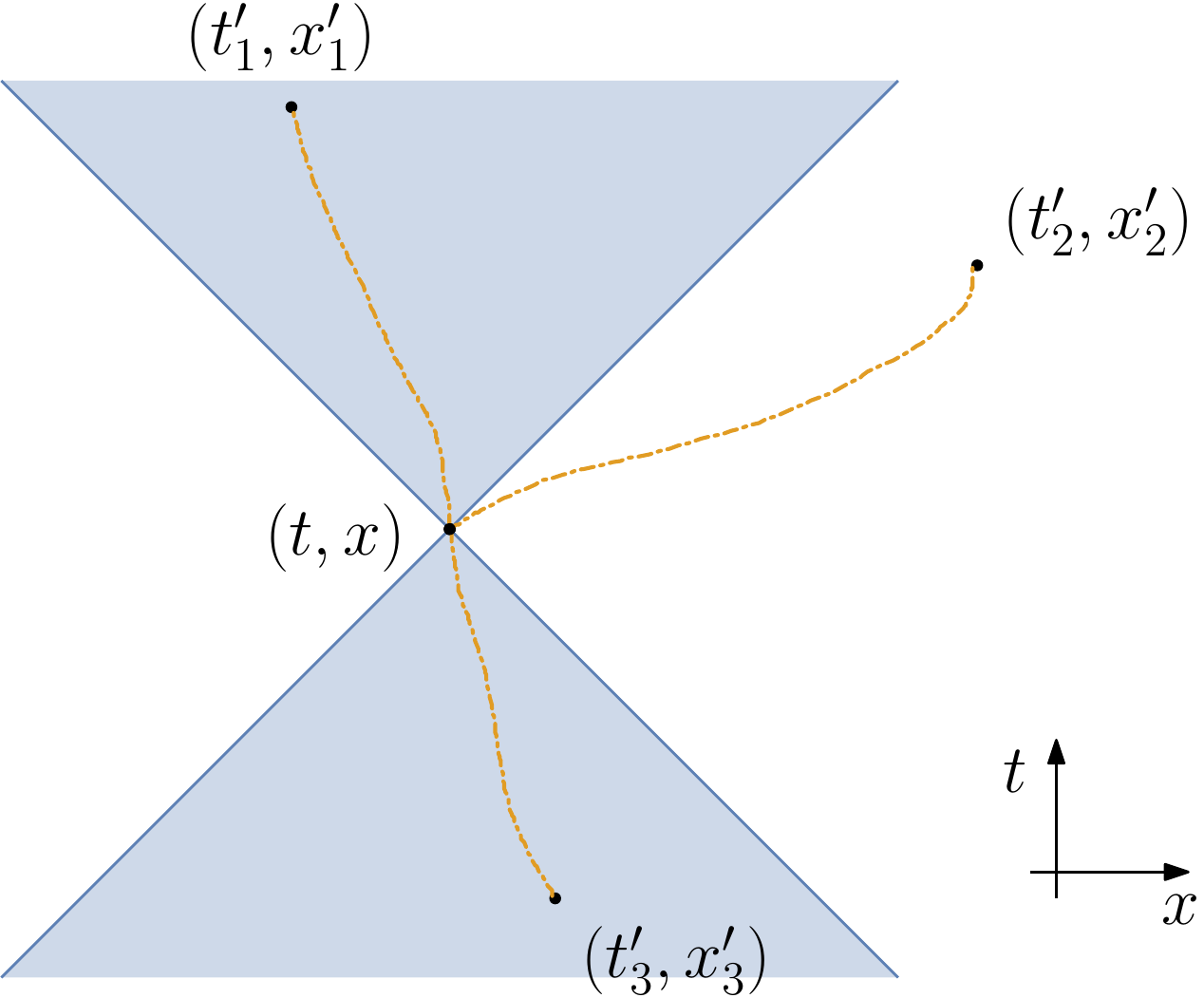}
    \caption[Spacetime interpretation of a free Green function.]{Consider the free Green function $G(t'-t,x'-x)$ and let us suppose that there exists a field fluctuation at $(t,x)$. The Green function then lets us determine how this fluctuation, to linear order, propagates through empty spacetime. The retarded Green function is non-vanishing if and only if $(t'_1,x'_1)$ lies in the causal future of $(t,x)$. The advanced Green function, however, will be non-vanishing if and only if $(t'_3, x'_3)$ lies in the causal past of $(t,x)$. All causal Green functions vanish identically for spacelike separated events, such as $(t,x)$ and $(t'_2, x'_2)$.}
    \label{fig:ch2:spacetime-interpretation-gf}
\end{figure}

These considerations allow for the following interpretation of the Green function $G(t'-t,x'-x)$, where we assume that the correct boundary conditions for the physical problem under consideration have been specified: if an event $(t,x)$ describes the location of a small perturbation, then the event $(t',x')$ corresponds to a possible observation of this perturbation, if and only if $G(t'-t,x'-x)$ is non-zero. There are of course different Green functions, depending on the $i\epsilon$-prescription employed in the complex $\omega$-plane, and we visualize this principle for the retarded and advanced Green function in Fig.~\ref{fig:ch2:spacetime-interpretation-gf}.

Sometimes it is also useful to consider the purely temporal Fourier transformation of a Green function, and for the sake of this example we will consider the two-dimensional case:
\begin{align}
\label{eq:ch2:green-function-fourier-2}
G_\omega(x'-x) = \int\limits_{-\infty}^\infty \frac{\dd k}{2\pi} e^{+ik(x'-x)} G_{\omega,k} \, .
\end{align}
This is useful because it allows us to discuss a mode decomposition of the Green function into frequencies. For this reason we may understand the Green function as a kind of superposition of propagating plane waves. In this case it is more useful to think of $k$ as a complex variable instead of $\omega$, and to perform the contour integration in the complex $k$-plane. In order not to derail the discussion too much we refer to appendix \ref{app:2d-massive-green-functions} for the calculational details. Then, the temporal Fourier transforms of the retarded, advanced, and Feynman Green function are
\begin{align}
G^\text{R}_\omega(x'-x) &= +\frac{i}{2\omega}e^{+i\omega|x'-x|} \, , \\
G^\text{A}_\omega(x'-x) &= -\frac{i}{2\omega}e^{-i\omega|x'-x|} \, , \\
G^\text{F}_\omega(x'-x) &= +\frac{i}{2|\omega|}e^{+i|\omega(x'-x)|} \, .
\end{align}
Let us understand why the above expression for $G^\text{R}_\omega(x'-x)$ is indeed the retarded Green function. To that end, recall that each Fourier mode of the retarded Green function in spacetime, $G^\text{R}(t'-t,x'-x)$, modulo normalization, is given by the plane wave
\begin{align}
e^{-i\omega(t'-t)} e^{+i\omega|x'-x|} = e^{-i\omega( t'-t - |x'-x| )} \, .
\end{align}
Identifying $t$ and $x$ as the spacetime location of a perturbation, we can consider them fixed, say, for a $\delta$-shaped perturbation $\sim\delta(t-t_0)\delta(x-x_0)$. Only $t'$ and $x'$ are variables, and they describe the directions in which the perturbations travel. Surfaces of constant phase are hence given by
\begin{align}
\text{const.} = -i\omega( t'-t_0 - |x'-x_0|) \quad \Rightarrow \quad \begin{cases}
\displaystyle x' > x_0 : \quad \frac{\dd x'}{\dd t'} &= +1 > 0 \, , \\[0.8em]
\displaystyle x' < x_0 : \quad \frac{\dd x'}{\dd t'} &= -1 < 0 \, .
\end{cases}
\end{align}
This means that the perturbation, while travelling forward in time, moves away from the source at $x_0$ in both directions. For the advanced Green function the only difference (besides the sign in the normalization factor) lies in the sign of the exponent. Going through a similar calculation one finds that for the advanced Green function the perturbations travel \emph{towards} the initial perturbation, corresponding to the time reversal that one may apply to the free retarded Green function to arrive at the advanced Green function.

The Feynman Green function, on the other hand, behaves the same as the retarded Green function for positive frequencies. Because the Feynman Green function only depends on the modulus of the frequency there is one important distinction: negative frequencies are propagated forwards in time as well, whereas the retarded Green function, by construction, propagates them backwards in time.

\subsection{Homogeneous Green functions}
For completeness let us also address the temporal Fourier transform of the homogeneous Green functions which play an important role in quantum theory. They may also be evaluated using contour integration, but unlike in the case of the inhomogeneous Green functions their contours are compact; see appendix \ref{app:green-functions-homogeneous} for more details. They take the form
\begin{align}
G^{(1)}_\omega(x'-x) = \frac{\cos(\omega x)}{|\omega|}\theta(\omega^2) \, , \\
G^+_\omega(x'-x) = \frac{\cos(\omega x)}{|\omega|}\theta(+\omega) \, , \\
G^-_\omega(x'-x) = \frac{\cos(\omega x)}{|\omega|}\theta(-\omega) \, .
\end{align}
The Wightman functions $G^+_\omega$ and $G^-_\omega$ propagate positive and negative frequencies, and their sum is the Hadamard\index{Green function!Hadamard} function. One may verify that the Hadamard function can be derived from the Feynman Green function via
\begin{align}
G^{(1)}_\omega(x'-x) = 2\Im\left[ \mathcal{G}^\text{F}_\omega(x'-x) \right] \, .
\end{align}
Let us emphasize again that in the free, non-local case these Green functions remain unchanged because they are solutions to homogeneous equations.

\section{Asymptotic causality condition on Green functions}
\label{sec:ch2:asymptotic-causality}
As we will see in the remainder of this thesis, in non-local theories it is very difficult to define causality in a local sense. If non-local effects are present in a spacetime region, a local light cone may not have physical significance. This of course directly affects the understanding of Green functions in non-local theories.

In his book ``Dynamical Theory of Groups and Fields'' \cite{DeWitt:1965}, DeWitt briefly addresses this issue. He defines a notion of \emph{asymptotic causality} that any physical Green function should satisfy. In the later sections we will see that the Green functions of ghost-free theories fall under this category.

As a reminder, in the above we have shown that a free causal Green function encountered in a local theory should satisfy
\begin{align}
G(t'-t,x'-x) = 0 \quad \text{if~~} t>t' \, .
\end{align}
In a non-local theory, DeWitt generalizes this condition to
\begin{align}
\mathcal{G}(t'-t,x'-x) \rightarrow 0 \quad \text{as~~} t'-t \rightarrow \infty \, .
\end{align}
One may think of this relation as a necessary consistency condition to recover macroscopic causality at some length scale. We will demonstrate later that the Green functions encountered in ghost-free, infinite-derivative theories do indeed satisfy this condition, both in the free and interacting case. We refer to Chs.~\ref{ch:ch5} and \ref{ch:ch6} for a detailed treatise with explicit examples.

\section{Causality from analyticity: non-local case}
\label{sec:ch2:causality-from-analyticity-non-local}
Let us now consider again a two-dimensional example, but this time with a non-local, infinite-derivative operator. Let us call solutions of the equation
\begin{align}
f(\Box) \Box \mathcal{G}(t'-t,x'-x) = -\delta(t'-t)\delta(x'-x) \, , \quad f(0) = 1 \, ,
\end{align}
non-local Green functions. The function $f(\Box)$ is non-polynomial and features a convergent series expansion subject to the constraint $f(0) = 1$. In the remainder of this thesis we will focus on functions of the form
\begin{align}
f(\Box) = \exp\left[(-\ell^2\Box)^N\right] \, , \quad \ell>0 \, , \quad N \in \mathbb{N} \, ,
\end{align}
and let us, for simplicity, in this section focus on the case of $N=1$. The Green function can again be represented as a double Fourier transform,
\begin{align}
\mathcal{G}(t'-t,x'-x) = \int\limits_{-\infty}^\infty \frac{\dd\omega}{2\pi} e^{-i\omega(t'-t)} \int\limits_{-\infty}^\infty \frac{\dd k}{2\pi} e^{+ik(x'-x)} \mathcal{G}_{\omega,k} \, , \quad \mathcal{G}_{\omega,k} = (-1)\frac{\exp\left[(\omega^2-k^2)\ell^2\right]}{\omega^2-k^2}
\end{align}
Clearly, the function $\mathcal{G}_{\omega,k}$ has poles in the complex plane at $\omega=\pm k$, which coincide with the local theory as the absence of new poles is what gave rise to ``ghost-free'' attribute to those theories.

That being said, due to the exponential factor it is impossible to perform a contour integration in the complex $k$-plane to evaluate the full integral since the exponential diverges in the directions $k \rightarrow \pm i\infty$. In the cases of higher $N$, these directions can also be at an angle in the complex plane. These divergences are \emph{essential singularities} and not mere poles, and for that reason one cannot employ contour integration to solve the above integral. Moreover, the presence of these essential singularities poses an enormous obstacle towards Wick rotation methods\index{Wick rotation} in the full non-local theory.

However, there is a simple way to solve this problem. Let us define
\begin{align}
\mathcal{G}_{\omega,k} = G{}_{\omega,k} + \Delta\mathcal{G}_{\omega,k} \, , \quad G_{\omega,k} = \frac{-1}{\omega^2-k^2} \, , \quad \Delta\mathcal{G}_{\omega,k} = \frac{1-\exp\left[(\omega^2-k^2)\ell^2\right]}{\omega^2-k^2} \, .
\end{align}
Then, the double Fourier transform of $G_{\omega,k}$ can be performed identical to the local theory. The term $\Delta\mathcal{G}_{\omega,k}$ should be understood as a non-local modification. Its double Fourier transform is given by the following integral:
\begin{align}
\Delta\mathcal{G}(t'-t,x'-x) = \int\limits_{-\infty}^\infty \frac{\dd\omega}{2\pi} e^{-i\omega(t'-t)} \int\limits_{-\infty}^\infty \frac{\dd k}{2\pi} e^{+ik(x'-x)} \Delta\mathcal{G}_{\omega,k}
\end{align}
The integrand $\Delta\mathcal{G}_{\omega,k}$ does not have any poles: the two simple poles are absent for any function that satisfies $f(0) = 1$, which is guaranteed in the class of non-local field theories studied here.

Does this integral always exist? The answer is: it depends. In the above case we inserted the choice $N=1$, and in this case the temporal part of the Fourier transform is unbounded. However, in purely static and some stationary situations when the temporal part of the Green function can be discarded, these integrals exist and provide interesting insights. We will make use of these static Green functions extensively in Chs.~\ref{ch:ch3}--\ref{ch:ch5}. Whenever the temporal part is required, or a summation over frequency modes needs to be performed (see, e.g., the case of quantum field theory in Ch.~\ref{ch:ch6}), only the case of even $N$ is permissible.

Because integration is performed over a Lorentz-invariant integrand that is purely a function of $\omega^2-k^2$ one can employ a transformation to hyperbolic coordinates to simplify these integrals substantially, not dissimilar to the case of spherical coordinates in Euclidean geometries. However, due to the non-compactness of the Lorentz group one needs more than one coordinate patch to cover the entirety of Minkowski space. We refer the reader to the work of DeWitt--Morette \textit{et al.} \cite{Morette:2002} for explicit representations of these integrals in two and higher dimensions.

Let us rewrite the above integration (and include a mass $m>0$ as a two-dimensional regulator)
\begin{align}
\Delta\mathcal{G}(t,x) = \int\limits_{\mathbb{R}^{1,1}} \frac{\dd^2 q}{(2\pi)^2} e^{iq{}_\mu x{}^\mu} A(\ts{q}^2) \, , \quad A(\ts{q}^2) = -\frac{1-\exp\left[-\ell^{2N}(\ts{q}^2+m^2)^N\right]}{\ts{q}^2+m^2} \, ,
\end{align}
where we defined $q{}^\mu = (\omega,k)$ and $\ts{q}^2=-\omega^2+k^2$. Then, using the results of \cite{Morette:2002} one finds
\begin{align}
\Delta\mathcal{G}(t,0) &= -\frac{1}{2\pi} \int\limits_0^\infty \dd s A(-s^2) s Y_0(st) + \frac{1}{\pi^2} \int\limits_0^\infty \dd s A(s^2) s K_0 (st) \, , \\
\Delta\mathcal{G}(0,x) &= \frac{1}{\pi^2} \int\limits_0^\infty \dd s A(-s^2) s K_0 (sx) - \frac{1}{2\pi} \int\limits_0^\infty \dd s A(s^2) s Y_0(sx) \, .
\end{align}
For $t>0$, $x>0$, and even $N$ the above integrals can be evaluated numerically since the integrands are regular everywhere in the integration domain and approach zero as $s\rightarrow\pm\infty$ sufficiently fast. One can also show that
\begin{align}
\label{eq:ch2:deltaG-tx-relation}
\Delta\mathcal{G}(0,u) = - \Delta\mathcal{G}(u,0) \, .
\end{align}

\begin{figure}[!htb]%
    \centering
    \includegraphics[width=0.7\textwidth]{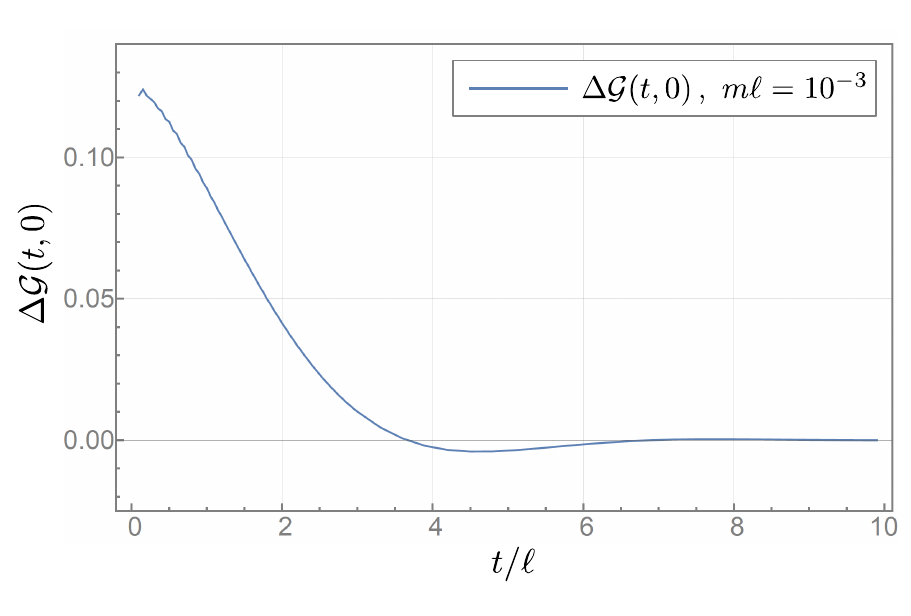}\\[-15pt]
    \caption[Non-local modification of the Green function in $\mathrm{GF_2}$ theory.]{We evaluate the non-local modification $\Delta\mathcal{G}(t,x)$ numerically for $\mathrm{GF_2}$ theory (mass parameter $m\ell = 10^{-3}$). Equation~\eqref{eq:ch2:deltaG-tx-relation} implies that the spacelike and timelike directions behave similarly. The diagram shows that $\Delta\mathcal{G}(t,0)$ decreases rapidly in time, and hence a similar property holds for the decrease of $\Delta\mathcal{G}(0,x)$ in space. We will address the null direction separately below.}
    \label{fig:ch2:deltaG-tx}
\end{figure}

\begin{figure}
    \centering
    \includegraphics[width=0.47\textwidth]{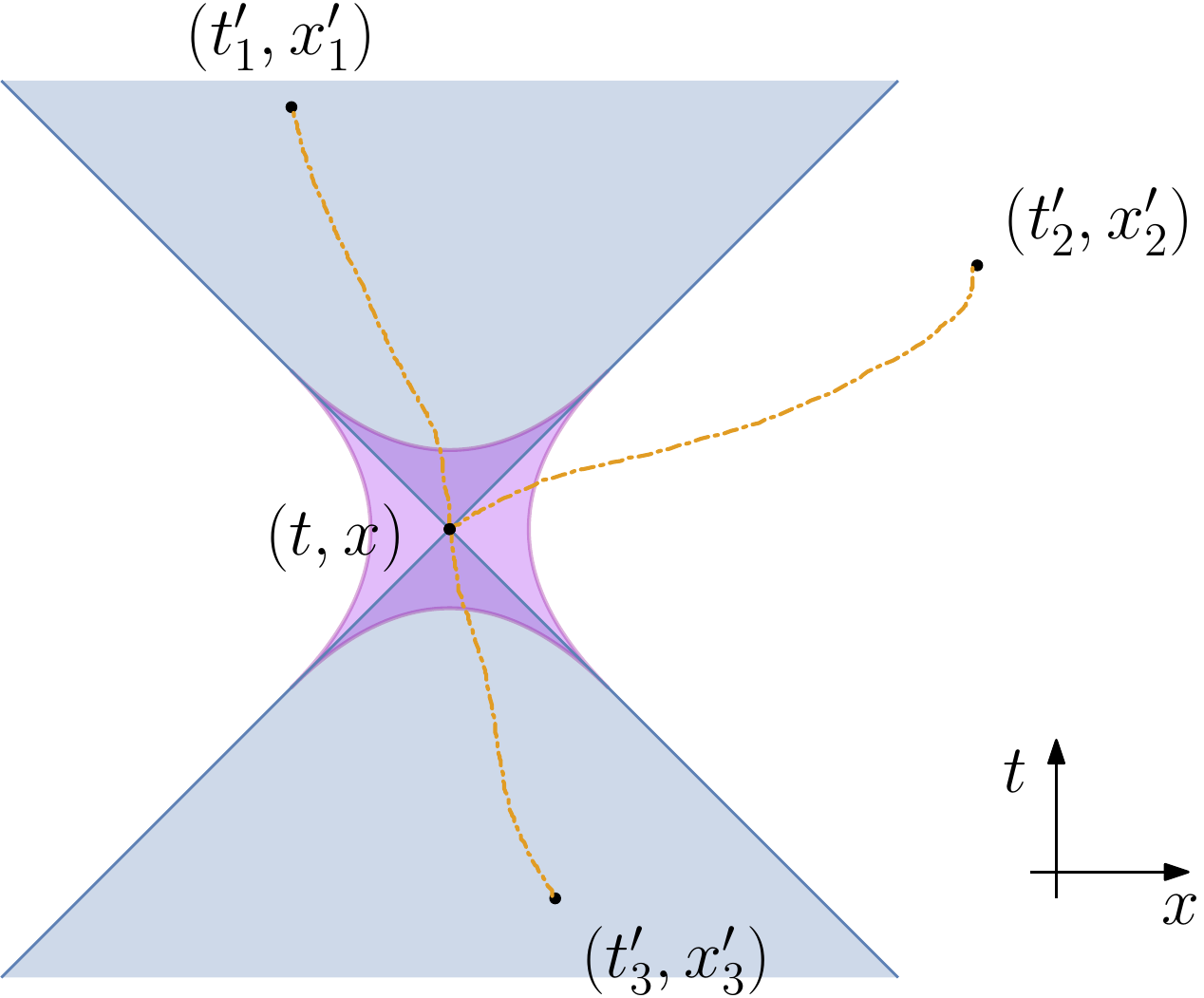}
    \caption[Spacetime interpretation of a non-local free Green function.]{Consider the non-local free Green function $\mathcal{G}(t'-t,x'-x)$ and let us suppose that there exists a field fluctuation at $(t,x)$. At large distances, the non-local Green function mimics the causal behavior of the local Green function. At ``small scales,'' however, where $|-(t'-t)^2+(x'-x)^2|\sim\ell^2$, non-causal effects may appear. Note that this ``small scale,'' highlighted as the shaded hyperboloid in the above figure, in fact has infinite extension in the null directions and is only a small scale in the purely timelike or spacelike directions. At this point, this is only to be understood as a qualitative picture, and we will devote a substantial part of this thesis to understanding this better.}
    \label{fig:ch2:spacetime-interpretation-gf-non-local}
\end{figure}

See Fig.~\ref{fig:ch2:deltaG-tx} for a plot of the non-local modification in $\mathrm{GF_2}$ theory for $N=2$. Evidently, the non-local modification decrease equally fast in both temporal and spatial distances. However, it is manifestly non-zero outside the light cone, thereby violating microcausality; see Fig.~\ref{fig:ch2:spacetime-interpretation-gf-non-local} for a visualization. DeWitt's condition of asymptotic causality, however, is satisfied. In particular, this class of non-local field theories comes with a length scale $\ell$, such that we can write (for even $N$)
\begin{align}
\Delta\mathcal{G}(t,0) = 0 \quad \text{if~~} t \gg \ell \, , \quad
\Delta\mathcal{G}(0,x) = 0 \quad \text{if~~} x \gg \ell \, .
\end{align}
We should point out that in the framework of special relativity there is no such thing as a ``small spacetime volume'' on which non-locality becomes important. This is due to the indefinite signature of the spacetime metric: If one restricts an area by demanding
\begin{align}
|-(t'-t)^2+(x'-x)^2|\sim\ell^2
\end{align}
it corresponds to spatial and temporal distance of $\mathcal{O}(\ell)$, but corresponds to infinite extension in spacetime along the null directions. In this thesis we will primarily deal with static situations, where we can always effectively ignore the temporal direction, and hence the notion of ``small scales'' is sensible.

For the remainder of this section we will collect a few analytical results we were able to find.

\subsection{Explicit expressions}
\label{sec:deltaG-explicit-expressions}
Let us now consider the purely temporal Fourier transform of the non-local modification term in one spatial dimension,
\begin{align}
\label{eq:ch2:deltaG-omega}
\Delta\mathcal{G}_\omega(x) = \int\limits_{-\infty}^\infty \frac{\dd q}{2\pi} \cos(qx)A_\omega(q) \, , \quad A_\omega(q) = \frac{1-\exp\left[-\ell^{2N}(q^2-\omega^2)^N\right]}{\omega^2-q^2} \, .
\end{align}
In the case of $N=1$ this can be evaluated explicitly by expressing the above as a double integral:
\begin{align}
\Delta\mathcal{G}_\omega(x) = \int\limits_{-\infty}^\infty \frac{\dd q}{2\pi} e^{iqx} \frac{e^{-\ell^2(q^2-\omega^2)}-1}{q^2-\omega^2} = -\int\limits_0^{\ell^2}\dd s \int\limits_{-\infty}^\infty \frac{\dd k}{2\pi} e^{iqx} e^{-s(q^2-\omega^2)} \, .
\end{align}
By commuting the integrals and evaluating the Gaussian $s$-integral first one arrives at
\begin{align}
\mathcal{G}^\text{R}_\omega(x) &= \frac{i}{4\omega} \left[ e^{i\omega x} Y(x) + e^{-i\omega x}Y(-x) \right] \, , \quad
Y(x) = 1 + \text{erf}\left( i\omega\ell + \frac{x}{2\ell} \right) \, , \\
\Delta\mathcal{G}_\omega(x) &= -\frac{1}{2\omega}\text{erfi}(\omega\ell) \, ,
\end{align}
where $\text{erfi}(x) = -i\text{erf}(ix)$ denotes the imaginary error function \cite{Olver:2010}. One may verify that in the limit of $x/\ell\rightarrow \infty$ one has $Y(x) = 1 + \text{sgn}(x)$, which implies $\mathcal{G}^\text{R}_\omega(x)\rightarrow G^\text{R}_\omega(x)$. This guarantees that in the limit of vanishing non-locality we recover the local theory, as well as the identical asymptotic behavior of the non-local retarded Green function. This is consistent with our previous considerations where we showed that the causal properties of a Green function stem from the analytic behavior of its local part because its non-local modification term does not have any poles in the complex plane at finite radius.

For $N=2$ such a general expression cannot be found, but instead we are able to compute $\Delta\mathcal{G}_\omega(0)$ explicitly. Since this discussion is a bit more technical we refer to appendix \ref{app:deltaG-0-gf2} for the calculational details. The result for $N=2$ in one spatial dimension is
\begin{align}
\Delta\mathcal{G}_\omega(0) = \frac{\sqrt{2}\omega^2\ell^3}{6\Gamma\left(\tfrac34\right)} {}_2 F{}_2\left( \tfrac34, \tfrac54;~ \tfrac32, \tfrac74;~-\omega^4\ell^4\right) - \frac{\Gamma\left(\tfrac34\right)\ell}{\pi} {}_2 F{}_2\left(\tfrac14,\tfrac34;~\tfrac12,\tfrac54;~-\omega^4\ell^4\right) \, .
\end{align}

\subsection{General asymptotics}
At this point, there is one additional insight that can be taken for general $N$. 
One may integrate by parts to obtain
\begin{align}
\Delta\mathcal{G}_\omega(x) = -\frac{1}{x} \int\limits_{-\infty}^\infty \frac{\dd q}{2\pi} \sin(qx) \frac{\dd A_\omega(q)}{\dd q} \, ,
\end{align}
where the boundary terms vanish both for even $N$ and odd $N$, since the spatial part is always decreasing. These integrations by part can be repeated ad infinitum, yielding another factor of $1/x$ each time, and for that reason prove that the non-local modification $\Delta\mathcal{G}_\omega(x)$ decreases faster than any power in $x$ for each given Fourier mode $\omega$.

The important insight of this section is the following: macroscopic causality is dictated purely by the analytic properties of the local Green function in Fourier space; the analytic properties of the non-local modification do not play any role. At length scales comparable to the scale of non-locality $\ell$, however, non-local Green functions violate the causality conditions, since they are non-vanishing outside the future light cone. Let us note that this property is intricately connected with $f(0) = 1$, which in turn guarantees that asymptotically one recovers the local theory.

\section{Non-local Green function contributions: some results}
An analytic treatment of the non-local modification term $\Delta\mathcal{G}(t,x)$ is difficult. While the case $N=1$ is easier to treat due to its effectively Gaussian appearance, the convergence is not guaranteed because of the divergence in the temporal Fourier transform. The closest alternative, $N=2$, is numerically well-behaved, but analytically more challenging due to the quartic exponential expression. Here we would like to present some intermediate results that may prove helpful in the future to understand the case $N=2$ better.

Let us start again with the representation of the non-local modification for $d$ spatial dimensions,
\begin{align}
\Delta\mathcal{G}(t,\ts{x}) &= \int\limits_0^\infty \frac{\dd\omega}{\pi} \cos(\omega t) \int\limits_{\mathbbm{R}^d} \frac{\dd^d k}{(2\pi)^d} e^{i\ts{k}\cdot\ts{x}} \frac{1-e^{-\ell^4(\ts{k}^2-\omega^2)^2}}{\omega^2-\ts{k}^2} \\
\label{eq:ch2:deltaG-representation}
&= \frac{1}{2\pi^2 x} \int\limits_0^\infty \dd\omega \cos(\omega t) \int \limits_0^\infty \dd k k^{d-2} \sin(k x) \frac{1-e^{-\ell^4(k^2-\omega^2)^2}}{\omega^2-k^2} \, ,
\end{align}
where we denote $x=|\ts{x}|$ and $k=|\ts{k}|$. There is the following identity:
\begin{align}
\frac{1-e^{-\ell^4p^2}}{p} = \frac{1}{\sqrt{\pi}\ell^2} \int\limits_0^\infty \dd y e^{-y^2/(4\ell^2)} \int\limits_0^y \dd z \sin(p z) \, , \quad p \in \mathbb{R} \, .
\end{align}
Inserting this into \eqref{eq:ch2:deltaG-representation} one finds
\begin{align}
\Delta\mathcal{G}(t,x) &= \frac{1}{2\pi^{5/2}\ell^2x} \int\limits_0^\infty \dd y e^{-y^2/(4\ell^2)} \int\limits_0^y I(t,x,z) \, , \\
I(t,x,z) &= I_1 I_2 - I_3 I_4 \, , 
\end{align}
where we defined the following regularized integrals:
\begin{alignat}{4}
I_1 &= \lim\limits_{\alpha\rightarrow 0} \, \int\limits_0^\infty \dd\omega \, e^{-\alpha\omega} \cos(\omega t)\sin(\omega^2 z) \, , \hspace{40pt} &
I_3 &= \lim\limits_{\alpha\rightarrow 0} \, \int\limits_0^\infty \dd\omega \, e^{-\alpha\omega} \cos(\omega t)\cos(\omega^2 z) \, , \\
I_2 &= \lim\limits_{\alpha\rightarrow 0} \, \int\limits_0^\infty \dd k  \, e^{-\alpha k}k^{d-2} \sin(k x)\cos(k^2 z) \, ,  &
I_4 &= \lim\limits_{\alpha\rightarrow 0} \, \int\limits_0^\infty \dd k  \, e^{-\alpha k}k^{d-2} \sin(k x)\sin(k^2 z) \, .
\end{alignat}

\subsection{Four-dimensional case}
In the case of $d=3$ one can show
\begin{align}
\Delta\mathcal{G}(t,x) = \frac{-1}{16\pi^{3/2}\ell^2} \int\limits_0^\infty \dd y e^{-y^2/(4\ell^2)} \int\limits_0^y \frac{\dd z}{z^2} \cos\left(\frac{t^2-x^2}{4z}\right) \, ,
\end{align}
where one now has to evaluate
\begin{align}
\int\limits_0^y \frac{\dd z}{z^2} \cos\left(\frac{t^2-x^2}{4z}\right) = \frac{4}{x^2-t^2} \left[ \sin\left(\frac{t^2-x^2}{4y}\right) - S_0 \right] \, , \quad S_0 = \lim\limits_{z\rightarrow 0} \sin\left(\frac{t^2-x^2}{4z}\right) \, .
\end{align}
Leaving aside the limiting procedure for $S_0$ one can calculate
\begin{align}
\Delta\mathcal{G}(t,x) &= \frac{1}{4\pi^{3/2}\ell^2(t^2-x^2)} \int\limits_0^\infty \dd y e^{-y^2/(4\ell^4)} \left[ \sin\left(\frac{t^2-x^2}{4y}\right) - S_0 \right] \\
&= \frac{1}{64\pi\ell^2} G^{20}_{03}\left( \frac{(t^2-x^2)^2}{256\ell^4} \left| \begin{matrix} \\ 0,0, -\tfrac12 \end{matrix} \right. \right) - \frac{S_0}{4\pi(t^2-x^2)} \, , \nonumber
\end{align}
where $G$ denotes the Meijer G-function with the parameters $n=0$, $p=0$, $m=2$, $q=3$ as well as
\begin{align}
b_1 = b_2 = 0 \, , \quad b_3 = -\frac12 \, .
\end{align}
The asymptotics of the Meijer G-function with the above values are \cite{Olver:2010}
\begin{align}
G^{20}_{03}(x) &\approx \begin{cases} 
\displaystyle -\frac{2}{\sqrt{\pi}}\left( 2\gamma - \Psi\left(\tfrac32\right) + \log x \right) & \text{ for } x \ll 1 \, , \\[10pt]
\displaystyle -\frac{2}{\sqrt{3x}} \Im \exp\left[ -3 e^{i\pi/3} x^{1/3} \right] & \text{ for } x \gg 1 \, ,
\end{cases} \nonumber \\[10pt]
\gamma &\approx 0.57722\dots \, , \quad \Psi\left(\tfrac32\right) \approx 0.03649 \dots \, .
\end{align}
From these asymptotics it is clear that for large distances, that is, either $t/\ell \gg 1$ or $x/\ell \gg 1$, the non-local modification decreases, in accordance with DeWitt's notion of asymptotic causality.

\subsection{Two-dimensional case}
Also the two-dimensional case can be treated to some extent. In this case one finds
\begin{align}
\Delta\mathcal{G}(t,x) &= \frac{1}{\sqrt{\pi}\ell^2} \int\limits_0^\infty \dd y e^{-y^2/(4\ell^2)}\int\limits_0^y \dd z ( \tilde{I}_1\tilde{I}_2 - \tilde{I}_3\tilde{I}_4 ) \\
&= \frac{1}{4\pi^{3/2}\ell^2} \int\limits_0^\infty \dd y e^{-y^2/(4\ell^2)} \int\limits_0^y \frac{\dd z}{z} \sin\left( \frac{t^2-x^2}{4z} + m^2 z \right) \, , 
\end{align}
where we used
\begin{align}
\tilde{I}_1 &= \int\limits_0^\infty \frac{\dd \omega}{\pi} \cos(\omega t) \cos(\omega^2 z) 
= \frac{\cos[t^2/(4z)] + \sin[t^2/(4z)]}{2\sqrt{2\pi z}} \, , \\
\tilde{I}_2 &= \int\limits_0^\infty \frac{\dd q}{\pi} \cos(q x) \sin[(q^2+m^2) z] 
= \frac{\cos[x^2/(4z)-m^2 z] - \sin[x^2/(4z)-m^2 z]}{2\sqrt{2\pi z}} \, , \\
\tilde{I}_3 &= \int\limits_0^\infty \frac{\dd \omega}{\pi} \cos(\omega t) \sin(\omega^2 z) 
= \frac{\cos[t^2/(4z)] - \sin[t^2/(4z)]}{2\sqrt{2\pi z}} \, , \\
\tilde{I}_4 &= \int\limits_0^\infty \frac{\dd q}{\pi} \cos(q x) \cos[(q^2+m^2) z] 
= \frac{\cos[x^2/(4z)-m^2 z] + \sin[x^2/(4z)-m^2 z]}{2\sqrt{2\pi z}} \, .
\end{align}
Again, we introduced the mass $m>0$ as a two-dimensional infrared regulator. On the light cone (and hence, by Poincar\'e invariance, in the coincidence limit) one can calculate
\begin{align}
\Delta\mathcal{G}(0,0) &= \frac{1}{4\pi\sqrt{\pi}\ell^2} \int\limits_0^\infty \dd y e^{-y^2/(4\ell^4)} \int\limits_0^y \dd z \frac{\sin m^2 z }{z} 
= \frac{1}{4\pi\sqrt{\pi}\ell^2} \int\limits_0^\infty \dd y e^{-y^2/(4\ell^4)} \, \text{Si}(m^2y) \\
&= \frac{m^2\ell^2}{2\pi\sqrt{\pi}} \, {}_2F{}_{2}\left( \tfrac12,1; \tfrac32,\tfrac32; -m^4\ell^4 \right)
\end{align}
In particular, for $m=0$ we find that $\Delta\mathcal{G}(0,0) = 0$. The shape of the generalized hypergeometric function with the above values behaves roughly like $e^{-m\ell/2}$.

In the massless case, $m=0$, we can calculate the non-local modification analytically. Defining the spacelike distance $\tau^2 := t^2-x^2$ for convenience, we make use of the identity
\begin{align}
\int\limits_0^y \frac{\dd z}{z} \sin\left(\frac{\tau^2}{4z}\right) = \text{sgn}\left(\tau^2\right) \frac{\pi}{2} - \text{Si}\left(\frac{\tau^2}{4y}\right) \, , \quad \tau^2 \not= 0 \, .
\end{align}
Then, one arrives at
\begin{align}
\Delta\mathcal{G}(t,x) = \text{sgn}(\tau^2)\left[ \frac18 - \frac{\tau^2}{128\pi\ell^2} G{}^{21}_{14}\left( \left. +\frac{\tau^4}{256\ell^4} \right| \begin{matrix} \tfrac12 &  \\ 0, & 0, -\tfrac12,-\tfrac12 \end{matrix} \right) \right] \, .
\end{align}
On the light cone one can calculate directly, also in the massless limit,
\begin{align}
\Delta\mathcal{G}(0,0) = 0 \, .
\end{align}
Even though these partial results only allow us an incomplete study of the non-local modification term of the Green function, we list them here in the hope that they will prove useful in the future. For similar expressions, also involving the Meijer G-function, we refer to the review article on ghost-free infinite-derivative quantum field theory by Buoninfante \textit{et al.} \cite{Buoninfante:2018mre}.

\section{Static Green functions}
\label{sec:ch2:static}
Let us now leave spacetime and focus on purely spatial Green functions. As we will see in the next chapter, spatial Green functions are entirely sufficient to discuss static and stationary situations where retardation effects are absent. While the use of these Green functions is quite straightforward, their derivation involves a few useful techniques which we would like to present in this section, before turning to applications in the next chapter.

To be more concrete, we will focus on spatial Green functions of the Laplace operator $\lap$ and functions thereof, $f(\lap)$. The fundamental difference of treating spatial Green functions of the Laplace operator $\lap$ as compared to spacetime Green functions of the d'Alembert operator $\Box$ lies in the fact that the Laplace operator is elliptical, whereas the d'Alembert operator is hyperbolic. This means that zero mode solutions of the form $\Box\phi=0$ and $\lap\psi=0$ have very different behavior. While $\phi$ may be a propagating wave, harmonic functions $\psi$, on a non-compact manifold, can often be excluded straightforwardly by fixing appropriate boundary conditions (since they are polynomial functions in flat space, for example).

In a mathematical language one could say that the kernel of the Laplace operator is empty, provided one forbids polynomially increasing or decreasing harmonic solutions by fixing the boundary conditions. This is very convenient because it implies that the Fourier representations for the Green functions do not need to be regulated: they can be calculated straightforwardly. This fact has become apparent already in the previous sections, because the convergence for the Fourier representation of spatial Green functions is always guaranteed.

For these reasons, spatial Green functions are easier to treat than the time-dependent spacetime Green functions. Let us now focus on the class of $\mathrm{GF_N}$ theories where
\begin{align}
f(\lap) = \exp[(-\ell^2\lap)^N] \, , \quad \ell>0 \, , \quad N \in \mathbb{N} \, .
\end{align}
Clearly this choice satisfies $f(0) = 1$. A spatial Green function satisfies this relation:
\begin{align}
\label{eq:ch2:static-green-function}
f(\lap)\lap \mathcal{G}_d(\ts{x}'-\ts{x}) = -\delta^{(d)}(\ts{x}'-\ts{x})
\end{align}
Here and in what follows we shall work in $d$-dimensional flat space, and $\ts{x}$ denotes the collection of all spatial coordinates. The Green function is only a function of the difference of $\ts{x}$ and $\ts{x}'$ due to the translational symmetry of empty flat space, and we introduced the subscript ``$d$'' to keep track of the dimensions. The Fourier representation of this Green function takes the form
\begin{align}
\mathcal{G}_d(\ts{x}'-\ts{x}) = \int\frac{\dd^d k}{(2\pi)^d} e^{+i\ts{k}\cdot(\ts{x}'-\ts{x})} \mathcal{G}_{d,k} \, , \quad \mathcal{G}_{d,k} = \frac{e^{-(k^2\ell^2)^N}}{k^2} \, ,
\end{align}
where $k = |\ts{k}|$. It is now useful to perform this Fourier transformation in spherical momentum coordinates instead, wherein one may choose
\begin{align}
\ts{k}\cdot(\ts{x}'-\ts{x}) = kr\cos\theta \, , \quad r = |\ts{x}'-\ts{x}| \, .
\end{align}
From the above consideration it is also clear that the Green function can be expressed as a function of $r$ alone, which follows from the spherical symmetry of empty flat space around the origin. Here and in what follows, in order to keep the notation somewhat manageable, we shall use $\mathcal{G}_d(r)$ and $\mathcal{G}_d(\ts{x}'-\ts{x})$ interchangeably, that is, we will use the same symbol both for the Green function with a $d$-dimensional vectorial argument as well as for the Green function with the radial argument. Because the distance $\ts{x}'-\ts{x}$ only enters the free Green function $\mathcal{G}_d$ via the absolute value this abbreviation is unique.

Introducing $d$-dimensional spherical momentum coordinates $\{k,\theta,\varphi_1,\dots,\varphi_{d-2}\}$ one finds
\begin{align}
\mathcal{G}_d(r) &= \frac{A_{d-2}}{(2\pi)^d} \int\limits_0^\infty \dd k k^{d-3} e^{-(k^2\ell^2)^N} \int\limits_0^\pi \dd\theta \sin^{d-2}\theta e^{ikr\cos\theta} \\
&= \frac{A_{d-2}}{(2\pi)^d} \int\limits_0^\infty \dd k k^{d-3} e^{-(k^2\ell^2)^N} \int\limits_0^\pi \dd\theta \sin^{d-2}\theta \cos(kr\cos\theta) \, ,
\end{align}
where $A_{d-2}$ is the surface of the $(d-2)$-sphere,
\begin{align}
A_{d-2} = 2\frac{\pi^{(d-1)/2}}{\Gamma\left(\tfrac{d-1}{2}\right)} \, .
\end{align}
Using now the following representation of the Bessel function (Eq.~(10.9.4) in Ref.~\cite{Olver:2010}),
\begin{align}
J_\nu(z) = \frac{\left(\tfrac{z}{2}\right)^\nu}{\sqrt{\pi}\Gamma\left(\nu+\tfrac12\right)} \int\limits_0^\pi \dd\theta \sin^{2\nu}\theta\cos(z\cos\theta) \, ,
\end{align}
identifying $d-2=2\nu$, and performing a variable substitution $\zeta=kr$, we finally arrive at
\begin{align}
\label{eq:ch2:bessel-representation-1}
\mathcal{G}_d(r) = \frac{1}{(2\pi)^{d/2}r^{d-2}} \int\limits_0^\infty \dd\zeta \zeta^{\frac{d-4}{2}} e^{-(z\ell/r)^{2N}} J_{\frac{d}{2}-1}(\zeta) \, .
\end{align}
This integral can be evaluated analytically for a wide range of $d \ge 3$ and $N \ge 1$. For other functions $f(\lap)$ the integral representation takes the more general form
\begin{align}
\mathcal{G}_d(r) = \frac{1}{(2\pi)^{d/2}r^{d-2}} \int\limits_0^\infty \dd\zeta \zeta^{\frac{d-4}{2}} \frac{1}{f(-\zeta^2/r^2)} J_{\frac{d}{2}-1}(\zeta) \, .
\end{align}
There are some interesting observations which we want to address next.

\subsection{Higher and lower dimensions}
First, one may wonder what happens in the case $d=2$? To that end, note that the Green function $\mathcal{G}_d(r)$ satisfies the following recursion relations:
\begin{align}
\label{eq:ch2:recursion}
\mathcal{G}_d(r) = -2\pi\int\limits_{r_0}^r \dd\tilde{r} \, \tilde{r} \mathcal{G}_{d+2}(\tilde{r}) \, , \quad
\mathcal{G}_{d+2}(r) = -\frac{1}{2\pi r} \frac{\partial \mathcal{G}_d(r)}{\partial r} \, .
\end{align}
This happens because of spherical symmetry: the surfaces of the $(d-2)$-sphere and the $d$-sphere are related recursively as well. There is an ambiguity in the above formulas regarding the lower integration bound $r_0$. While in the cases $d\ge3$ one may set $r_0 = \infty$ because the Green functions vanish asymptotically, in the case of $d=2$ this is not permissible. This problem is well known and stems from the fact that the two-dimensional Green function is dimensionless and hence requires an ad hoc reference scale. This will become more clear in explicit examples, such as the gravitational field of a cosmic string in Ch.~\ref{ch:ch3}. With that caveat out of the way, the recursion relations \eqref{eq:ch2:recursion} can be used to determine the Green functions in arbitrarily high dimensions with only two ``seed Green functions.''

\subsection{Local limit}
Second, one may wonder what happens in the case of $\ell\rightarrow 0$, that is, in the local case. From the condition $f(0) = 1$ it follows immediately that
\begin{align}
\mathcal{G}_d(r) = \frac{1}{(2\pi)^{d/2}r^{d-2}} \int\limits_0^\infty \dd\zeta \zeta^{\frac{d-4}{2}} J_{\frac{d}{2}-1}(\zeta) \, .
\end{align}
This integral, while being just a dimension-dependent factor in the local limit, is only well-defined for $d=3,4$. In higher dimensions it needs to be regulated and one finds for all $d \ge 3$ the following result:
\begin{align}
\lim\limits_{\epsilon\rightarrow 0} \int\limits_0^\infty \dd\zeta e^{-\epsilon\zeta} \zeta^{\tfrac{d-4}{2}} J_{\tfrac{d}{2}-1}(\zeta) = 2^{d/2-2}\Gamma\left(\tfrac{d}{2}-1\right) \, .
\end{align}
Interestingly, for $\ell>0$ the scale of non-locality automatically regularizes this integral, and one may simply calculate it for any $d \ge 3$ with $\ell>0$ and then take the limit $\ell\rightarrow 0$. One may wonder what the origin of this divergence is. The integrand is perfectly regular at $\zeta=0$ and $d\ge 3$ because
\begin{align}
\zeta^{\tfrac{d-4}{2}} J_{\tfrac{d}{2}-1}(\zeta) = \frac{2^{1-d/2}}{\Gamma\left(\tfrac{d}{2}\right)} \zeta^{d-3} + \mathcal{O}(\zeta^{d-1}) \quad \text{for~~} \zeta \ll 1 \, .
\end{align}
Conversely, for large arguments $\zeta$ the Bessel function exhibits oscillatory behavior, and together with the polynomially growing prefactor these oscillations are no longer decreasing as $\zeta\rightarrow\infty$ and $d \ge 5$. Since $\zeta=kr$, and the divergence occurs for any radius variable $r$, we conclude that the divergence in $d \ge 5$ must stem from large momenta $\ts{k}$.

The local limit then takes the following form for any $d \ge 3$:
\begin{align}
G_d(r) = \lim\limits_{\ell\rightarrow 0} \mathcal{G}_d(r) = \frac{\Gamma\left(\tfrac{d}{2}-1\right)}{4 \pi^{d/2}} \frac{1}{r^{d-2}} \, .
\end{align}
Because $f(0) = 1$ the local limit also corresponds to the large-distance limit, and hence we see that the non-local modification $\ell>0$ is not relevant at length scales much larger than $\ell$, giving yet another argument in favour of DeWitt's asymptotic causality condition already mentioned in Sec.~\ref{sec:ch2:asymptotic-causality}. Without time it is of course difficult to speak of causality, but what we have proven here is that the influence of the non-local modification decreases rapidly with distance.

\subsection{Coincidence limit}
\label{sec:ch2:coincidence}
Third, after discussing the local limit and the asymptotics $r/\ell\rightarrow \infty$, let us also take a look at the coincidence limit $r\rightarrow 0$. Inserting the transformation $\eta=z^2/r^2$ into Eq.~\eqref{eq:ch2:bessel-representation-1} one finds yet another representation of the static Green function in terms of the Bessel function \cite{Frolov:2015usa},
\begin{align}
\label{eq:ch2:bessel-representation-2}
\mathcal{G}_d(r) = \frac{1}{4\pi} \int\limits_0^\infty \dd\eta \frac{1}{f(-\eta)\eta} \left(\frac{\sqrt{\eta}}{2\pi r}\right)^{\frac{d}{2}-1} J_{\tfrac{d}{2}-1}(\sqrt{\eta}r) \, .
\end{align}
Using again the expansion for small arguments $r\rightarrow 0$ one finds
\begin{align}
\label{eq:ch2:coincidence-limit}
\mathcal{G}_d(r\rightarrow 0) &\approx c_0 + c_2 r^2 + \mathcal{O}\left(r^4\right) \, , \\
c_0 &= \frac{1}{(4\pi)^{d/2}\Gamma\left(\tfrac{d}{2}\right)} \int\limits_0^\infty \dd \eta \, \eta^{\tfrac{d}{2}-2} \frac{1}{f(-\eta)} \, , \\
c_2 &= -\frac{1}{4(4\pi)^{d/2}\Gamma\left(\tfrac{d}{2}+1\right)} \int\limits_0^\infty \dd \eta \, \eta^{\tfrac{d}{2}-1} \frac{1}{f(-\eta)} \, .
\end{align}
It is clear that in the local theory when $f\equiv 1$ both $c_0$ and $c_2$ diverge, explaining the standard divergence of fields at the origin of a $\delta$-distributional source. Note, however, that all non-zero $f$ that also have sufficiently decreasing asymptotics for large arguments regularize this integral and hence regularize the physical fields at the origin!

Let us focus again on the $\mathrm{GF_N}$ case when
\begin{align}
f(-\eta) = \exp\left[(\eta\ell^2)^N\right] \, .
\end{align}
Then one obtains, for any $N \ge 1$ and $d \ge 3$, the following finite coefficients:
\begin{align}
c_0 = \frac{\Gamma\left(\tfrac{d-2}{2N}\right)}{(4\pi)^{d/2}N\Gamma\left(\tfrac{d}{2}\right)\ell^{d-2}} \, , \qquad c_2 = -\frac{\Gamma\left(\tfrac{d}{2N}\right)}{4(4\pi)^{d/2}N\Gamma\left(\tfrac{d}{2}+1\right)\ell^d} \, .
\end{align}
This constitutes a fairly general proof of the regularity of the static Green functions at the origin for any $\mathrm{GF_N}$ theory. It is clear that similar properties hold for other non-local theories with similar asymptotic in the function $f$.

Even though it is not the main focus of this thesis we would like to remark that certain higher-derivative theories, where, for example, $f(\lap) = 1 + (-\ell^2\lap)^N$, share this property. However, in this case the regularity only holds if one chooses a suitably high $N$ given a dimension $d$ due to the merely polynomial falloff behavior.

Last, we would like to comment on the absence of the linear $c_1$-term in the $r$-expansion \eqref{eq:ch2:coincidence-limit}: it implies that the first derivative is differentiable at $r=0$, which is important because $r$ is a radius variable; compare also the considerations in Ref.~\cite{Frolov:2015bta} in this context.

\subsection{Explicit expressions}
\label{sec:ch2:explicit}
\begin{figure}[!hbt]
  \centering
  \includegraphics[width=\textwidth]{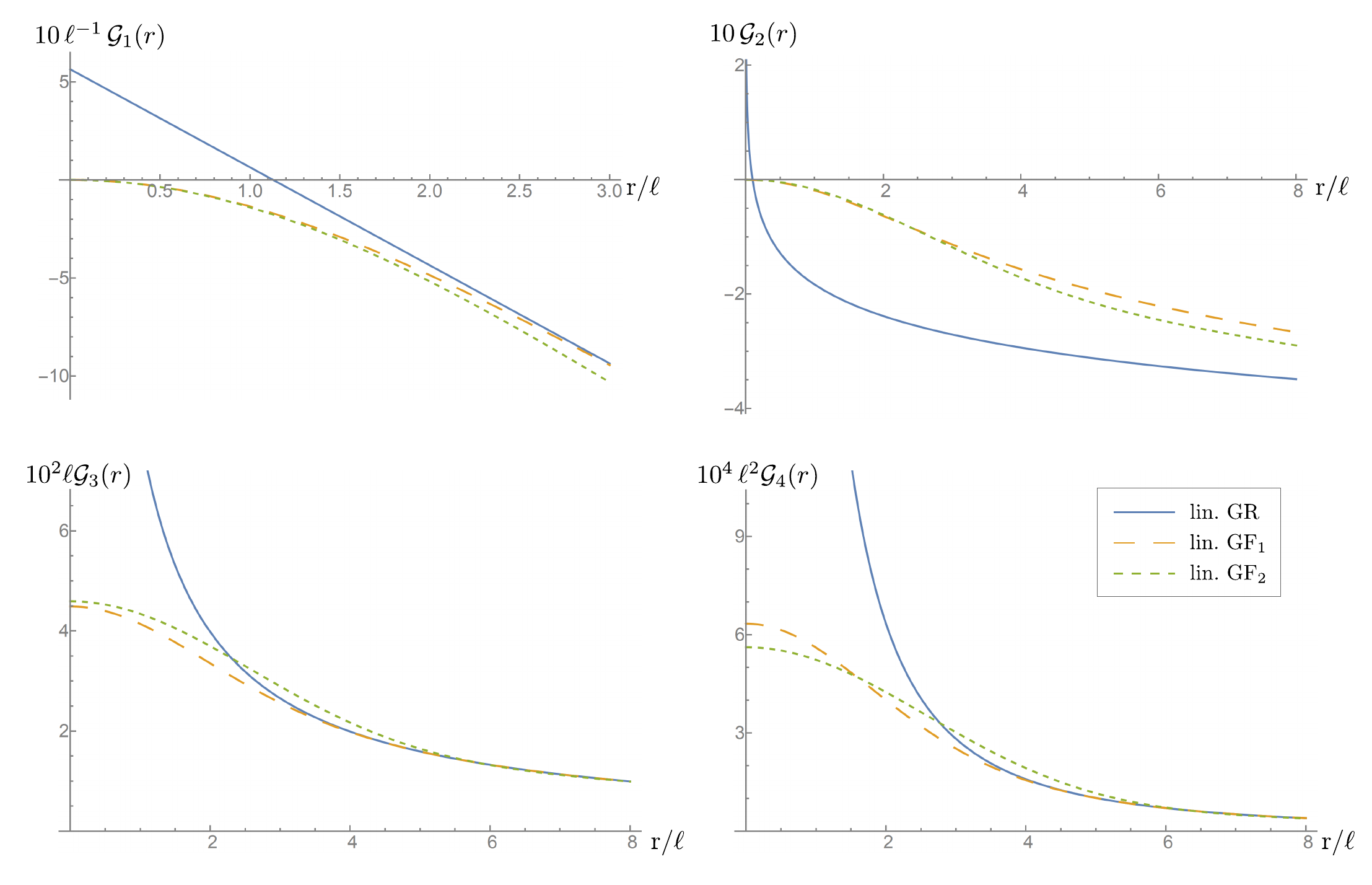}
  \caption[Local and non-local static Green functions.]{We plot the dimensionless Green functions $\ell^{d-2}\,\mathcal{G}_d(r)$ for the local case of linearized General Relativity as well as for the non-local cases of linearized  $\mathrm{GF_1}$ and $\mathrm{GF_2}$ theory for the dimensions $d=1,2,3,4$. At scales much larger than the scale of non-locality, $r/\ell \gg 1$, all three functions agree. At short distances, $r/\ell \ll 1$, they differ quite strongly: while the local Green functions are singular at $r=0$ for $d>1$, all non-local Green functions are finite and well-behaved.}
  \label{fig:ch2:visualization-gf1-gf2}
\end{figure}

It will be useful for the rest of this thesis to have exact expressions for these static Green functions at our disposal. In this subsection, for notational compactness, let us denote the static Green functions of $\mathrm{GF_N}$ theory in $d$-dimensional Euclidean space as $\mathcal{G}^N_d(r)$, and their local counterparts as $G_d(r)$. We find the following expressions:
\begin{align}
G{}_1(r) &= -\frac{r}{2} \, , \\
\mathcal{G}_1^1(r) &= -\frac r2 \text{erf}\left(\frac{r}{2\ell}\right) - \ell\frac{\exp{\left[-r^2/(4\ell^2)\right]} - 1}{\sqrt{\pi}} \, , \\
\mathcal{G}_1^2(r) &= -\frac{\ell}{\pi} \Big\{ \hspace{6pt} 2\Gamma(\tfrac14) y\, {}_1\!F\!{}_3\left( \tfrac14;~ \tfrac34,\tfrac54,\tfrac32;~ y^2 \right) + \Gamma(\tfrac34) \Big[ {}_1\!F\!{}_3\left( -\tfrac14;~ \tfrac14,\tfrac12,\tfrac34;~ y^2 \right) - 1 \Big] \Big\} \, \\
G{}_2(r) &= -\frac{1}{2\pi}\log\left(\frac{r}{r_0}\right) \, , \\
\mathcal{G}_2^1(r) &= \frac{1}{4\pi} \left[ \text{Ei}\left( -\frac{r^2}{4\ell^2} \right) - 2\log \left( \frac{r}{r_0} \right) \right] \, , \\
\mathcal{G}_2^2(r) &= -\frac{y}{2\pi} \Big[ \hspace{4pt} \sqrt{\pi}\, {}_1\!F\!{}_3\left(\tfrac12;~1,\tfrac32,\tfrac32;~y^2\right) - y\, {}_2 \!F\!{}_4\left(1,1;~\tfrac32,\tfrac32,2,2;~ y^2 \right) \Big] \, , \\
G_3(r) &= \frac{1}{4\pi r} \, , \\
\mathcal{G}_3^1(r) &= \frac{\text{erf}[r/(2\ell)]}{4\pi r} \, , \\
\mathcal{G}_3^2(r) &= \frac{1}{6\pi^2\ell}\Big[ 3 \Gamma\!\left(\tfrac54\right) {}_1\!F\!{}_3\left( \tfrac14;~ \tfrac12,\tfrac34,\tfrac54;~ y^2 \right) - 2y\Gamma\!\left(\tfrac34\right) {}_1\!F\!{}_3\left( \tfrac34;~ \tfrac54, \tfrac32, \tfrac74;~ y^2 \right) \Big] \, , \\
G_4(r) &= \frac{1}{4\pi^2 r^2} \, , \\
\mathcal{G}_4^1(r) &= \frac{1 - \exp\left[-r^2/(4\ell^2)\right]}{4\pi^2 r^2} \, , \\
\mathcal{G}_4^2(r) &= \frac{1}{64\pi^2 y\ell^2}\Big[ 1 - {}_0\!F\!{}_2\left( \tfrac12,\tfrac12;~ y^2 \right) + 2\sqrt{\pi} y\, {}_0\!F\!{}_2\left( 1, \tfrac32;~ y^2 \right) \Big] \, ,
\end{align}
where we used the abbreviation $y =( r/4\ell)^2$, $\gamma = 0.577\dots$ is the Euler--Mascheroni constant,\footnote{At the time of writing this thesis the author is not sure whether there already exists a Don Page mnemonic for this quantity to a precision of at least six digits. Addendum: It is possible to approximate $\gamma\approx 228/395$ to one part in 1,200,000 while simultaneously endowing it with a suitable mnemonic \cite{Page:2020a}.} and ${}_a F_{b}$ denotes the hypergeometric function \cite{Olver:2010}. Moreover, $\text{erf}(x)$ is the error function, and $\text{Ei}(x)$ denotes the exponential integral,
\begin{align}
\text{erf}(x) = \frac{2}{\sqrt{\pi}}\int\limits_0^x \dd z e^{-z^2} \, , \qquad
\text{Ei}(-x) = -E_1(x) = -\int\limits_x^\infty \dd z \frac{e^{-z}}{z} \quad \text{for~~} x > 0 \, .
\end{align}
Let us also comment that in the cases $N=1$ and $N=2$ it is possible to give the static Green functions $\mathcal{G}^N_d(r)$ in a closed form \cite{Frolov:2015usa} for $d\ge3$,
\begin{align}
\mathcal{G}^1_d(r) &= \frac{\gamma\left(\frac d2 - 1, \tfrac{r^2}{4\ell^2}\right)}{4\pi^{d/2}r^{d-2}} \, , \\
\mathcal{G}^2_d(r) &= \frac{2^{2 - \tfrac{3d}{2}}\pi^{\tfrac{1-d}{2}}}{d(d-2)\ell^{d-2}} \Big[ \frac{d}{\Gamma\left(\frac d4\right)} {}_1\!F\!{}_3\left( \tfrac d4 - \tfrac 12;~ \tfrac 12, \tfrac d4 , \tfrac d4 + \tfrac 12;~ y^2 \right) -\frac{2(d-2) y}{\Gamma\left(\tfrac d4 + \tfrac 12 \right)} {}_1\!F\!{}_3\left( \tfrac d4;~ \tfrac 32, \tfrac d4 + 1, \tfrac d4 + \tfrac 12;~ y^2 \right) \Big] \, , \nonumber
\end{align}
where $\gamma(s,x)$ denotes the lower incomplete gamma function \cite{Olver:2010},
\begin{align}
\gamma(s,x) &:= \int\limits_0^x z^{s-1} e^{-z} \dd z \, .
\end{align}
We chose to display these quantities in the middle of the thesis over banishing them to the appendix since they play a central role for many of the forthcoming results to be presented in the following chapters. We display the dimensionless Green functions $\ell^{d-2}\mathcal{G}_d(r)$ as a function of dimensionless distance $r/\ell$ in Fig.~\ref{fig:ch2:visualization-gf1-gf2}, showing that they are manifestly regular at $r=0$ and coincide with the local expressions for $r\gg\ell$.

\subsection{Heat kernel representation of static Green functions}
\label{sec:ch2:heat-kernel}
Last, we would like to mention a useful representation of non-local static Green functions in terms of the heat kernel in imaginary time. This heat kernel satisfies
\begin{align}
\label{eq:ch2:heat-equations}
\lap K_d (\ts{x}|\tau) = -i\partial_\tau K_d(\ts{x}|\tau) \, , \quad \lim\limits_{\tau\rightarrow 0} K_d(\ts{x}|\tau) = \delta^{(d)}(\ts{x}) \, , \quad \lim\limits_{\tau\rightarrow\pm\infty} K_d(\ts{x}|\tau) = 0 \, .
\end{align}
The explicit representation in $d$ spatial dimensions is
\begin{align}
K_d(\ts{x}|\tau) = \frac{1}{(4\pi i \tau)^{d/2}} \exp\left(\frac{i \ts{x}^2}{4\tau}\right) \, .
\end{align}
We will now prove that one can write the static $d$-dimensional Green function as
\begin{align}
\mathcal{G}_d(r) = \frac{1}{2\pi} \int\limits_{-\infty}^\infty \frac{\dd\eta}{f(-\eta\ell^2)\eta} \int\limits_{-\infty}^\infty \dd\tau K_d(\ts{r}|\tau) e^{i\eta\tau} \, .
\end{align}
To prove this result, let us define an object $\mathcal{K}_d(\ts{x}|\tau)$ as a solution of
\begin{align}
\label{eq:ch2:ktilde}
f(\lap) \mathcal{K}_d(\ts{x}|\tau) = iK_d(\ts{x}|\tau) \, .
\end{align}
Then the Green function $\mathcal{G}_d(\ts{x})$ can be written as
\begin{align}
\mathcal{G}_d(\ts{x}) = \int\limits_0^\infty \dd\tau \mathcal{K}_d(\ts{x}|\tau) \, .
\end{align}
To see this, simply insert the above representation into the definition of the non-local Green function \eqref{eq:ch2:static-green-function}, make use of \eqref{eq:ch2:heat-equations}, integrate by parts, and again use \eqref{eq:ch2:heat-equations}. Let us now introduce the imaginary-time Fourier transform of the object $\mathcal{K}_d(\ts{x}|\tau)$ such that
\begin{align}
\widetilde{\mathcal{K}}_d(\ts{x}|\omega) = \int\limits_{-\infty}^\infty \dd\tau e^{i\omega\tau}\mathcal{K}_d(\ts{x}|\tau) \, , \qquad
\mathcal{K}_d(\ts{x}|\tau) = \int\limits_{-\infty}^\infty \frac{\dd\omega}{2\pi} e^{-i\omega\tau} \widetilde{\mathcal{K}}(\ts{x}|\omega) \, .
\end{align}
Then we can write
\begin{align}
\mathcal{G}_d(\ts{x}) &= \int\limits_0^\infty \dd\tau \int\limits_{-\infty}^\infty \frac{\dd\omega}{2\pi} \int\limits_{-\infty}^\infty \dd\tau' e^{-i\omega(\tau-\tau')} \mathcal{K}_d(\ts{x}|\tau') \\
&= \int\limits_0^\infty \dd\tau \int\limits_{-\infty}^\infty \frac{\dd\omega}{2\pi} \int\limits_{-\infty}^\infty \dd\tau' e^{-i\omega(\tau-\tau')} \frac{i}{f(\lap)}K_d(\ts{x}|\tau') \\
&= \int\limits_0^\infty \dd\tau \int\limits_{-\infty}^\infty \frac{\dd\omega}{2\pi} \int\limits_{-\infty}^\infty \dd\tau' e^{-i\omega(\tau-\tau')} \frac{i}{f(-i\partial_\tau)}K_d(\ts{x}|\tau') \\
&= \int\limits_0^\infty \dd\tau \int\limits_{-\infty}^\infty \frac{\dd\omega}{2\pi} \int\limits_{-\infty}^\infty \dd\tau' e^{-i\omega(\tau-\tau')} \frac{i}{f(-\omega)}K_d(\ts{x}|\tau') \, .
\end{align}
First we used the definition \eqref{eq:ch2:ktilde}, then made use of the heat equations \eqref{eq:ch2:heat-equations}, and integrated by parts. The boundary contributions vanish due to \eqref{eq:ch2:heat-equations}. The remaining integral over $\tau$ can be regulated via
\begin{align}
\int\limits_0^\infty \dd\tau e^{-i\omega\tau} \equiv \lim\limits_{\epsilon\rightarrow 0}\int\limits_0^\infty \dd\tau e^{-i(\omega-i\epsilon)\tau} = \frac{-i}{\omega} \, .
\end{align}
We then finally obtain the desired representation
\begin{align}
\mathcal{G}_d(\ts{x}) &= \int\limits_{-\infty}^\infty \frac{\dd\omega}{2\pi} \int\limits_{-\infty}^\infty \dd\tau' e^{i\omega\tau'} \frac{1}{\omega f(-\omega)}K_d(\ts{x}|\tau')\, .
\end{align}
As we will see later, this representation is very useful because it isolates the spatial dependence into the $d$-dimensional heat kernel, which has a Gaussian form. For this reason, all spatial coordinates appear as symmetric factors.

\section{Concluding remarks}
In this chapter we have developed the idea of non-local Green functions, both in spacetime and in the Fourier domain. In general, non-local Green functions violate causality inside a non-local hyperbolic region, and it is useful to define the non-local modification of a Green function as the difference between the non-local version and the local version. We have shown that this difference is regular in the Fourier domain and does not introduce any new poles. For this reason the global causal properties of the Green functions are predicted by their local part alone, and we demonstrate explicitly that ghost-free non-local Green functions obey DeWitt's asymptotic causality condition.

While the region where non-locality becomes important is indeed compact in purely timelike and spacelike directions, it may extend infinitely far along the null directions. For this reason it is instructive to evaluate the non-local modification term on the light cone, and we were able to show in some examples that its value on the light cone is proportional to the mass parameter of the theory under consideration.

If the situation is time-independent, the considerations simplify drastically. The effects of non-locality become confined to a compact region in space alone, and it is possible to construct corresponding spatial Green functions explicitly in a wide range of theories. They have two important properties. First, they are finite and regular in the coincidence limit, unlike in the local case, where they diverge. And second, they asymptotically coincide with their local counterparts, which guarantees that non-local modifications tend to decrease with spatial distance.

%%%%%%%%%%%%%%%%%%%%%%%%%%%%%%%%%%%%%%%%%%%%%%%%%%%%%%%%%%%%%%%%%%%%%%%%%%%%%%%%%%%%%%%%%%%%%%%%%%%
%
% Chapter: Static & stationary solutions in weak-field gravity
%
\chapter{Static and stationary solutions in weak-field gravity}
\label{ch:ch3}
\textit{Using the previous results on static Green functions we will demonstrate in this chapter how we can utilize them to construct the gravitational field of static and stationary mass distributions in the weak-field limit of non-local ghost-free gravity. This chapter is based on Refs.~\cite{Boos:2018bxf,Boos:2018bhd,Boos:2020kgj,Boos:2020ccj}.}

\section{Weak-field limit of non-local gravity}
\label{sec:ch3:weak-field}

We are interested in the dynamics of weak-field non-local gravity around a flat Minkowski background. Considering metrics of the form $g{}_{\mu\nu} = \eta{}_{\mu\nu} + \epsilon h{}_{\mu\nu}$, where $\eta{}_{\mu\nu}$ is the Minkowski metric in Cartesian coordinates, $h{}_{\mu\nu}$ describes the weak gravitational field, and $\epsilon \ll 1$, it is sufficient to consider an action that is at most quadratic in the perturbation and hence of $\mathcal{O}(\epsilon^2)$. This guarantees that after the variation the field equations are of linear order in the perturbation $h{}_{\mu\nu}$.

A suitable parity-even and torsion-free action for non-local infinite-derivative gravity is at most quadratic in curvature \cite{Biswas:2011ar} and can be written in terms of a non-local operator $\mathcal{O}^{\mu\nu\rho\sigma}_{\alpha\beta\gamma\delta}$ in the form
\begin{align}
\label{eq:ch3:action-idg}
S[g{}_{\mu\nu}] = \frac{1}{2\kappa} \int \sqrt{-g} \dd^D x \left( R + \frac12 R{}_{\mu\nu\rho\sigma} \mathcal{O}^{\mu\nu\rho\sigma}_{\alpha\beta\gamma\delta} R{}^{\alpha\beta\gamma\delta} \right) \, .
\end{align}
As is well known, see \cite{Stephani:2003}, there only exist three quadratic curvature invariants proportional to $C{}_{\mu\nu\rho\sigma}C{}^{\mu\nu\rho\sigma}$, $R{}_{\mu\nu}R{}^{\mu\nu}$ and $R^2$, where $C{}_{\mu\nu\rho\sigma}$ is the Weyl tensor, $R{}_{\mu\nu}$ is the Ricci tensor, and $R$ is the Ricci scalar, in which case the operator $\mathcal{O}^{\mu\nu\rho\sigma}_{\alpha\beta\gamma\delta}$ can be constructed purely from the metric $g{}_{\mu\nu}$ and the Kronecker symbol $\delta{}^\mu_\nu$. In non-local theories, however, the operator may also include an arbitrary amount of covariant derivatives, either as ``free indices'' $\nabla{}_\mu$ or contracted into covariant d'Alembert operators $g{}^{\mu\nu}\nabla{}_\mu\nabla{}_\nu$. For studies including non-vanishing torsion we refer to Ref.~\cite{delaCruz-Dombriz:2018aal}.

It is important to stress that not all possible derivative terms are physically relevant at leading order (that is, quadratic in $\epsilon$ at the level of the action, or linear in $\epsilon$ at the level of the field equations). For example, is it possible to commute two covariant derivatives between the two curvature tensors, since the correction term is proportional to the Riemann curvature (or a contraction thereof), which generates a higher-order term in curvature that leaves the linearized field equations invariant. Moreover one may always employ the Bianchi identities $R{}_{\mu\nu[\rho\sigma;\lambda]} = 0$.

For a detailed consideration of all possible quadratic terms we refer to Ref.~\cite{Biswas:2011ar}, where it is proven that the operator $\mathcal{O}^{\mu\nu\rho\sigma}_{\alpha\beta\gamma\delta}$ can be written in terms of covariant derivatives and six independent functions of the d'Alembert operator, which, after applying the aforementioned commutator of covariant derivatives, reduces the amount of free functions to only three.

During the derivations it is helpful to keep the covariant formalism for a general metric $g{}_{\mu\nu}$ and only at the end, when the action is at the final form, expand to quadratic order in $\epsilon$ using $g{}_{\mu\nu} = \eta{}_{\mu\nu} + \epsilon h{}_{\mu\nu}$. Then, one can demonstrate that the number of independent functions reduces even more to only two \cite{Frolov:2015usa}, and the final weak-field action takes the form \cite{Biswas:2011ar}
\begin{align}
\label{eq:ch3:action}
S\left[h{}_{\mu\nu}\right] &= \frac{1}{2\kappa} \int \dd^D x \Big( \hspace{18pt} \frac12 h^{\mu\nu}\,a(\Box)\Box\,h_{\mu\nu}-h^{\mu\nu}\,a(\Box)\partial_{\mu}\partial_{\alpha}\,h^{\alpha}{}_{\nu} + h^{\mu\nu}\, c(\Box)\partial_{\mu}\partial_{\nu} h - \frac12 h\,c(\Box)\Box h \nonumber \\
&\hspace{79pt} + \frac12 h^{\mu\nu}\,\frac{a(\Box)-c(\Box)}{\Box}\partial_{\mu}\partial_{\nu}\partial_{\alpha}\partial_{\beta}\,h^{\alpha\beta}\Big) \, ,
\end{align}
where from now on we shall drop the coefficient $\epsilon$ and assume $h{}_{\mu\nu}\ll 1$. This action may be seen as a non-local generalization of the quadratic action proposed by van Nieuwenhuizen\cite{VanNieuwenhuizen:1973fi}. The two expressions $a(\Box)$ and $c(\Box)$ are non-zero functions of the d'Alembert operator (``form factors'') that satisfy
\begin{align}
a(0) = c(0) = 1 \, .
\end{align}
 As we have already seen in Ch.~\ref{ch:ch2}, and will see in more detail in this chapter as well, this constraint guarantees that one recovers linearized General Relativity at large distances. The resulting field equations are
\begin{align}
\begin{split}
\label{eq:ch3:eom}
a(\Box)\big[\Box\,h_{\mu\nu}-\partial_{\sigma}\big(\partial_{\nu}\,h_{\mu}{}^{\sigma} +\partial_{\mu}h_{\nu}{}^{\sigma}\big)\big] &+ c(\Box)\big[\eta_{\mu\nu}\big(\partial_{\rho}\partial_{\sigma}h^{\rho\sigma}-\Box h\big)+\partial_{\mu}\partial_{\nu}h \big]\\
& +\frac{a(\Box)-c(\Box)}{\Box}\partial_{\mu}\partial_{\nu}\partial_{\rho}\partial_{\sigma}h^{\rho\sigma}=-2\kappa  T_{\mu\nu} \, ,
\end{split}
\end{align}
where $T{}_{\mu\nu}$ is the energy-momentum tensor. Note that the energy-momentum tensor conservation, $\partial{}^\mu T{}_{\mu\nu} = 0$, is consistent with the field equations, as it should be. In the time-independent case one obtains
\begin{align}
\begin{split}
\label{eq:ch3:eom-stationary}
a(\lap)\big[\lap\,h_{\mu\nu}-\partial_{\sigma}\big(\partial_{\nu}\,h_{\mu}{}^{\sigma} +\partial_{\mu}h_{\nu}{}^{\sigma}\big)\big] &+ c(\lap)\big[\eta_{\mu\nu}\big(\partial_{\rho}\partial_{\sigma}h^{\rho\sigma}-\lap h\big)+\partial_{\mu}\partial_{\nu}h \big]\\
& +\frac{a(\lap)-c(\lap)}{\lap}\partial_{\mu}\partial_{\nu}\partial_{\rho}\partial_{\sigma}h^{\rho\sigma}=-2\kappa  T_{\mu\nu} \, ,
\end{split}
\end{align}
where we replaced $\Box\rightarrow\lap$, and we shall employ this equation in the following sections in order to find static or stationary solutions.

\subsection{Gauge freedom}
The action of non-local ghost-free gravity is generally covariant because the non-local modification enters via a scalar form factor containing the d'Alembert operator ``$\Box$'' sandwiched between curvature tensors. After the linearization this general covariance becomes a gauge freedom in $h{}_{\mu\nu}$ that takes the form (see also the related discussion in DeWitt \cite{DeWittChristensen:2011})
\begin{align}
\begin{split}
\label{eq:ch3:hmunu-gauge}
h{}_{\mu\nu} \rightarrow h{}_{\mu\nu}' &= h{}_{\mu\nu} + \delta h{}_{\mu\nu} \, , \quad \delta h{}_{\mu\nu} = \partial{}_\mu \epsilon{}_\nu + \partial{}_\nu\epsilon{}_\mu \, , \\
h \rightarrow h' &= h + \delta h \, , \hspace{43pt} \delta h = 2 \partial{}_\mu \epsilon{}^\mu \, .
\end{split}
\end{align}
Here, $\epsilon{}^\mu$ is a vector field that parametrizes an infinitesimal diffeomorphism. The linearized Riemann tensor as the ``field strength tensor'' is invariant under this transformation,
\begin{align}
\delta R{}_{\mu\nu\rho\sigma} &= \partial{}_\nu \partial{}_{[\rho} \delta h{}_{\sigma]\mu} - \partial{}_\mu \partial{} _{[\rho}\delta h{}_{\sigma]\nu} = 0 \, .
\end{align}
One may also check that the transformation \eqref{eq:ch3:hmunu-gauge} leaves the left-hand side of the field equations \eqref{eq:ch3:eom} invariant, which is another reason we may think of $\epsilon{}^\mu$ as a gauge parameter.

Sometimes, however, it is useful to fix the gauge in order to simplify the field equations, which in turn have then lost their gauge invariance. For this reason, gauge-fixing conditions manifestly break gauge invariance. In the present context, let us consider the expression
\begin{align}
X{}_\nu(\lambda) = \partial{}^\mu h{}_{\mu\nu} - \lambda \partial{}_\nu h \, , \quad \lambda \in \mathbb{R} \, .
\end{align}
Let us now demand that we perform a gauge transformation that sets this expression to zero,
\begin{align}
\label{eq:ch3:gauge-fixing}
X{}_\nu(\lambda) \rightarrow X{}_\nu'(\lambda) = \partial{}^\mu h{}_{\mu\nu} - \lambda \partial{}_\nu h + \Box \epsilon{}_\nu + (1-2\lambda)\partial{}_\nu \partial{}^\mu \epsilon{}_\mu = 0 \, .
\end{align}
We may interpret this as a differential equation for $\epsilon{}^\mu$, where the $h$-dependent terms appear as sources. Since we are interested in perturbations around flat Minkowski spacetime we may utilize the Fourier representations
\begin{align}
\tilde{\epsilon}_\mu(\ts{k}) = \int \dd^D x e^{-i\ts{k}\cdot\ts{x}} \epsilon_\mu(\ts{x}) \, , \quad
\tilde{j}_\nu(\ts{k}) = \int \dd^D x e^{-i\ts{k}\cdot\ts{x}} \left( \partial{}^\mu h{}_{\mu\nu} - \lambda \partial{}_\nu h \right) \, ,
\end{align}
such that the gauge fixing condition \eqref{eq:ch3:gauge-fixing} becomes an algebraic relation in momentum space,
\begin{align}
\left[ \ts{k}^2\eta{}_{\mu\nu} + (1-2\lambda)k{}_\mu k{}_\nu \right]\tilde{\epsilon}{}^\nu(\ts{k}) = \tilde{j}_\mu(\ts{k}) \quad \overset{\lambda\not=1}{\Rightarrow} \quad \tilde{\epsilon}^\mu(\ts{k}) = \frac{1}{\ts{k}^2}\left[ \eta{}^{\mu\nu} - \frac{1-2\lambda}{2(1-\lambda)} \frac{k{}^\mu k{}^\nu}{\ts{k}^2} \right] \tilde{j}_\nu(\ts{k}) \, .
\end{align}
This expression is well-defined for $\lambda\not=1$. Suppose, for the sake of the argument, that this Fourier transform exists and we can hence find an $\epsilon{}^\mu$ that solves Eq.~\eqref{eq:ch3:gauge-fixing}, given a choice $\lambda\not=1$ and a field configuration $h{}_{\mu\nu}$. Then, the field equations \eqref{eq:ch3:eom} become
\begin{align}
a(\Box)\Box h{}_{\mu\nu} - (1-\lambda)\eta{}_{\mu\nu} c(\Box)\Box h + \Big[ (1-\lambda)c(\Box) - \lambda a(\Box)\Big] \partial{}_\mu \partial{}_\nu h ~\overset{*}{=}\, -2\kappa T{}_{\mu\nu} \, ,
\end{align}
where we have denoted the equality as ``$\overset{*}{=}$'' since it only holds true in the specific $\lambda$-gauge. Let us now introduce
\begin{align}
\hat{h}{}_{\mu\nu} = h{}_{\mu\nu} - \lambda h \eta{}_{\mu\nu} \, , \quad h{}_{\mu\nu} = \hat{h}{}_{\mu\nu} - \frac{\lambda}{\lambda D-1} \hat{h}\eta{}_{\mu\nu} \, , \quad h = \frac{1}{1-\lambda D} \hat{h} \, ,
\end{align}
such that $\partial{}^\mu \hat{h}{}_{\mu\nu} \overset{*}{=} 0$. Then, the field equations take the simplified form
\begin{align}
a(\Box)\Box\hat{h}{}_{\mu\nu} + \frac{(1-\lambda)c(\Box)-\lambda a(\Box)}{\lambda D - 1}\Big( \eta{}_{\mu\nu}\Box - \partial{}_\mu\partial{}_\nu \Big) \hat{h} = -2\kappa T{}_{\mu\nu} \, .
\end{align}
In the case of $a(\Box) = c(\Box)$ they simplify even further and become
\begin{align}
a(\Box)\left[ \Box\hat{h}{}_{\mu\nu} + \frac{1-2\lambda}{\lambda D - 1}\Big( \eta{}_{\mu\nu}\Box - \partial{}_\mu\partial{}_\nu \Big) \hat{h} \right] = -2\kappa T{}_{\mu\nu} \, .
\end{align}
From these considerations, purely at the level of the field equations, it becomes clear that the choice $\lambda=\tfrac12$ is preferable. The differential equation for the suitable $\epsilon{}^\mu$ then also simplifies:
\begin{align}
\label{eq:ch3-dedonder}
X{}_\nu\left(\tfrac12\right) = \partial{}^\mu h{}_{\mu\nu} - \frac12 \partial{}_\nu h + \Box \epsilon{}_\nu = 0 \, .
\end{align}
In fact, this partial differential equation is formally equivalent to the Maxwell equations for a ``vector potential'' $\epsilon{}_\mu$ in the ``Lorenz gauge'' $\partial{}^\mu \epsilon{}_\mu=0$ for the current $j{}_\nu \sim \partial{}^\mu h{}_{\mu\nu} - \frac12 \partial{}_\nu h$. This current, by means of the Lorenz condition, is conserved. In other words, we can be confident that for any given $h{}_{\mu\nu}$ we can find a function $\epsilon{}^\mu$ that guarantees Eq.~\eqref{eq:ch3-dedonder}, which is sometimes also referred to as a ``De\,Donder gauge condition.''

The final equations, in the case $a(\Box)=c(\Box)$ then take the form
\begin{align}
\label{eq:ch3-eom-hat-a=c}
a(\Box) \Box\hat{h}{}_{\mu\nu} = -2\kappa T{}_{\mu\nu} \, ,
\end{align}
where the energy-momentum conservation emerges as a consequence of the gauge choice $\partial{}^\mu \hat{h}{}_{\mu\nu} = 0$. This representation has the advantage that the solution can be given immediately as
\begin{align}
\hat{h}{}_{\mu\nu}(\ts{x}) = \hat{h}{}^0_{\mu\nu}(\ts{x}) + 2\kappa\int\dd^d y \, \mathcal{G}_d(\ts{x}-\ts{y}) T{}_{\mu\nu}(\ts{y}) \, ,
\end{align}
provided the situation allows for a scalar Green function acting trivially on the tensorial structure, and where $\hat{h}{}^0_{\mu\nu}$ is the homogeneous solution consistent with the imposed boundary conditions.

\section{Staticity and stationarity}

In this chapter we will be interested purely in static and stationary situations. A geometry $g{}_{\mu\nu}$ is said to be \emph{stationary}\index{stationarity!definition of} if it possesses a timelike Killing vector $\ts{\xi}$ such that \cite{Poisson:2004,Frolov:2011}
\begin{align}
\mathcal{L}_\ts{\xi} g{}_{\mu\nu} = \xi{}^\rho\partial{}_\rho g{}_{\mu\nu} + (\partial{}_\mu \xi{}^\rho)g{}_{\rho\nu} + (\partial{}_\nu \xi{}^\rho)g{}_{\mu\rho} = 0 \, .
\end{align}
In the present case we are interested in the special geometry $g{}_{\mu\nu} = \eta{}_{\mu\nu} + h{}_{\mu\nu}$, where $h{}_{\mu\nu}$ is a gravitational perturbation and $\eta{}_{\mu\nu}$ is the Minkowski metric in Cartesian coordinates $\{t,\ts{x}\}$. Let us parametrize the timelike Killing vector as $\ts{\xi} = \ts{\xi}_0 + \ts{\zeta}$, where $\ts{\xi}_0 = \partial_t$ is the timelike Killing vector of Minkowski space, and $\ts{\zeta}$ describes a possible deviation due to the gravitational perturbation $h{}_{\mu\nu}$. Then, to linear order in the perturbations, we obtain
\begin{align}
\mathcal{L}_{\ts{\xi}} g{}_{\mu\nu} = \mathcal{L}_{\ts{\xi}_0} h{}_{\mu\nu} + \zeta_{(\mu,\nu)} \, .
\end{align}
Provided the gravitational perturbations satisfy $\mathcal{L}_{\ts{\xi}_0} h{}_{\mu\nu}=0$ we may then choose $\ts{\zeta}=0$, rendering $\ts{\xi}_0$ a Killing vector of the perturbed geometry as well. Inserting $\xi_0^\mu = \delta^\mu_t$ one obtains
\begin{align}
\mathcal{L}_{\ts{\xi}_0} h{}_{\mu\nu} = \partial_t h{}_{\mu\nu} = 0 \, .
\end{align}
For this reason any time-independent perturbation $h{}_{\mu\nu}$ yields a stationary geometry. If the timelike Killing vector is also hypersurface orthogonal, the geometry is \emph{static} \cite{Poisson:2004,Frolov:2011}:\index{staticity!definition of}
\begin{align}
\xi{}_{[\mu;\nu} \xi{}_{\rho]} = 0 \, ,
\end{align}
where the semicolon denotes the covariant derivative. Introducing the 1-form $\xi = \xi{}_\mu \dd x{}^\mu$, and using the fact that in Riemannian geometries with vanishing torsion\index{torsion!vanishing} the Levi--Civita connection is symmetric in a coordinate frame, $\Gamma{}^\mu{}_{\nu\rho} = \Gamma{}^\mu{}_{\rho\nu}$, the above condition is equivalent to
\begin{align}
\dd \xi \wedge \xi = 0 \, ,
\end{align}
where ``$\dd$'' denotes the exterior derivative\index{exterior derivative}. Let us now substitute the timelike Killing vector $\xi^\mu_0 = \delta^\mu_t$ such that $\xi = \xi_0 = \xi_0^\mu (\eta{}_{\mu\nu} + h{}_{\mu\nu}) \dd x{}^\nu$, and parametrize $h_{tt} = \phi$ as well as $A = h_{t\alpha}\dd x{}^\alpha = A_\alpha\dd x{}^\alpha$, where $\alpha$ is a spatial index. Then, to linear order, we find
\begin{align}
\dd \xi_0 \wedge \xi_0 = -\dd A \wedge \dd t \, .
\end{align}
This implies that geometries with $A=0$ are static to leading order in the perturbation.

\subsection{Overview of the rest of the chapter}
In what follows, we will discuss the gravitational field of point particles, extended brane-like objects, rotating objects in general, as well as rotating point particles, rotating strings, and rotating p-branes as specific examples. When deriving these solutions we will make heavy use of the previously derived Green functions, see Ch.~\ref{ch:ch2}, and will focus on three major aspects:
\begin{itemize}
\item[(i)] Regularity of the solutions.
\item[(ii)] Asymptotics of the solutions at large distances, and relation to linearized solutions obtained within General Relativity. In particular, how does the choice of the functions $a(\Box)$ and $c(\Box)$ influence the asymptotics?
\item[(iii)] Whenever possible, we will provide solutions that are valid in any number of spacetime dimensions.
\end{itemize}
Instead of starting with the most general ansatz (which, by linearity, is just a superposition of separate solutions), we shall discuss first the simplest case of an isolated point particle and discuss the resulting spherically symmetric and static gravitational field. Then, we shall move on to extended brane-like objects which are still static but feature a reduced spatial symmetry that includes rotational symmetry inside the brane and translational symmetry along the brane. The cosmic string is a special example of that. After that, we will move on to the stationary case and discuss rotating objects, first in some generality, and then we shall focus on the examples of the rotating point particle and the rotating cosmic string. And finally, we will provide the expressions for rotating $p$-branes in any number of dimensions.

\section{Gravitational sources}
\label{sec:ch3:sources}
Before addressing the gravitational field let us briefly discuss the matter sources of the gravitational field. In order to properly distinguish between spacetime indices and spatial indices, let us introduce the following notation (see also the list of symbols on p.~\pageref{ListOfSymbols}):
\begin{itemize}
\item Greek letters from the beginning of the alphabet ($\alpha,\beta,\gamma,\dots$) denote purely spatial indices.\\[-1.4\baselineskip]
\item $D$-dimensional Cartesian spacetime coordinates can be written as $x{}^\mu = (t,x{}^\alpha)$.\\[-1.4\baselineskip]
\item When the index structure is not important we will denote $x{}^\alpha = \ts{x}$ for brevity.
\end{itemize}
In this chapter, we will talk about the sources of a stationary gravitational field that have the energy-momentum tensor
\begin{align}
\label{eq:ch3:tmunu-stationary}
T{}_{\mu\nu} = \rho(\ts{x})\delta{}^t_\mu\delta{}^t_\nu + \delta{}^t_{(\mu} \delta{}^\alpha_{\nu)} \frac{\partial}{\partial x{}^\beta} j{}_\alpha{}^\beta(\ts{x}) \, .
\end{align}
The function $\rho(\ts{x})$ describes the matter density, and the antisymmetric tensor field $j{}_{\alpha\beta}(\ts{x}) = -j{}_{\beta\alpha}(\ts{x})$ parametrizes the angular momentum density. We shall assume that $j{}_{\alpha\beta}(\ts{x}) \rightarrow 0$ sufficiently fast for $|\ts{x}|\rightarrow\infty$, but leave $\rho(\ts{x})$ unrestricted. Moreover, as demanded by stationarity, we assume that both of these functions are independent of time. Then it is straightforward to check that this energy-momentum tensor is indeed conserved,
\begin{align}
\partial{}^\mu T{}_{\mu\nu} = \delta{}^t_\nu \left[ \frac{\partial \rho(\ts{x})}{\partial t} + \frac12 \frac{\partial^2}{\partial x{}^\alpha \partial x{}^\beta} j{}^{\alpha\beta}(\ts{x}) \right] + \frac12 \delta{}^\alpha_\nu \frac{\partial^2}{\partial x{}^\beta \partial t} j{}_\alpha{}^\beta(\ts{x})  = 0 \, .
\end{align}
The timelike component vanishes due to the time-independence of $\rho(\ts{x})$ and the antisymmetry of $j{}_{\alpha\beta}(\ts{x})$, and the spacelike component vanishes due to the time-independence of $j{}_{\alpha\beta}(\ts{x})$.

Let us now relate the components of the energy-momentum tensor to conserved charges consistent with the isometries of underlying spacetime. We denote the generators of spacetime translations and spatial rotations as
\begin{align}
\ts{\xi}_{(\mu)} = \xi{}^\nu_{(\mu)}\partial{}_\nu \,\overset{*}{=} \partial{}_\mu \, , \quad
\ts{\zeta}_{(\alpha\beta)} = \zeta{}^\nu_{(\alpha\beta)} \partial{}_\nu \,\overset{*}{=} x{}_\alpha \partial{}_\beta - x{}_\beta\partial{}_\alpha \, ,
\end{align}
where in the equality ``$\,\overset{*}{=}$'' we have inserted the real-space representation of the generators
\begin{align}
\xi{}^\nu_{(\mu)} = \delta{}^\nu_\mu \, , \quad \zeta{}^\nu_{(\alpha\beta)} = 2\delta{}^\nu_{[\beta} x{}_{\alpha]} \, .
\end{align}
Let us now fix an observer $\ts{u} = \partial_t$ and calculate the conserved charges associated to the spacetime isometries $\{\ts{\xi}_{(\mu)}, \ts{\zeta}_{(\alpha\beta)}\}$ on the congruence $\ts{u}$. They take the form
\begin{align}
P_\mu &= \int \dd^d x T_{\rho\nu} u{}^\rho \xi{}^\nu_{(\mu)} = \int \dd^d x T{}_{t\mu} \, , \\
J_{\alpha\beta} &= \int \dd^d x T{}_{\rho\nu} u{}^\rho \xi{}^\nu_{(\alpha\beta)} = \int \dd^d x \left( x{}_\alpha T_{t\beta} - x{}_\beta T_{t\alpha} \right)
\end{align}
For the energy-momentum tensor \eqref{eq:ch3:tmunu-stationary} one finds
\begin{align}
P{}_\mu &= \delta{}^t_\mu \int\dd^d x \rho(\ts{x}) \, , \\
J{}_{\alpha\beta} &= \int \dd^d x j_{\alpha\beta}(\ts{x}) + \frac12 \int\dd^d x \frac{\partial}{\partial{} x{}^\gamma} \left( x{}_\alpha j{}_\beta{}^\gamma - x{}_\beta j{}_\alpha{}^\gamma \right) = \int \dd^d x j_{\alpha\beta}(\ts{x}) \, .
\end{align}
In the second equality we have integrated by parts and in the third equality we made have assumed that $j{}_{\alpha\beta}(\ts{x}) \rightarrow 0$ sufficiently fast for $|\ts{x}|\rightarrow\infty$. Moreover, $P_t$ admits the physical interpretation as the total mass $m$ of the physical system. Last, the fact that $P_\alpha=0$ implies that we study the gravitational field in the center of mass frame.

\section{Point particles}
A point particle of mass $m$ in $d$ spatial dimensions has the energy-momentum tensor
\begin{align}
T{}_{\mu\nu} = m\delta{}^t_\mu \delta{}^t_\nu \delta{}^{(d)}(\ts{r}) \, ,
\end{align}
where $\ts{r}=(x^1,\dots,x^d)$ denotes the collection of spatial Cartesian coordinates in $d$ dimensions. In accordance with the spherical symmetry of the energy-momentum we make the following ansatz:
\begin{align}
\label{eq:ch3:particle-ansatz}
\dd s^2 = -[1+\phi(r)]\dd t^2 + [1+\psi(r)] \dd\ts{r}^2 \, , \quad \dd\ts{r}^2 = \sum\limits_{\alpha=1}^d (\dd x^\alpha)^2 \, , \quad r^2 = \sum\limits_{\alpha=1}^d (x^\alpha)^2 \, .
\end{align}
Inserting this metric into the field equations \eqref{eq:ch3:eom} and replacing $\Box\rightarrow\lap$ yields
\begin{align}
(c-a)\lap\phi + (d-1)c\lap\psi &= -2\kappa m\delta{}^{(d)}(\ts{r}) \, , \\
(\delta_{ij}\lap - \partial_i\partial_j) \left\{[a-(d-1)c]\psi - c\phi\right\} &= 0 \, ,
\end{align}
where we have suppressed the arguments of the functions for better readability. In order to solve the homogeneous equation, let us assume that
\begin{align}
c(\lap) = (1+\alpha)a(\lap) \, , \quad \alpha \not= -1 \, , \qquad \psi = \frac{1+\alpha}{1-(d-1)(1+\alpha)} \phi \, .
\end{align}
Then, the inhomogeneous equation becomes
\begin{align}
\label{eq:ch3:eom-inh-pp}
a(\lap)\lap \phi(r) = -2 \frac{d-2+\alpha(d-1)}{1-d(1+\alpha)}\kappa m \delta{}^{(d)}(\ts{r}) \, .
\end{align}
We recognize that this equation is solved by a static Green function $\mathcal{G}_d(r)$ solving in turn
\begin{align}
a(\lap)\lap\mathcal{G}_d(r) = -\delta{}^{(d)}(\ts{r}) \, ,
\end{align}
which we already discussed in Ch.~\ref{ch:ch2}. The gravitational field of a point particle is hence
\begin{align}
\label{eq:ch3:particle-solution}
\phi(r) = 2 \frac{d-2+\alpha(d-1)}{1-d(1+\alpha)}\kappa m \mathcal{G}_d(r) \, , \quad \psi(r) = 2 \frac{1+\alpha}{-1+d(1+\alpha)} \kappa m \mathcal{G}_d(r) \, .
\end{align}
From Sec.~\ref{sec:ch2:explicit} we know the explicit expressions for these Green functions for the choice
\begin{align}
a(\lap) = \exp[(-\ell^2\lap)^N] \, , \quad N=1,2 \, ,
\end{align}
which correspond to $\mathrm{GF_1}$ and $\mathrm{GF_2}$ theory, respectively.

\subsection{Regularity}
From the considerations presented in Sec.~\ref{sec:ch2:coincidence} it is clear that the resulting metric \eqref{eq:ch3:particle-ansatz} together with the solutions \eqref{eq:ch3:particle-solution} is completely regular at $r=0$, in stark contrast to the metric obtained in linearized General Relativity. This is the first of many examples where non-locality regularizes gravitational singularities at the linear level. Moreover, as we have shown in Sec.~\ref{sec:ch2:coincidence}, since the Green functions behave as
\begin{align}
\mathcal{G}_d(r\rightarrow 0) = c_0 + c_2 r^2 + \mathcal{O}(r^4) \, , \quad c_2 < 0 \, ,
\end{align}
there is also no conical singularity at $r=0$; rather, the behavior is similar to that of non-singular black hole metrics \cite{Frolov:2016pav,Simpson:2019mud} in what is also referred to as a ``de\,Sitter core.'' A more rigorous proof of the regularity of the metric \eqref{eq:ch3:particle-solution} involves the calculation of curvature invariants. We will postpone that until the next section, when we will discuss extended, brane-like objects.

\subsection{Asymptotics}
On perturbed Minkowski spacetime with a timelike Killing vector $\ts{\xi}=\partial_t$ one can define the Newtonian potential as
\begin{align}
\Phi_\text{N} = -\frac12\left( 1 + \ts{\xi}\cdot\ts{\xi} \right) = \frac{\phi}{2} \, ,
\end{align}
which for $d=3$ dimensions, for small values of $\alpha$, and in the local limit reduces to
\begin{align}
\Phi_\text{N}^{d=3} = -\frac{\kappa m}{2} \frac{1+2\alpha}{1+\frac{3}{2}\alpha} G_3(r) \approx -\frac{G m}{r} \left(1+\frac{\alpha}{2} \right) \, ,
\end{align}
which is only the correct Newtonian limit if $\alpha=0$, provided we do not want to change the value of the gravitational constant.

One possible interpretation of this fact is that $\alpha\not=0$ introduces new gravitational degrees of freedom, 
which can be made rigorous at the level of the propagator around Minkowski space, and we refer to Biswas \textit{et al.} \cite{Biswas:2011ar} for these calculations. That being said, from now on we shall set $\alpha=0$ in our future considerations.

\section{Friedel oscillations around point particles}
\label{sec:ch3:friedel}
Before moving on to extended objects we would like to study the properties of the gravitational field of point particles in a bit more detail. Utilizing the timelike Killing vector $\ts{\xi} = \partial_t$ we may construct an effective energy density by projecting out the purely timelike direction of the linearized field equations \eqref{eq:ch3-eom-hat-a=c}, giving rise to the following expression:
\begin{align}
\label{eq:ch3:energy-density}
\rho_\text{eff}(r) \equiv \kappa T{}^\text{eff}_{\mu\nu}\xi{}^\mu\xi{}^\nu \equiv \frac{1}{a(\lap)} \kappa T{}_{\mu\nu}\xi{}^\mu\xi{}^\nu = -\frac{\lap \phi}{2(d-2)} \, .
\end{align}
The interpretation of this quantity as an effective energy density is completely analogous to the considerations presented in Sec.~\ref{sec:ch1:on-shell-off-shell}. In fact, since the energy density of a point particle is proportional to the $d$-dimensional $\delta$-function, the effective energy density is just the rescaled integral kernel of the non-local ghost-free form factor,
\begin{align}
\rho_\text{eff}(r) = \frac{\kappa m}{a(\lap)} \delta^{(d)}(\ts{r}) \equiv \kappa m K_d(r) \, ,
\end{align}
which follows immediately from Eq.~\eqref{eq:ch3:eom-inh-pp}; compare also Sec.~\ref{sec:ch1:kernel}. With the explicit form of the solution $\phi(r)$ available via the Green functions from Sec.~\ref{sec:ch2:explicit}, it is straightforward to evaluate the above expressions using the identity
\begin{align}
\lap \phi(r) = \frac{1}{r^{d-1}} \partial_r \left[ r^{d-1} \partial_r \phi(r) \right] \, .
\end{align}
For simplicity we will focus on the four-dimensional case $d=3$ and the ghost-free theories $\mathrm{GF_1}$, $\mathrm{GF_2}$, and $\mathrm{GF_3}$.\footnote{We do not list the exact expressions for the relevant $\mathrm{GF_3}$ Green function here and refer instead to Ref.~\cite{Boos:2018bhd}.} Displayed in a logarithmic plot, see Fig.~\ref{fig:ch3:energy-density-gf-log}, it becomes apparent that the effective energy density oscillates in the cases of $\mathrm{GF_2}$ and $\mathrm{GF_3}$ theory, and does \emph{not} fluctuate in the case of $\mathrm{GF_1}$ theory. It should be noted that fluctuations to \emph{negative values} do occur as well. In the local case, by means of the local Einstein equations, the energy density is just a $\delta$-function.

\subsection{Oscillations in higher-derivative gravity}
\begin{figure}[!htb]
\centering
\subfloat[Higher-derivative theories $N=1,2,3$.]
{
    \includegraphics[width=0.5\textwidth]{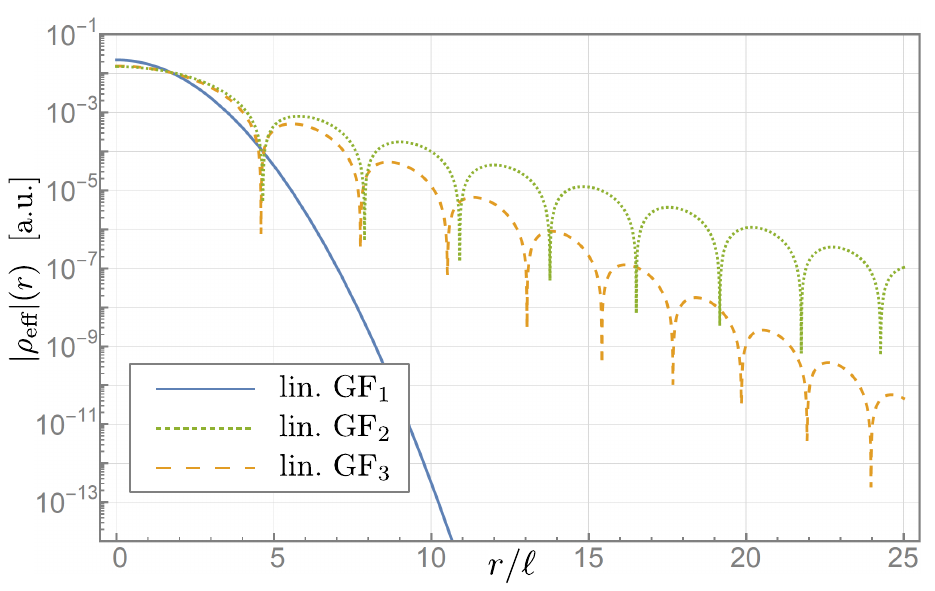}
	\label{fig:ch3:energy-density-gf-log}
}
\subfloat[Infinite-derivative theories $N=1,2,3$.]
{
    \includegraphics[width=0.5\textwidth]{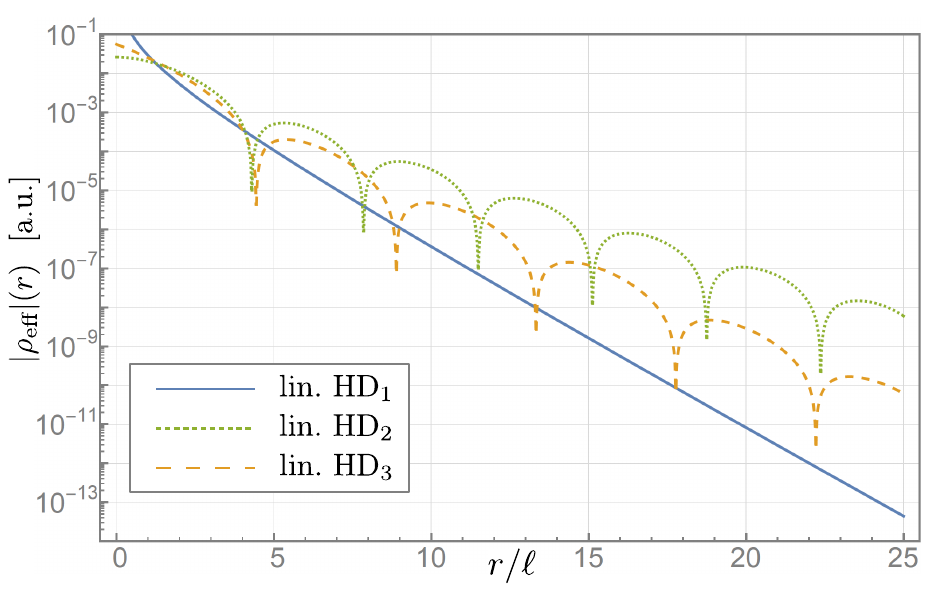}
    \label{fig:ch3:energy-density-hd-log}
}
\caption[Effective energy densities in higher-derivative and non-local theories.]{Absolute value of the effective energy densities $\rho_\text{eff}(r)$ in higher-derivative theories (left) as well as non-local theories (right), plotted in arbitrary units on a logarithmic vertical axis over the dimensionless distance $r/\ell$. Both undergo periodic spatial oscillations for $N \ge 2$ and change their signs periodically. In this logarithmic plot the sign change corresponds to a vertical slope.}
\label{fig:ch3:energy-density-log}
\end{figure}

Typically, these oscillations are observed in higher-derivative theories of gravity \cite{Accioly:2016etf,Accioly:2016qeb,Perivolaropoulos:2016ucs,Giacchini:2016xns}. Let us understand how they come about there by truncating our non-local form factors at first order. This way, we can introduce a new class of higher-derivative theories that we shall dub $\mathrm{HD_N}$ theories, in close resemblance to our non-local ghost-free $\mathrm{GF_N}$ theories:
\begin{alignat}{3}
\mathrm{HD_N}:& \qquad f(\lap) = 1 + (-\ell^2\lap)^N \, , \quad && N \in \mathbb{N} \, , \\
\mathrm{GF_N}:& \qquad f(\lap) = \exp\left[(-\ell^2\lap)^N\right] \, , \quad && N \in \mathbb{N} \, .
\end{alignat}
While we leave aside the physical relevancy of these higher-derivative theories, which is an interesting topic of its own, we merely would like to demonstrate at this point that these oscillations also occur in higher-derivative theories. This is rather straightforward and can be done quickly making use of the Green function representation \eqref{eq:ch2:bessel-representation-1} subject to the higher-derivative form factors. After employing \eqref{eq:ch3:particle-solution} one finds the following potentials for linearized higher-derivative gravity:
\begin{align}
\mathrm{HD_1} :& \quad \phi(r) = -\frac{2Gm}{r} \left(1-e^{-r/\ell}\right) \, , \\
\mathrm{HD_2} :& \quad \phi(r) = -\frac{2Gm}{r} \left\{ 1-e^{-r/(\sqrt{2}\ell)}\cos\left[r/(\sqrt{2}\ell)\right] \right\} \, , \\
\mathrm{HD_3} :& \quad \phi(r) = -\frac{2Gm}{r} \left\{ 1-\tfrac13 e^{-r/\ell} -\tfrac23 e^{-r/(2\ell)}\cos\left[\sqrt{3}r/(2\ell)\right] \right\} \, .
\end{align}
Then, one can calculate the energy density \eqref{eq:ch3:energy-density} and compare it to the non-local case. The result can be seen in Fig.~\ref{fig:ch3:energy-density-hd-log}. Similar to the non-local case, the higher-derivative theories also produce oscillations. For the cases $\mathrm{HD_2}$ and $\mathrm{HD_3}$ these oscillations are even visible at the level of the potentials since they include manifestly oscillatory terms via trigonometric functions.

The higher-derivative form factors, while satisfying $f(0) = 1$ and thereby reproducing the correct asymptotic behavior, are non-zero for imaginary values of momentum:
\begin{align}
1 + (k^2\ell^2)^N = 0 \quad \text{if} \quad k^2\ell^2 = \sqrt[N]{-1} \, .
\end{align}
These complex roots are thought to cause the observed oscillations. Our conclusions cast some doubt on that because the oscillations, albeit in a slightly different magnitude, still occur in the case of exponential form factors that are non-local and everywhere manifestly non-zero.

\begin{figure}
\centering
\subfloat
{
    \bgroup
	\def\arraystretch{1.2}
	\footnotesize
	\begin{tabular}{rccrcc}
	$r/\ell$ & $\delta_2/(2\ell)$ & ~~ & $r/\ell$ & $\delta_3/(2\ell)$ \\ \hline
	 4.59 & 3.16 &&  4.65 & 3.23 \\
	 7.75 & 2.76 &&  7.88 & 3.02 \\
	10.51 & 2.53 && 10.90 & 2.86 \\
	13.04 & 2.38 && 13.76 & 2.75 \\
	15.42 & 2.26 && 16.51 & 2.66 \\
	17.68 & 2.17 && 19.17 & 2.58 \\
	19.85 & ---  && 21.75 & ---
	\end{tabular}
	\egroup
}
\qquad
\subfloat{\adjustbox{raise=-6pc}{\includegraphics[width=0.5\textwidth]{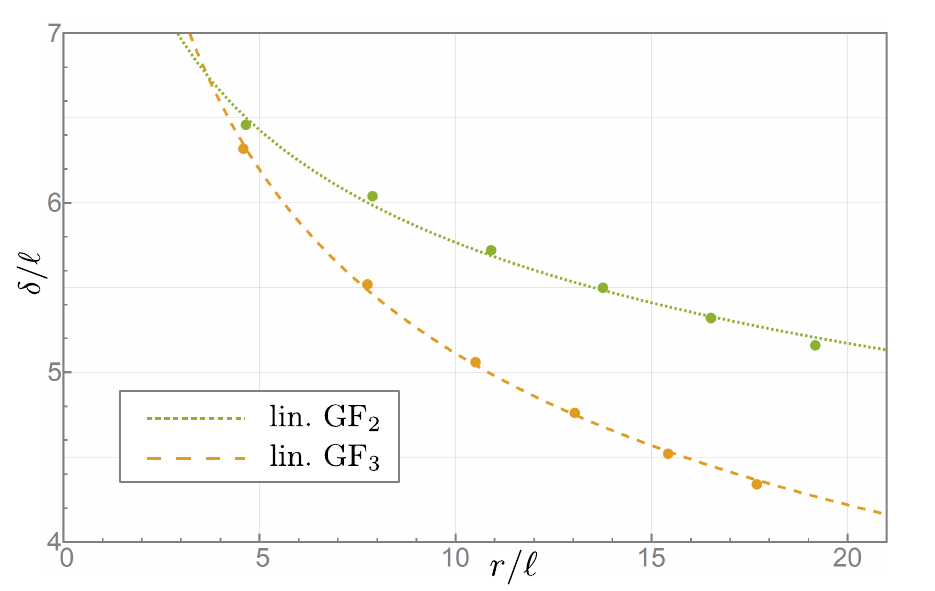}}}
\caption[Wavelength of spatial Friedel oscillations for $\mathrm{GF_2}$ and $\mathrm{GF_3}$ theory.]{Evaluating the zeros of the energy density it is possible to read off half the corresponding wavelength $\delta_N$. These wavelength are not constant but depend on the position: they decrease with increasing distance $r/\ell$, and to first approximation this can be described by simple power laws (even though there still exist subleading oscillations around those power laws, as the above plot shows).}
\label{fig:ch3:wavelengths}
\end{figure}

Some numerical investigations into the oscillation wavelengths in the non-local theories $\mathrm{GF_2}$ and $\mathrm{GF_2}$ can be found in Fig.~\ref{fig:ch3:wavelengths}, which reveals a spatial dependence of the oscillations that roughly follow simple power laws,
\begin{align}
\mathrm{GF_2}: \quad \frac{\delta_2}{\ell} \sim 9.68 \left(\frac{r}{\ell}\right)^{-0.28} \, , \quad
\mathrm{GF_3}: \quad \frac{\delta_3}{\ell} \sim 8.28 \left(\frac{r}{\ell}\right)^{-0.16} \, ,
\end{align}
where we denote the dimensionless wavelength for $\mathrm{GF_N}$ theory as $\delta_N/\ell$. At this point we are unable to provide a physical explanation for this behavior, but hope to revisit this phenomenon at some time.

%\todo{Are the roots of unity aligned with the essential divergence directions of the exponential form factors? If yes, then let us mention it here.}

\subsection{Physical interpretation: an attempt}
In the context of condensed matter physics one encounters the following phenomenon: place an electron, an electric impurity, inside a cold metal. Then one can calculate the effective potential around the electron, and one finds that the resulting potential oscillates in space, related to screening and anti-screening effects of the surrounding electrons \cite{Altland:2006}, and these oscillations are called Friedel oscillations \cite{Friedel:1952,Friedel:1954,Friedel:1958}. We cannot help but notice the similarity in the present context: placing a gravitational impurity (point particle) in the Minkowski vacuum, the effective potential is not of the standard Newtonian form but involves spatial oscillations.

We may interpret the local energy density as an effective density for the following reason: we take it to be to timelike component of the Einstein tensor, which can be interpreted as the energy density of a smeared matter distribution. Hence the local energy density is an effective energy density that knows about the spreading of sharp, $\delta$-shaped matter distributions due to the presence of non-locality. Moreover, the framework of non-local ghost-free gravity may be viewed as an effective field theory obtained from integrating out UV degrees of freedom.

It would be rewarding to perform a detailed quantum-field-theoretical calculation that also results in visible oscillations in the effective potential. While we will not entertain these calculations further in this thesis, there is a potential similarity to the oscillations encountered in the vacuum polarization around a $\delta$-shaped potential in non-local ghost-free quantum field theory that we will discuss later in Ch.~\ref{ch:ch6}, where we show that the difference between the local and non-local vacuum polarization fluctuates from negative to positive values with increasing distance.

\section{Extended objects: $p$-branes}
\label{sec:ch3:p-branes}
Having demonstrated that the gravitational field of point particles is regularized due to the presence of non-locality, a natural extension is now to look at static mass distributions with a finite extension. Since the divergences stem from infinitely thin $\delta$-sources, we are interested in calculating the gravitational field of $p$-branes, that is, $p$-dimensional infinitely thin sheets of matter.

A point particle is a special case of a $p$-brane, namely, $p=0$. For a string (which is, spatially, a one-dimensional object), one has instead $p=1$. In general, $D$ spacetime dimensions can be written as
\begin{align}
D = p + m + 1 \, .
\end{align}
Let us adopt the coordinates in Minkowski space to this notation write
\begin{align}
x{}^\mu = (t, x{}^\alpha) = (t, z^a, y^i) \, , \quad a=1,\dots,p, \quad i=1,\dots,m \, .
\end{align}
The Minkowski metric then takes the form
\begin{align}
\dd s^2 = -\dd t^2 + \sum\limits_{a=1}^p (\dd z^a)^2 + \sum\limits_{i=1}^m (\dd y^i)^2 \, ,
\end{align}
where Latin indices from the beginning of the alphabet label the brane coordinates $z^a$, and Latin indices from the middle of the alphabet label the ambient coordinates $y^i$. One may think of $p$ as the spatial extension of the brane, and of $m$ as the spatial codimension of the brane. In other words, the brane will be located at $\{y^i=0\}$ for $i=1,\dots,m$.

We parametrize energy-momentum tensor of a $p$-brane with a surface tension $\epsilon>0$ as
\begin{align}
\label{eq:ch3:tmunu-p-brane}
T{}_{\mu\nu} = \epsilon\left( \delta{}^t_\mu\delta{}^t_\nu - \sum\limits_{a=1}^p \delta{}^a_\mu\delta{}^a_\nu \right) \prod\limits_{i=1}^m \delta\left(y^i\right) \, .
\end{align}
The physical dimensions of $\epsilon$ are
\begin{align}
[\epsilon] = \frac{\text{M}}{\text{L}^p} \, ,
\end{align}
and hence it is clear that for a particle, $p=0$, $\epsilon$ corresponds to its mass, and that for a string, $p=1$, $\epsilon$ is a line density, and so on. Let us briefly comment on the total mass of such a system. It is given by
\begin{align}
m = \int\dd^d z \, \epsilon \prod\limits_{i=1}^m \delta\left(y^i\right) = \int\dd^p z \, \epsilon = \infty \, .
\end{align}
Clearly this expression diverges for $p>0$, but this is to be expected since we introduced an infinitely extended object as a source. As we will see, however, the gravitational field does not depend on that quantity and only the ``line density'' $\epsilon$ enters.

The metric describing $p$-branes is often called a warped geometry of the form \cite{Dabholkar:1990yf,Stelle:1998xg}
\begin{align}
\dd s^2 = f(y^i)\dd\sigma^2(t,z^a) + \dd\gamma^2(y^i) \, .
\end{align}
In the context of our linearized weak-field ansatz, we re-parametrize this metric as
\begin{align}
\label{eq:ch3:warped-ansatz}
\dd s^2 = [1+u(r)]\left[-\dd t^2 + \sum\limits_{a=1}^p (\dd z^a)^2 \right] + [1+v(r)] \sum\limits_{i=1}^m (\dd y^i)^2 \, , \quad r^2 = \sum\limits_{i=1}^m (y^i)^2 \, .
\end{align}
This geometry, as we will explain now, is consistent with the isometries encoded in the tensorial structure of the energy-momentum tensor:
\begin{itemize}
\item The Poincar\'e symmetry $P(1,p)$ in the $tz^a$-sector is an invariance ``inside'' the $p$-brane. Boosts and rotations inside the brane do not change the geometry, since it is a static, homogeneous object that extends to both spacelike and timelike infinity. In the case of a point particle, $p=0$, the Poincar\'e symmetry just becomes temporal translation invariance corresponding to staticity.

\item The rotational $O(m)$ symmetry in the $y^i$-sector reflects the rotational invariance ``around'' the brane in the ambient space. For a particle it corresponds to the usual spherical symmetry around it. For a cosmic string, however, this captures the axisymmetry around it in four spacetime dimensions. In the case of $m=1$ it becomes a discrete reflection symmetry from one side of the brane to the other, in any number of dimensions.
\end{itemize}
Inserting the warped geometry \eqref{eq:ch3:warped-ansatz} into the time-independent, linearized field equations \eqref{eq:ch3:eom-stationary} gives the following set of equations:
\begin{align}
\big[ (p+1)c(\lap) - a(\lap) \big] \lap u + (m-1)c(\lap) \lap v &= -2\kappa\epsilon \prod\limits_{i=1}^m \delta(y_i) \, , \\
\left( \delta_{ij} \lap - \partial_i \partial_j \right)\left\{ \big[a(\lap) - (m-1)c(\lap)\big] v - (p+1)c(\lap)u \right\} &= 0 \, .
\end{align}
While they can be solved for the case of $c=(1+\alpha)a$, see Ref.~\cite{Boos:2018bxf}, here we will focus on the simpler case $a=c$ that produces the same Newtonian limit as General Relativity. Then, the equations simplify and become
\begin{align}
a(\lap)\lap\left[ pu + (m-1)v \right] &= -2\kappa\epsilon \prod\limits_{i=1}^m \delta(y^i) \, , \\
a(\lap)\left( \delta_{ij} \lap - \partial_i \partial_j \right)\left[ (2-m)v - (p+1)u \right] &= 0 \, .
\end{align}
The homogeneous equation is solved by
\begin{align}
\label{eq:ch2:p-brane-homogeneous-solution}
v = \frac{p+1}{2-m} u \, ,
\end{align}
such that the remaining, inhomogeneous equation takes the final form
\begin{align}
a(\lap)\lap u = -\frac{2-m}{p+m-1} 2\kappa\epsilon \prod\limits_{i=1}^m \delta(y^i) \, .
\end{align}
In the limiting case of $p=0$ and $m=d$ one recovers the field equation for a point particle, compare to Eq.~\eqref{eq:ch3:eom-inh-pp}, as it should be. The solution of this equation is given by the $m$-dimensional static Green function $\mathcal{G}_m$ and takes the form
\begin{align}
\label{eq:ch2:p-brane-solution}
u(r) = \frac{2-m}{p+m-1} 2\kappa\epsilon \mathcal{G}_m(r) \, , \quad v(r) = \frac{p+1}{p+m-1} 2\kappa\epsilon \mathcal{G}_m(r) \, .
\end{align}
Observe that the radial dependence of both $u(r)$ and $v(r)$ is universal, in the sense that it does not depend on the dimension $d=m+p$ of space. Rather, it depends solely on the codimension $m$, or, in other words, on the dimensionality of the transverse space.

\subsection{Regularity}
As already demonstrated earlier in Sec.~\ref{sec:ch2:static}, the static non-local Green functions are manifestly regular at $r=0$. For this reason, the resulting metric \eqref{eq:ch3:warped-ansatz} with the functions \eqref{eq:ch2:p-brane-solution} is also a regular metric at $r=0$.

\subsection{Curvature expressions}
\label{sec:ch3:p-brane-curvature}
Let us now turn towards a more detailed analysis of the curvature. To leading order in the metric perturbation $h{}_{\mu\nu}$ the Riemann curvature tensor, the Ricci tensor, and the Ricci scalar are given by
\begin{align}
R{}^\mu{}_{\nu\rho\sigma} &= \partial{}_\nu \partial{}_{[\rho} h{}_{\sigma]}{}^\mu - \partial{}^\mu \partial{} _{[\rho}h{}_{\sigma]\nu} \, , \\
R{}_{\mu\nu} &= R{}^\rho{}_{\mu\rho\nu} = \partial{}_\rho \partial{}_{(\mu} h{}_{\nu)}{}^\rho - \frac12 \left(\partial{}_\mu\partial{}_\nu h + \Box h{}_{\mu\nu} \right) \, , \\
R &= R{}^\rho{}_\rho = \partial{}_\rho \partial{}_\sigma h{}^{\rho\sigma} - \Box h \, .
\end{align}
As in the case of the point particle, it can be useful to consider the quantity
\begin{align}
-\kappa \rho_\text{eff} = \left( R{}_{\mu\nu} - \frac12 R g{}_{\mu\nu} \right) \xi{}^\mu \xi{}^\nu = \frac{m+p-1}{2(1+p)} R \, , \quad \ts{\xi} = \partial_t \, ,
\end{align}
which justifies the interpretation of the Ricci scalar as some sort of rescaled effective energy density. With that in mind, let us introduce new indices that are tailored to the warped geometry \eqref{eq:ch3:warped-ansatz}. Collecting the coordinates from the $tz^a$-sector in one index $\overline{a}$ such that $\overline{a}=0,1,\dots,p$, let us denote the Minkowski metric on this sector as $\eta{}_{\overline{a}\overline{b}}$, and keep the Latin indices from the middle of the alphabet for the $y^i$-sector. We will denote derivatives as $\partial_i = \partial/\partial y^i$ and $\partial_a = \partial/\partial z^a$ for brevity. Then, substituting the on-shell condition \eqref{eq:ch2:p-brane-homogeneous-solution}, we can write the curvature tensor components as
\begin{align}
R{}_{\overline{a}\overline{b}\overline{c}\overline{d}} = 0 \, , \quad
R{}_{\overline{a}i\overline{b}j} = \frac{m-2}{2(1+p)} \eta{}_{\overline{a}\overline{b}} \partial_i \partial_j v \, , \quad
R{}_{ijkl} = 2\partial{}_{[i} \delta{}_{j][k} \partial_{l]} v \, .
\end{align}
The Ricci tensor and Ricci scalar take the form
\begin{align}
R_{\overline{a}\overline{b}} = \frac{m-2}{2(1+p)} \eta{}_{\overline{a}\overline{b}} \lap v \, , \quad
R{}_{ij} = -\frac12 \delta{}_{ij} \lap v \, , \quad R = -\lap v \, .
\end{align}
The Weyl tensor can be written as
\begin{align}
C{}_{\mu\nu\rho\sigma} = R{}_{\mu\nu\rho\sigma} - \frac{2}{D-2}\left( \eta{}_{\mu[\rho} R{}_{\sigma]\nu} - \eta{}_{\nu[\rho} R{}_{\sigma]\mu} \right) + \frac{2}{(D-1)(D-2)} R \eta{}_{\mu[\rho} \eta{}_{\sigma]\nu} \, .
\end{align}
Its components are
\begin{align}
\label{eq:ch3:weyl-1}
C{}_{\overline{a}\overline{b}\overline{c}\overline{d}} &= \frac{1-m}{(1+p)(m+p)} 2\eta{}_{\overline{a}[\overline{c}} \eta{}_{\overline{d}]\overline b} \lap v \, , \\
\label{eq:ch3:weyl-2}
C{}_{\overline{a}i\overline{b}j} &= \frac{m-2}{2(1+p)} \eta{}_{\overline{a}\overline{b}} \partial_i \partial_j v + \frac{2-m+p}{2(1+p)(m+p)} \eta{}_{\overline{a}\overline{b}}\delta{}_{ij} \lap v \, , \\
\label{eq:ch3:weyl-3}
C{}_{ijkl} &= 2 \partial{}_{[i} \delta{}_{j][k} \partial{}_{l]} v + \frac{2}{m+p}  \delta{}_{i[k} \delta{}_{l]j} \lap v \, .
\end{align}
We are also interested in quadratic curvature invariants, and, as is well known \cite{Stephani:2003}, the above expression for the Weyl tensor is not the most convenient form. This is because the tensorial Ricci term and the scalar Ricci term are not orthogonal in the sense of an irreducible decomposition. It is more convenient, for that reason, to introduce the tracefree Ricci tensor
\begin{align}
\cancel{R}{}_{\mu\nu} = R{}_{\mu\nu} - \frac{1}{D} R \eta{}_{\mu\nu} \, , \quad \cancel{R}{}^\rho{}_\rho = 0 \, , \quad D = m + p + 1 \, .
\end{align}
Its non-vanishing components are
\begin{align}
\cancel{R}_{\overline{a}\overline{b}} = \frac{m(m+p-1)}{2(1+p)(m+p+1)} \eta{}_{\overline{a}\overline{b}} \lap v \, , \quad
\cancel{R}_{ij} = \frac{1-m-p}{2(m+p+1)} \delta{}_{ij} \lap v \, .
\end{align}
The Weyl tensor can then be written as
\begin{align}
C{}_{\mu\nu\rho\sigma} = R{}_{\mu\nu\rho\sigma} - \frac{2}{D-2}\left( \eta{}_{\mu[\rho} \cancel{R}{}_{\sigma]\nu} - \eta{}_{\nu[\rho} \cancel{R}{}_{\sigma]\mu} \right) - \frac{2}{D(D-1)} R \eta{}_{\mu[\rho} \eta{}_{\sigma]\nu} \, .
\end{align}
The square of the Weyl tensor is then the square of each individual piece, which follows from the tracelessness of the Weyl tensor, as well as the tracelessness of the tracefree Ricci tensor \cite{Stephani:2003,Boos:2014hua}. 

\begin{figure}[!hbt]
    \centering
    \subfloat{{ \includegraphics[width=0.42\textwidth]{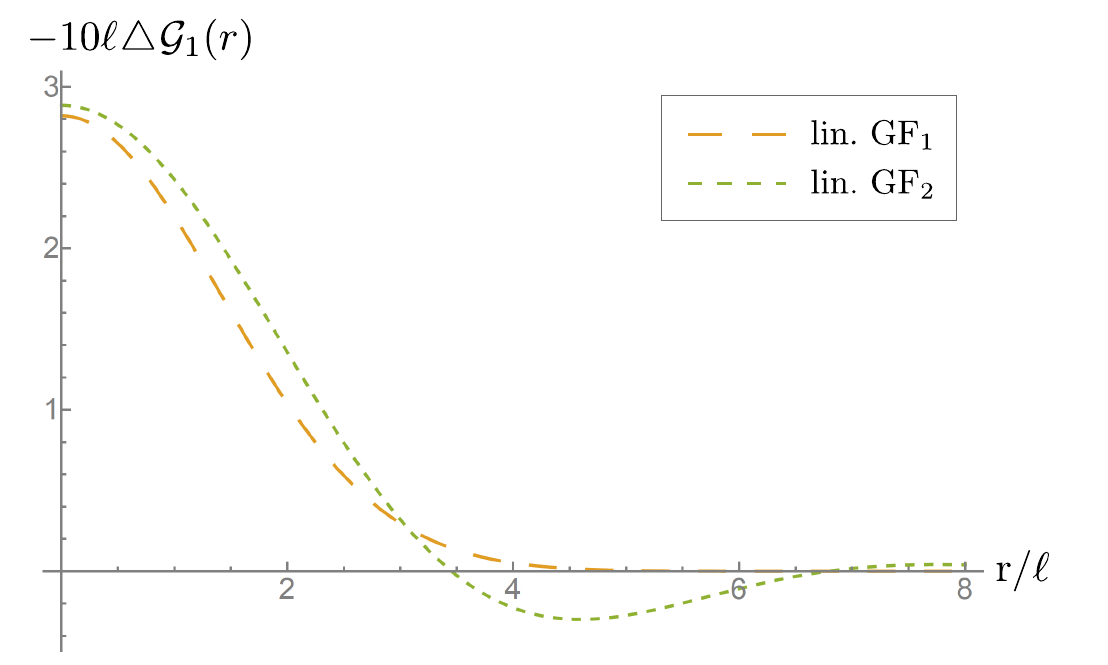} }}%
    \qquad
    \subfloat{{ \includegraphics[width=0.42\textwidth]{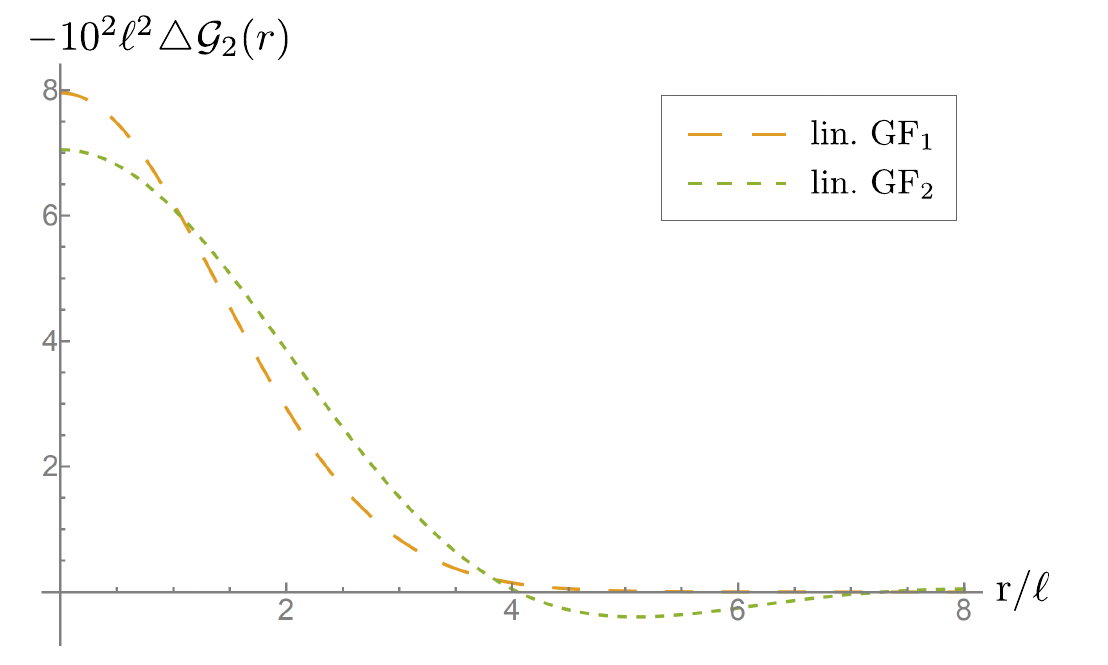} }}%

    \subfloat{{ \includegraphics[width=0.42\textwidth]{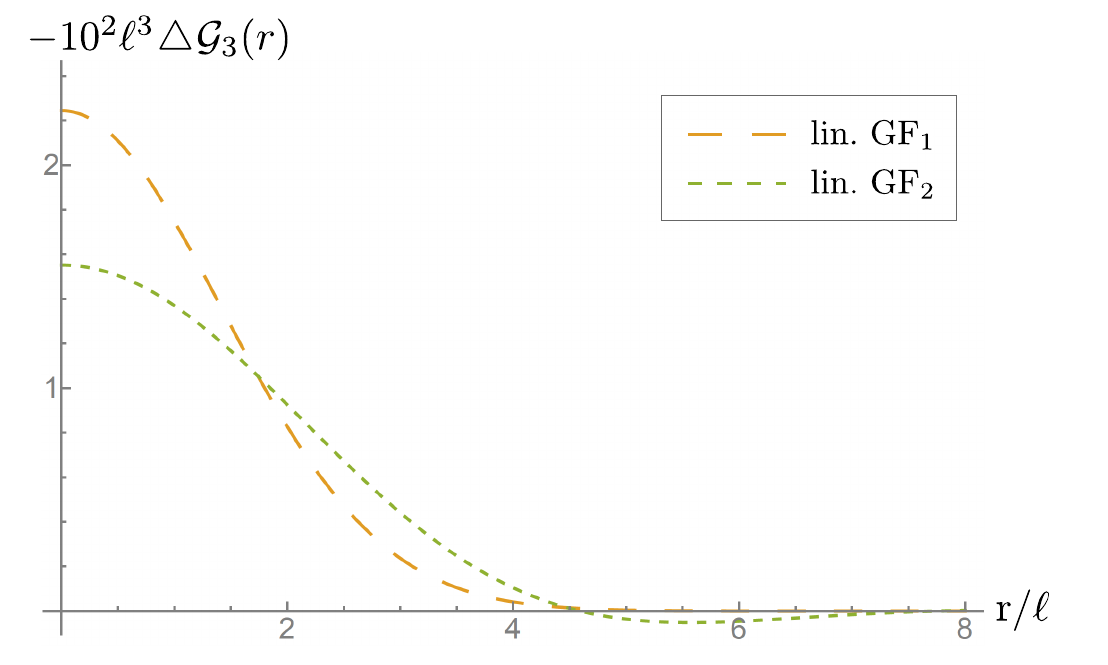} }}
    \qquad
    \subfloat{{ \includegraphics[width=0.42\textwidth]{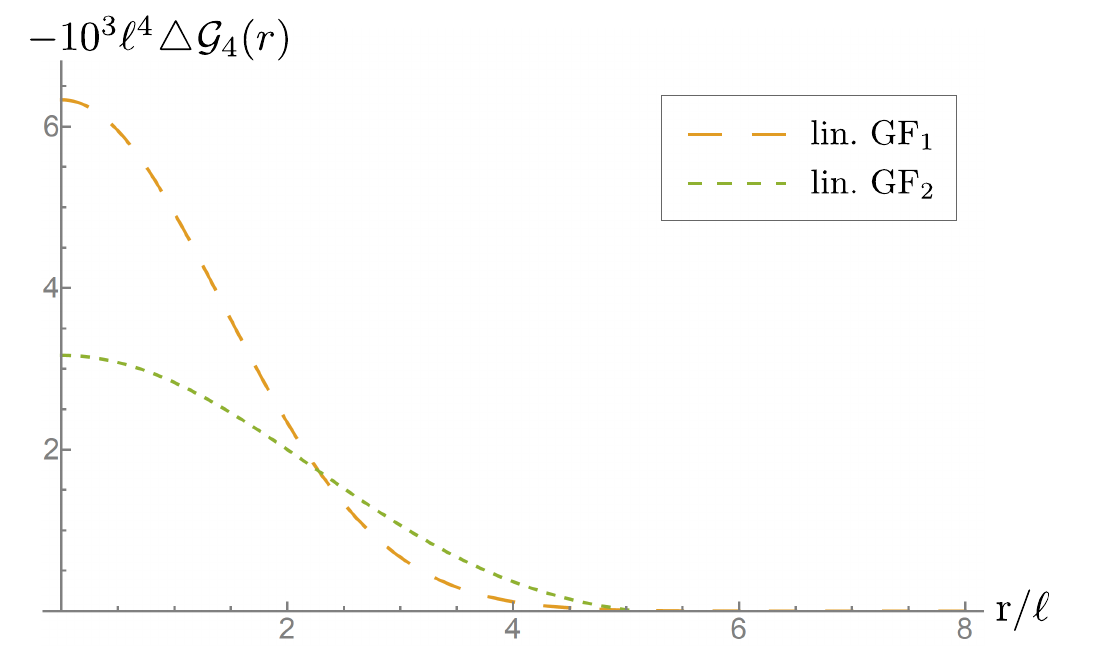} }}%
    \caption[Linear curvature invariants.]{Dimensionless linear curvature invariant $I_m$ evaluated for $\mathrm{GF_1}$ and $\mathrm{GF_2}$ theory in the cases $m=1,2,3,4$. It is finite and regular at $r/\ell=0$.}
    \label{fig:ch3:visualization-i}
\end{figure}

\begin{figure}[!hbt]
    \centering
    \subfloat{{ \includegraphics[width=0.42\textwidth]{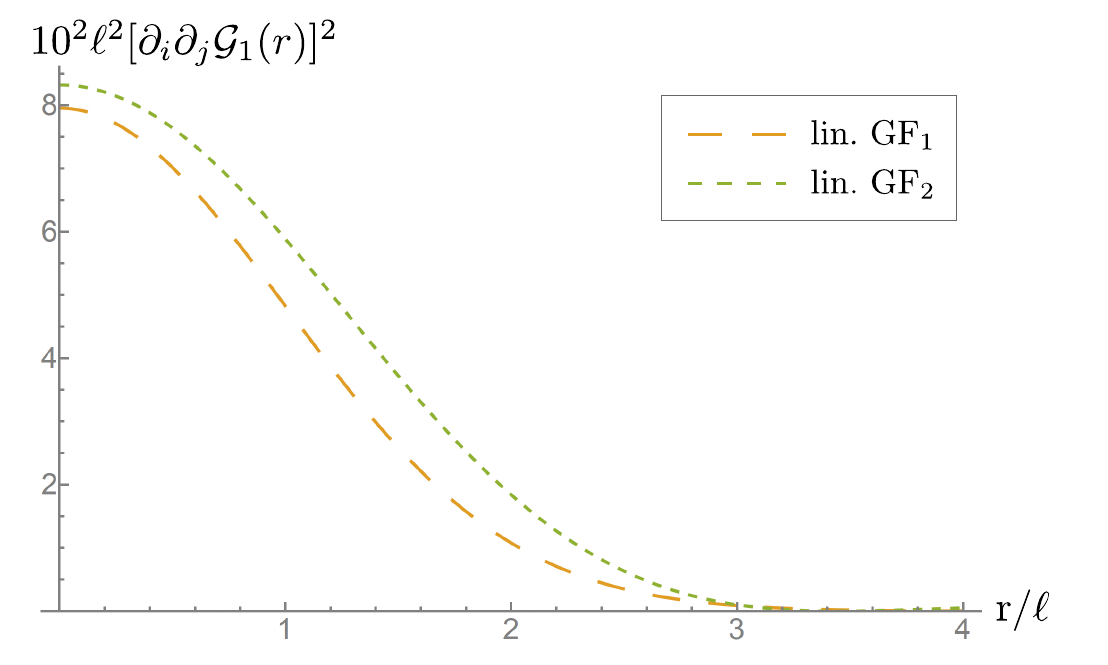} }}%
    \qquad
    \subfloat{{ \includegraphics[width=0.42\textwidth]{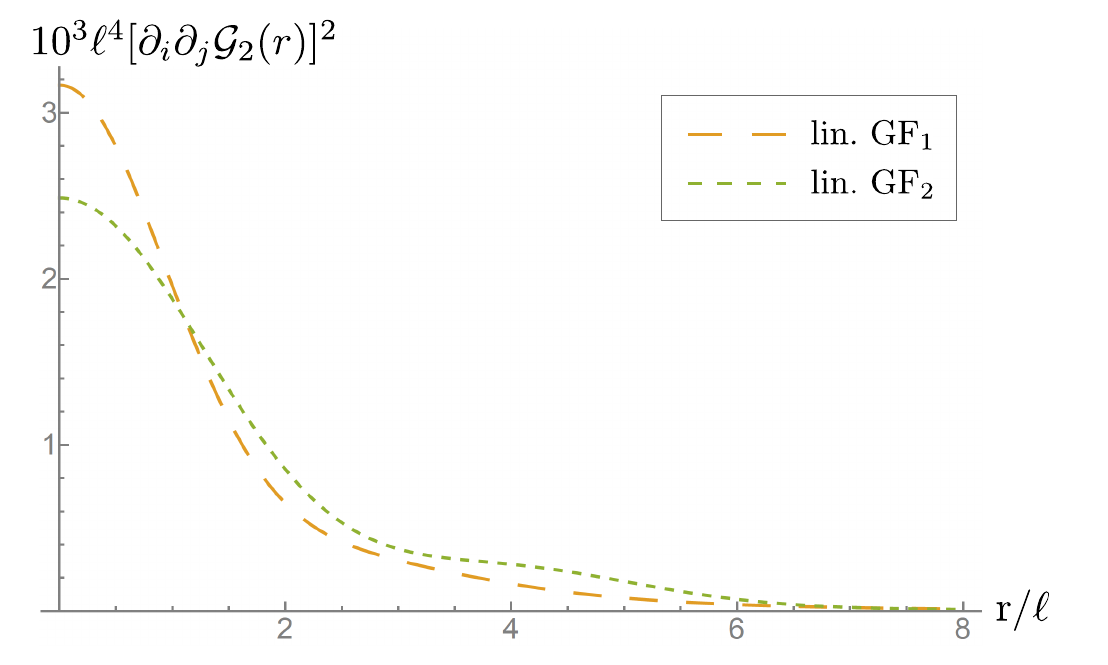} }}%

    \subfloat{{ \includegraphics[width=0.42\textwidth]{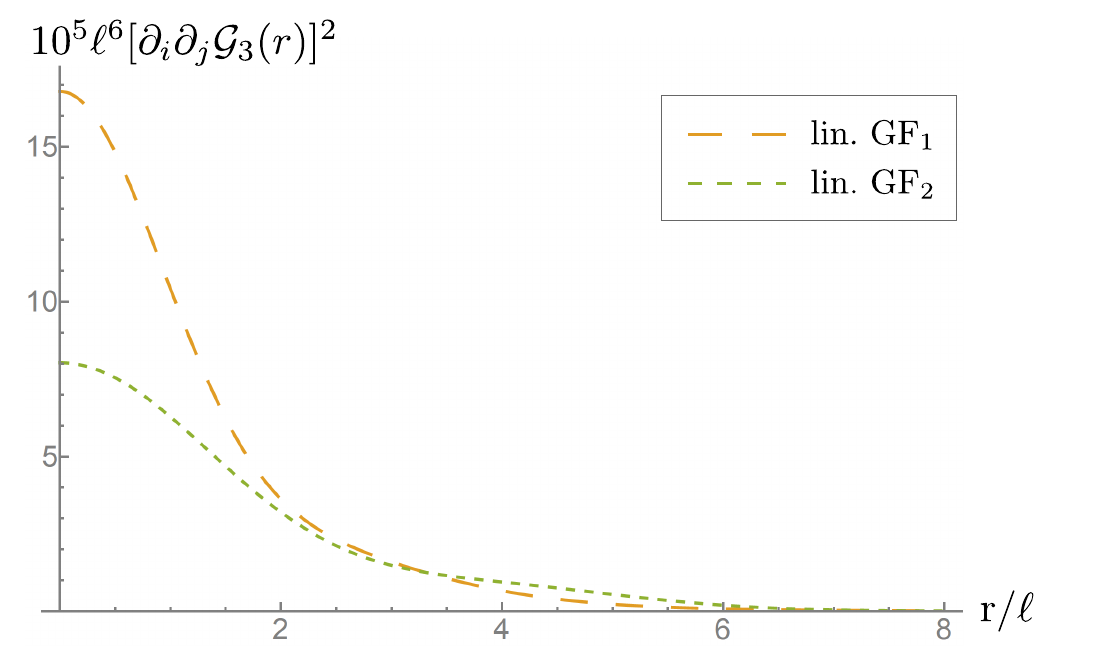} }}
    \qquad
    \subfloat{{ \includegraphics[width=0.42\textwidth]{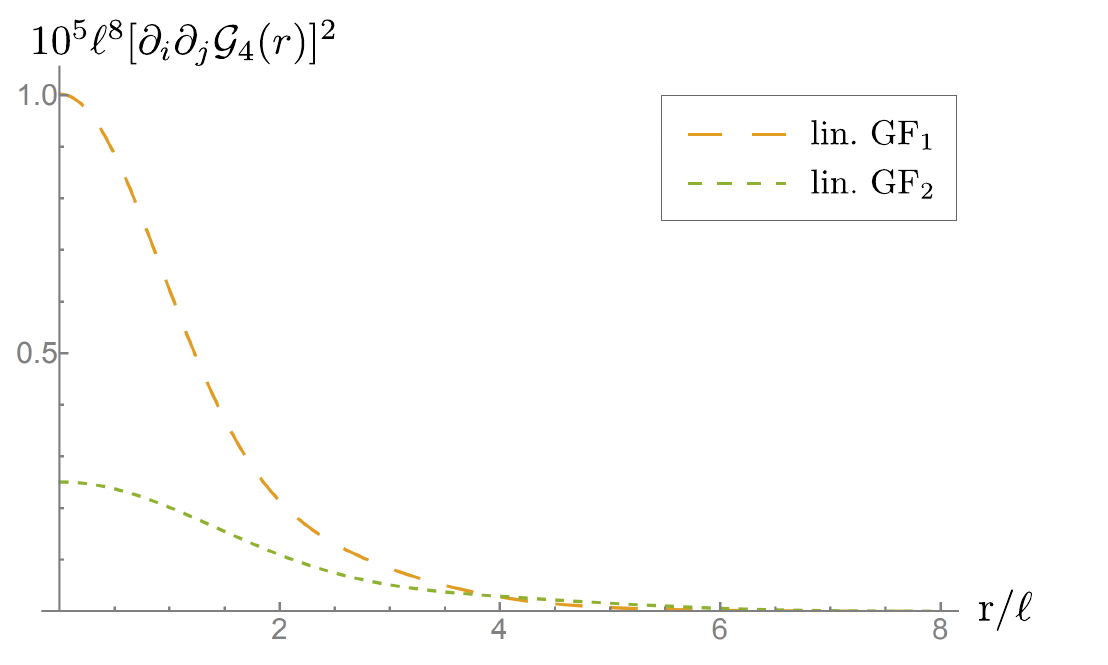} }}%
    \caption[Quadratic curvature invariants.]{Dimensionless quadratic curvature invariant $J_m$ evaluated for $\mathrm{GF_1}$ and $\mathrm{GF_2}$ theory in the cases $m=1,2,3,4$. It is finite and regular at $r/\ell=0$.}
    \label{fig:ch3:visualization-j}
\end{figure}

Then, the ``orthogonal'' quadratic curvature invariants are
\begin{align}
C^2 &= C{}_{\mu\nu\rho\sigma} C{}^{\mu\nu\rho\sigma} = \frac{p^2-m^2+3m+p-2}{(1+p)(m+p)} (\lap v)^2 + \frac{(m-2)(m+p-1)}{1+p} (\partial_i\partial_j v)(\partial^i \partial^j v) \, , \\
\cancel{R}^2 &= \cancel{R}_{\mu\nu} \cancel{R}^{\mu\nu} = \frac{m(m+p-1)^2}{(1+p)(m+p+1)} (\lap v)^2 \, .
\end{align}
Another quadratic invariant is the four-dimensional Chern--Pontryagin pseudoscalar which however vanishes for the warped geometry \eqref{eq:ch3:warped-ansatz},
\begin{align}
\mathcal{P} = \frac12 \epsilon{}^{\mu\nu}{}_{\alpha\beta} C{}^{\alpha\beta}{}_{\rho\sigma} C{}^{\rho\sigma}{}_{\mu\nu} = 0 \, .
\end{align}
Let us notice that there are essentially only two types of expressions in the curvature invariants: a square of the Laplace operator, $(\lap v)^2$, and a square of the double divergence, $(\partial_i\partial_j v)(\partial^i \partial^j v)$. For this reason let us define the dimensionless invariants
\begin{align}
\label{eq:ch3:p-brane-invariants}
I_m \equiv -\ell^m \lap \mathcal{G}_m(r) \, , \quad
J_m \equiv \ell^{2m} \left( \partial_i\partial_j \mathcal{G}_m(r) \right)\left( \partial{}^i \partial_j \mathcal{G}_m(r) \right) \, .
\end{align}
$I_m$ is directly proportional to the Ricci scalar, and $J_m$ can be expressed as a linear combination of the two ``orthogonal'' quadratic curvature invariants. Due to the $O(m)$ symmetry around the $p$-brane, the metric function $v$ is a function of the radius alone, $v=v(r)$. Making use of the $m$-dimensional spherical identities for any function $f=f(r)$,
\begin{align}
\lap f = f'' + (m-1)\frac{f'}{r} \, , \quad
(\partial_i\partial_j f)(\partial^i \partial^j f) = (f'')^2 + (m-1)\left(\frac{f'}{r}\right)^2 \, ,
\end{align}
we can now employ the recursion relations for static Green functions \eqref{eq:ch2:recursion} and rewrite \eqref{eq:ch3:p-brane-invariants} as
\begin{align}
I_m &= 2\pi \ell^m \Big[ m \mathcal{G}_{m+2}(r) - 2\pi r^2 \mathcal{G}_{m+4}(r) \Big] \, , \\
J_m &= 4\pi^2 \ell^{2m} \Big\{ m \left[\mathcal{G}_{m+2}(r)\right]^2 - 4\pi r^2 \mathcal{G}_{m+2}(r) \mathcal{G}_{m+4}(r) + 4\pi^2 r^4 \left[\mathcal{G}_{m+4}(r)\right]^2 \Big\} \, .
\end{align}
Since each Green function $\mathcal{G}_m(r)$ is finite at $r=0$, the above relations prove that also the linear and quadratic curvature invariants are finite at the origin. See Figs.~\ref{fig:ch3:visualization-i} and \ref{fig:ch3:visualization-j} for plots of the linear and quadratic curvature invariants for $\mathrm{GF_1}$ and $\mathrm{GF_2}$ theory in the cases of $m=1,2,3,4$.

\subsection{Concrete examples}
Having the general expressions for both the metric and the curvature invariants readily available, we would now like to focus on four important subclasses:
\begin{itemize}
\item The point particle ($p=0$ and $m=D-1$).
\item A cosmic string ($p=1$ and $m=D-2$) in $D=4$.
\item A domain wall ($m=1$ and $p=D-2$).
\item ``Angle deficit configurations'' ($m=2$, $p=D-3$).
\end{itemize}

\subsubsection{The point particle, revisited}
It has been observed that the Weyl tensor vanishes at the location of a point particle in four spacetime dimensions in non-local gravity \cite{Buoninfante:2018xiw}. Here we can generalize this result to any number of dimensions, and, in fact, any ghost-free theory that belongs to the $\mathrm{GF_N}$ class. In the present context, a point particle corresponds to $p=0$ and hence the metric functions describing a point particle in $d$ dimensions are
\begin{align}
u(r) = \frac{2-d}{d-1} 2\kappa\epsilon \mathcal{G}_d(r) \, , \quad v(r) = \frac{1}{d-1} 2\kappa\epsilon \mathcal{G}_d(r) \, .
\end{align}
Note that we have proven in Sec.~\ref{sec:ch2:static} that the non-local, static Green functions behave as
\begin{align}
\mathcal{G}_d(r\rightarrow 0) \approx c_0 + c_2 r^2 + \mathcal{O}\left(r^4\right) \, .
\end{align}
The absence of the linear term should be noted. Consequently, since the metric function $v(r)$ is a multiple of the static Green function, a similar identity holds for the metric function $v(r)$ itself. Then one can calculate, to leading order,
\begin{align}
\label{eq:ch3:smoothness-r=0}
\partial_i \partial_j v(r) = \frac{1}{d} \delta{}_{ij} \lap v + \mathcal{O}(r^2) \, .
\end{align}
Spherical symmetry and smoothness of the metric at the center $r=0$ (mathematically, the smoothness is encoded in the absence of the linear term, which is sometimes also referred to as ``regularity'') imply the absence of a tensorial structure in the curvature in proximity to $r=0$. Let us verify this explicitly by inserting \eqref{eq:ch3:smoothness-r=0} into the Weyl tensor:
\begin{align}
C{}_{\overline{a}\overline{b}\overline{c}\overline{d}}(r\rightarrow 0) &\approx 0 \, , \\
C{}_{\overline{a}i\overline{b}j}(r\rightarrow 0) &\approx \frac{(m-1)p}{(m+p)(1+p)m} \eta{}_{\overline{a}\overline{b}} \delta{}_{ij} \lap v \, , \\
C{}_{ijkl}(r\rightarrow 0) &\approx \frac{-p}{m(m+p)}2 \delta{}_{i[k} \delta{}_{l]j} \lap v \, .
\end{align}
The above relations imply that $C{}_{\mu\nu\rho\sigma}=0$ at $r=0$ if $p=0$. In other words, as long as point particles have a regular gravitational field (meaning the absence of a linear term close to $r=0$) their Weyl tensor vanishes at $r=0$ for any $\mathrm{GF_N}$ theory. For the special case $p=D-2$ (``domain walls'') the Weyl tensor also vanishes at $r=0$: If one inserts $p=D-2$ (which implies $m=1$) then $C_{\bar{a}i\bar{b}j}=0$. But $C_{ijkl}=0$ as well in this case because the transverse space is one-dimensional. In fact, for $p=D-2$ the Weyl tensor vanishes everywhere due to conformal flatness, see Sec.~\ref{sec:ch3:domain-wall}

%There is a general argument that the Weyl tensor needs to vanish at $r=0$ for spherically symmetric and regular geometries... \todo{only at $r=0$ or everywhere?}

In the case of a static, regular geometry and $p=0$ one may also argue from a different perspective.\footnote{We thank Don Page for pointing this out to us.} Let us focus on the case $d=3$ and denote the Riemann normal coordinates at $r=0$ as $\{t,x,y,z\}$, such that $\dd s^2 = -\dd t^2 + \dd x^2 + \dd y^2 + \dd z^2$ at $r=0$ with deviations of order $\mathcal{O}(r^2)$ due to the assumed regularity. Then, by rotational symmetry around $r=0$ one has $C_{txtx} = C_{tyty} = C_{tztz}$. Since the Weyl tensor is tracefree and $C_{tttt} = 0$ by symmetry, this implies that $C_{txtx}=0$ at $r=0$. A similar argument follows from the $xx$-component of the tracefree condition, $C_{xyxy}=C_{xzxz}=0$, and so on. Last, the rotational symmetry forbids non-vanishing components where a spatial index appears only once, $C_{txty}$, which hence have to vanish as well. Consequently, for a smooth geometry at $r=0$, the Weyl tensor has to vanish at $r=0$. Similar considerations hold true in higher-dimensional spacetimes, but only for $p=0$ since the argument employs spherical symmetry.

We cannot help to notice the conceptual similarity to fixed points in quantum field theory, where order parameters diverge and the physical situation no longer has a scale: since a zero Weyl tensor implies conformal invariance, this also seems to be the case as one approaches non-local gravitational objects. However, one should be careful with taking this result all too seriously: It may very well be that close to the gravitational object, although regular and well-behaved, the linear theory is no longer sufficient. It remains to be seen what happens to these properties in the full, non-linear theory.

\subsubsection{A cosmic string}
\label{eq:ch3:cosmic-string-prelude}
Let us focus on the cosmic string of four-dimensional spacetime, $D=4$. Hence we have $m=2$ and $p=1$, and the metric functions take the form
\begin{align}
u(r) = 0 \, , \quad v(r) = 2\kappa\epsilon \mathcal{G}_2(r) \, .
\end{align}
Defining the Newtonian gravitational potential as
\begin{align}
\Phi_\text{N} := -\frac12\left( g{}_{\mu\nu}\xi{}^\mu\xi{}^\nu + 1 \right) = \frac12 u \, , \quad \ts{\xi} = \partial_t \, ,
\end{align}
we see that it vanishes for a cosmic string where $u=0$. Because cosmic strings are interesting objects \cite{Kibble:1976sj,Vilenkin:1981zs,Hindmarsh:1994re} of some hypothetical physical significance in cosmology, we will devote an entire section to the study of their gravitational field in linearized, non-local gravity after these more general examples for $p$-branes.

\subsubsection{A domain wall}
\label{sec:ch3:domain-wall}
Domain walls, per definition, have only one orthogonal direction, $m=1$, such that $p=D-2=d-1$ and the metric functions are
\begin{align}
u(r) = \frac{2\kappa\epsilon}{d-1} \mathcal{G}_1(r) \, , \quad v(r) = \frac{2d\kappa\epsilon}{d-1}  \mathcal{G}_1(r) \, .
\end{align}
The resulting metric, irrespective of the underlying gravitational theory, is conformally flat because the introduction of a new radial variable $y'$ according to
\begin{align}
\sqrt{1+v(y)}\dd y = \sqrt{1+u(y')}\dd y'
\end{align}
transforms the metric into $g{}_{\mu\nu} = (1+u)\eta{}_{\mu\nu}$. This can also be seen at the level of the vanishing of the Weyl tensor by inserting the one-dimensional identity 
\begin{align}
\partial_i \partial_j v = \delta{}_{ij} \lap v
\end{align}
together with $m=1$ into the expressions \eqref{eq:ch3:weyl-1}--\eqref{eq:ch3:weyl-3}. In the context of General Relativity the vanishing of the Weyl tensor has far-reaching implications: since the Ricci tensor is algebraically linked to the energy-momentum tensor by means of the field equations, spacetime outside of the domain wall is locally flat. This is not true in non-local ghost-free gravity, where the Ricci curvature is non-zero in the domain wall's vicinity, before decreasing at large radial distances and approaching the zero value encountered in linearized General Relativity asymptotically.

\subsubsection{``Angle deficit configurations''}
As we saw in the section on the cosmic string, whenever $m=2$ the Newtonian gravitational potential vanishes, $\Phi_\text{N} = 0$. In higher dimensions we may still have $m=2$, but the brane itself may be higher-dimensional and no longer correspond to a string because $p=D-3=d-2$. The metric functions take the form
\begin{align}
u(r) = 0 \, , \quad v(r) = 2\kappa\epsilon \mathcal{G}_2(r) \, .
\end{align}
The metric functions are identical with that of a cosmic string, but the resulting metric is not due to the presence of a different dimensionality of the brane. As is well known, cosmic strings produce an angle deficit \cite{Kibble:1976sj,Vilenkin:1981zs,Hindmarsh:1994re}. We may introduce higher-dimensional equivalents of this scenario and refer to them as ``angle deficit configurations,'' and we define them as spacetimes with $m=2$.

Then, looking at the $y^1 y^2$-sector alone (that is, imagine you are on the brane and keep your position fixed), the transverse geometry is two-dimensional. That geometry possesses an angle deficit around the point $y^1=y^2=0$, except that $\{y^1=y^2=0\}$ no longer describes a line in space (or a sheet in spacetime) as in the case with the cosmic string, but a $(D-3)$-brane.

\section{Geometry of a cosmic string in non-local gravity}
Using the results of Sec.~\ref{eq:ch3:cosmic-string-prelude}, the metric of a cosmic string in four dimensions, in the linear approximation, may be written as (compare also Ref.~\cite{Boos:2020kgj} and Kolar \& Mazumdar \cite{Kolar:2020bpo})
\begin{align}
\dd s^2 &= -\dd t^2 + \dd z^2 + v(\rho) (\dd x^2 + \dd y^2) = -\dd t^2 + \dd z^2 + \left[1+v(\rho)\right]\left( \dd\rho^2 + \rho^2 \dd \varphi^2 \right) \, , \\
v(\rho) &= 2\kappa\mu \mathcal{G}_2(\rho) \, .
\end{align}
where the string extends along the $z$-axis, and we introduced polar coordinates $\{\rho,\varphi\}$ such that $x=\rho\cos\varphi$ and $y=\rho\sin\varphi$. We have also replaced $\epsilon$ by $\mu$, which is the symbol more frequently employed to denote the string line density (or string tension). Concretely, using Sec.~\ref{sec:ch2:static}, we find in linearized General Relativity and linearized $\mathrm{GF_1}$ theory
\begin{alignat}{5}
\text{GR} : && \quad v(\rho) &= -8G\mu \ln\left( \frac{\rho}{\rho_0} \right) \, , \\
\mathrm{GF_1} : && \quad v(\rho) &= +4G\mu \left[ \text{Ei}\left(-\frac{\rho^2}{4\ell^2}\right) - 2\ln\left(\frac{\rho}{\rho_0}\right) \right] \, .
\end{alignat}
Here, ``$\text{Ei}$'' denotes the exponential integral \cite{Olver:2010}
\begin{align}
\text{Ei}(-x) = -E_1(x) = -\int\limits_x^\infty \dd z \frac{e^-z}{z} \, , \quad x > 0 \, .
\end{align}

\subsection{No distributional curvature}
The Ricci scalar of this geometry is
\begin{align}
R = -\lap v = -2\kappa\mu \lap\mathcal{G}_2(\rho) = 4\kappa\mu K_2(\rho|\ell) \, , \quad K_2(\rho|\ell) = \frac{1}{4\pi\ell^2} e^{-\tfrac{\rho^2}{4\ell^2}} \, ,
\end{align}
where $K_2(\rho|\ell)$ is heat kernel of two dimensional space, satisfying $K_2(\rho|\ell\rightarrow 0) = \delta(\rho)/2\pi\rho$. Unlike in General Relativity, the Ricci curvature is not distributional, but smoothly distributed around the $z$-axis. Only in the limiting case $\ell\rightarrow 0$ it becomes sharply concentrated. The Ricci tensor and Weyl tensor are also non-vanishing in the non-local case, with the quadratic curvature invariants taking the form
\begin{align}
C^2 = \frac13 (\lap v)^2 + 0 \times \left(\partial_i\partial_j v\right)\left(\partial^i \partial^j v\right) \, , \quad \cancel{R}^2 = (\lap v)^2 \, .
\end{align}
The tensorial part does not contribute in the case $m=2$, and hence the only relevant functions describing the curvature are the two-dimensional heat kernel and its square.

At far distances, however, when $\rho/\ell \rightarrow \infty$, the heat kernel approaches zero and one recovers the standard locally flat spacetime describing a cosmic string as already found in General Relativity. This does not come as a surprise but is an expected result that is inherited from the properties of static, non-local Green functions as discussed in Sec.~\ref{sec:ch2:static}.

\subsection{Angle deficit}
At the linear level, let us now prove that cosmic strings indeed mediate an angle deficit around the $z$-axis (for positive $\mu$). Defining the angle deficit as
\begin{align}
\delta\varphi = 2\pi - \frac{C(\rho)}{R(\rho)} \, ,
\end{align}
where $C(\rho)$ is the proper circumference of a circle with coordinate radius $\rho$, and $R(\rho)$ is the proper radius. At linear order in $\mu$ they take the form
\begin{align}
C(\rho) = \left[1 + \frac{v(\rho)}{2}\right] 2\pi \rho \, , \quad
R(\rho) = \int\limits_0^\rho \dd\rho' \left[1 + \frac{v(\rho')}{2}\right]
\end{align}
The integral can be taken analytically for $\mathrm{GF_1}$ theory and at leading order one finds
\begin{align}
\delta\varphi(\rho) = 8\pi G \mu \left[ 1-\frac{\sqrt{\pi}\ell\text{erf}\left(\frac{\rho}{2\ell}\right)}{\rho} \right]
\end{align}
Unlike in linearized General Relativity, the angle deficit is now a function of the radial distance $\rho$. At large distances or vanishing scale of non-locality, $\rho/\ell\rightarrow\infty$, one recovers the standard result from linearized General Relativity, $\delta\varphi = 8\pi G \mu$. For small distances, however, the behavior is very different, and at $\rho=0$ the angle deficit even vanishes,
\begin{align}
\lim\limits_{\rho\rightarrow 0} \delta\varphi(\rho) = 0 \, .
\end{align}
Because the angle deficit grows quadratically at the origin, and not linearly, there is some hope that the resulting geometry will be regular at the origin $\rho=0$. In order to understand that better, let us isometrically embed the spacelike surface described by $\dd t = \dd z = 0$.

\subsection{A tale of two cones}
The two-dimensional spacelike surface $\{\dd t = \dd z = 0\}$, call it $\Sigma$, is parametrized by the line element
\begin{align}
\dd\Sigma^2 = \left[1+v(\rho)\right]\left( \dd\rho^2 + \rho^2 \dd \varphi^2 \right) \, ,
\end{align}
and it is rotationally symmetric. This means, provided $\dd v/\dd\rho \leq 0$, it is possible to find a coordinate $z=z(\rho')$ such that this line element can be embedded in $\mathbb{R}^3$,
\begin{align}
\dd\Sigma_{\text{3D}}^2 = \dd z^2 + \dd\rho'^2 + \rho'^2\dd\varphi^2 \, .
\end{align}
To find the appropriate transformation one can fix the $\varphi$-sector by requiring
\begin{align}
\label{eq:ch3:rho-rhoprime}
\rho'^2 = \left[1+v(\rho)\right]\rho^2 \, ,
\end{align}
which is an implicit equation that gives $\rho = \rho(\rho')$. At the linear level one may approximate
\begin{align}
\rho' = \left[1+\frac{v(\rho)}{2}\right]\rho \, .
\end{align}
Then we fix the function $z(\rho')$ by identifying
\begin{align}
\left[1+v(\rho)\right]\dd\rho^2 = \left[ 1 + \left( \frac{\dd z}{\dd \rho'} \right)^2 \right] \dd\rho'^2 = \dd \rho'^2 + \dd z^2 \, .
\end{align}
Integrating this relation and using the intermediate formula
\begin{align}
\left[1+v(\rho)\right]\dd\rho^2 = \left[ 1 - v'(\rho)\rho \right] \dd\rho'^2
\end{align}
we obtain, to leading order in $v$,
\begin{align}
z(\rho') &= \int\limits_0^{\rho'} \dd\tilde{\rho}' \sqrt{-\frac{\dd v(\rho)}{\dd\rho}\rho}
= \sqrt{8\mu G}\int\limits_0^{\rho'} \dd\tilde{\rho}' \sqrt{4\pi^2 \mathcal{G}_4(\rho)\rho^2} \, .
\end{align}
where we have made use of the recursion formula \eqref{eq:ch2:recursion}, and $\rho=\rho(\tilde{\rho}')$ by means of \eqref{eq:ch3:rho-rhoprime}. For $\mathrm{GF_1}$ theory one obtains the explicit expression
\begin{align}
z(\rho') &= \sqrt{8\mu G} \int\limits_0^{\rho'} \dd\tilde{\rho}' \sqrt{1 - \exp\left[-\frac{\rho^2}{4\ell^2}\right]} \, .
\end{align}
This integral is very difficult to treat analytically due to the implicit dependence of the integrand on $\tilde{\rho}'$, which is why we implemented a numerical method. First, it given a value $\rho'$ a numerical root solver finds the appropriate $\rho$. In a second step, during the numerical integration, this function is called at every point of the integration. Since $\rho/\ell \ll 1$ in our example the effect is very small and we trust our numerics.

\begin{figure}[!htb]
\centering
\subfloat[Linearized General Relativity.]
{
    \includegraphics[width=0.49\textwidth]{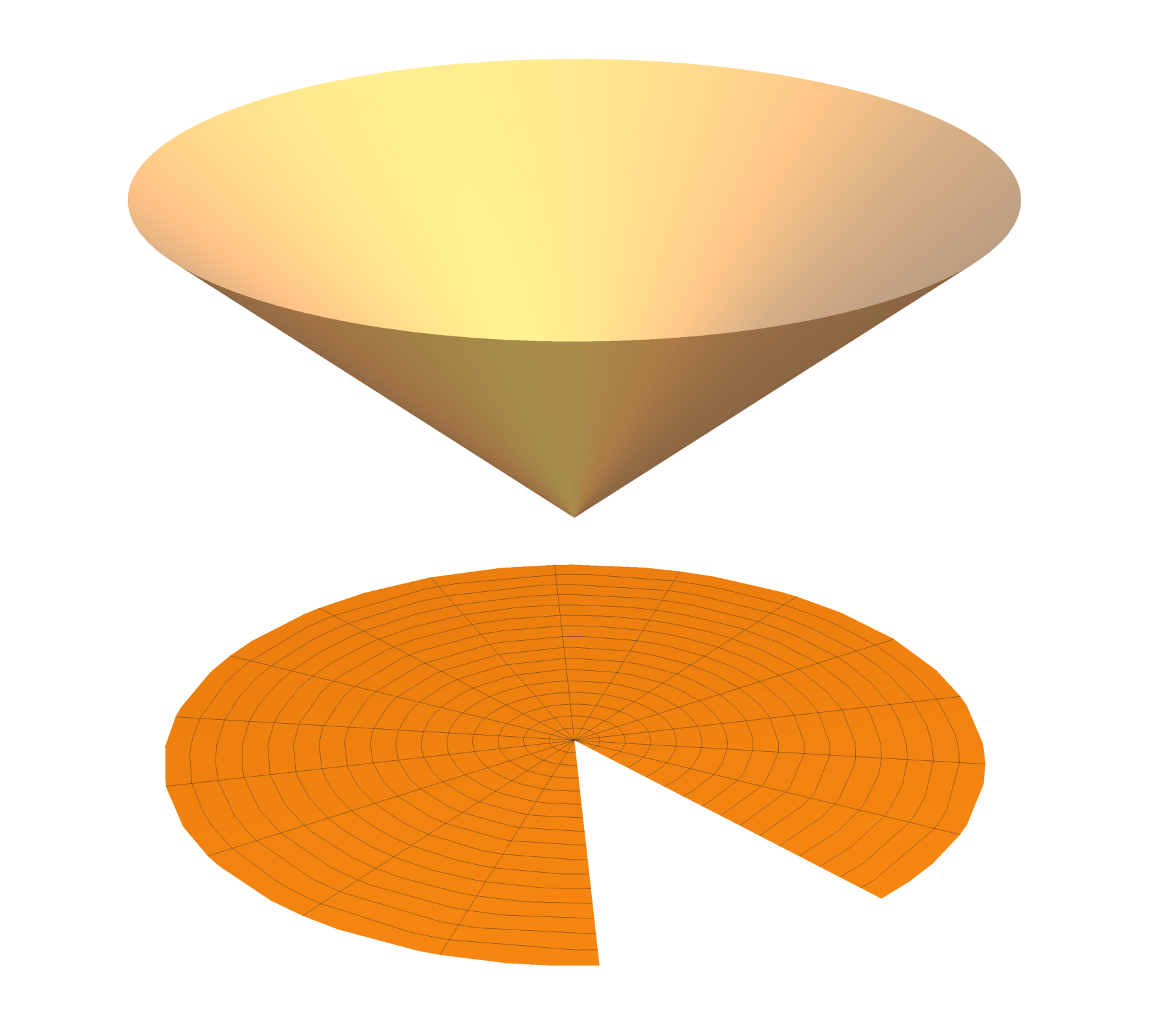}
}
\subfloat[Linearized non-local gravity.]
{
    \includegraphics[width=0.49\textwidth]{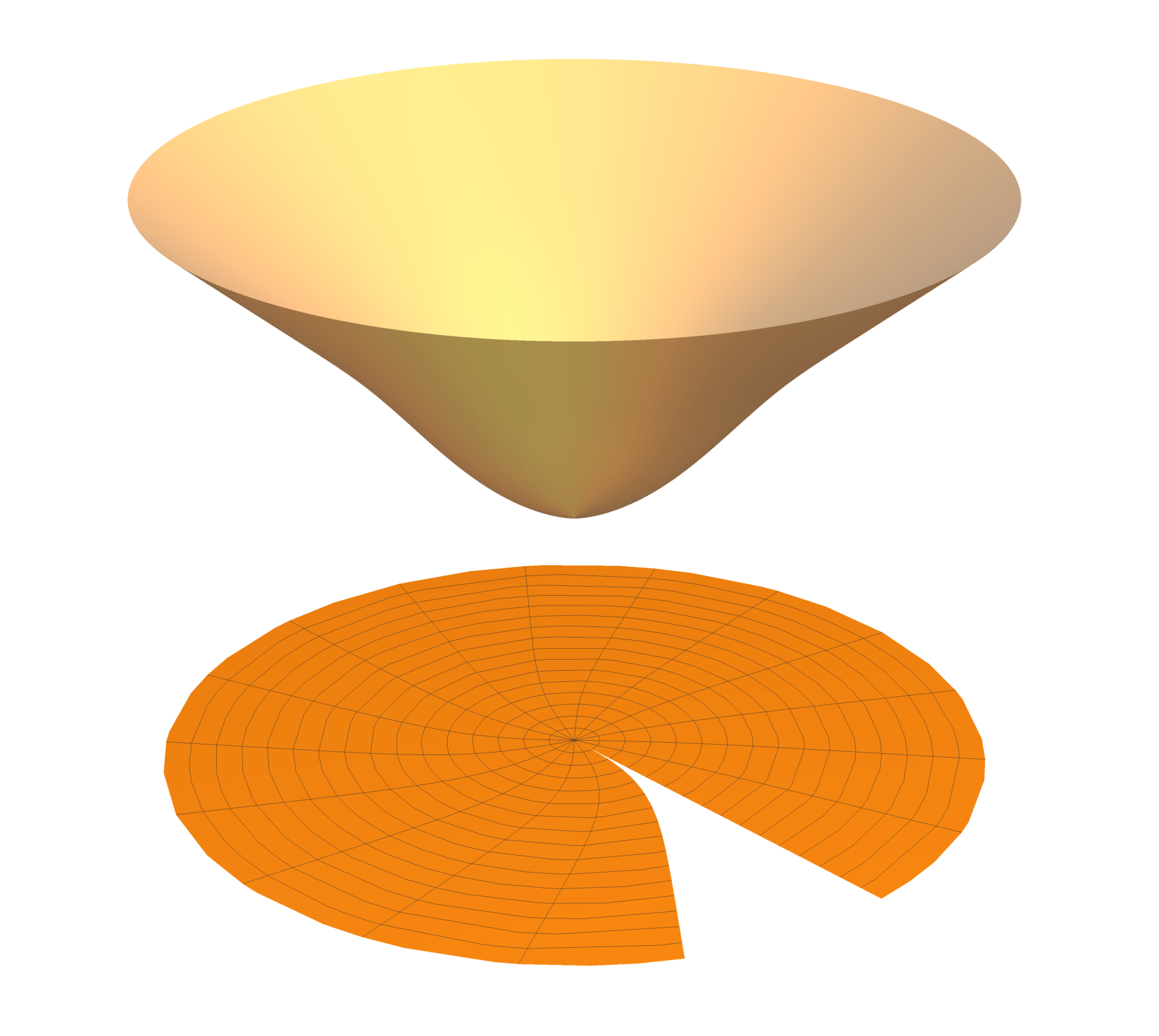}
}
\caption[Transverse spatial geometry of a cosmic string.]{We display an isometric embedding of the transverse geometry of a cosmic string in three-dimensional Euclidean space for (a) General Relativity and (b) non-local $\mathrm{GF_1}$ theory. Whereas this geometry is a cone for General Relativity, it becomes a smoothed cone in the ghost-free theory. Below, we show the ``unrolled geometry,'' which, in the case of General Relativity, is simply a plane with a constant angle deficit. In case of $\mathrm{GF_1}$ theory it is instead a plane with a growing angle deficit that asymptotically becomes constant. We used the parameters $8\mu G=0.7, \ell=0.3\rho_0$, and the circle's coordinate radius is chosen to be $2\rho_0$. }
\label{fig:cones}
\end{figure}

In the limiting case of $\ell\rightarrow 0$ the entire situation becomes much simpler: one may simply perform a coordinate transformation $\rho\rightarrow\rho''$ \cite{Vilenkin:1981zs},
\begin{align}
(1-8\mu G)\rho''^2 = \left[1-8G\mu\ln\left(\frac{\rho}{\rho_0}\right)\right]\rho^2 \, ,
\end{align}
that renders the geometry, at leading order in $\mu G$, to be flat space with a conical deficit:
\begin{align}
\dd s^2 = \dd\rho''^2  + \rho''^2(1-8\mu G)\dd\varphi^2 \approx \dd\rho''^2  + \rho''^2 \left[(1-4\mu G)\dd\varphi\right]^2 \, .
\end{align}
The embedding function $z(\rho'')$ can be found by elementary trigonometry and for small $\delta\varphi\ll 1$ is
\begin{align}
z(\rho'') = \sqrt{\frac{\delta\varphi}{\pi}}\rho'' = \sqrt{8\mu G} \rho'' \, .
\end{align}
We visualize these two geometries in Fig.~\ref{fig:cones}, along with the ``unrolled'' geometry that is either a plane with a wedge cut out (General Relativity) or a plane with a slowly growing angle deficit ($\mathrm{GF_1}$ theory). The resulting geometry in non-local $\mathrm{GF_1}$ theory is that of a cone with smoothed tip, with a curvature radius of $\mathcal{O}(\ell)$.

\section{Stationary rotating objects (general solution)}
Let us now move on to stationary rotating objects. As already mentioned in Sec.~\ref{sec:ch3:sources}, the form of the energy-momentum tensor is
\begin{align}
\tag{\ref*{eq:ch3:tmunu-stationary}}
T{}_{\mu\nu} = \rho(\ts{x})\delta{}^t_\mu\delta{}^t_\nu + \delta{}^t_{(\mu}\delta{}^\alpha_{\nu)} \frac{\partial}{\partial x{}^\beta} j{}_\alpha{}^\beta(\ts{x}) \, ,
\end{align}
where $\rho(\ts{x})$ is the time-independent matter density, and $j{}_{\alpha\beta}(\ts{x}) = -j{}_{\beta\alpha}(\ts{x})$ is the time-independent angular momentum density. Its indices are purely spatial, meaning that in the $D$-dimensional language one may write
\begin{align}
j{}_{\mu\nu}(\ts{x}) = \begin{pmatrix} 0 & 0 & \dots & 0 \\ 0 & & & \\ \vdots & & j{}_{\alpha\beta}(\ts{x}) & \\ 0 & & & \end{pmatrix} \, .
\end{align}
This form implies the absence of boost charges, consistent with the stationarity of the physical scenario. As a metric ansatz we parametrize
\begin{align}
\dd s^2 = -\left[1+\phi(\ts{x})\right]\dd t^2 + \left[1+\psi(\ts{x})\right] \dd \ts{x}^2 + 2 A{}_\alpha(\ts{x}) \dd x{}^\alpha \dd t
\end{align}
Here, $\phi(\ts{x})$ and $\psi(\ts{x})$ are two static potentials, and $A{}_\alpha(\ts{x})$ is a gravitomagnetic potential that encompasses the stationary rotation. Inserting this ansatz into the stationary field equations \eqref{eq:ch3:eom-stationary} for the choice $a(\Box) = c(\Box)$ yields
\begin{align}
(d-1)a(\lap)\lap\psi &= -2\kappa \rho(\ts{x}) \, , \\
(\delta_{ij}\lap - \partial_i\partial_j) a(\lap) \left[(2-d)\psi - \phi\right] &= 0 \, , \\
a(\lap)\left( \lap A{}_\alpha - \partial{}_\alpha \partial{}^\beta A{}_\beta \right) &= -\kappa \partial{}_\beta j{}_\alpha{}^\beta(\ts{x}) \, .
\end{align}
At the linear level it is helpful to perform the infinitesimal coordinate transformation $t \rightarrow t + f(\ts{x})$ which induces the transformation
\begin{align}
A_\alpha \rightarrow A_\alpha - \partial_\alpha f(\ts{x}) \, .
\end{align}
We can find a suitable function $f$ to impose the \emph{Lorenz gauge condition} $\partial^\alpha A{}_\alpha = 0$ such that the field equations become
\begin{align}
(d-1)a(\lap)\lap\psi &= -2\kappa \rho(\ts{x}) \, , \\
(\delta_{ij}\lap - \partial_i\partial_j) a(\lap) \left[(2-d)\psi - \phi\right] &= 0 \, , \\
a(\lap) \lap A{}_\alpha &= -\kappa \partial{}_\beta j{}_\alpha{}^\beta(\ts{x}) \, .
\end{align}
Invoking the Lorenz gauge condition has the substantial advantage that we may also utilize the static Green function method to find a solution for $A{}_\alpha(\ts{x})$. The homogeneous equation, as in the case of the point particle already discussed above, is solved by
\begin{align}
\psi(\ts{x}) = \frac{1}{2-d} \phi(\ts{x}) \, .
\end{align}
Then, the remaining inhomogeneous equations have the solutions
\begin{align}
\label{eq:ch3:phi-psi-rotating}
\phi(\ts{x}) = -2\kappa \frac{d-2}{d-1} \int\dd^d y \, \rho(\ts{y}) \mathcal{G}_d(\ts{x}-\ts{y}) \, , \quad
A_\alpha(\ts{x}) = \kappa \int \dd^d y \, j{}_\alpha{}^\beta(\ts{y}) \frac{\partial \mathcal{G}_d(\ts{x}-\ts{y})}{\partial x{}^\beta} \, .
\end{align}
Recall that the static Green function $\mathcal{G}_d(\ts{x})$ is only a function of the absolute value of its argument, $\mathcal{G}_d(|\ts{x}|)$. While the expression is straightforward for $\phi(\ts{x})$, let us derive the expression for $A_\alpha(\ts{x})$ explicitly by formally inverting the differential equation:
\begin{align}
A_\alpha(\ts{x}) &= -\frac{\kappa}{a(\lap)\lap} \frac{\partial}{\partial x{}^\beta} j{}_\alpha{}^\beta(\ts{x}) = -\kappa \frac{\partial}{\partial x{}^\beta} \frac{1}{a(\lap)\lap} j{}_\alpha{}^\beta(\ts{x}) \\
&= -\kappa \int\dd^d y \, j{}_\alpha{}^\beta(\ts{y}) \frac{\partial}{\partial x{}^\beta} \left[ \frac{1}{a(\lap)\lap} \delta(\ts{x}-\ts{y}) \right] \\
&= +\kappa \int\dd^d y \, j{}_\alpha{}^\beta(\ts{y}) \frac{\partial \mathcal{G}_d(\ts{x}-\ts{y})}{\partial x{}^\beta}
\end{align}
At this point it is also straightforward to verify that $A{}_\alpha$ indeed satisfies the Lorenz gauge condition,
\begin{align}
\partial{}^\alpha A_\alpha(\ts{x}) = \kappa \int \dd^d y \, j{}^{\alpha\beta}(\ts{y}) \frac{\partial^2 \mathcal{G}_d(\ts{x}-\ts{y})}{\partial x{}^\alpha \partial x{}^\beta} = 0 \, ,
\end{align}
where the last identity follows from the antisymmetry of $j{}_{\alpha\beta}(\ts{x})$.

Making use of the recursion formula \eqref{eq:ch2:recursion} we can rewrite the expression of the gravitomagnetic potential in terms of a higher-dimensional static Green function,
\begin{align}
A_\alpha(\ts{x}) = -2\pi\kappa \int \dd^d y \, j{}_{\alpha\beta}(\ts{y}) (x^\beta-y^\beta) \mathcal{G}_{d+2}(\ts{x}-\ts{y}) \, .
\end{align}
It is sometimes also useful to work with the gravitomagnetic potential 1-form defined via
\begin{align}
\label{eq:ch3:a:1-form}
\ts{A}(\ts{x}) = A{}_\alpha(\ts{x}) \dd x{}^\alpha \, .
\end{align}
With these expressions given in a general form we may now construct stationary rotating gravitational fields in the weak-field limit of non-local ghost-free gravity.

\section{Angular momentum in higher dimensions}
Let us briefly discuss the properties of the angular momentum tensor density $j{}_{\alpha\beta}(\ts{x}) = -j{}_{\beta\alpha}(\ts{x})$. As a purely spatial antisymmetric tensor of rank 2 we can also think of it as a 2-form
\begin{align}
\ts{j} = \frac12 j{}_{\alpha\beta}(\ts{x}) \dd x{}^\alpha \wedge \dd x{}^\beta \, .
\end{align}
Only in three spatial dimensions 2-forms are dual to 1-forms, which in turn are equivalent to vectors if a metric is present. For this reason a 2-form is the appropriate generalization of the angular momentum pseudo-vector to higher dimensions.

\begin{figure}
  \centering
  \includegraphics[width=0.5\textwidth]{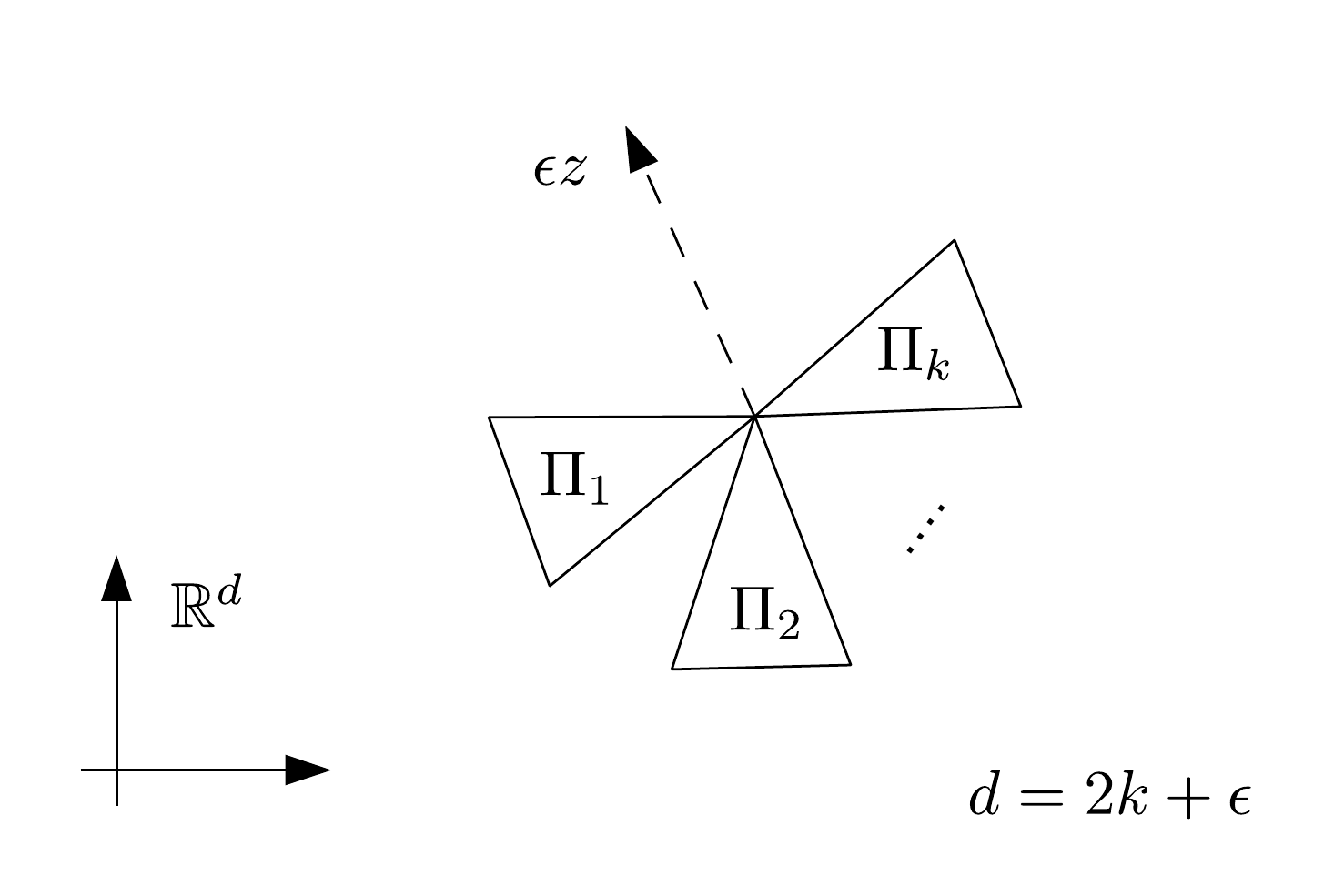}
  \caption[Darboux decomposition of $d$-dimensional space.]{In the Darboux decomposition of $d$-dimensional Euclidean space there are $k$ orthogonal Darboux planes $\Pi_A$ labelled by $A=1,\dots,k$ where $d=2k+\epsilon$. The additional $z$-direction only exists if $d$ is odd, or, equivalently, when $\epsilon=1$. In the above diagram we visualize the Darboux planes by wedges and the $z$-axis by a dashed line.}
  \label{fig:ch3:darboux-1}
\end{figure}

Any antisymmetric matrix can be brought to its so-called \emph{Darboux form} and for more details one may consult the literature; see e.g.~\cite{Greub:1981,Dreiner:2008tw} and references therein. Here we will sketch the main idea of the Darboux decomposition as applied to the antisymmetric matrix $j{}_{\alpha\beta} = -j{}_{\beta\alpha}$, where $\alpha,\beta$ are spatial indices and may be raised and lowered at will using a flat metric. In even dimensions, that is, when $d=2k$ with $k\in\mathbb{N}$, the Darboux theorem states that for this matrix there exist pairs of vectors $\{m_A^\alpha, \hat{m}_A^\alpha\}$ with $A=1,\dots,k$ such that
\begin{align}
j{}^\alpha{}_\beta m{}^\beta_A = +j_A \hat{m}^\alpha_A \, , \quad j{}^\alpha{}_\beta \hat{m}{}^\beta_A = -j_A m^\alpha_A \, , \quad j_A \ge 0 \, .
\end{align}
In odd dimensions, $d=2k+\epsilon$, there exists one additional vector $n^\alpha$ such that
\begin{align}
j{}^\alpha{}_\beta n{}^\beta = 0 \, .
\end{align}
We can show that these vectors $\{m_A^\alpha, \hat{m}_A^\alpha,\epsilon n{}^\alpha\}$ are orthogonal. To that end, introduce a symmetric matrix $\ts{Q} = \ts{j}{}^T \ts{j}$ such that
\begin{align}
Q{}^\alpha{}_\beta &= j{}_\beta{}^\gamma j{}^\alpha{}_\gamma \, , \\
j{}^\alpha{}_\beta m{}^\beta_A &= -j_A \, j{}^\alpha{}_\gamma \hat{m}^\gamma_A = j_A^2 m{}^\alpha_A \, , \\
j{}^\alpha{}_\beta \hat{m}{}^\beta_A &= +j_A \, j{}^\alpha{}_\gamma m^\gamma_A = j_A^2 \hat{m}{}^\alpha_A \, .
\end{align}
That is, for each $A$ the vectors $\{m_A^\alpha, \hat{m}_A^\alpha\}$ are eigenvectors of $Q$ with the same eigenvalue. Moreover, the above relations imply that $m_A^\alpha$ and $\hat{m}_A^\alpha$ are orthogonal for the same $A$ (provided that $j_A \not= 0$):
\begin{align}
m{}^\alpha_A \hat{m}_{A\alpha} = -\frac{1}{j_A^2} j{}^\alpha{}_\beta \hat{m}^\beta_A j{}_\alpha{}^\gamma m{}_{A\gamma} = -\frac{1}{j_A^2} Q{}^\gamma{}_\beta \hat{m}^\beta_A m{}_{A\gamma} = -\hat{m}{}^\gamma_A m{}_{A\gamma} = -m{}^\alpha_A \hat{m}{}_{A\alpha} \, .
\end{align}
The overlap between $n{}^\alpha$ and any of the $\{m_A^\alpha, \hat{m}_A^\alpha\}$ also vanishes if $j_A\not=0$. As a result, the antisymmetric matrix $j{}_{\alpha\beta}$ can always be brought to the form
\begin{align}
\ts{j}&\hat{=} \begin{pmatrix}
0 & j_1 & & & \dots & & & 0 \\
-j_1 & 0 & & & & & & \\
& & 0 & j_2 & & & & & \\
& & -j_2 & 0 & & & & \\
\vdots & & & & \ddots & & & \\
& & & & & 0 & j_k & \\
& & & & & -j_k & 0 & \\
0 & & & & & & & 0
\end{pmatrix} \, ,
\end{align}
where the last line, entirely composed of zeros, only exists in odd dimensions when $\epsilon=1$. Given any such matrix $j{}_{\alpha\beta}$ there exists a coordinate system in which the matrix takes the above form, and our previous considerations show that this coordinate system is indeed spanned by the vectors $\{m{}^\alpha_A,\hat{m}^\alpha_A, \epsilon n{}^\alpha\}$. Let us define coordinates $\{y_A, \hat{y}_A, \epsilon z\}$ such that
\begin{align}
m{}^\alpha_A \partial_\alpha = \partial_{y_A} \, , \quad 
\hat{m}{}^\alpha_A \partial_\alpha = \partial_{\hat{y}_A} \, \quad
n{}^\alpha \partial_\alpha = \partial_z \, .
\end{align}
Moreover, we define a \emph{Darboux plane} $\Pi_A$ to be the two-dimensional surface spanned by two vectors $\{m{}^\alpha_A,\hat{m}^\alpha_A\}$ such that
\begin{align}
\Pi_A = \text{span}\left\{\partial_{y_A} , \partial_{\hat{y}_A} \right\} \, .
\end{align}
It is clear that Darboux planes are only defined up to an orthogonal transformation (that is, a rotation or a reflection) inside that plane. Provided we choose the orientation in a right-handed fashion, the volume element in each Darboux plane is
\begin{align}
\epsilon{}^A_{\alpha\beta} = 2 \delta{}^A_{[\alpha} \delta{}^{\hat{A}}_{\beta]} \, , \quad \left( \epsilon{}^A_{\alpha\beta} \right) = \begin{pmatrix} ~0 & 1 \\ -1 & 0 \end{pmatrix} \, .
\end{align}
Collecting these results we can write the antisymmetric matrix $j{}_{\alpha\beta}$ as
\begin{align}
j{}_{\alpha\beta} = \sum\limits_{A=1}^k j_A \epsilon{}^A_{\alpha\beta} \, .
\end{align}
For a convenient visual depiction of Darboux planes in $d$-dimensional space see Fig.~\ref{fig:ch3:darboux-1}.

To make the rotational ambiguity in each of the Darboux planes manifest, we may also introduce polar coordinates $\{\rho_A,\varphi_A\}$ in each plane such that
\begin{align}
y_A = \rho_A\cos\varphi_A \, , \quad \hat{y}_A = \rho_A\sin\varphi_A \, .
\end{align}
It is important to emphasize that these discussions concern an antisymmetric matrix with constant coefficients. If we are instead interested in an antisymmetric tensor field of rank 2, we may of course still perform the same Darboux decomposition, but there is no guarantee that the notion of Darboux planes remains the same as one moves through space. In other words: a general treatise of an angular momentum density would be very interesting, but must necessarily involve additional assumptions on the spatial behavior of that density and can therefore not be purely algebraic in nature. In the context of symplectic mechanics this is usually phrased as ``Darboux theorem.'' We will revisit this issue in Ch.~\ref{ch:ch4} when discussing extended ``pencils'' of matter, but for now we shall restrict ourselves to a simpler case of specially aligned angular momentum densities where the tensorial structure decouples from the spatial dependence.

\section{Spinning point particles}
If the tensorial and spatial structures factorize, the angular momentum density takes the form
\begin{align}
j{}_{\alpha\beta}(\ts{x}) = j{}_{\alpha\beta} f(\ts{x}) \, ,
\end{align}
where $f(\ts{x})$ is a dimensionless function proportional to the matter density and $j{}_{\alpha\beta}$ is a constant matrix that describes the angular momentum density. While we may not be able to describe all rotating matter with the above factorized angular momentum density it is general enough to construct the gravitational field of point particles with intrinsic angular momentum as well as strings and higher-dimensional $p$-branes. In the case of localized ``thin'' objects the matter density $\rho(\ts{x})$ becomes a $\delta$-function which leads to further simplifications because then, as we will see, the solutions are directly proportional to the relevant static Green functions.

Let us finish this introduction by introducing a few helpful definitions. First, it is useful to introduce a $d$-dimensional radial variable according to
\begin{align}
r^2 = \varepsilon z^2 + \sum\limits_{A=1}^k \rho_A^2 \, .
\end{align}
The gravitomagnetic potential includes terms of the form $j{}_{\alpha\beta}x^\alpha\dd x{}^\beta$, which, in the Darboux formalism combined with polar Darboux coordinates, take a rather simple form:
\begin{align}
j{}_{\alpha\beta}x{}^\alpha\dd x{}^\beta = \sum\limits_{A=1}^k j_A \epsilon{}^A_{\alpha\beta}x{}^\alpha\dd x{}^\beta = \sum\limits_{A=1}^k j_A \rho_A^2 \dd\varphi_A \, .
\end{align}
We are now fully equipped to study concrete examples of rotating objects in linearized, non-local ghost-free gravity.

\subsection{Spinning point particle in four spacetime dimensions}
Let us begin with the simplest example: a point particle in four spacetime dimensions. In $d=3$ spatial dimensions one has $k=1$ and $\varepsilon=1$, which means that there exists one Darboux plane for the angular momentum and one independent $z$-axis. Because there exists only one Darboux plane we shall suppress the Darboux plane index in what follows, use the coordinates $x$ and $y$ inside the Darboux plane, and call the angular momentum eigenvalue $j$. Then, the angular momentum density takes the simple form
\begin{align}
j{}_{\alpha\beta}(\ts{x}) = 2j \delta{}^x_{[\alpha} \delta{}^y_{\beta]} \delta{}^{(3)}(\ts{x})
\end{align}
Inserting this expression into \eqref{eq:ch3:phi-psi-rotating} as well as \eqref{eq:ch3:a:1-form} one finds
\begin{align}
\dd s^2 &= -\left[1+\phi(r)\right]\dd t^2 + \left[1+\psi(r)\right] \dd \ts{x}^2 + 2 \ts{A}(\ts{x}) \dd t \, , \\
\phi(r) &= -\kappa m\mathcal{G}_3(r) \, , \quad
\psi(r) = \kappa m\mathcal{G}_3(r) \, , \quad
\ts{A}(\ts{x}) = 2\pi\kappa \mathcal{G}_5(r) j \rho^2 \dd\varphi \, .
\end{align}
The geometry of a slowly-rotating point particle in four-dimensional General Relativity, also called Thirring--Lense metric \cite{Thirring:1918,Lense:1918,Thirring:1921,Mashhoon:1984fj,Pfister:2007}, arises in the limit $\ell\rightarrow 0$ from this expression,
\begin{align}
\phi(r) &= -\frac{2Gm}{r} \, , \quad
\psi(r) = \frac{2Gm}{r} \, , \quad
\ts{A}(\ts{x}) = \frac{2Gj}{r} \sin^2\theta \dd\varphi \, .
\end{align}
where we inserted $\rho = r\sin\theta$, and $\theta$ is the standard polar angle. This metric also corresponds to the linearized limit of the well-known Kerr metric describing a rotating black hole \cite{Kerr:1963ud,Frolov:2011}.

\subsection{Spinning point particle in higher dimensions}
\label{sec:ch3-rotating-particle-higher-dimensions}
As seen above, a particle in $d=2k+\epsilon$ spatial dimensions has $k$ independent Darboux planes $\Pi_A$ and hence $k$ independent angular momentum eigenvalues $j_k$. Its energy-momentum tensor is \eqref{eq:ch3:tmunu-stationary} with the angular momentum density
\begin{align}
j{}_{\alpha\beta}(\ts{x}) = \delta{}^{(d)}(\ts{x}) \sum\limits_{A=1}^k j_A \epsilon{}^A_{\alpha\beta}
\end{align}
and as per Eqs.~\eqref{eq:ch3:phi-psi-rotating} and \eqref{eq:ch3:a:1-form} the resulting metric is
\begin{align}
\dd s^2 &= -\left[1+\phi(r)\right]\dd t^2 + \left[1+\psi(r)\right] \dd \ts{x}^2 + 2 \ts{A}(\ts{x}) \dd t \, , \\
\phi(r) &= -2\kappa m \frac{d-2}{d-1} \mathcal{G}_d(r) \, , \quad
\psi(r) = \frac{2\kappa m}{d-1} \mathcal{G}_d(r) \, , \quad
\ts{A}(\ts{x}) = 2\pi\kappa \mathcal{G}_{d+2}(r) \sum\limits_{A=1}^k j_A \rho_A^2 \dd\varphi_A \, .
\end{align}
At large distances where the non-local modification no longer plays a role one recovers the linearized form of the Myers--Perry solution \cite{Myers:1986} for $d\ge3$:
\begin{align}
\phi(r) \sim \frac{\Gamma\left(\tfrac{d}{2}\right)}{(d-1)\pi^{\tfrac{d}{2}}} \frac{\kappa m}{r^{d-2}} \, , \quad
\psi(r) \sim -\frac{\phi(r)}{d-2} \, , \quad
\ts{A}(\ts{x}) \sim -\frac{\Gamma\left(\tfrac{d}{2}\right)}{2\pi^{\tfrac{d}{2}}} \frac{\kappa}{r^d} \sum\limits_{A=1}^k j_A \rho_A^2 \dd\varphi_A \, .
\end{align}

\section{Spinning strings and $p$-branes}
\label{sec:ch3:rotating-p-branes}
Before we discuss extended spinning objects, let us briefly clarify the nature of rotation treated in this section in order to avoid possible confusion. The angular momentum density $j{}_{\alpha\beta}$ can, in principle, be used to describe both \emph{orbital} and \emph{intrinsic} angular momentum. Orbital angular momentum is well known in classical mechanics, and a point particle at a position $\ts{r} = r_\alpha \dd x{}^\alpha$ and a linear momentum of $\ts{p} = p_\alpha \dd x{}^\alpha$ has the orbital angular momentum
\begin{align}
\ts{j} = \ts{r} \wedge \ts{p} \, .
\end{align}
In three spatial dimensions the above may be recast into the perhaps more familiar vectorial expression $\vec{j} = \vec{r} \times \vec{p}$. At any rate, if one considers extended, rotating objects that have non-vanishing orbital angular momentum one encounters the problem that at some point the local angular velocity will exceed the speed of light. Moreover, the translational symmetry parallel to that extended object will be broken due to the existence of a preferred ``center'' where the rotation axes pierce the object.

In these sections we are interested in purely \emph{intrinsic} angular momentum. One may think of the classical analogue of spin, or perhaps of a Weyssenhoff-type fluid with non-vanishing spin density \cite{Weyssenhoff:1947}. This type of angular momentum does not require the definition of a rotational center, consistent with our linearized weak-field ansatz.

There is another potential issue that arose when we integrated by parts in Sec.~\ref{sec:ch3:sources} while discussing the form of the stationary energy-momentum tensor in Eq.~\eqref{eq:ch3:tmunu-stationary}. In particular, we required that the following surface integral $\mathfrak{S}$ vanishes:
\begin{align}
\label{eq:ch3:am-surface-1}
\mathfrak{S} = \frac12 \int\dd^d x \frac{\partial}{\partial{} x{}^\gamma} \left( x{}_\alpha j{}_\beta{}^\gamma - x{}_\beta j{}_\alpha{}^\gamma \right) \, .
\end{align}
For compact point particles this is satisfied trivially since $j{}_{\alpha\beta} \sim \delta{}^{(d)}(\ts{x})$. For extended objects, however, it may no longer be the case. Consider, for example, a cosmic string along the $z$-axis: its angular momentum density certainly does not vanish as $z\rightarrow\pm\infty$.

\begin{figure}
  \centering
  \includegraphics[width=0.5\textwidth]{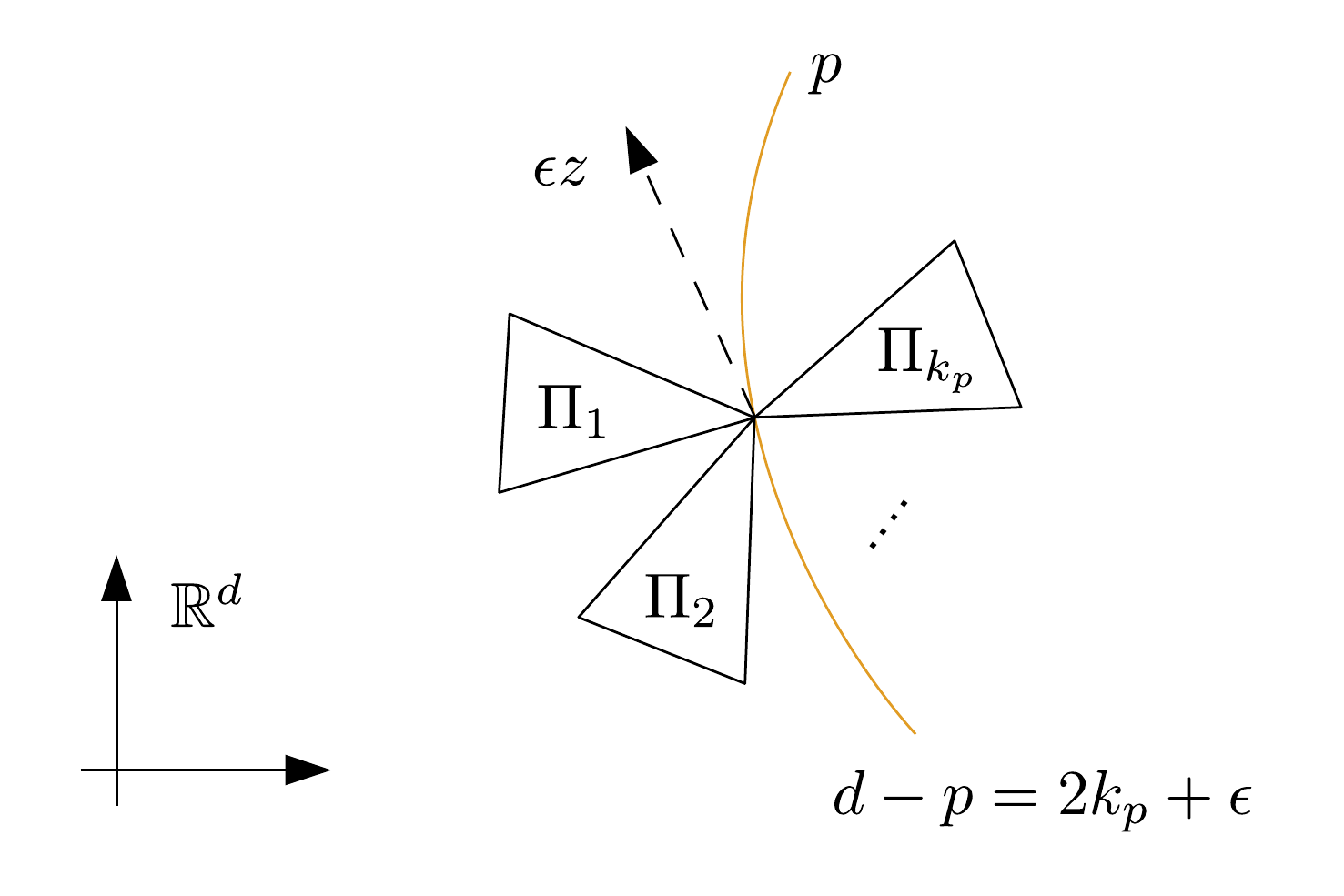}
  \caption[Darboux decomposition of $d$-dimensional space in presence of a $p$-brane.]{In the presence of a $p$-brane one may perform a Darboux decomposition of the $(d-p)$-dimensional transverse Euclidean space. Given a $p$, there are $k_p$ orthogonal Darboux planes $\Pi_A$ labelled by $A=1,\dots,k_p$ where $d-p=2k_p+\epsilon$. The additional $z$-direction only exists if $d-p$ is odd, or, equivalently, when $\epsilon=1$. In the above diagram we visualize the Darboux planes by wedges, the $z$-axis by a dashed line, and the $p$-brane by a curved line.}
  \label{fig:ch3:darboux-2}
\end{figure}

Let us briefly recall the notation employed in the description of $p$-branes. In $d=m+p$ spatial dimensions, a spatial coordinate $x{}^\alpha$ is split into $p$ directions $z^a$ along the $p$-brane and $m$ directions $y^i$ transverse to the brane, $x{}^\alpha = (z^a, y^i)$. In what follows we will assume that the angular momentum is aligned with the $p$-brane such that $j_{ab}=0$ and $j{}_{ij}\not=0$. In other words, we assume that the spatial part of the angular momentum density is block diagonal,
\begin{align}
j{}_{\alpha\beta}(\ts{x}) = \begin{pmatrix}
0 & \dots & 0 & 0 & \dots & 0 \\
\vdots & \ddots & \vdots  & & & \\
0 & \dots & 0 & & & \\
\vdots & & & & j{}_{ij}(\ts{x}) & \\
0 & & & & &
\end{pmatrix} \, .
\end{align}
For this reason the above surface term becomes
\begin{align}
\label{eq:ch3:am-surface-2}
\mathfrak{S} = \frac12 \int\dd^d x \frac{\partial}{\partial{} y{}^k} \left( y{}_i j{}_j{}^k - y{}_j j{}_i{}^k \right) = 0 \, .
\end{align}
The potentially diverging contributions stem from $y_i \rightarrow \infty$, that is, away from the $p$-brane which is located at $y^i = 0$. But away from the $p$-brane the angular momentum density vanishes, and hence $\mathfrak{S} = 0$. What if we instead had non-vanishing angular momentum density components $j_{ab}\not=0$? Then, looking at Eq.~\eqref{eq:ch3:am-surface-1}, there would also be contributions from $z^a\rightarrow\pm\infty$ irrespective of the values of $y^i$, leading to possible divergences. For these reasons we can still think of $j{}_{ij}$ as the proper angular momentum density such that for $p$-branes we find
\begin{align}
J{}_{ij} = \int\dd^p z \, j{}_{ij} \, .
\end{align}
We can also perform a Darboux decomposition of $j_{ij}$ in higher dimensions. Let us define
\begin{align}
d - p = m = 2k_p + \epsilon \, ,
\end{align}
and see Fig.~\ref{fig:ch3:darboux-2} for a visual representation. In other words: one may perform a Darboux decomposition purely in transverse space. Again, one may label the the individual planes as $\Pi_A$ with $A=1,\dots,k_p$ and call their angular momentum eigenvalues $j_A$. The total angular momentum is
\begin{align}
J{}_{ij} = \int\dd^p z \, \sum\limits_{A=1}^{k_p} \epsilon{}^A_{ij} j_A \, .
\end{align}
In this case, the physical dimension of $j_A$ is of course dimension-dependent as it corresponds to an angular momentum line density in the case of $p=1$ and to its higher-dimensional generalization for $p>1$. In our case  $j_A = \text{const.}$ such that the total angular momentum diverges, but this is to be expected since the object itself is infinitely large. In that sense it mimics the divergence of the total mass $m$ in the context of the static $p$-branes discussed in Sec.~\ref{sec:ch3:p-branes}, and just as before, we will see that the total angular momentum does not appear in the resulting gravitational field.

We are now ready to construct the gravitational field of rotating $p$-branes in higher dimensions. The energy-momentum tensor is that of a static $p$-brane \eqref{eq:ch3:tmunu-p-brane} combined with a purely transverse angular momentum density,
\begin{align}
T{}_{\mu\nu} = \left[ \epsilon\left( \delta{}^t_\mu\delta{}^t_\nu - \sum\limits_{a=1}^p \delta{}^a_\mu\delta{}^a_\nu \right) + \delta{}^t_{(\mu}\delta{}^i_{\nu)} j{}_i{}^k \partial_k \right] \prod\limits_{i=1}^m \delta(y^i)  \, ,
\end{align}
where the shorthand $\partial_i$ denotes partial differentiation with respect to $y^i$. Our ansatz is the superposition of a warped geometry with a gravitomagnetic potential $\ts{A} = A_i\dd y^i$,
\begin{align}
\label{eq:ch3:warped-ansatz-rotating}
\dd s^2 = [1+u(r)]\left[-\dd t^2 + \sum\limits_{a=1}^p (\dd z^a)^2 \right] + [1+v(r)] \sum\limits_{i=1}^m (\dd y^i)^2 + 2 \ts{A}(\ts{y})\dd t \, , \quad r^2 = \sum\limits_{i=1}^m (y^i)^2 \, ,
\end{align}
where $\ts{y} = (y^i)$ collects all transverse coordinates. Let us again focus on the case $a(\Box) = c(\Box)$ such that the equations of motion take the simple form
\begin{align}
\tag{\ref*{eq:ch3-eom-hat-a=c}}
a(\Box)\Box\hat{h}{}_{\mu\nu} = -2\kappa T{}_{\mu\nu} \, ,
\end{align}
where we impose the De\,Donder gauge $\partial{}^\mu \hat{h}{}_{\mu\nu} = 0$. After inserting our ansatz \eqref{eq:ch3:warped-ansatz-rotating} the field equations reduce to
\begin{align}
a(\lap) \lap \left( \frac{p-1}{2} u + \frac{m}{2} v \right) &= -2\kappa\epsilon \prod\limits_{i=1}^m \delta(y^i) \, , \\
a(\lap) \lap \left( \frac{2-m}{2} v - \frac{p+1}{2} u \right) &= 0 \, , \\
a(\lap) \lap A{}_i &= -\kappa j{}_i{}^k \partial{}_k \prod\limits_{i=1}^m \delta(y^i) \, .
\end{align}
One may verify that the ansatz \eqref{eq:ch3:warped-ansatz-rotating} together with the above field equations is indeed consistent with the De\,Donder gauge. As before, the homogeneous equation implies
\begin{align}
v(r) = \frac{p+1}{2-m} u(r) \, ,
\end{align}
such that the solution for $u(r)$ and $\ts{A}(\ts{y})$ becomes
\begin{align}
u(r) = \frac{2-m}{p+m-1} 2\kappa\epsilon \mathcal{G}_m(r) \, , \quad
\ts{A}(\ts{y}) = 2\pi\kappa j{}_{ij} y^i \dd y{}^j \mathcal{G}_{m+2}(r) \, .
\end{align}
In the limit $p=0$ one has $m=d$ and we recover the rotating point particle already discussed in Sec.~\ref{sec:ch3-rotating-particle-higher-dimensions}. Moreover, we may again introduce polar coordinates $\{\rho_A,\varphi_A\}$ in each Darboux plane such that
\begin{align}
\ts{A}(\ts{y}) = 2\pi\kappa \mathcal{G}_{m+2}(r) \sum j_A \rho_A^2 \dd\varphi_A \, ,
\end{align}
with the additional relations
\begin{align}
r^2 = \epsilon z^2 + \sum\limits_{A=1}^{k_p} \rho_A^2 \, , \quad \sum\limits_{i=1}^m(\dd y^i)^2 = \epsilon \dd z^2 + \sum\limits_{A=1}^{k_p}\left(\dd\rho_A^2 + \rho_A^2\dd\varphi_A^2 \right) \, .
\end{align}
These expressions are quite general and describe a $p$-brane in any number of spacetime dimensions, endowed with a purely transverse angular momentum density.

As we mentioned above, in the case $p=0$ one recovers the rotating point particle. If one sets instead $j_A=0$ one recovers the static $p$-brane solutions discussed earlier. In that sense the above expressions present the most general solution constructed in the context of this thesis.

In some cases, however, it is not possible to construct transversely-rotating solutions. For example the case of a domain wall in four spacetime dimensions ($d=3$ and $p=2$). In that case $d-p = 1$ such that $k_p=0$ and $\epsilon=1$: there are not enough spatial dimensions left for the $p$-brane to have intrinsic angular momentum perpendicular to itself.

\subsection{Cosmic string in four dimensions}
Let us focus on the only case that is relevant to our four-dimensional Universe: a cosmic string ($p=1,d=3,m=2$). In that case there is only one Darboux plane, $k_p=1$, and no additional $z$-axis. Let us call the angular momentum eigenvalue $j$ and denote the polar coordinates in the Darboux plane as $\{\rho,\varphi\}$, and---to be more in line with traditional notation---denote the string surface tension as $\mu$ instead of $\epsilon$. The gravitational field of a rotating cosmic string is then
\begin{align}
\dd s^2 &= -\dd t^2 + \dd z^2 + [1+v(\rho)] (\dd \rho^2 + \rho^2\dd\varphi^2) + 2 \ts{A}(\ts{y})\dd t \, , \\
u(\rho) &= 0 \, , \quad
v(\rho) = 2\kappa\mu\mathcal{G}_2(\rho) \, , \quad
\ts{A}(\ts{y}) = 2\pi\kappa \mathcal{G}_2(\rho) j \rho \dd\varphi \, .
\end{align}
In the above we did not make any distinction between the variable $\rho$ and $r$ since for $k_p=1$ and $\epsilon=0$ they coincide. Let us comment on a possible source of confusion: we identified the direction of the cosmic string with the $z$-direction, but this $z$-axis is completely independent of that $z$-axis encountered in the Darboux decomposition. If anything, this constitutes an informal proof of the hypothesis that the number of symbols in the Greek and Latin alphabet are saturated roughly half-way through a typical doctoral thesis.

\subsection{Angle deficit configurations}
``Angle deficit configurations'' in higher dimensions, as we defined them above, are solutions where $m=2$ such that the Newtonian potential vanishes. In the static case the metric function $v(r)$ coincides with the geometry of a cosmic string and the only difference lies in the appearance of additional spatial directions. This remains true for rotating angle deficit configurations in higher dimensions since the gravitomagnetic potential $\ts{A}(\ts{y})$ still only has one angular momentum eigenvalue due to $m=2$. The metric is
\begin{align}
\dd s^2 &= -\dd t^2 + \sum\limits_{a=1}^{d-2} (\dd z^a)^2 + [1+v(\rho)] (\dd \rho^2 + \rho^2\dd\varphi^2) + 2 \ts{A}(\ts{y})\dd t \, , \\
u(\rho) &= 0 \, , \quad
v(\rho) = 2\kappa\epsilon\mathcal{G}_2(\rho) \, , \quad
\ts{A}(\ts{y}) = 2\pi\kappa \mathcal{G}_2(\rho) j \rho \dd\varphi \, .
\end{align}

\subsection{Curvature expressions}
With the curvature expressions for a static $p$-brane well known, see Sec.~\ref{sec:ch3:p-brane-curvature}, one may wonder how the presence of the gravitomagnetic potential $A_i(\ts{y})$ affects the curvature. Due to linearity, at this point we do not have to reproduce the previous results but merely list the newly arising terms. The Riemann tensor and Ricci tensor pick up the new components
\begin{align}
R{}_{tijk} = \partial{}_i \partial{}_{[j} A{}_{k]} \, , \quad R{}_{ti} = \frac12\left( \partial_i \partial^k A{}_k - \lap A{}_i \right) \overset{*}{=} -\frac12 \lap A{}_i \, ,
\end{align}
where in the last equality we have invoked the Lorenz gauge condition $\partial^k A{}_k = 0$. Because the new components of the Ricci tensor are off-diagonal the Ricci scalar and therefore the tracefree Ricci tensor are unaffected. The Weyl tensor, however, has two new non-vanishing components:
\begin{align}
C_{tabi} &= \frac{1}{d-1} \eta{}_{ab} R{}_{ti} \overset{*}{=} \frac{1}{2(1-d)} \eta{}_{ab} \lap A{}_i \, , \\
C{}_{tijk} &= \partial{}_i \partial{}_{[j} A{}_{k]} + \frac{1}{d-1}\left( \eta{}_{ij}R{}_{tk} - \eta{}_{ik}R{}_{tj} \right) \overset{*}{=} \partial{}_i \partial{}_{[j} A{}_{k]} + \frac{1}{2(1-d)}\left( \eta{}_{ij}\lap A{}_k - \eta{}_{ik} \lap A{}_j \right) \, .
\end{align}
The quadratic invariants then take the form
\begin{align}
C^2 &= C{}_{\mu\nu\rho\sigma} C{}^{\mu\nu\rho\sigma} = \frac{p^2-m^2+3m+p-2}{(1+p)(m+p)} (\lap v)^2 + \frac{(m-2)(m+p-1)}{1+p} (\partial_i\partial_j v)(\partial^i \partial^j v) \\
&\hspace{85pt} + 4 \left(\partial_i \partial_{[j} A{}_{k]}\right)\left(\partial^i \partial^{[j} A{}^{k]}\right) - \frac{2m+3p}{2(m+p+1)^2} (\lap A{}_k)(\lap A{}^k) \, , \\
\cancel{R}^2 &= \cancel{R}_{\mu\nu} \cancel{R}^{\mu\nu} = \frac{m(m+p-1)^2}{(1+p)(m+p+1)} (\lap v)^2 \, .
\end{align}	
This expression could be further simplified by utilizing the explicit representations for $v$ and $A_i$ in terms of the static Green functions, and then making use of the recursion relations that express the derivatives in terms of higher-dimensional Green functions, but we will omit this rather lengthy study here.

In four dimensions ($d=3$) the one may also consider the Chern--Pontryagin pseudoscalar
\begin{align}
\mathcal{P} &= \frac12 \epsilon{}^{\mu\nu}{}_{\alpha\beta} C{}^{\alpha\beta}{}_{\rho\sigma} C{}^{\rho\sigma}{}_{\mu\nu} \, .
\end{align}
For $p=0$ this invariant takes the form (in the Lorenz gauge $\partial^i A_i = 0$)
\begin{align}
\mathcal{P} &= 2 \epsilon{}^{ti}{}_{jk} C{}^{jk}{}_{lm} C{}^{lm}{}_{ti} = 2 \epsilon{}^{ti}{}_{jk} \left[ 2 \partial{}^{[j} \delta{}^{k][l} \partial{}^{m]} v + \frac23  \delta{}^{j[l} \delta{}^{m]k} \lap v \right]\left[ \partial{}_i \partial{}_{[l} A{}_{m]} - \frac{\eta{}_{il}\lap A{}_m - \eta{}_{im} \lap A{}_l}{4} \right] \nonumber \\
&= 2\epsilon{}^t{}_{ijk} (\partial{}^i \partial_l A{}^j)( \partial^k \partial^l v ) \, .
\end{align}
The form of this invariant could have been guessed from first principles, except for the prefactor. It is the only combination that can be formed that is quadratic in the second derivatives of the gravitational potentials that also includes the $\epsilon$-symbol. We could in principle also evaluate this expression by similar methods as presented above in the non-rotating case, but we shall omit these studies at this point and perhaps revisit them some other time.

For $p=1$ the Chern--Pontryagin pseudoscalar vanishes identically since both possible terms vanish: $\epsilon{}^{tc}{}_{ta}C{}^{ta}{}_{bi}C{}^{bi}{}_{tc} = 0$ because $t$ appears twice in the $\epsilon$-symbol, and $\epsilon{}^{cj}{}_{bi}C{}^{bi}{}_{ta}C{}^{ta}{}_{cj} = 0$ because the indices $a,b$ are one-dimensional and hence the $\epsilon$-symbol vanishes. In the case of $p=2$ we cannot define orthogonal rotation since the transverse space is one-dimensional in $d=2$, and hence $\ts{A}=0$.

\section{Concluding remarks}
In this chapter we have constructed a fairly general class of stationary geometries that are supported by brane-like objects in any number of dimensions. For simplicity we have assumed a particular form of the angular momentum density, even though, by linearity, the methods presented in this chapter should be sufficient to generate the metrics for more complicated objects.

All gravitational fields were expressed in terms of static non-local Green functions. These Green functions have the property that they are finite in the coincidence limit, as well as regular (in the sense that there is no conical deficit; see the discussions in Ch.~\ref{ch:ch2}). Moreover, these Green functions asymptotically coincide with those encountered in linearized General Relativity, which guarantees that in the large-distance limit all solutions described in this chapter coincide with their corresponding solutions from linearized General Relativity.

As already indicated in Ch.~\ref{ch:ch1}, the reason for this behavior is simple: on may think of these solutions in two equivalent ways: As solutions of linearized non-local ghost-free gravity with $\delta$-shaped matter sources, or as solutions of linearized General Relativity, where the $\delta$-shaped matter sources have been smoothed out.

For this reason it is important to move beyond the linear description, which unfortunately is connected with considerable calculational complications. Only when a fully non-linear solution to the non-local ghost-free gravity equations has been found we can truly evaluate the role of the linearized solutions described in the present chapter. Nevertheless we hope that the considerations presented here prove useful for that purpose.

%%%%%%%%%%%%%%%%%%%%%%%%%%%%%%%%%%%%%%%%%%%%%%%%%%%%%%%%%%%%%%%%%%%%%%%%%%%%%%%%%%%%%%%%%%%%%%%%%%%
%
% Chapter: Ultrarelativistic objects
%
\chapter{Ultrarelativistic objects}
\label{ch:ch4}
\textit{Aichelburg and Sexl constructed the gravitational field of ultrarelativistic objects from linearized solutions in General Relativity. In this chapter we will generalize this method to the linearized solutions of non-local ghost-free gravity and obtain the gravitational field of ultrarelativistic, rotating particles (gyratons) as well as their higher-dimensional generations (rotating particles and rotating branes in arbitrary dimensions). This chapter is based on Ref.~\cite{Boos:2020ccj} whereas the material in Sec.~\ref{sec:ch4:gyratonic-branes} has not yet been submitted.}

\section{The Aichelburg--Sexl metric and the Penrose limit}

The gravitational field of ultrarelativistic objects has been studied for a rather long time. Tolman, Ehrenfest, and Podolski have shown that the gravitational force between a beam of light and a massless particle moving in the same direction vanishes \cite{Tolman:1931}. Later, Bonnor studied extensively the gravitational field of null fluids and spinning pencils of light \cite{Bonnor:1969a,Bonnor:1969b,Bonnor:1977}, see also Refs.~\cite{Mitskievic:1989,Lynden-Bell:2017vax,Schneiter:2018}. These studies have been generalized to higher dimensions \cite{Frolov:2005in,Frolov:2005zq}, to incldue electric charge \cite{Frolov:2005ja}, in other asymptotic geometries \cite{Frolov:2005ww,Kadlecova:2016irj} as well as in supergravity \cite{Frolov:2006va}. For more information on gyratons and their geometric classifications see Refs.~\cite{Kadlecova:2009qu,Krtous:2012qa,Podolsky:2014lpa,Podolsky:2018oov}.

Instead of solving the Einstein--Maxwell equations to obtain the gravitational field of light it is also possible to consider instead an initial ``seed'' metric, such as the Schwarzschild black hole, and then perform a Lorentz boost, and this method goes back to Aichelburg and Sexl \cite{Aichelburg:1971}. In a second step one may then take the limit of the boost velocity approaching the speed of light. Since a Lorentz transformation does not remain regular in this limit, the limit has to be extended by additional assumptions that can counteract this pathology. As Penrose demonstrated a few years later \cite{Penrose:1976}, this limit may be defined much more generally, and hence we shall refer to the limiting procedure as the \emph{Penrose limit}.

In four spacetime dimensions there exists another avenue: instead of starting with the full, non-linear solution of Einstein's equations one may utilize the weak-field approximation of that solution as well. For example, it may be verified (see below) that the linearized Schwarzschild metric, when boosted to the speed of light in the correct Penrose limit, coincides with the Aichelburg--Sexl metric. This prescription is particularly fruitful in the context of mini black hole formation from the collision of ultrarelativistic particles \cite{Eardley:2002re,Yoshino:2002tx,Giddings:2004xy,Yoshino:2005hi,Yoshino:2006dp,Yoshino:2007ph}; for applications in non-local ghost-free gravity see Refs.~\cite{Frolov:2015bta,Frolov:2015usa}.

Na\"ively this fact can be understood as follows: to leading order, the spacetime curvature of a point particle has the form $\mathcal{R}\sim m/r^3$. In the Penrose limit the product $m\gamma$ is kept fixed while $\gamma\rightarrow\infty$, which in turn implies that the parameter $m$ is very small. This justifies that only leading order contributions in both the curvature, the metric, and the Levi--Civita connection need to be taken into account, since all higher order terms vanish in the limit $\gamma\rightarrow\infty$. For this reason one may hope that the same features remain true if one extends the studies to non-local gravity as described by infinite-derivative form factors: it may be possible that the gyratons obtained from performing the Penrose limit on a linearized weak-field solution of infinite-derivative gravity are in fact also exact solutions of the full, non-linear field equations of infinite-derivative gravity.

However, perhaps surprisingly, it has been shown that within General Relativity this mechanism only works in four spacetime dimensions \cite{Frolov:2011,Frolov:2005in,Frolov:2005zq}. In higher-dimensional General Relativity, exact gyraton solutions may contain terms quadratic in the angular momentum, which vanish in their linear approximation. In other words: Boosted weak-field solutions in spacetime dimensions higher than four may no longer solve the full, non-linear field equations.

In the following sections we will proceed as follows. First we will boost a rotating weak-field ``seed metric'' as obtained in linearized infinite-derivative gravity to a velocity $\beta$. In a second step we will perform the suitable Penrose limit which involves criteria on its mass $m$ and its angular momentum $j$ and demonstrate that in the limiting case of vanishing non-locality and vanishing angular momentum we recover the Aichelburg--Sexl metric. In a third step, we will examine the properties of the resulting rotating ultrarelativistic gyraton metrics in four and higher dimensions, and also discuss a possible generalization towards spinning ultrarelativistic $p$-branes.

\section{Gyratons}

Let us now focus on the spinning solutions we obtained in Ch.~\ref{ch:ch3}. In particular, for later convenience, we are interested in the gravitational field of a ``pencil,'' that is, an object with finite extension $L$ in one spatial direction, while being infinitely thin in all other spatial directions. We shall assume that the pencil lives in $d$-dimensional Euclidean space.

\subsection{Geometrical setup}
\begin{figure}
  \centering
  \includegraphics[width=0.5\textwidth]{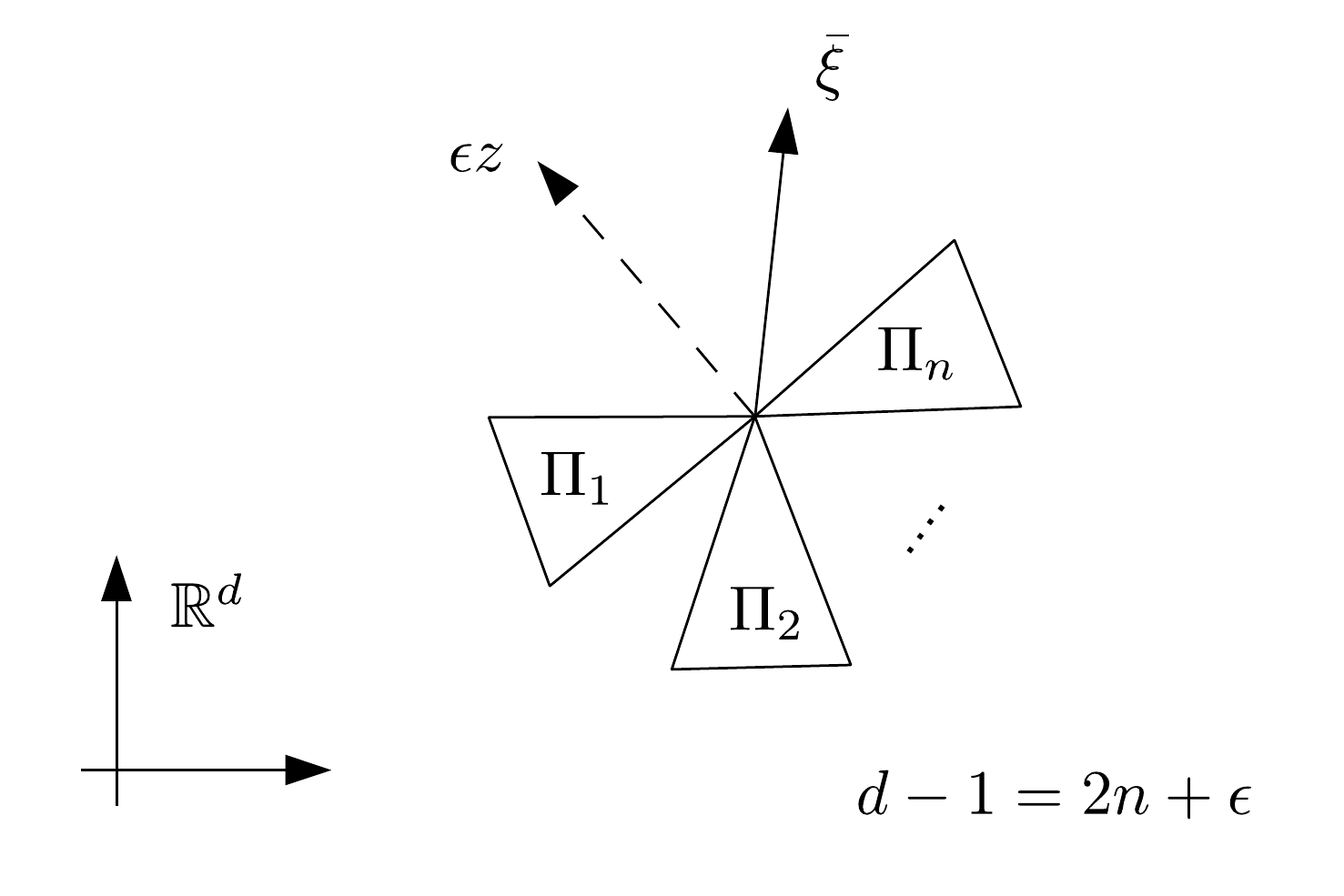}
  \caption[Darboux decomposition of $(d-1)$-dimensional space.]{Ignoring the $\bar{\xi}$-direction, one may perform a Darboux decomposition of the remaining $(d-1)$-dimensional Euclidean space. There are $n$ orthogonal Darboux planes $\Pi_A$ labelled by $A=1,\dots,n$ where $d-1=2n+\epsilon$. The additional $z$-direction only exists if $d-1$ is odd, or, equivalently, when $\epsilon=1$. In the above diagram we visualize the Darboux planes by wedges, the $z$-axis by a dashed line, and the boost direction $\bar{\xi}$ by a straight line.}
  \label{fig:ch4:darboux-3}
\end{figure}

It is helpful to consider two frames in this scenario, and we shall call these frames $\bar{S}$ and $S$. In the $\bar{S}$-frame the pencil is at rest, and in the $S$-frame it moves with a constant velocity $\beta$ along the direction of its spatial extension. These two frames are related by a Lorentz boost, and the coordinates in each frame are
\begin{align}
\bar{X}{}^\mu = (\bar{t}, \bar{\xi}, x^i_\perp) \, , \quad X{}^\mu = (t, \xi, x^i_\perp) \, .
\end{align}
While $\bar{t}$ and $t$ denote the time coordinates in the frames $\bar{S}$ and $S$, respectively, $\bar{\xi}$ denotes the direction of the spatial extension of the pencil in its own rest frame, and $\xi$ is the corresponding coordinate after the boost. There are also $i=1,\dots,d-1$ transverse coordinates $x_\perp^i$ for which we choose the same symbol in both frames since their values are unaffected by the boost in the $\bar{\xi}$-direction. The Minkowski line element, in both frames, takes the form
\begin{align}
\dd s^2 = -\dd \bar{t}^2 + \dd\bar{\xi}^2 + \dd\ts{x}_\perp^2 = -\dd t^2 + \dd \xi^2 + \dd\ts{x}_\perp^2 \, , \quad \dd\ts{x}_\perp^2 = \sum\limits_{i=1}^{d-1} (\dd x_\perp^i)^2 = \delta{}_{ij}\dd x_\perp^i \dd x_\perp^j \, .
\end{align}
In what follows, we shall denote all quantities associated with the rest frame $\bar{S}$ with a bar, and all quantities related to the $S$-frame in which the pencil is moving without a bar. Let us now briefly specify the orthogonal coordinates in $\bar{S}$ further. We write
\begin{align}
x_\perp^i = (y_A, \hat{y}_A, \epsilon z) \, , \quad A = 1, \dots, n \, , \quad n = \left\lfloor \frac{d-1}{2} \right\rfloor \, , \quad d = 2n+1+\epsilon \, .
\end{align}
This means that the transverse, $(d-1)$-dimensional space is split into $n$ Darboux planes $\Pi_A$, each plane spanned by a pair of coordinates $\{y_A, \hat{y}_A\}$ with $A=1,\dots,n$ labelling the independent planes. If $(d-1)$ is odd, there is also an additional $z$-axis. This is quite similar to our previous considerations in Ch.~\ref{ch:ch3} and we visualize this Darboux decomposition in Fig.~\ref{fig:ch4:darboux-3} for the present case. In other words: the Darboux decomposition performed in this section is identical to that of the previous chapter, provided we ignore the $\bar{\xi}$-direction.

Let us now specify the energy-momentum tensor for the pencil in the $\bar{S}$-frame:
\begin{align}
\label{eq:ch4:tmunu-pencil}
T_{\mu\nu}=\left[ \delta^{\bar{t}}_{\mu} \delta^{\bar{t}}_{\nu} \bar{\lambda}(\bar{\xi})+\sum\limits_{A=1}^n \left( \bar{j}_A(\bar{\xi}) \delta^{\bar{t}}_{(\mu} \epsilon^{A~j} _{\nu)}\frac{\partial}{\partial x_\perp^j}\right) \right]\delta^{(d-1)}(\ts{x}_{\perp})\, ,
\end{align}
In the above, $\bar{\lambda}(\bar{\xi})$ denotes the line density of the gravitational pencil along the $\bar{\xi}$-direction and has dimensions of mass per length, and $\bar{j}{}_A(\bar{\xi})$ denotes the line density of angular momentum for a given Darboux plane $\Pi_A$ with dimensions of angular momentum per length. The quantity $\epsilon{}^A_{ij}$ denotes the surface element for the Darboux plane $\Pi_A$, as already employed in Ch.~\ref{ch:ch3}, such that
\begin{align}
\bar{j}{}_{ij}(\bar{\xi}) = \sum\limits_{A=1}^n \bar{j}_A(\bar{\xi}) \epsilon{}^A_{ij} \, .
\end{align}
The total energy-momentum tensor is proportional to the transverse $\delta$-function $\delta{}^{(d-1)}(\ts{x}_\perp)$ since the pencil has zero width in the transverse direction.

We assume that the pencil has a finite length $\bar{L}$ such that both $\bar{\lambda}(\bar{\xi})=0$ and $ \bar{j}_A(\bar{\xi})=0$ for $\bar{\xi}\not\in(0,\bar{L})$. The mass and the angular momentum of such a pencil are
\begin{align}
\label{eq:ch4:m-j-def}
\bar{m}=\int\limits_{-\infty}^\infty \dd\bar{\xi} \, \bar{\lambda}(\bar{\xi})\, , \quad
\bar{J}_{i j}=\int\limits_{-\infty}^\infty \dd\bar{\xi} \, \bar{j}_{ij}(\bar{\xi})\, , \quad
\bar{j}_{ij}(\bar{\xi})=\sum\limits_{A=1}^n \epsilon^A_{ij} \bar{j}_A(\bar{\xi})\, ,
\end{align}
Note that the above energy-momentum tensor implies that the angular momentum is transverse to the $\bar{\xi}$ direction since $j{}_{\alpha\beta}\delta{}^\alpha_{\bar{\xi}} = j{}_{\beta\alpha}\delta{}^\alpha_{\bar{\xi}} = 0$. In other words, the angular momentum density tensor $j{}_{\alpha\beta}$ is trivial in the $\bar{t}\bar{\xi}$-sector. In four dimensions this is easier to visualize, and the above conditions correspond to the angular momentum pseudovector being aligned with the boost direction $\bar{\xi}$ of the pencil. Expressed as a 2-form this implies that the angular momentum density 2-form is perpendicular to the $\bar{\xi}$-direction.

While it may be possible to consider the gravitational field of spinning ultrarelativistic objects with angular momentum components in the direction of motion, this choice of only transverse angular momentum density is convenient because it implies that the tensorial structure of the angular momentum 2-form $j{}_{\alpha\beta}$ transforms trivially under the Lorentz boost in the $\bar{\xi}$-direction.

That being said, we may now use Eqs.~\eqref{eq:ch3:phi-psi-rotating}--\eqref{eq:ch3:a:1-form} of the previous chapter to find the gravitational field of the pencil-shaped energy-momentum of Eq.~\eqref{eq:ch4:tmunu-pencil}. The transverse $\delta$-function collapses most integrals and one finds
\begin{align}
h{}_{\mu\nu}\dd\bar{X}^\mu\dd\bar{X}^\nu &= \bar{\phi} \left[\dd \bar{t}^2+{1\over d-2} (\dd\bar{\xi}^{\,2}+\dd\ts{x}_{\perp}^2)\right]+2 \bar{A}_{i}\dd x_{\perp}^{i}\dd \bar{t}\, , \label{eq:ch4:sol-h} \\
\bar{\phi}(\bar{\xi},x_{\perp}^{i}) &= 2\kappa \frac{d-2}{d-1} \int\limits_{-\infty}^\infty \dd\bar{\xi}' \, \bar{\lambda}(\bar{\xi}')\mathcal{G}_d(\bar{r})\, , \label{eq:ch4:sol-phi}  \\
\bar{A}_{i}(\bar{\xi},x_{\perp}^{i}) &= -2\pi \kappa \int\limits_{-\infty}^\infty \dd\bar{\xi}' \, \bar{j}_{ij}(\bar{\xi}') x_\perp^j \mathcal{G}_{d+2}(\bar{r})\, , \label{eq:ch4:sol-a}
\end{align}
where we defined the auxiliary expression
\begin{align}
\bar{r}^2 = \left(\bar{\xi} - \bar{\xi}'\right)^2 + \delta{}_{ij} x_\perp^i x_\perp^j \, .
\end{align}
This gravitational field is the starting point for constructing a rotating gyraton solution in infinite-derivative gravity, and the free parameters are given by the line densities $\bar{\lambda}(\bar{\xi})$ and $\bar{j}_A(\bar{\xi})$. Sometimes it is also useful to write
\begin{align}
\bar{j}_{ij}(\bar{\xi}) \dd x{}_\perp^i x{}_\perp^j = -\sum\limits_{A=1}^n \bar{j}_A(\bar{\xi}) \rho_A^2 \dd\varphi_A \, ,
\end{align}
where $\{\rho_A, \varphi_A\}$ are polar coordinates in the Darboux plane $\Pi_A$ as already introduced in Ch.~\ref{ch:ch3}.

\subsection{Boost and Penrose limit}
We have found the gravitational field of a pencil in an arbitrary number of dimensions for a wide range of linearized infinite-derivative gravity theories, and we can now proceed to boost the solution to a velocity $\beta$, before taking the limit $\beta\rightarrow 1$ in a suitable Penrose limit.

\begin{figure}
  \centering
  \includegraphics[width=0.5\textwidth]{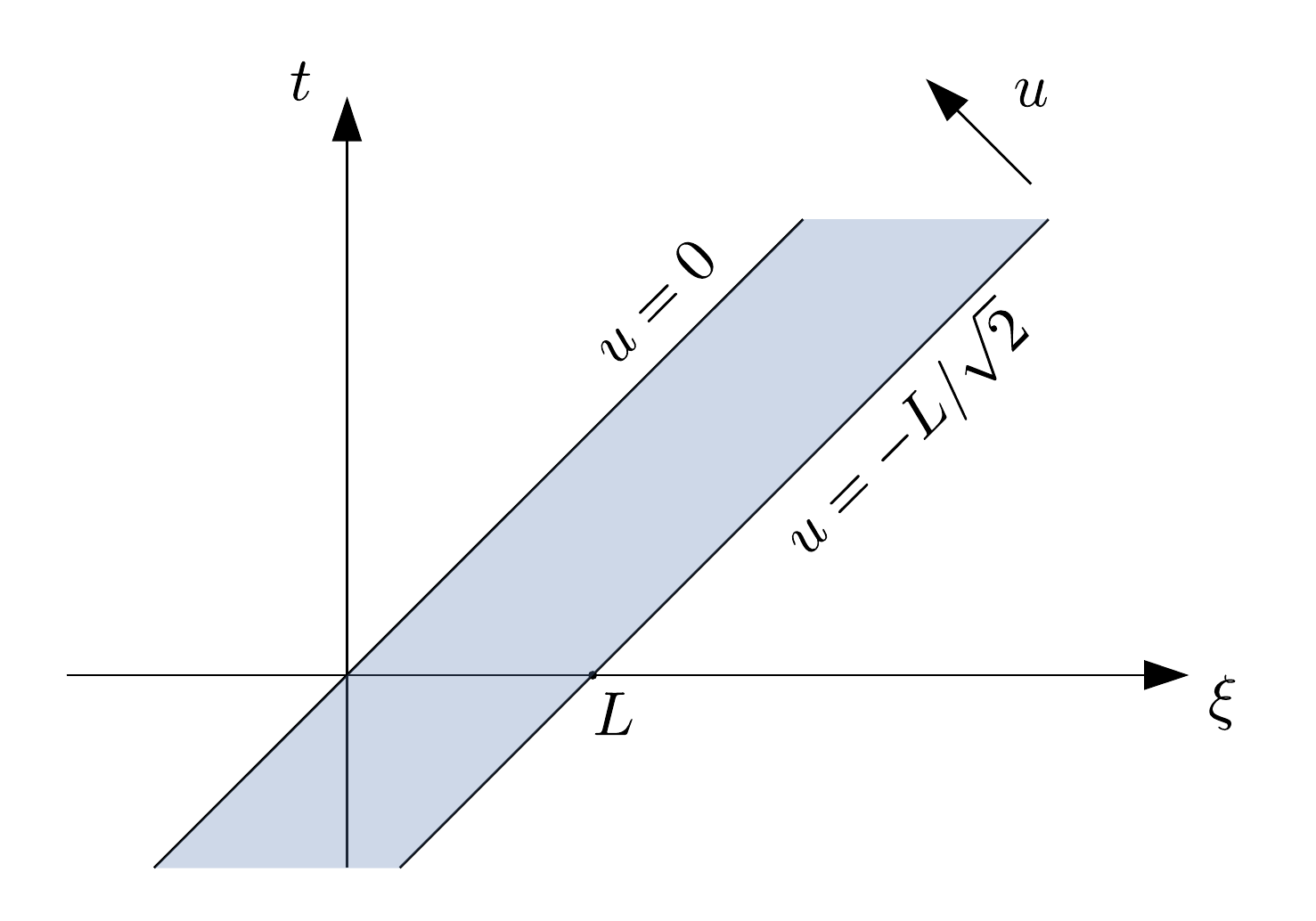}
  \caption[Motion of an ultrarelativistic object in Minkowski space.]{A pencil of length $L$ moving with the speed of light. Its energy density, viewed as a function of retarded time $u$, is non-zero in the interval $u\in[-L/\sqrt{2}, 0]$.}
  \label{fig:ch4:fig-u}
\end{figure}

During this limit we assume that $\bar{m}\gamma$ and $\bar{L}/\gamma$ remain constant, where
\begin{align}
\gamma = \frac{1}{\sqrt{1-\beta^2}} \, .
\end{align}
Let us first remark on the scaling behavior of the mass and angular momentum line densities as given in Eq.~\eqref{eq:ch4:m-j-def} which imply
\begin{align}
\label{eq:ch4:line-densities}
\bar{\lambda} = \frac{\dd \bar{m}}{\dd \bar{\xi}} \, , \quad \bar{j}_{ij} = \frac{\dd \bar{J}_{ij}}{\dd \bar{\xi}} \, .
\end{align}
The Penrose limit consists of two assumptions during the limiting process of $\beta\rightarrow 1$: first, we assume that the energy $\gamma\bar{m}$ remains constant. Second, we assume that the ratio $\bar{L}/\gamma$ remains constant as well. Then, Eq.~\eqref{eq:ch4:line-densities} implies the following scaling behavior during the Penrose limit:
\begin{align}
\label{eq:ch4:line-densities-scaling}
\lambda \sim \gamma^2 \bar{\lambda} \, , \quad j_{ij} \sim \gamma \bar{j}_{ij} \, .
\end{align}
The scaling weights are different because the angular momentum is transverse to the boost direction and hence it is not required to demand a similar condition to the energy. To move forward, let us now parametrize the boost in the $\bar{\xi}$-direction as
\begin{align}
\label{eq:ch4:boost}
\overline{t} = \gamma\left( t - \beta\xi \right) \, , \quad
\overline{\xi} = \gamma\left( \xi - \beta t \right) \, .
\end{align}
For fixed $\xi$ in the new frame $S$ one has $\bar{\xi} = -\gamma\beta t + \text{const.}$, which implies that the pencil moves into the positive $\xi$-direction. Let us now introduce retarded and advanced null coordinates,
\begin{align}
u = \frac{t - \xi}{\sqrt{2}} \, , \quad v = \frac{t + \xi}{\sqrt{2}} \, .
\end{align}
In terms of the $\bar{S}$-coordinates one has
\begin{align}
\overline{t} = {\gamma\over \sqrt{2}}[(1+\beta) u+(1-\beta)v] \, , \quad
\overline{\xi} = {\gamma\over \sqrt{2}}[-(1+\beta) u+(1-\beta)v] \, ,
\end{align}
and in the ultrarelativistic limit, $\beta\rightarrow 1$, this becomes
\begin{align}
\overline{t} \to\sqrt{2}\gamma u \, , \quad \overline{\xi} \to -\sqrt{2}\gamma u \, ,
\end{align}
which in turn implies that the matter distribution has a finite duration $u\in[-L/\sqrt{2},0]$; we visualize this in Fig.~\ref{fig:ch4:fig-u}. Based on the scaling properties \eqref{eq:ch4:line-densities-scaling} we define the mass and angular momentum line densities in the $S$ frame as
\begin{align}
\label{eq:pr-limit-mass}
\lambda(u) &= \lim \limits_{\gamma\rightarrow\infty} \sqrt{2} \gamma^2 \, \overline{\lambda}(-\sqrt{2}\gamma u) \, , \\
\label{eq:pr-limit-angular-momentum}
j_{ij}(u) &= \lim\limits_{\gamma\rightarrow\infty} \sqrt{2}\gamma \, \overline{j}_{ij}(-\sqrt{2}\gamma u) \, , \quad j_A(u) = \lim\limits_{\gamma\rightarrow\infty} \sqrt{2}\gamma \, \bar{j}_A(-\sqrt{2}\gamma u) \, .
\end{align}

These precise scaling relations imply that the following expressions for the total mass $m$ and the total angular momentum $J_{ij}$ remain constant,
\begin{align}
m &= \gamma \, \overline{m} = \gamma \int\limits_{-\infty}^\infty \dd\overline{\xi} \, \overline{\lambda}(\overline{\xi}) = \int\limits_{-\infty}^\infty \dd u \, \lambda(u) = \text{const} \, , \\
J_{ij} &= \bar{J}_{ij} = \int\limits_{-\infty}^\infty\dd\overline{\xi} \, \overline{j}_{ij}(\overline{\xi}) = \int\limits_{-\infty}^\infty \dd u j{}_{ij}(u) = \text{const} \, .
\end{align}
Applying the coordinate transformation \eqref{eq:ch4:boost} to the metric \eqref{eq:ch4:sol-h} under the scaling assumptions \eqref{eq:pr-limit-mass}--\eqref{eq:pr-limit-angular-momentum} then gives the gyraton\index{gyraton!metric in ghost-free gravity} metric
\begin{align}
\label{eq:gyraton-1}
\ts{g} &= \left(\eta{}_{\mu\nu} + h{}_{\mu\nu}\right) \dd X{}^\mu \dd X{}^\nu = -2\dd u \dd v + \phi \dd u^2 + 2 A{}_i\dd x{}^i_\perp \dd u + \dd \ts{x}_\perp^2 \, , \\
\phi &= \lim\limits_{\gamma\rightarrow\infty} 2 \gamma^2 \frac{d-1}{d-2} \bar{\phi} \, , \quad A_i = \lim\limits_{\gamma\rightarrow\infty} \sqrt{2} \gamma \bar{A}_i \, .
\end{align}
Here, $\phi$ and $A_i$ denote the gravitational potentials in the frame $S$ after the Penrose limit. Their precise form can be derived conveniently by using Eqs.~\eqref{eq:ch4:sol-phi}--\eqref{eq:ch4:sol-a} in conjunction with the following Green function representation derived in Sec.~\ref{sec:ch2:heat-kernel},
\begin{align}
\mathcal{G}_d(r) = \frac{1}{2\pi}\int\limits_{-\infty}^\infty \frac{\dd\eta}{a(-\eta\ell^2)\eta} \int\limits_{-\infty}^\infty \dd\tau \, K_d(r|\tau) \, e^{i\eta\tau} \, , \quad
K_d(r|\tau) = \frac{1}{(4\pi i \tau)^{\tfrac{d}{2}}} e^{i\tfrac{r^2}{4\tau}} \, , \quad a(z) = e^{(-z)^N} \, ,
\end{align}
which is valid for linearized $\mathrm{GF_N}$ theories with $N=1,2,\dots$ (in the case of linearized General Relativity one sets $\ell=0$ and hence $a=1$). The above representation is useful because radial distance enters quadratically in the exponent of the $d$-dimensional heat kernel $K_d(\bar{r}|\tau)$, where
\begin{align}
\bar{r}^2 = (\bar{\xi}-\bar{\xi}')^2 + r_\perp^2 = 2\gamma^2(u-u')^2 + r_\perp^2 \, , \quad r_\perp^2 = \delta_{ij}x_\perp^i x_\perp^j \, .
\end{align}
Then one can make use of the identity (see \cite{Shankar:1994})
\begin{align}
\delta(u) = \lim\limits_{\epsilon\rightarrow 0} \frac{1}{\sqrt{2\pi i\epsilon}}e^{i\tfrac{u^2}{2\epsilon}} \, ,
\end{align}
where we identify $\epsilon=\tau/\gamma^2$, yielding the universal relation
\begin{align}
\label{eq:ch4:g-gamma-limit}
\lim\limits_{\gamma\rightarrow\infty} \gamma \, \mathcal{G}_d(\bar{r}) &= \frac{1}{\sqrt{2}} \mathcal{G}_{d-1}(r_\perp) \delta(u-u') \, .
\end{align}
This relation is true for any linearized non-local ghost-free theory as well as linearized General Relativity, and is valid in any number of dimensions. It is an interesting relation because it relates the static $d$-dimensional Green function to the $(d-1)$-dimensional static Green function, multiplied by a $\delta$-function in the retarded time that collapses the $u'$-integral in Eqs.~\eqref{eq:ch4:sol-phi}--\eqref{eq:ch4:sol-a}. We find
\begin{align}
\label{eq:ch4:phi-final}
\phi(u,r_\perp) = 2\sqrt{2} \kappa \lambda(u) \mathcal{G}_{d-1}(r_\perp) \, , \quad A_i(u,x_\perp^i) = -2\pi\kappa j{}_{ij}(u) x^j_\perp \mathcal{G}_{d+1}(r_\perp) \,  .
\end{align}
Again, one may employ the Darboux decomposition of the angular momentum density such that
\begin{align}
\label{eq:ch4:a-final}
A_i(\ts{x}_\perp) \dd x{}^i_\perp = 2\pi\kappa \mathcal{G}_{d+1}(r_\perp) \sum\limits_{A=1}^n j_A(u) \rho_A^2 \dd \varphi_A \, .
\end{align}
The gyraton metric of a rotating pencil in an arbitrary number of dimensions in linearized ghost-free gravity is hence given by \eqref{eq:gyraton-1} in conjunction with Eqs.~\eqref{eq:ch4:phi-final}--\eqref{eq:ch4:a-final}.

Let us now study some explicit examples under the knowledge of the static Green functions presented in great detail in Sec.~\ref{sec:ch2:static}. We have already argued there that all non-local static Green functions $\mathcal{G}_d(r)$ are regular as $r\rightarrow 0$ and for that reason the gyratons presented in this chapter are regular as well. Let us note, however, that were it not for the rescaling in the $\bar{\xi}$-direction during the Penrose limit the resulting metric would be singular as it would contain a $\delta$-function in retarded time, corresponding to a relativistic shock wave. This can be shown from setting $\bar{\lambda}(\bar{\xi}) =\bar{m} \delta(\bar{\xi})$ in the above considerations. The importance of these results lies in the fact that the metric is now also regular in the \emph{transverse} space as $r_\perp \rightarrow 0$, which is an effect generated by the presence of non-locality.

In order to study these metrics in a bit more detail it is useful to introduce the gravitomagnetic charge\index{gravitomagnetic charge} $Q$ that is related to the gravitomagnetic potential $\ts{A}$ as follows:
\begin{align}
Q = \int\limits_\mathcal{A} \ts{F} = \oint\limits_{\partial \mathcal{A}} \ts{A} = 4\pi^2\kappa \mathcal{G}_{d+1}(r_\perp) \sum\limits_{A=1}^n j_A(u) \rho_A^2 \, ,
\end{align}
where $\mathcal{A}$ denotes an area of radii $\rho_A = \text{const}$. Let us now turn to some concrete examples.

\subsection{Gyraton solutions in $d=3$}
As a warm-up, let us reproduce well-known results from $(3+1)$-dimensional General Relativity \cite{Bonnor:1969a,Bonnor:1969b,Aichelburg:1971,Frolov:2005in,Frolov:2005zq}. Here $d=3$ and hence the transverse space is two-dimensional, $n=1$ and $\epsilon=0$. For this reason there is only one Darboux plane. Calling the angular momentum eigenfunction $j(u)$ and introducing the polar coordinates $\{\rho,\varphi\}$ we may identify $r_\perp^2 = \rho^2$.

\subsubsection{General Relativity}
In General Relativity $\ell=0$ and the relevant static Green functions take the form\index{Aichelburg--Sexl metric}\index{gyraton!metric in General Relativity}
\begin{align}
G_2(r) = -\frac{1}{2\pi} \log(r) \, , \quad
G_4(r) = \frac{1}{4\pi^2r^2} \, .
\end{align}
Then, the gravitational potential and the gravitomagnetic potential are
\begin{align}
\phi(u,\rho) &= -\frac{\sqrt{2}\kappa\lambda(u)}{2\pi}\log(\rho) \, , \quad \ts{A}(u) = \frac{\kappa j(u)}{2\pi} \dd \varphi \, .
\end{align}
In the limiting case of $j(u)=0$ we reproduce the Aichelburg--Sexl metric \cite{Aichelburg:1971,Frolov:2011}. For $j(u)\not=0$, on a $u=\text{const.}$ slice, the gravitomagnetic potential is an exact form in transverse space such that
\begin{align}
\ts{F} = \dd \ts{A} = 0 \quad \text{for} \quad \rho > 0 \quad \text{and} \quad u=\text{const.} \, .
\end{align}
This implies that the gravitomagnetic charge is just a function of time,
\begin{align}
Q_0 = \kappa j(u) \, .
\end{align}

\subsubsection{Ghost-free infinite-derivative gravity}
\begin{figure}
  \centering
  \includegraphics[width=0.8\textwidth]{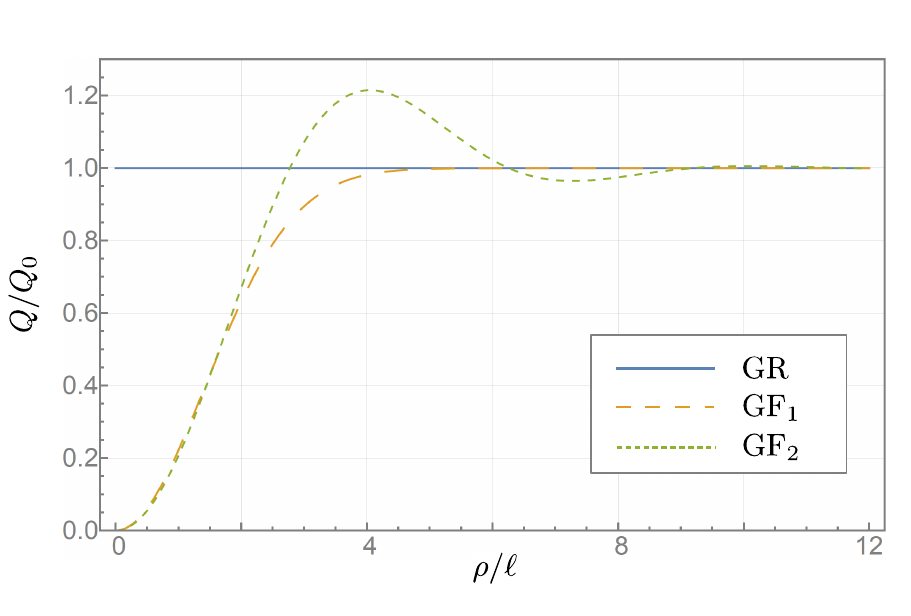}
  \caption[Gravitomagnetic charges in local and non-local gravitational theories.]{For a fixed retarded time $u$ the gravitomagnetic charge of a rotating gyraton in four dimensional spacetime for General Relativity is constant in space; in the non-local $\mathrm{GF_1}$ and $\mathrm{GF_2}$ theories, however, the gravitomagnetic charge is a function of transverse distance $\rho$. In this plot we have normalized all charges to the constant expression encountered in General Relativity at any given fixed time $u = \text{const}$. It is interesting to note that in $\mathrm{GF_2}$ theory the charge oscillates, whereas it approaches the constant asymptotic value much more smoothly in $\mathrm{GF_1}$ theory.}
  \label{fig:ch4:fig-q}
\end{figure}

Even though our general expressions hold for arbitrary ghost-free $\mathrm{GF_N}$ theories, for simplicity we will focus on $N=1$ and $N=2$. In $\mathrm{GF_1}$ theory one finds
\begin{align}
\mathcal{G}{}_2(r) &= - \frac{1}{4\pi} \text{Ein}\left( \frac{r^2}{4\ell^2} \right) \, ,
\end{align}
which results in
\begin{align}
\phi(u,\rho) &= -\frac{\sqrt{2}\kappa \lambda(u)}{2\pi}\text{Ein}\left(\frac{\rho^2}{4\ell^2} \right) \, , \quad
\ts{A}(u,\ts{x}_\perp) = \frac{\kappa j(u)}{2\pi}\left[ 1 - \exp\left(-\frac{r_\perp^2}{4\ell^2} \right) \right] \dd \varphi \, .
\end{align}
The gravitomagnetic charge now depends on the transverse distance,
\begin{align}
Q_1(\rho) = \kappa j(u) \left[ 1 - \exp\left(-\frac{\rho^2}{4\ell^2}\right) \right] \, ,
\end{align}
It is clear that in the limiting case of $\ell=0$ one recovers the results for General Relativity. Similarly, for $\mathrm{GF_2}$ theory we obtain
\begin{align}
\begin{split}
\mathcal{G}{}_2(r) &= -\frac{y}{2\pi} \Big[ \hspace{4pt} \sqrt{\pi}\, {}_1\!F\!{}_3\left(\tfrac12;~1,\tfrac32,\tfrac32;~y^2\right) - y\, {}_2 \!F\!{}_4\left(1,1;~\tfrac32,\tfrac32,2,2;~ y^2 \right) \Big] \, ,
\end{split}
\end{align}
where $y=\rho^2/(16\ell^2)$. The gravitomagnetic charge is
\begin{align}
\begin{split}
Q_2(\rho) &= \kappa j(u) \Big[ 1 - {}_0 F{}_2\left( \tfrac12,\tfrac12; y^2 \right) - 2\sqrt{\pi}y {}_0 F{}_2 \left( 1, \tfrac32; y^2 \right) \Big] \, .
\end{split}
\end{align}
For a visualization of the charges see Fig.~\ref{fig:ch4:fig-q}.

\subsection{Curvature}
Given the fact that the gyraton metric is regular one might wonder whether that fact is reflected in the curvature invariants as well. For the sake of simplicity, let us focus again on the four-dimensional case and use Cartesian coordinates in the Darboux plane such that $x=\rho\cos\varphi$ and $y=\rho\sin\varphi$ and the metric takes the form
\begin{align}
\label{eq:ch4:gyraton-metric-4d}
\ts{g} &= -2\dd u \dd v + \phi(u,x,y) \dd u^2 + \dd x^2 + \dd y^2 + 2 \left[ A_x(u,x,y)\dd x + A_y(u,x,y)\dd y \right] \dd u  \, .
\end{align}
Let us treat this as a full, non-linear metric. One may show that the null vector field $\ts{k}=\partial_v$ is covariantly constant,
\begin{align}
\nabla_\nu k{}^\mu = 0 \, .
\end{align}
which implies that the metric \eqref{eq:ch4:gyraton-metric-4d} corresponds to a pp-wave \cite{Stephani:2003}. This is true irrespective of the form of the functions $\phi$, $A_x$ and $A_y$, provided one does not change their functional dependence. This is of course not unexpected expected since already Aichelburg and Sexl proved this in their original paper \cite{Aichelburg:1971} in the non-rotating case of $\ts{A}=0$, where they moreover prove that the resulting metric is of Petrov type N, as opposed to Petrov type D for the initial, unboosted metric. This fact in itself is also interesting since a coordinate transformation does not change the Petrov type. The Penrose limit, however, is of course singular in the limit of $\beta\rightarrow 1$ and for that reason the change in the algebraic type is possible.

One may also show that all polynomial curvature invariants vanish, as is typical for pp-waves,
\begin{align}
R = R{}_{\mu\nu}R{}^{\mu\nu} = R{}_{\mu\nu\rho\sigma}R{}^{\mu\nu\rho\sigma} = 0 \, .
\end{align}
This fact is interesting since it may imply that the gyraton metric \eqref{eq:ch4:gyraton-metric-4d} might be a solution of the full, non-linear non-local field equations, and hence it would be interesting to study this further in the future; see also Ref.~\cite{Kilicarslan:2019njc} in this context.

\subsection{Gyraton solutions in $d\ge 4$}
\label{sec:ch4:gyratons-higher-dimensions}
Last, let us demonstrate how to construct gyraton solutions in higher dimensions. First, in the case of $d=4$ one finds again one Darboux plane, but now $\epsilon=1$ such that there is also an additional $z$-coordinate. The metric functions then take the form
\begin{align}
\phi &= 2\sqrt{2}\kappa\lambda(u)\mathcal{G}_3(r_\perp) \, , \quad A_i \dd x_\perp^i = -\frac{\kappa}{r_\perp}\frac{\dd}{\dd r_\perp}\mathcal{G}_3(r_\perp) j(u) \rho^2 \dd \varphi \, , \quad r_\perp^2 = \rho^2 + z^2 \, .
\end{align}
For higher dimensions we omit exact expressions but rather specify an algorithmic procedure of how to generate them. It consists of three steps.

First, after specifying the number of spatial dimensions $d$, one may determine the number of Darboux planes $n$, and whenever $d-1$ is odd there exists an additional $z$-axis. Because the original, unboosted metric is axisymmetric around the $\bar{\xi}$-direction it is useful to introduce at this point the polar coordinates $\{\rho_A,\varphi_A\}$ for each Darboux plane. Moreover, this construction is unique if one chooses the direction of the polar angles $\varphi_A$ to be right-handed with respect to the $\bar{\xi}$-direction.

Second, one may introduce the transverse radial variable $r_\perp$ such that
\begin{align}
r_\perp^2 = \sum\limits_{A=1}^n \rho_A^2 + \epsilon z^2 \, .
\end{align}
Substituting this radius variable into Eqs.~\eqref{eq:ch4:phi-final} and \eqref{eq:ch4:a-final} in the metric \eqref{eq:gyraton-1}, one now specifies the functional form of the static Green functions $\mathcal{G}_{d-1}(r_\perp)$ and $\mathcal{G}_{d+1}(r_\perp)$, which can be derived from the expressions provided in Sec.~\ref{sec:ch2:static} in any number of dimensions in conjunction with the recursion relations \eqref{eq:ch2:recursion}.

Last, one may want to consider a specific mass line density $\bar{\lambda}(\bar{\xi})$ and angular momentum line density $\bar{j}_{ij}(\bar{\xi})$ in the original rest frame $\bar{S}$, in which case Eqs.~\eqref{eq:pr-limit-mass}--\eqref{eq:pr-limit-angular-momentum} provide the correct prescriptions to find the line densities $\lambda(u)$ and $j{}_{ij}(u)$ in retarded time in the frame $S$.

Realistic gyratons are not infinitely thin in the transverse direction, but by linearity one may modify the ``pencil'' energy-momentum tensor \eqref{eq:ch4:tmunu-pencil} to include a transverse density profile as well, mutatis mutandis. In that case, however, it is necessary to to impose certain rigidity conditions on the Darboux form of the angular momentum tensor, and for that reason we won't discuss this more complicated case in more detail at this point.

\section{Gyratonic $p$-branes}
\label{sec:ch4:gyratonic-branes}
With the gyraton metrics derived from a rotating pencil readily available, a natural extension is to consider higher-dimensional, extended objects as gravitational sources: $p$-branes. In this concluding section we will construct the gravitational field of rotating ultrarelativistic (``gyratonic'') $p$-branes in close similarity to that of the gyratons discussed in the previous section.

\begin{figure}
  \centering
  \includegraphics[width=0.5\textwidth]{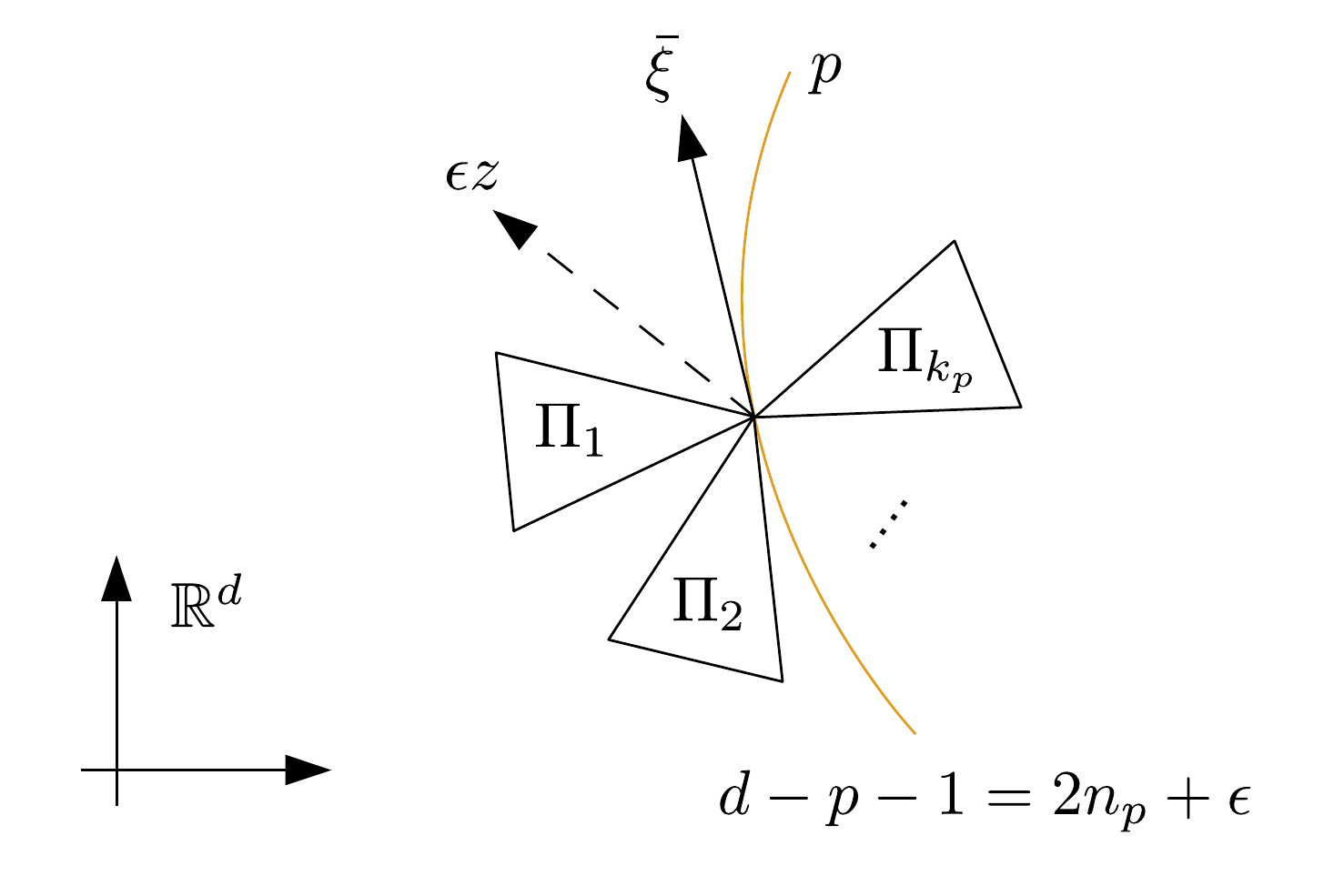}
  \caption[Darboux decomposition of $(d-1)$-dimensional space in presence of a $p$-brane.]{Ignoring the $\bar{\xi}$-direction, one may perform a Darboux decomposition of the $(d-p-1)$-dimensional transverse Euclidean space. Given a $p$, there are $n_p$ orthogonal Darboux planes $\Pi_A$ labelled by $A=1,\dots,n_p$ where $d-p-1=2n_p+\epsilon$. The additional $z$-direction only exists if $d-p-1$ is odd, or, equivalently, when $\epsilon=1$. In the above diagram we visualize the Darboux planes by wedges, the $\bar{\xi}$-axis by a solid line, the $z$-axis by a dashed line, and the $p$-brane by a curved line.}
  \label{fig:ch4:darboux-4}
\end{figure}

To begin with, the energy momentum tensor now takes the form
\begin{align}
\label{eq:ch4:tmunu-pencil-brane}
T_{\mu\nu}=\left\{ \bar{\epsilon}(\bar{\xi}) \left[ \delta^{\bar{t}}_{\mu} \delta^{\bar{t}}_{\nu} - \sum\limits_{a=1}^p \delta{}^a_\mu\delta{}^a_\nu \right] +\sum\limits_{A=1}^n \left( \bar{j}_A(\bar{\xi}) \delta^{\bar{t}}_{(\mu} \epsilon^{A~j} _{\nu)}\frac{\partial}{\partial y^j}\right) \right\} \prod\limits_{i=1}^m \delta\left(y^i\right) \, ,
\end{align}
which is a combination of the $p$-brane energy-momentum tensor \eqref{eq:ch3:tmunu-p-brane} endowed with an additional transverse angular momentum density \eqref{eq:ch4:tmunu-pencil}. This deserves some explanation. First, we denote the coordinates in the pre-boost reference system $\bar{S}$ as
\begin{align}
\bar{X}{}^\mu = (\bar{t}, \bar{\xi}, z^a, y^i) \, , \quad a=1,\dots, p \, , \quad i=1,\dots,m = d-p-1 \, .
\end{align}
The considerations are otherwise identical to those of Sec.~\ref{sec:ch3:rotating-p-branes}, with the exception that we single out one preferred $\bar{\xi}$-direction. The remaining space is split between the $m$-dimensional transverse space and the $p$-dimensional space along the brane; see Fig.~\ref{fig:ch4:darboux-4} for a visualization. The main conceptual difference lies in the fact that the transverse space is now $(d-p-1)$-dimensional, since we also subtract the $\bar{\xi}$-direction in which the $p$-brane will be boosted. We call the number of Darboux planes $n_p$ such that
\begin{align}
\label{eq:ch4:np-definition}
d-p-1 = 2n_p + \epsilon \, .
\end{align}

Moreover, we assume that the angular momentum lies entirely in the transverse space, that is, there are no components in either the $\bar{\xi}$-direction or the $z^a$-direction,
\begin{align}
\label{eq:ch3:j-constraints}
j{}_{\bar{\xi}\mu} = j{}_{\mu\bar{\xi}} = j{}_{a\mu} = j{}_{\mu a} = 0 \, .
\end{align}
And last, the line density $\bar{\epsilon}(\bar{\xi})$ is assumed to be non-zero only inside an interval $\bar{\xi}\in[0,\bar{L}]$. In this sense this corresponds to the generalization of a ``pencil'' to the case of a $p$-brane: it is a $p$-brane that has a finite thickness in the boost direction $\bar{\xi}$. This assumption is made in order to define a proper Penrose limit by keeping the resulting boosted $p$-brane of finite width, in complete analogy to the Penrose limit of the pencil discussed in the previous section.

\subsection{$p$-brane metric}
By linearity, we choose the corresponding metric ansatz to be similar to the rotating $p$-brane ansatz \eqref{eq:ch3:warped-ansatz-rotating} and stipulate
\begin{align}
\dd s^2 &= \left[1+\bar{\phi}(\bar{r})\right]\left[-\dd \bar{t}^2 + \sum\limits_{a=1}^p (\dd z^a)^2 \right] + \left[1+\bar{\psi}(\bar{r})\right] \left[ \sum\limits_{i=1}^m (\dd y^i)^2 + \dd\bar{\xi}^2 \right] + 2 \bar{\ts{A}}(\ts{y})\dd \bar{t} \, , \\
\bar{r}^2 &= r_\perp^2 + \bar{\xi}^2 \, , \quad r_\perp^2 = \sum\limits_{i=1}^m (y^i)^2 \, , \quad \bar{\ts{A}}(\ts{y}) = \bar{A}_i(\ts{y})\dd y{}^i \, .
\end{align}
The solution is given by
\begin{align}
\bar{\psi}(\bar{r}) &= \frac{p+1}{2-m} \bar{\phi}(\bar{r}) \, , \quad
\bar{\phi}(\bar{r}) = \frac{2-m}{p+m-1} 2\kappa \int \limits_{-\infty}^\infty \dd \bar{\xi}' \, \bar{\epsilon}(\bar{\xi}') \mathcal{G}_{m+1}(\bar{r}') \, , \\
\bar{A}_i(\bar{\ts{r}}) &= -2\pi\kappa \int\limits_{-\infty}^\infty \dd\bar{\xi}' \, \bar{j}_{ij}(\bar{\xi}') y^j \mathcal{G}_{m+3}(\bar{r}') \, , \quad
\bar{r}'^2 = \left(\bar{\xi} - \bar{\xi}'\right)^2 + r_\perp^2 \, ,
\end{align}
where the static Green function in $(m+1)$ dimensions appears in the gravitoelectric potential $\bar{\phi}$ because $d-p = m+1$. In this notation we can also introduce the Darboux planes
\begin{align}
r_\perp^2 = \epsilon z^2 + \sum\limits_{A=1}^{n_p} \rho_A^2 \, , \quad \sum\limits_{i=1}^m(\dd y^i)^2 = \epsilon \dd z^2 + \sum\limits_{A=1}^{n_p}\left(\dd\rho_A^2 + \rho_A^2\dd\varphi_A^2 \right) \, .
\end{align}

\subsection{Boost and Penrose limit}
We can now again perform the boost from the $\bar{S}$ frame to the $S$ frame such that
\begin{align}
\overline{t} = {\gamma\over \sqrt{2}}[(1+\beta) u+(1-\beta)v] \, , \quad
\overline{\xi} = {\gamma\over \sqrt{2}}[-(1+\beta) u+(1-\beta)v] \, .
\end{align}
The scaling relations for the $p$-brane's energy density and angular momentum density are generalized from the pencil-shaped gyraton as follows:
\begin{align}
\epsilon(u) &= \lim \limits_{\gamma\rightarrow\infty} \sqrt{2} \gamma^2 \, \overline{\epsilon}(-\sqrt{2}\gamma u) \, , \\
j_{ij}(u) &= \lim\limits_{\gamma\rightarrow\infty} \sqrt{2}\gamma \, \overline{j}_{ij}(-\sqrt{2}\gamma u) \, , \quad j_A(u) = \lim\limits_{\gamma\rightarrow\infty} \sqrt{2}\gamma \, \bar{j}_A(-\sqrt{2}\gamma u) \, .
\end{align}
This constitutes a physically sensible generalization of the Penrose limit to $p$-branes since the boost is again performed along the $\bar{\xi}$-direction, as we assume that the $p$-brane has a finite thickness $\bar{L}$ along that direction. The only change in these calculations is the form of the original metric, which is that of a warped geometry. After the boost one obtains
\begin{align}
\label{eq:ch4:gyratonic-p-brane}
\ts{g} &= \left(\eta{}_{\mu\nu} + h{}_{\mu\nu}\right) \dd X{}^\mu \dd X{}^\nu = -2\dd u \dd v + \phi \dd u^2 + 2 A{}_i\dd y{}^i \dd u + \sum\limits_{a=1}^p (\dd z^a)^2 + \sum\limits_{i=1}^m (\dd y^i)^2 \, , \\
\phi &= \lim\limits_{\gamma\rightarrow\infty} 2 \gamma^2 \frac{m+p-1}{2-m} \bar{\phi} \, , \quad A_i = \lim\limits_{\gamma\rightarrow\infty} \sqrt{2} \gamma \bar{A}_i \, .
\end{align}
We may again perform the limiting case of $\gamma\rightarrow\infty$ using the universal relation \eqref{eq:ch4:g-gamma-limit} such that
\begin{align}
\label{eq:ch4:p-brane-phi-final}
\phi(u,r_\perp) = 2\sqrt{2} \kappa \epsilon(u) \mathcal{G}_m(r_\perp) \, , \quad A_i(u,y^i) = -2\pi\kappa j{}_{ij}(u) y^j \mathcal{G}_{m+2}(r_\perp) \, , \quad r_\perp^2 = \sum\limits_{i=1}^m (y^i)^2 \, .
\end{align}
Using the $n_p$ independent Darboux planes we can also rewrite the gravitomagnetic potential as
\begin{align}
\label{eq:ch4:p-brane-a-final}
A_i(\ts{r}_\perp) \dd y{}^i = 2\pi\kappa \mathcal{G}_{m+2}(r_\perp) \sum\limits_{A=1}^{n_p} j_A(u) \rho_A^2 \dd \varphi_A \, , \quad r_\perp^2 = \sum\limits_{A=1}^{n_p} \rho_A^2 + \epsilon z^2 \, .
\end{align}
This result is quite interesting, since it implies that the presence of the $p$-brane does not affect the functional form of the metric since Eq.~\eqref{eq:ch4:gyratonic-p-brane} is functionally identical to \eqref{eq:gyraton-1}. The reason is that the $z^a$-sector remains unaffected by the boost in the $\bar{\xi}$-section, and for this reason the corresponding metric components tend to zero in the ultrarelativistic limit. This is mathematically identical to the $y^i$-sector as well as to the entire transverse $x_\perp^i$-sector in the case of the pencil-shaped gyratons described in the previous section, which also vanish asymptotically in the Penrose limit.

The difference between the gyratonic $p$-brane metric and the gyraton metric lies in the functional dependence of the metric functions, as well as in the precise metric functions. Whereas in the case of a pencil gyraton the transverse distance is a $(d-1)$-dimensional distance (because only the $\xi$-direction does not play a role), for a gyratonic $p$-brane the transverse direction is $(d-1-p)$-dimensional since not only the $\xi$-direction disappears but also the $p$ spatial directions along the brane do not enter due to the translational isometry of the original $p$-brane metric in that direction. For that reason the metric functions for the pencil are $\mathcal{G}_{d-1}(r_\perp)$ and $\mathcal{G}_{d+1}(r_\perp)$, whereas for the gyratonic $p$-brane one has instead $\mathcal{G}_m(r_\perp)$ and $\mathcal{G}_{m+2}(r_\perp)$. In the limiting case of $p=0$ the $z^a$-sector collapses and one has $m=d-1$, thereby recovering the results of the pencil-shaped gyraton.

\subsection{Examples in $d=3$?}
Let us consider the simplest example of a $p$-brane, that is, a cosmic string with $p=1$ in $(3+1)$-dimensional spacetime. As it turns out, it is impossible to construct a \emph{rotating} string in $d=3$ subject to \eqref{eq:ch3:j-constraints} because $k_p=0$ and $\epsilon=1$ as per Eq.~\eqref{eq:ch4:np-definition}. This implies that $m=1$ and hence the example is quite degenerate: there is not a single Darboux plane, and hence we cannot define a rotation parameter. Moreover, the transverse space is only one-dimensional, and hence
\begin{align}
\ts{g} &= -2\dd u \dd v + \phi \dd u^2 + (\dd z^1)^2 + \dd z^2 \, , \quad \phi(u,z) = 2\sqrt{2} \kappa \epsilon(u) \mathcal{G}_1(z) \, .
\end{align}
Here $z^1$ is the direction along the brane and $z$ is the only transverse coordinate.

\subsection{Example in $d=4$}
Instead, let us focus on the simplest non-trivial case: a cosmic string ($p=1$) in $(4+1)$-dimensional spacetime. Indeed, for $d=4$ the transverse space of such a string is two-dimensional such that there is one Darboux plane with an angular momentum profile $j(u)$ and coordinates $\{\rho,\varphi\}$. The metric and the gravitational potentials are then
\begin{align}
\ts{g} &= -2\dd u \dd v + \phi \dd u^2 + 2 \ts{A} \dd u + (\dd z^1)^2 + \dd\rho^2 + \rho^2\dd\varphi^2 \, , \\
\phi(u,\rho) &= 2\sqrt{2} \kappa \epsilon(u) \mathcal{G}_2(\rho) \, , \quad A_i(\ts{r}_\perp) \dd y{}^i = 2\pi\kappa \mathcal{G}_4(\rho) j(u) \rho^2 \dd \varphi \, .
\end{align}
In this case the $z^1$-direction denotes the direction along the $1$-brane and should not be confused with the additional $z$-coordinate that may exist if the transverse space is odd-dimensional.

\subsection{Higher dimensions}
Similar to what we elucidated in Sec.~\ref{sec:ch4:gyratons-higher-dimensions}, with Eqs.~\eqref{eq:ch4:gyratonic-p-brane}--\eqref{eq:ch4:p-brane-a-final} one can algorithmically generate rotating gyratonic $p$-branes in an arbitrary number of dimensions for linearized General Relativity as well as a wide range of linearized non-local ghost-free gravity theories.

\section{Concluding remarks}
In this chapter we have constructed the gravitational field of ultrarelativistic objects by performing a suitable Penrose limit for weak-field solutions found in Ch.~\ref{ch:ch3}. While the obtained solutions have the same asymptotic behavior the gyraton solutions encountered in General Relativity, a striking difference is their regularity: the gravitational field is finite at the location of the infinitely thin matter source. Moreover, the gyraton metrics include the well known Aichelburg--Sexl metric as a special limiting case. In a last step, we have applied a generalized Penrose limit to rotating $p$-branes and thereby constructed objects that we refer to as gyratonic $p$-branes.

An interesting questions remains: are the obtained metrics also solutions to the full non-linear non-local field equations? It seems possible since in the Penrose limit the effective mass is scaled with the inverse of the Lorentz boost parameter, making it arbitrarily small. We hope to address this question in more detail inside our future light cone.

%%%%%%%%%%%%%%%%%%%%%%%%%%%%%%%%%%%%%%%%%%%%%%%%%%%%%%%%%%%%%%%%%%%%%%%%%%%%%%%%%%%%%%%%%%%%%%%%%%%
%
% Chapter: Quantum-mechanical scattering
%
\chapter{Quantum-mechanical scattering}
\label{ch:ch5}
\textit{Turning towards quantum aspects of non-locality, we will focus on the quantum-mechanical scattering on a $\delta$-shaped potential. Using asymptotic properties of the non-local modification terms we will extract the scattering coefficients and analyze how the presence of non-locality affects their properties. This chapter is based on Ref.~\cite{Boos:2018kir}.}

\section{Introduction}
Scattering experiments are fascinating probes for the quantum aspects of Nature, with applications ranging to state-of-the-art particle colliders searching for new physics beyond the Standard Model. In this chapter we will focus on the quantum-mechanical scattering problem of a plane wave in the presence of a $\delta$-shaped potential.

While it is straightforward in local quantum mechanics to solve this problem using the standard language of scattering states in a Hilbert space, in non-local quantum theory this is not so simple: the main problem lies in the fact that the Hilbert space is difficult to define since the Hamiltonian operator contains infinitely many derivatives. For this reason we will follow a different avenue. Namely, if we are only interested in asymptotic properties of the scattering problem (such as the scattering coefficients) we may use the fact that in a static situation the effects of non-locality are limited to a finite spatial domain of $\mathcal{O}(\ell)$, where $\ell>0$ is the scale of non-locality; see also the relevant discussions on the initial value problem in Sec.~\ref{sec:ch1:initial-value} as well as the asymptotic causality criterion in Sec.~\ref{sec:ch2:asymptotic-causality}. Asymptotically we therefore need to recover the same scattering states as the ones encountered in local quantum theory, with their amplitudes possibly depending on the presence of non-locality. In what follows we will describe how we can use this property to extract the non-local scattering coefficients.

\section{A non-local scalar field in quantum mechanics}
We begin with a non-local modification of the Klein--Gordon equation describing the dynamics of a real scalar field $\phi$ in presence of a potential $V$,
\begin{align}
\label{eq:ch5:non-local-klein-gordon}
\big[\mathcal{D} - V(\ts{x}) \big] \varphi(t,\ts{x}) = 0 \, .
\end{align}
Here, $\mathcal{D}$ is a differential operator given by
\begin{align}
\mathcal{D} = \exp\left[(-\ell^2\Box)^N\right] \Box \, , \quad \ell > 0 \, , \quad N \in \mathbb{N} \, .
\end{align}
In the local limit $\ell\rightarrow 0$ one has $\mathcal{D}=\Box$ and hence recovers the local Klein--Gordon equation. Since we assume the potential to be time-independent we give up Lorentz invariance and will limit our considerations to the frame where the situation is static. Then, we may perform a temporal Fourier decomposition of the field $\varphi$ such that
\begin{align}
\label{eq:ch5:phi-fourier}
\varphi(t,x) = \int\limits_{-\infty}^\infty \frac{\dd\omega}{2\pi} e^{-i\omega t} \varphi_\omega(\ts{x}) \, .
\end{align}
Since $\varphi$ is a real-valued field, the complex Fourier components satisfy $\varphi_{-\omega} = \varphi_\omega^\ast$. Inserting the decomposition \eqref{eq:ch5:phi-fourier} into \eqref{eq:ch5:non-local-klein-gordon} yields a differential equation for each Fourier mode,
\begin{align}
\label{eq:ch5:non-local-klein-gordon-fourier}
\big[ \mathcal{D}_\omega - V(\ts{x}) \big] \varphi_\omega(\ts{x}) = 0 \, ,
\end{align}
where the differential operator $\mathcal{D}_\omega$ is given by
\begin{align}
\mathcal{D}_\omega = \exp\left[(-\ell^2\lap - \omega^2\ell^2)^N\right] (\lap + \omega^2) \, .
\end{align}
It is our goal to find the scattering coefficients for each Fourier mode off of the potential $V(\ts{x})$, and we will demonstrate how to find them exactly for the simple case of a $\delta$-shaped potential
\begin{align}
V(\ts{x}) = \lambda\delta(\ts{x}) \, .
\end{align}
In what follows, we will first discuss the Lippmann--Schwinger method for solving scattering problems of the above form. Then, we will apply this method to find the scattering coefficients for each Fourier mode in the local case where $\ell=0$ and we may check our results against the literature. In a last step, we will generalize the calculations to the non-local case, and for calculational simplicity we will focus on the case of $N=1$. As argued in Sec.~\ref{sec:ch2:causality-from-analyticity-non-local}, the case of $N=1$ is well-defined as long as no summation over $\omega$ is performed. For this reason we will restrict ourselves to the analysis of the scattering coefficients for each Fourier mode, but it should be noted that our methods can also be used to derive the full, real-space scattering coefficients for any wave packet, albeit in the mathematically more involved non-local $\mathrm{GF_2}$ theory with $N=2$.

\section{Lippmann--Schwinger method}
Suppose the operator $\mathcal{D}_\omega$ has a Green function $G_\omega(\ts{x}'-\ts{x})$ such that
\begin{align}
\mathcal{D}_\omega G_\omega(\ts{x}'-\ts{x}) = - \delta(\ts{x}'-\ts{x}) \, .
\end{align}
We will later assume that $G_\omega(\ts{x}'-\ts{x})$ is the retarded Green function, but this formalism works for any Green function that solves the above equation. Moreover, let $\varphi_\omega^0(\ts{x})$ be a solution of the free equation
\begin{align}
\mathcal{D}_\omega \varphi_\omega^0(\ts{x}) = 0 \, .
\end{align}
The physical interpretation of the mode $\varphi_\omega^0(\ts{x})$ is that of a freely moving plane wave that parametrizes the incoming part of the scattering problem. Then, the exact solution for $\varphi_\omega(\ts{x})$ is
\begin{align}
\varphi_\omega(\ts{x}) = \varphi^0_\omega(\ts{x}) - \int\dd\ts{y} G_\omega(\ts{y}-\ts{x}) V(\ts{y}) \varphi_\omega(\ts{y}) \, ,
\end{align}
which is sometimes called the Lippmann--Schwinger equation\cite{Lippmann:1950}\index{Lippmann--Schwinger equation}. Indeed, the above solves the Klein--Gordon equation \eqref{eq:ch5:non-local-klein-gordon-fourier}:
\begin{align}
\begin{split}
\mathcal{D}_\omega \varphi_\omega(\ts{x}) &= \mathcal{D}_\omega \varphi^0_\omega(\ts{x}) - \int\dd\ts{y} \mathcal{D}_\omega G_\omega(\ts{y}-\ts{x}) V(\ts{y}) \varphi_\omega(\ts{y}) \\
&= 0 + \int\dd\ts{y} \delta(\ts{y}-\ts{x}) V(\ts{y}) \varphi_\omega(\ts{y}) = V(\ts{x}) \varphi_\omega(\ts{x}) \, .
\end{split}
\end{align}
However, the above Lippmann--Schwinger\index{Lippmann--Schwinger equation} integral is just a representation of the solution since it is an integral equation for $\varphi_\omega(\ts{x})$ that can only be solved if the integral kernel can be inverted. In general this may not always be possible, but then one may instead use the Lippmann--Schwinger representation for an iterative approximation scheme where
\begin{align}
\varphi^{(1)}_\omega(\ts{x}) &= \varphi^0_\omega(\ts{x}) - \int\dd\ts{y} G_\omega(\ts{y}-\ts{x}) V(\ts{y})  \varphi^0_\omega(\ts{y}) \, , \\
\varphi^{(2)}_\omega(\ts{x}) &= \varphi^0_\omega(\ts{x}) - \int\dd\ts{y} G_\omega(\ts{y}-\ts{x}) V(\ts{y})  \varphi^{(1)}_\omega(\ts{y}) \, , \\
\varphi^{(3)}_\omega(\ts{x}) &= \varphi^0_\omega(\ts{x}) - \int\dd\ts{y} G_\omega(\ts{y}-\ts{x}) V(\ts{y})  \varphi^{(2)}_\omega(\ts{y}) \, , \quad \dots \, ,
\end{align}
which is known as the Born series\index{Born series}. However, it is not always clear if and under which conditions this method is accurate, and in what follows we shall not follow this avenue. Instead, we will focus on a simple potential that allows for substantial simplifications.

\section{Transmission and reflection coefficients for a $\delta$-potential}
In the simple case of $V(\ts{x}) = \lambda\delta(\ts{x})$ with $\lambda>0$ it is possible to extract an exact solution from the Lippmann--Schwinger representation as the integral collapses,
\begin{align}
\varphi_\omega(\ts{x}) = \varphi^0_\omega(\ts{x}) - \lambda G_\omega(\ts{x}) \varphi_\omega(0) \, .
\end{align}
Provided $1 + \lambda G_\omega(0) \not=0$ we find the exact result
\begin{align}
\label{eq:ch5:phi-solution-local}
\varphi_\omega(\ts{x}) = \varphi^0_\omega(\ts{x}) - \frac{\lambda \varphi^0_\omega(0)}{1 + \lambda G_\omega(0)} G_\omega(\ts{x}) \, .
\end{align}

\subsection{Local case}
At this point we may choose our initial conditions and appropriate Green function. Let us study the well-known local case first \cite{Dirac:1947,Landau:1965,Tannoudji:1987,Sakurai:1994,Shankar:1994,Griffiths:1995}. For simplicity we will also restrict ourselves to the one-dimensional case such that the differential equation becomes
\begin{align}
\big[ \mathcal{D}_\omega - \lambda\delta(x) \big] \varphi_\omega(x) = 0 \, , \quad \mathcal{D}_\omega = \partial_x^2 + \omega^2 \, .
\end{align}
The Green function to the homogeneous equation with $\lambda=0$ is
\begin{align}
(\partial_x^2 + \omega^2) G_\omega(x'-x) = -\delta(x'-x) \, ,
\end{align}
where $G_\omega(x)$ is the partial Fourier transform of the free Green function of a scalar field we already discussed previously in Sec.~\ref{sec:ch2:physical-interpretation}. Choosing the free solution $\varphi_\omega^0(x)$ to be a right-moving plane wave we have
\begin{align}
\varphi^0_\omega(x) = e^{i\omega x} \, , \quad G_\omega(x) = G^\text{R}_\omega(x) = \frac{i}{2\omega}e^{i\omega|x|} \, , \quad \omega > 0 \, ,
\end{align}
where, as per our previous discussions in Sec.~\ref{ch:ch2}, the choice of $G_\omega(x)$ corresponds to the retarded Green function. Without loss of generality we restrict ourselves to a positive frequency Fourier mode. Inserting these choices into \eqref{eq:ch5:phi-solution-local} yields
\begin{align}
\varphi_\omega(x) = e^{i\omega x} - \frac{1}{1-2i\gamma} e^{i\omega|x|} = \begin{cases} \displaystyle \frac{-2i\gamma}{1-2i\gamma}e^{i\omega x} & \text{for~} x > 0 \, , \\[15pt] \displaystyle e^{i\omega x} - \frac{1}{1-2i\gamma} e^{-i\omega x} & \text{for~} x < 0 \, , \end{cases}
\hspace{45pt} \gamma = \frac{\omega}{\lambda} \, .
\end{align}
For $x>0$ the solution is purely given by transmitted, right-moving solution. For $x<0$, however, there solution consists of a superposition of both the original, right-moving plane wave as well as a left-moving reflected part. If we had chosen instead a left-moving ingoing wave, the situation would be mirrored, but it is clear by the symmetry of the situation that the results are identical under the transformation $x \rightarrow -x$.

\begin{figure}
  \centering
  \includegraphics[width=0.75\textwidth]{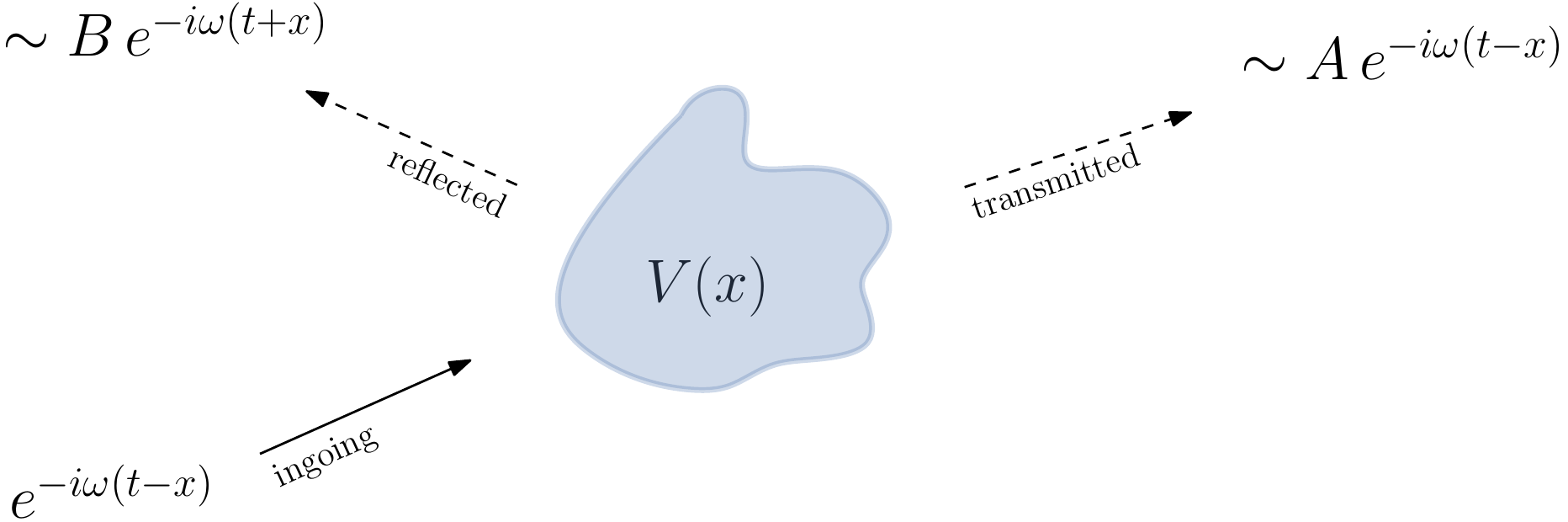}\vspace{10pt}
  \caption[Visualization of scattering coefficients.]{In a time-independent setting, the quantum-mechanical scattering coefficients can be conveniently derived by studying plane waves of a fixed frequency $\omega>0$. A free, ingoing wave $\sim e^{-i\omega(t-x)}$, after scattering at the linear level off of the static potential $V(x)$, can be decomposed into a reflected part $\sim B e^{-i\omega(t+x)}$ and a transmitted part $\sim A e^{-i\omega(t-x)}$, where $A$ and $B$ are complex numbers that depend on the properties of the potential $V(x)$. The transmission coefficient $T$ is given by $|A|^2$ and the reflection coefficient is $|B|^2$ with $|A|^2+|B|^2 = 1$ due to unitarity. Note that these coefficients depend on the frequency of the Fourier mode for which they are calculated.}
  \label{fig:ch5:scattering-sketch}
\end{figure}

We can now use these expressions to read off the transmission and reflection coefficients, see also Fig.~\ref{fig:ch5:scattering-sketch} for a visualization. The coefficients are
\begin{align}
t_\omega = -\frac{2i\gamma}{1-2i\gamma} \, , \quad r_\omega = -\frac{1}{1-2i\gamma}
\end{align}
Then, we can define $T_\omega$ and $R_\omega$ such that
\begin{align}
T_\omega = |t_\omega|^2 = \frac{4\gamma^2}{1+4\gamma^2} \, , \quad R_\omega = |r_\omega|^2 = \frac{1}{1+4\gamma^2} \, , \quad T_\omega + R_\omega = 1 \, .
\end{align}
It is now our goal to extend these computations to the non-local case of $\mathrm{GF_1}$ theory.

\subsection{Non-local case}
Let us now switch non-locality on, $\ell>0$, and consider the non-local Klein--Gordon equation
\begin{align}
\big[ \mathcal{D}_\omega - \lambda\delta(x) \big] \varphi_\omega(x) = 0 \, , \quad \mathcal{D}_\omega = \exp\left[(- \ell^2\partial_x^2 -\omega^2\ell^2)^N\right](\partial_x^2 + \omega^2) \, .
\end{align}
Formally we can again introduce a Green function
\begin{align}
\exp\left[(- \ell^2\partial_x^2 - \omega^2\ell^2)^N\right](\partial_x^2 + \omega^2) \mathcal{G}^\text{R}_\omega(x'-x) = -\delta(x'-x) \, .
\end{align}
From now on we will choose this Green function as the retarded Green function. As our previous considerations have shown, this criterion implies that its local part corresponds to the retarded Green function of the local theory. Decomposing this Green function as
\begin{align}
\mathcal{G}^\text{R}_\omega(x'-x) = G^\text{R}_\omega(x'-x) + \Delta\mathcal{G}_\omega(x'-x) \, ,
\end{align}
see also the relevant discussion in Sec.~\ref{sec:ch2:causality-from-analyticity-non-local}, we have
\begin{align}
\Delta\mathcal{G}_\omega(x) = \int\limits_{-\infty}^\infty \frac{\dd q}{2\pi} \cos(qx)A_\omega(q) \, , \quad A_\omega(q) = \frac{1-\exp\left[-\ell^{2N}(q^2-\omega^2)^N\right]}{\omega^2-q^2} \, .
\end{align}
As we have shown previously in Sec.~\ref{sec:ch2:causality-from-analyticity-non-local} this expression can be solved explicitly for $N=1$ but is difficult to treat analytically for higher values of $N$. However, as we will show now, the scattering coefficients can be determined without knowing the exact form of $\Delta\mathcal{G}_\omega(x)$.

Provided $1+\lambda\mathcal{G}_\omega(0)\not=0$ the full solution for $\varphi_\omega(x)$ takes the form
\begin{align}
\label{eq:ch5:phi-solution-nonlocal}
\varphi_\omega(x) = \varphi^0_\omega(x) - \Lambda_\omega \mathcal{G}^\text{R}_\omega(x) \, , \quad \Lambda_\omega = \frac{\lambda \varphi^0_\omega(0)}{1 + \lambda \mathcal{G}_\omega(0)} \, .
\end{align}
For the determination of the scattering coefficients the full form of $\varphi_\omega(x)$ is not required: its asymptotic form is sufficient. As we have previously shown in great detail that at large distances $x\rightarrow\infty$ we have $\mathcal{G}^\text{R}_\omega(x) \rightarrow G^\text{R}_\omega(x)$ we can write asymptotically
\begin{align}
\varphi_\omega(x\rightarrow\infty) = \varphi^0_\omega(x) - \Lambda_\omega G^\text{R}_\omega(x) \, , \quad \Lambda_\omega = \frac{\lambda \varphi^0_\omega(0)}{1 + \lambda \big[ G^\text{R}_\omega(0) + \Delta\mathcal{G}_\omega(0) \big]} \, .
\end{align}
As discussed in Sec.~\ref{sec:ch1:on-shell-off-shell}, homogeneous differential equations are insensitive to a non-local modification term. For this reason we may again choose the same free, right-moving plane wave $\varphi^0_\omega(x) = e^{i\omega x}$ for the non-local scattering problem. Since the local Green function is an outgoing plane wave, $G^\text{R}_\omega(x) =\tfrac{i}{2\omega} e^{i\omega|x|}$, the scattering coefficients can be read off and can be recast purely in terms of the non-local contribution $\Delta\mathcal{G}_\omega(0)$:
\begin{align}
\label{eq:ch4:scattering-non-local}
T_\omega = |t_\omega|^2 = \frac{4\gamma^2}{1+4\gamma^2} \, , \quad R_\omega = |r_\omega|^2 = \frac{1}{1+4\gamma^2} \, , \quad \gamma = \omega\left(\frac{1}{\lambda} + \Delta\mathcal{G}_\omega(0) \right) \, .
\end{align}
These general expressions holds true for a wide range of non-local theories discussed in this thesis.

\subsection{Properties of the scattering coefficients}
With the scattering coefficients available for both the local and non-local theory we are now ready to discuss their properties and isolate the effects that non-locality brings about.

The scattering coefficients depend solely on the value of the non-local modification at the location of the $\delta$-potential. This is not too surprising since the class of non-local ghost-free theories discussed in this thesis are constructed precisely in such a way that their asymptotic behavior coincides with that of the local theory.

Let us make some general remarks before moving on to concrete, analytical examples. It is well known that in the local case the scattering coefficients are invariant under $\lambda\rightarrow-\lambda$. Evidently, non-locality spoils that behavior and the two cases are now distinct. Second, in the non-local case the transmission coefficient may become zero at a finite frequency $\omega_\star$ leading to a complete reflection of any Fourier mode with that frequency,
\begin{align}
\lambda = -\frac{1}{\Delta\mathcal{G}_{\omega_\star}(0)} \, .
\end{align}
We will address this unexpected feature in more detail below. Third, in the limit of $\lambda\rightarrow\infty$, the transmission coefficient no longer vanishes. And last, the condition $1+\lambda\mathcal{G}_\omega(0)\not=0$ has been employed to find the exact representation of $\varphi_\omega(x)$, but there may be frequencies for which this inequality is violated. We will revisit these frequencies in a later section and motivate their interpretation as quasinormal modes of non-local quantum theory.

For the sake of brevity, in what follows let us focus solely on $\mathrm{GF_1}$ theory with $N=1$. Since the analysis is restricted to Fourier modes of a fixed frequency $\omega$, the situation is effectively static and therefore not subject to the temporal instabilities of $\mathrm{GF_1}$ theory. As it turns out, similar results as the ones presented in the following also hold for $\mathrm{GF_2}$ theory, but the analytical expressions become much more involved.

Using the results from Sec.~\ref{sec:ch2:causality-from-analyticity-non-local} for $\Delta\mathcal{G}_\omega(0)$ one has
\begin{align}
\mathcal{G}^\text{R}_\omega(0) = \frac{i}{2\omega}\left[ 1 + \text{erf}(i\omega\ell) \right] \, , \quad \gamma = \frac{\omega}{\lambda} - \frac12 \text{erfi}(\omega\ell) \, ,
\end{align}
where $\text{erfi}(x) = -i\text{erf}(ix)$ is the imaginary error function \cite{Olver:2010}, which is real-valued for $x\in\mathbb{R}$. In the case of $\ell\rightarrow 0$ we recover the local theory, $\gamma=\omega/\lambda$. Inserting this representation into Eq.~\eqref{eq:ch4:scattering-non-local} we can calculate the scattering coefficients for $\mathrm{GF_1}$ theory for arbitrary values of $\lambda$, $\omega$, and $\ell$. For visualization purposes it is useful to consider instead the dimensionless parameters $\lambda\ell$ and $\omega\ell$, where $\ell>0$ takes on the role of a normalization parameter, not dissimilar to the Planck length frequently employed in such considerations.

\begin{figure}
    \centering
    \subfloat{{\includegraphics[width=0.47\textwidth]{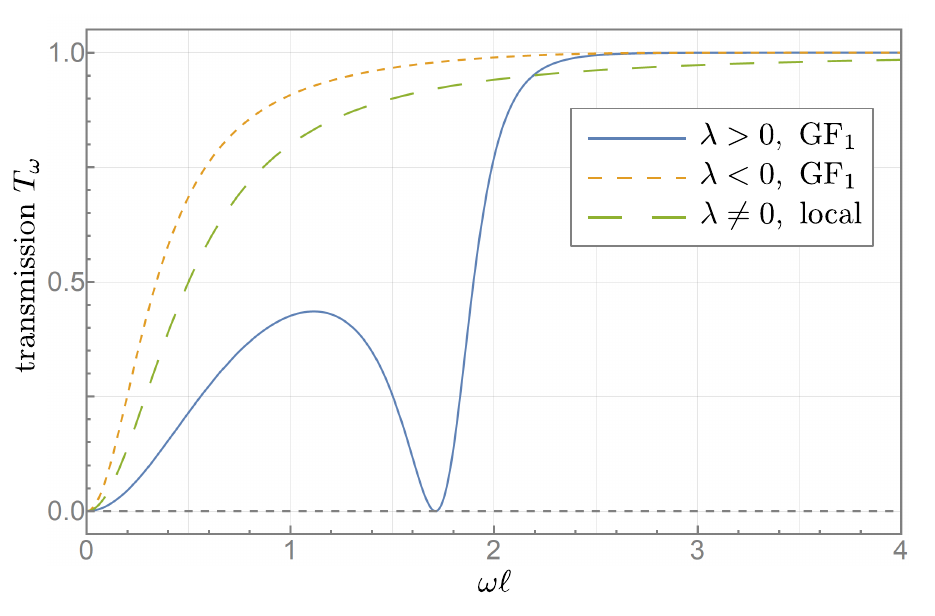} }}
    \subfloat{{\includegraphics[width=0.47\textwidth]{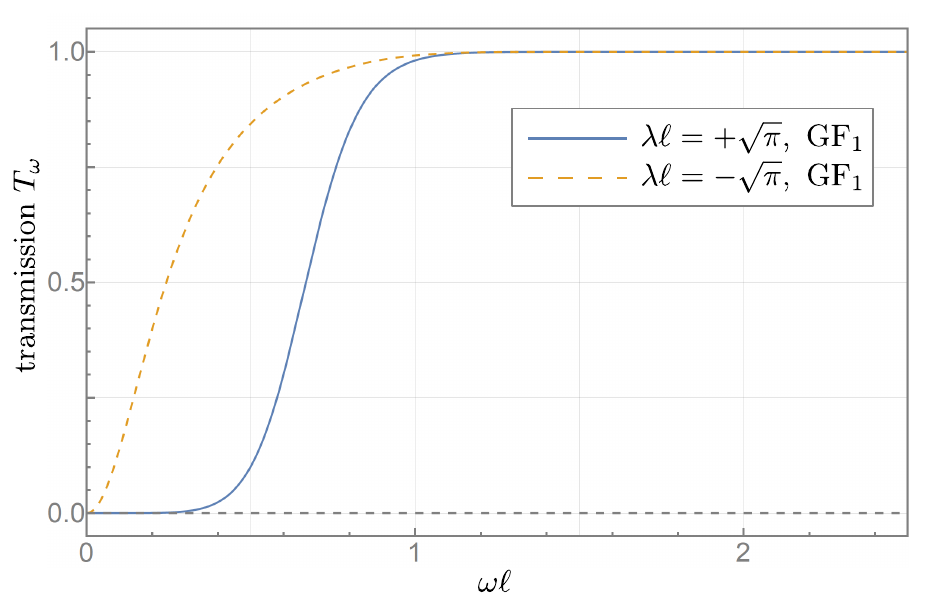} }}

    \caption[Transmission coefficient in the local theory and non-local $\mathrm{GF_1}$ theory.]{We visualize the dimensionless transmission coefficient $T_\omega$ as a function of the dimensionless parameter $\omega\ell$. Left: If the dimensionless coupling $\lambda\ell$ is in the critical regime, $0 < \lambda\ell < \sqrt{\pi}$, the transmission coefficient goes to zero at a critical frequency $\omega_\star>0$. Note that for a $\delta$-potential well ($\lambda<0$) this minimum does not exist. In the local case the transmission coefficient does not vanish anywhere except at zero frequency, and in this case $\ell$ just plays the role of a normalization parameter. Right: At the critical coupling $\lambda\ell = \sqrt{\pi}$, all frequencies $\omega\ell \lesssim 1$ are strongly suppressed. Again, for a $\delta$-potential well with $\lambda = -\sqrt{\pi}$, nothing special happens.}
    \label{fig:ch5:transmission}
\end{figure}

In the local case, where $\ell=0$, this normalization does not appear to make much sense. However, in that case we may still use this constant as an ad hoc normalization. For that reason, in what follows, the local case corresponds to the case $\Delta\mathcal{G}_\omega(0)=0$ and cannot simply be obtained by $\ell\rightarrow 0$. We hope that it is clear that this just corresponds to a particular choice of normalization. Find a graphical representation of the transmission coefficient $T{}_\omega$ as a function of the dimensionless frequency $\omega\ell$ in Fig.~\ref{fig:ch5:transmission}. Let us now address a number of interesting observations.

With a concrete form of the scattering coefficients at our disposal, we can now plot, say, the transmission coefficient as a function of the dimensionless frequency $\omega\ell$, see Fig.~\ref{fig:ch5:transmission}. As already mentioned above, non-locality lifts the $\lambda\rightarrow-\lambda$ degeneracy such that these cases have now distinct scattering coefficients.

\begin{figure}
    \centering
    \subfloat{{\includegraphics[width=0.47\textwidth]{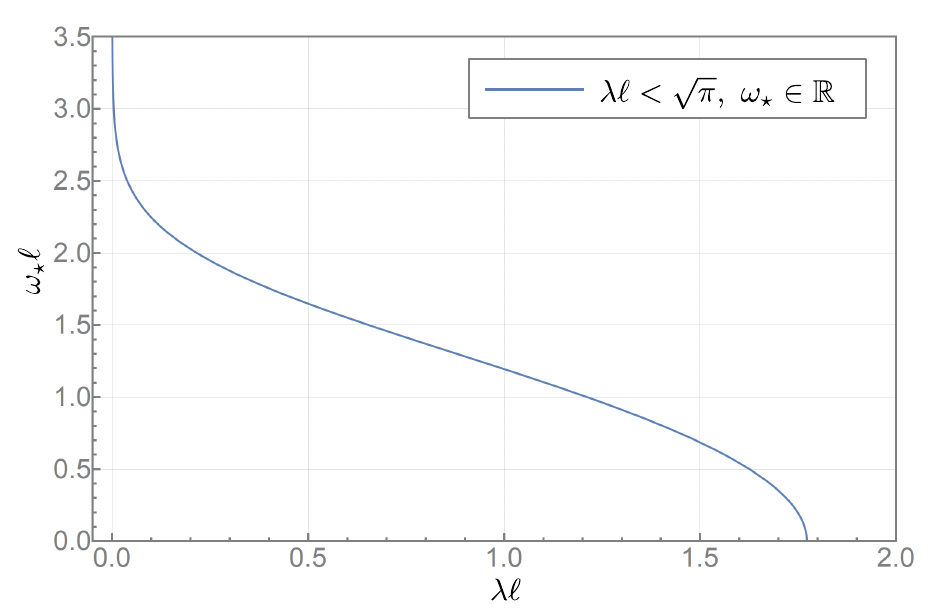} }}
    \subfloat{{\includegraphics[width=0.47\textwidth]{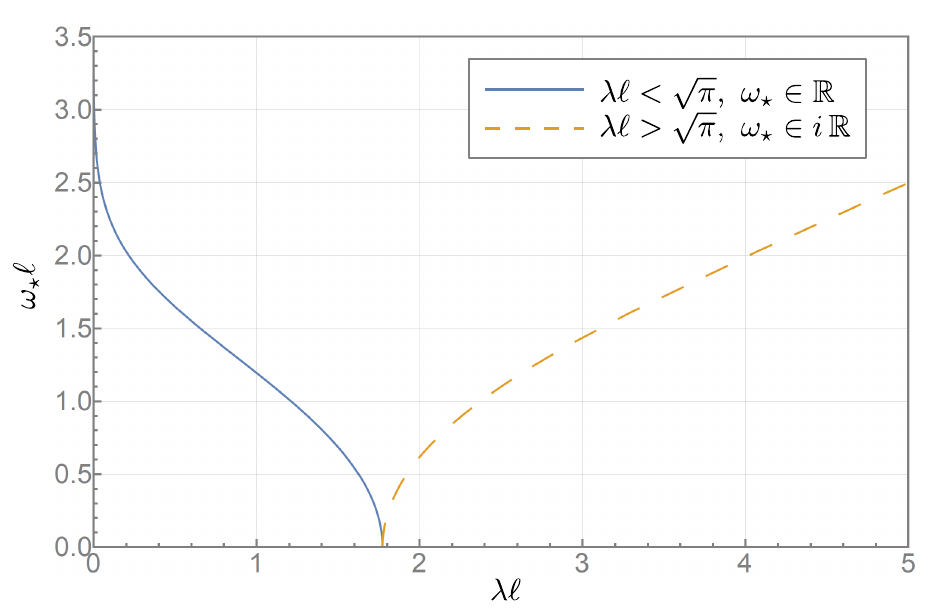} }}
    \caption[Critical frequency in non-local quantum scattering.]{These plots shows the dimensionless frequency $\omega_\star\ell$ as a function of $\lambda\ell$. For $\lambda\to 0$ the critical frequency diverges, but for $\lambda\ell \approx \mathcal{O}(1)$ the frequency $\omega_\star\ell$ is finite. If $\lambda\ell > \sqrt{\pi}$, see the dashed line in the right diagram, the critical frequency becomes imaginary.}
    \label{fig:ch5:critical-omega}
\end{figure}

We can now determine the critical frequency at which the transmission amplitude vanishes,
\begin{align}
\label{eq:ch5:omega-star}
\frac{\omega_\star}{\lambda} - \frac12 \text{erfi}(\omega_\star\ell) = 0 \, .
\end{align}
This equation is transcendent and, to the best of our knowledge, cannot be inverted analytically. However, it is possible to determine when such a solution $\omega_\star$ exists. To that end, we introduce the dimensionless quantities $\omega_\star\ell$ and $\lambda\ell$ such that
\begin{align}
\omega_\star\ell = \frac{\lambda\ell}{2} \text{erfi}(\omega_\star\ell) \, .
\end{align}
Given $\lambda\ell$, the above relation can be considered as a criterion for a straight line $\omega\ell$ to meet with the imaginary error function. As it turns out, this is not always possible because the imaginary error function is convex. It is sufficient to consider the expansion of the imaginary error function at $\omega\ell=0$,
\begin{align}
\text{erfi}(\omega_\star\ell) \approx \frac{2\omega_\star\ell}{\sqrt{\pi}} + \mathcal{O}\left[(\omega_\star\ell)^2 \right] \, .
\end{align}
For this reason, a critical frequency $\omega_\star$ only exists provided
\begin{align}
0 < \lambda\ell < \sqrt{\pi} \, .
\end{align}
In fact, one may show that if $\lambda\ell>\sqrt{\pi}$ the only solution of \eqref{eq:ch5:omega-star} is purely imaginary. Since it is difficult to find an analytical expression for $\omega_\star$ as a function of $\lambda$ we simply plotted $\lambda$ as a function of $\omega_\star$ and swapped the axes (``graphical inversion''); see Fig.~\ref{fig:ch5:critical-omega}. Thus, for a fixed $\lambda$ in this range, there always exists a critical frequency $\omega_\star$ for which the transmission amplitude is exactly zero, and hence the reflection coefficient exactly equals 1. This is an unexpected effect that arises purely due to the presence of non-locality. There is the following scaling relation:
\begin{align}
(\omega_\star-\omega)\ell = \sqrt{3}(1-\lambda\ell/\sqrt{\pi})^{1/2} \, .
\end{align}
That being said, let us try to understand this ``resonant reflection'' of a critical frequency $\omega_\star$. Employing the previously derived asymptotics in the non-local scattering problem we may write in the far-field regime
\begin{align}
\varphi_\omega(x\gg 1) = e^{i\omega x} - B_\omega e^{i\omega|x|} \, , \quad B = 1-i\left[ \frac{2\omega}{\lambda} - \text{erfi}(\omega\ell) \right] \, .
\end{align}
The wave consists of the ingoing plane wave in addition to two ``ringing'' modes that move away from the location of the potential barrier at $x=0$ with the asymptotic amplitude proportional to $B_\omega$. In the critical case one has $B_{\omega_\star} = 1$, which means that the ingoing wave and the transmitted wave cancel exactly. At the same time we need to recall that this is an asymptotic expansion, and hence the field $\varphi_{\omega_\star}$  penetrates the potential barrier a finite amount.

We cannot help but remark that this somewhat resembles high-reflection coatings in optics, where an enhanced reflection of a specific frequency can be generated via special coatings exhibiting a special index of refraction. In this sense this non-local scattering brings about new physics, at wavelengths comparable to the scale of non-locality, $\omega_\star\ell \sim 1$.

\section{Quasinormal modes}
\label{sec:ch5:qnm}
In our previous discussion we limited our consideration to scattering states that asymptotically correspond to plane waves. A necessary criterion to express the full solution in terms of plane waves was that the expression $1+\lambda\mathcal{G}_\omega(0)\not=0$. While scattering states have a continuous frequency $\omega$, let us now focus on \emph{bound states} which have discrete frequencies.

In order to discuss bound states we may again use the Lippmann--Schwinger method for a $\delta$-shaped potential,
\begin{align}
\varphi_\omega(x) = \varphi_\omega^0(x) - \lambda \varphi_\omega(0) \mathcal{G}_\omega(x) \, .
\end{align}
At this time, however, we assume that $\varphi_\omega(x)$ describes a bound state, and we may parametrize this by choosing the free solution to be vanishing, $\varphi_\omega^0(x) \equiv 0$. Then one obtains
\begin{align}
\varphi_\omega(x) = -\lambda \varphi_\omega(0) \mathcal{G}_\omega(x) \, .
\end{align}
Clearly, the consistency condition at $x=0$ can only be satisfied if
\begin{align}
1+\lambda\mathcal{G}_\omega(0) = 0 \, .
\end{align}
For a fixed potential strength $\lambda$ we may interpret the above condition as a condition on the frequency $\omega$, and because $\mathcal{G}_\omega(0)$ takes values in $\mathbb{C}$, the corresponding frequency is generally also complex. At this point we need to specify our definition of what constitutes a bound state. To that end, note that asymptotically when $x/\ell\rightarrow\infty$ the solution takes the form
\begin{align}
\varphi_\omega(x\rightarrow\infty) = -\frac{i\lambda\varphi_\omega(0)}{2\omega} e^{i\omega|x|} \, .
\end{align}
It is clear that only frequencies with a positive imaginary part are normalizable, and for that reason we define bound states to satisfy
\begin{align}
\varphi_\omega(x) = -\lambda \varphi_\omega(0) \mathcal{G}_\omega(x) \, , \quad 1+\lambda\mathcal{G}_\omega(0) = 0 \, , \quad \text{Im}(\omega) \ge 0 \, .
\end{align}
We refer to modes with frequencies subject to the above conditions as \emph{quasinormal modes}; for an in-depth review of quasinormal modes see \cite{Boonserm:2010px} and references therein.\index{quasinormal modes} In general, the condition $1+\lambda\mathcal{G}_\omega(0) = 0$ takes the form
\begin{align}
1+\lambda\left[ \frac{i}{2\omega} + \Delta\mathcal{G}_\omega(0) \right] = 0 \, .
\end{align}
In what follows, we will study the quasinormal modes both for the local theory and the non-local $\mathrm{GF_1}$ theory. Again, as in the calculation of the scattering coefficients, only the non-local modification term evaluated at the location of the potential (here at $x=0$) enters the results.

\subsection{Local case}
In the local case when $\Delta\mathcal{G}_\omega(0)=0$ there is only one quasinormal mode,
\begin{align}
\omega = -\frac{i\lambda}{2} \, .
\end{align}
The imaginary part of this frequency is positive if and only if $\lambda<0$. This corresponds to the well known fact that only an attractive $\delta$-potential ($\lambda<0$) has a normalizable bound state in local quantum theory, whereas a repulsive $\delta$-potential with $\lambda>0$ does not admit any such states.

\subsection{Non-local case}
\begin{figure}
    \centering
    \subfloat{{\includegraphics[width=0.47\textwidth]{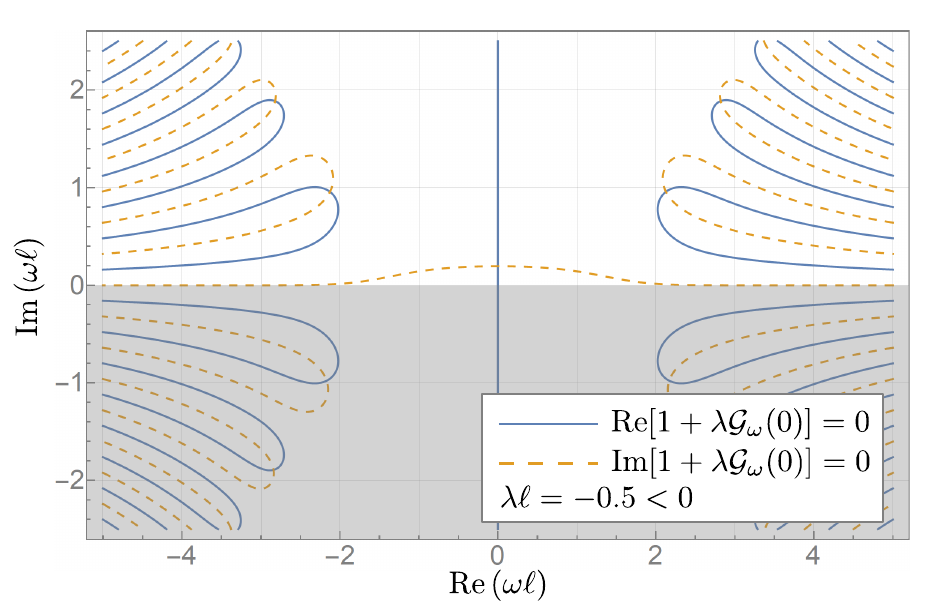} }}
    \subfloat{{\includegraphics[width=0.47\textwidth]{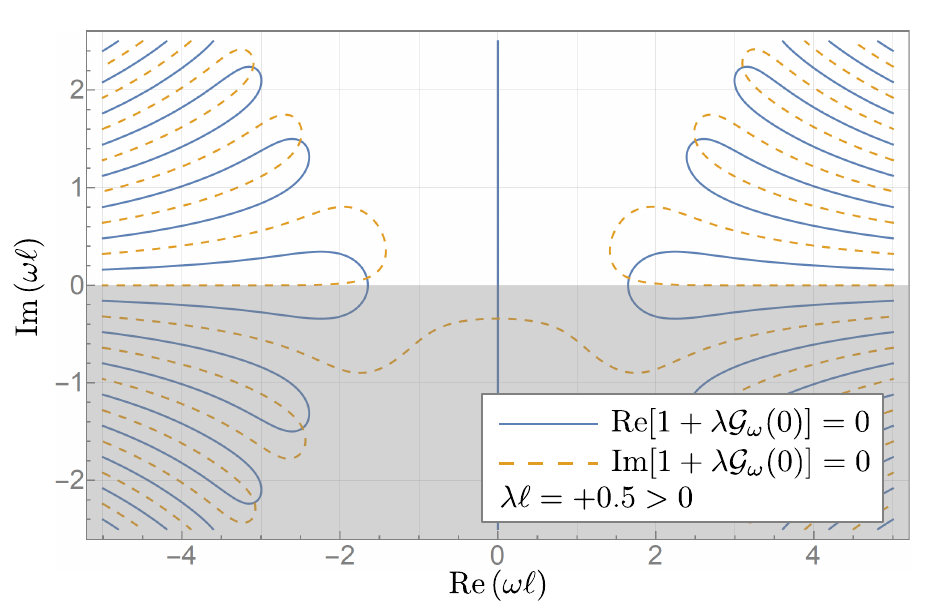} }}\\
    \subfloat{{\includegraphics[width=0.47\textwidth]{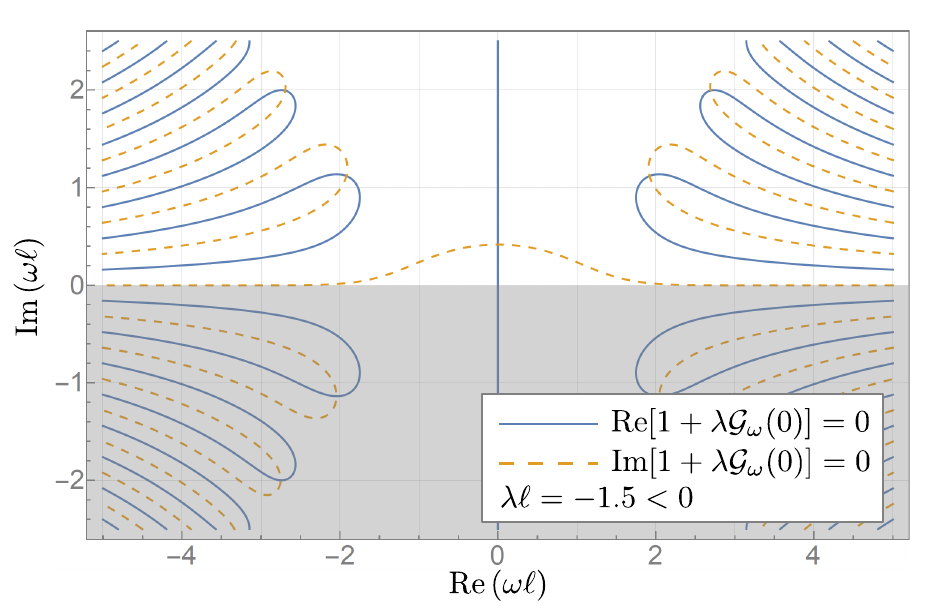} }}
    \subfloat{{\includegraphics[width=0.47\textwidth]{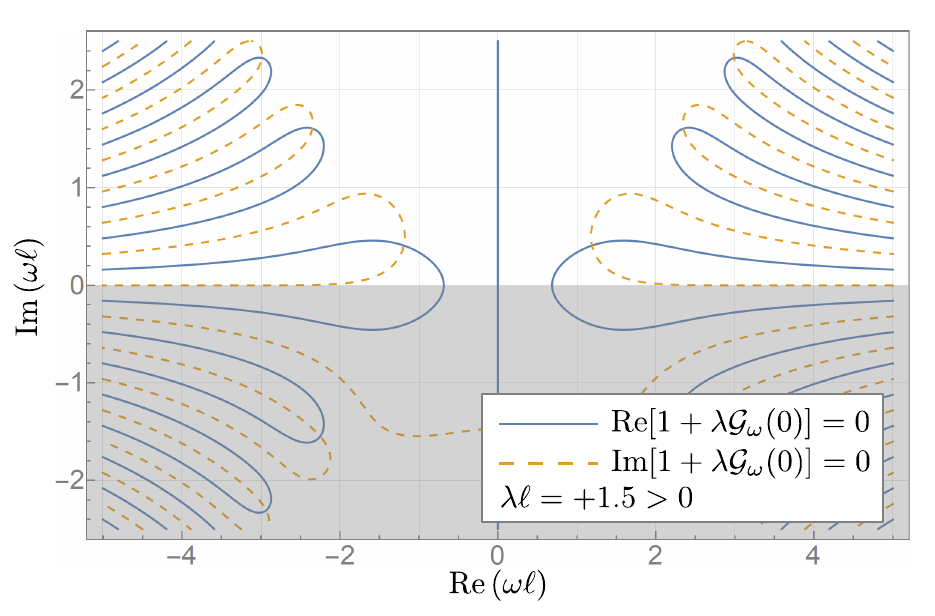} }}\\
    \subfloat{{\includegraphics[width=0.47\textwidth]{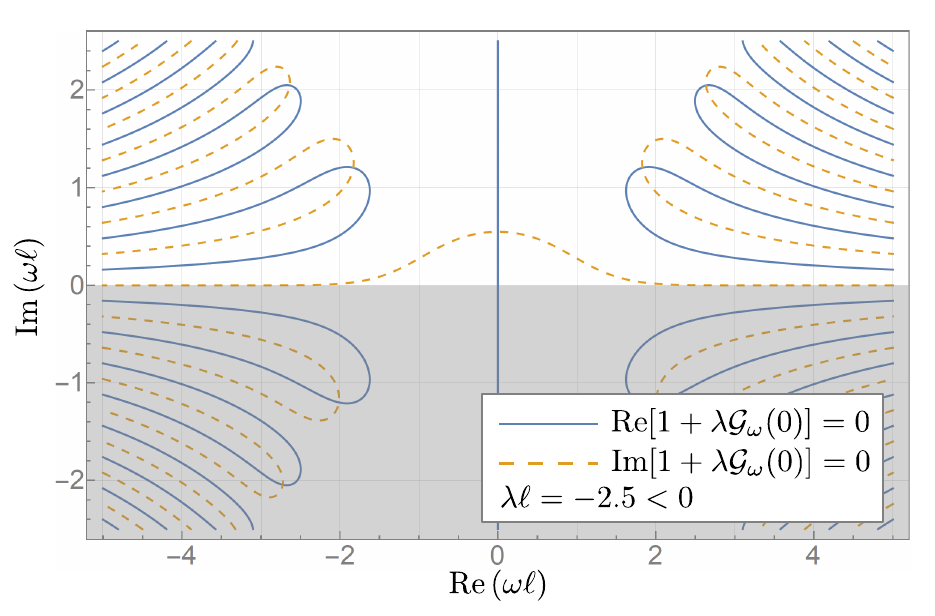} }}
    \subfloat{{\includegraphics[width=0.47\textwidth]{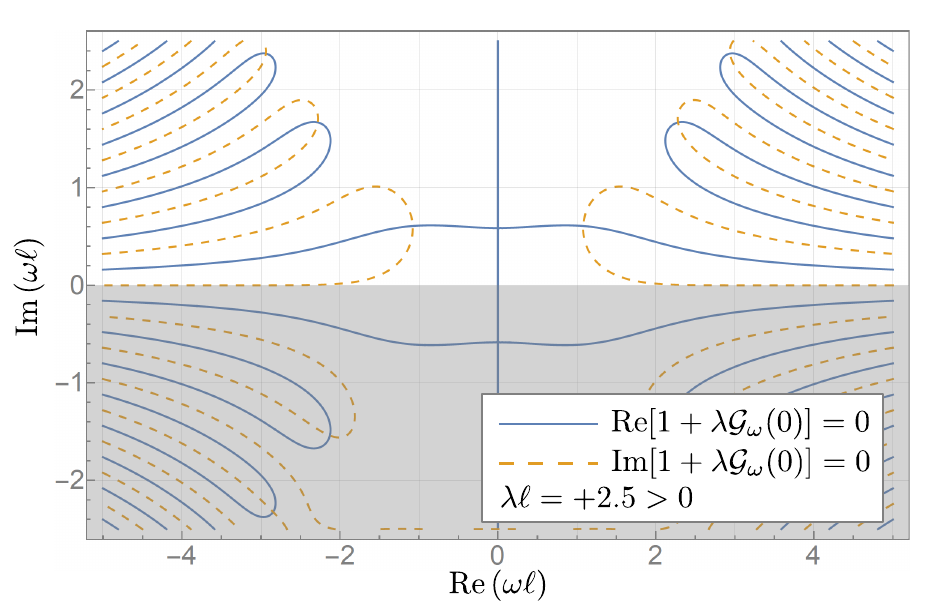} }}\\
    
    \caption[Quasinormal modes in non-local quantum scattering.]{We study the transcendental equation $1+\lambda\mathcal{G}_\omega(0)=0$ graphically for both negative (left) and positive (right) dimensionless couplings $\lambda\ell$. Each intersection point corresponds to a complex quasinormal mode $\omega\ell$, where we employ $\ell>0$ as a normalization factor. The shaded area contains un-normalizable modes with negative imaginary part. The central, purely imaginary mode corresponds to the well-known bound state $\omega=-i\lambda/2$ that is only bounded for $\lambda<0$, just as in the local theory. All other intersection points correspond to novel quasinormal modes that only exist in the non-local theory. }
    \label{fig:ch5:qnm}
\end{figure}

The situation is drastically different in the non-local theory. For $\mathrm{GF_1}$ theory, the bound state frequency criterion becomes
\begin{align}
\frac{\omega}{\lambda} + \frac{i}{2} - \frac12 \text{erfi}(\omega\ell) = 0 \, .
\end{align}
This is a transcendental equation and we decided to study its properties numerically. In particular, parametrizing $\omega=\omega_r + i \omega_i$ one may interpret this as two real-valued constraint equations on the two variables $\omega_r$ and $\omega_i$ such that it is possible to find the contour lines of this constraint in the $\omega_r\omega_i$-plane. We may use the scale of non-locality $\ell>0$ as a normalization parameter such that we measure both the frequency and potential strength in units of $\ell^{-1}$. Then, the only variable is the dimensionless potential strength $\lambda\ell$. In Fig.~\ref{fig:ch5:qnm} we plot the resulting contour lines of the following two real-valued equations:
\begin{align}
&\text{Re}\left\{ \frac{\omega_r + i \omega_y}{\lambda} + \frac{i}{2} - \frac12 \text{erfi}\big[(\omega_r + i \omega_i)\ell\big] \right\} = \frac{\omega_r}{\lambda} - \frac12 \text{Re} \Big\{ \text{erfi}\big[(\omega_r + i \omega_i)\ell\big] \Big\} = 0 \, , \\
&\text{Im}\left\{ \frac{\omega_r + i \omega_i}{\lambda} + \frac{i}{2} - \frac12 \text{erfi}\big[(\omega_r + i \omega_i)\ell\big] \right\} = \frac{\omega_i}{\lambda} + \frac12 - \frac12 \text{Im}\left\{ \text{erfi}\big[(\omega_r + i \omega_i)\ell\big] \right\} = 0 \, .
\end{align}
Wherever the contours intersect lies a solution $\{\omega_r,\omega_i\}$. There are a number of interesting observations: First, there exist normalizable bound states both for $\lambda>0$ and $\lambda<0$, in contrast to the local theory where there is only one bound state for $\lambda<0$. Second, the bound state of the local theory is shifted towards smaller imaginary arguments due to the presence of non-locality. And third, there is an entirely new set of quasinormal modes which are reflection-symmetric along the imaginary axis (that is, they feature equal, positive imaginary parts, but with real parts that are the negative of each other). They correspond to ingoing and outgoing modes, decreasing in amplitude as $e^{-\omega_i|x|}$, $\omega_i>0$, thereby justifying their designation as ``quasinormal modes.''

However, there are also several unnormalizable modes; in Fig.~\ref{fig:ch5:qnm} these are contained in the shaded regions. Their physical interpretation remains unclear, and we consider this study as a formal demonstration rather than physical fact. They \emph{might} indicate an instability in this particular class of non-local theories (see also the discussion on instabilities in $\mathrm{GF_1}$ theories in quantum field theory in Sec.~\ref{sec:ch6:stability}). On the other hand, their presence could also be an artefact of the insufficiency of a linear theory. Similar quasinormal modes have also been reported on in Ref.~\cite{Buoninfante:2019teo} in the context of a $\delta$-comb potential, which we would like to address next.

\section{Multiple $\delta$-potentials}
The above methods can be readily extended to $\delta$-comb potentials of the form
\begin{align}
V(x) = \sum\limits_{a=1}^n \lambda_a \delta(x-x_a) \, ,
\end{align}
where $\lambda_n$ is the strength and $x_a$ is the location of the $a$-th $\delta$-potential. A simple example is the double-barrier potential, but one may also think of extended lattice structures encountered in condensed matter physics where $n\gg 1$. In that case the function $\varphi_\omega(x)$ must satisfy
\begin{align}
\varphi_\omega(x) = \varphi^0_\omega(x) - \sum\limits_{a=1}^n \lambda_a \varphi_\omega(x_a) \mathcal{G}_\omega(x-x_a) \, .
\end{align}
In order to find the $n$ parameters $\varphi_\omega(x_a)$ one inserts $x=x_a$ for all $\delta$-barrier locations and solves the resulting system of equations, which may prove cumbersome for larger values of $n$,
\begin{align}
\varphi^0_\omega(x_b) = \sum\limits_{a=1}^n \left[ \delta_{ab} + \lambda_a\mathcal{G}_\omega(x_b-x_a) \right] \varphi_\omega(x_a) = \sum\limits_{a=1}^n M_{ab} \varphi_\omega(x_a) \, .
\end{align}
Formally, one then has
\begin{align}
\varphi_\omega(x) = \varphi^0_\omega(x) - \sum\limits_{a,b=1}^n \lambda_a M^{-1}_{ab} \varphi^0_\omega(x_b) \mathcal{G}_\omega(x-x_a) \, ,
\end{align}
where of course the challenge lies in finding the matrix $M^{-1}_{ab}$. For a specific choice of finite $n$ this can be done in a wide range of cases; for the explicit case of $n=2$, $\lambda_1=\lambda_2=\lambda$, and $x_1 = -x_2 = a$ we refer to Buoninfante \textit{et al.} \cite{Buoninfante:2019teo}.

\subsection{Quasinormal modes}
In the case of bound states the matrix $M_{ab}$ is not invertible such that each singular matrix $M_{ab}$ corresponds to a complex frequency $\omega$ that encodes a bound state (where, again, the sign of the imaginary part determines if that state is normalizable). This can be seen from choosing $\varphi^0_\omega(x)=0$ and setting $x=x_b$ such that
\begin{align}
\varphi_\omega(x_b) = - \sum\limits_{a=1}^n \lambda_a \varphi_\omega(x_a) \mathcal{G}_\omega(x_b-x_a) \qquad
\Leftrightarrow \qquad \sum\limits_{a=1}^n M_{ab} \varphi_\omega(x_b) = 0 \, .
\end{align}
For non-trivial bound state solutions we hence require $\text{det}\,M_{ab}=0$. In the limiting case of $n=1$ this reproduces our previous condition $1+\lambda\mathcal{G}_\omega(0)=0$.

In principle we can use these techniques to study excitations in one-dimensional systems whose structure may be modelled by $\delta$-comb potentials. Certain simplifications are to be expected when all $\lambda_a$ coincide and $x_a = a \times L$ for some lattice spacing $L>0$. It also seems possible to extend these studies to higher-dimensional systems whose potentials can be written as a product of $\delta$-functions. Last, it might also be interesting to study compact systems with periodic boundary conditions, $\varphi_\omega(x_n) = \varphi_\omega(x_1)$, and take the continuum limit $n\rightarrow\infty$ while keeping the size of the system $n\times L$ fixed. It would be interesting to study the quasinormal modes (and scattering coefficients) for extended bodies to further understand the role of non-locality in mesoscopic systems.

\section{Concluding remarks}
Using the asymptotic properties of non-local ghost-free quantum mechanics we were able to extract the scattering coefficients for a plane wave in the presence of a $\delta$-shaped potential. Moreover, using the Lippmann--Schwinger approach we demonstrated explicitly how the presence of non-locality enters these coefficients, with surprising consequences: the transmission vanishes for a critical frequency, implying complete reflection, provided the potential barrier is not too high. Moreover, non-locality destroys the symmetry of $\lambda\rightarrow-\lambda$ which is predicted by the local theory.

Since all our considerations were limited to a single Fourier mode of fixed energy, it would be interesting to extend these studies to wave packets. Since the non-local modification of the scattering coefficients depends on the dimensionless product of non-locality and energy $\omega\ell$ it is conceivable that the presence of non-locality will lead to a potentially observable difference in the dispersion relations for wave packets. It would also be interesting to study the effects of non-locality in periodic structures, such as crystals, and determine whether there are novel interactions between non-locality and phonons and hence observable consequences for lattice vibrations.

%%%%%%%%%%%%%%%%%%%%%%%%%%%%%%%%%%%%%%%%%%%%%%%%%%%%%%%%%%%%%%%%%%%%%%%%%%%%%%%%%%%%%%%%%%%%%%%%%%%
%
% Chapter: Vacuum polarization and the fluctuation-dissipation theorem
%
\chapter{Vacuum polarization and the fluctuation-dissipation theorem}
\label{ch:ch6}
\textit{We further develop the concept of interacting non-local Green functions and utilize them to determine the vacuum polarization of a non-local scalar quantum field around a $\delta$-shaped potential; we also address non-local thermal fluctuations and prove that they are compatible with the fluctuation-dissipation theorem. This chapter is based on Refs.~\cite{Boos:2019fbu,Boos:2019zml}.}

\section{Introduction}
The observation of zero-point fluctuations or ``vacuum polarization'' represents one of the key differences between classical fields and quantum fields. They are typically defined in terms of a subtraction scheme such that they vanish identically for a free field in empty spacetime. Around matter sources, conversely, they are non-vanishing. A well known example is the Casimir effect \cite{Casimir:1948a,Casimir:1948b,Lamoreaux:1996wh}, which has a long history \cite{Milton:2004ya,Bordag:2009}. Due to the presence of vacuum fluctuations in the vicinity of thin, conducting plates there is a net macroscopic force acting on them. Modeling the plates as $\delta$-shaped potentials, it is possible to extract the renormalized expectation value of the energy-momentum tensor and determine the forces explicitly \cite{Bordag:1992,Milton:2004vy,Munoz-Castaneda:2013yga}.

In the present chapter we would like to determine the vacuum polarization around a $\delta$-like potential in ghost-free quantum field theory. Having demonstrated that non-local ghost-free quantum mechanics admits physically sensible interpretations and scenarios we would now like to move on to non-local ghost-free quantum field theory. In particular, on a technical side, this involves the summation over temporal Fourier modes in order to generate real-space observables. Instead of relying on techniques from second quantization that in turn introduce creation and annihilation operators on a suitable Fock space, we will utilize the method of Green functions to extract physical predictions. In that context an important role is played not only by the causal propagators and their generalization to non-local ghost-free theories, but also by the Hadamard and Wightman functions in the presence of non-locality: while in the free case they coincide with their local counterparts, their interacting expressions are sensitive to the presence of non-locality which will lead to new observational signatures of non-locality.

\section{A model of a non-local scalar quantum field}
Even though the previous years have seen progress on general non-local quantum field theory, see Ref.~\cite{Buoninfante:2018mre} and references therein, we would like to examine a particular example for the sake of concreteness. To that end, in this chapter we will describe some properties of a non-local, massive real scalar quantum field in two-dimensional Minkowski spacetime. Since we will work in the Lorentzian setting we will not resort to path integral methods but rather employ the previously developed method of non-local Green functions.

In particular, we would like  to understand the vacuum polarization of a non-local quantum field $\varphi$ in the presence of a $\delta$-shaped potential. The action for such a non-local scalar field $\varphi$ can be written as
\begin{align}
S[\varphi] = \frac12 \int\dd^2 X\left( \varphi \mathcal{D} \varphi - V(X)\varphi^2 \right) \, ,
\end{align}
where $X{}^\mu = (t,x)$ describe the two-dimensional coordinates in Minkowski spacetime $\dd s^2 = -\dd t^2 + \dd x^2$, $V(X)$ represents a potential term, and $\mathcal{D}$ is a differential operator of the form
\begin{align}
\mathcal{D} = \exp\left[\ell^{2N}(-\Box+m^2)^N\right](\Box-m^2) \, , \quad \Box = -\partial_t^2 + \partial_x^2 \, .
\end{align}
We refer to Sec.~\ref{sec:ch1:action} for some considerations regarding non-local action principles as employed in this thesis. Here, $\ell>0$ is the scale of non-locality such that in the limiting case $\ell\rightarrow 0$ one reproduces the local theory, and $N=1,2,\dots$ is a positive integer. We call non-local ghost-free theories of the above type $\mathrm{GF_N}$ theories. It should be noted that for much of the following discussion the precise form of $\mathcal{D}$ is irrelevant since we shall express our findings purely in terms of the relevant Green function, but we decided to display the precise form of $\mathcal{D}$ at this point as well to provide some context.

In quantum field theory the field $\varphi$ is promoted to a field operator $\hat{\varphi}$, which in flat space can be expanded in Fourier modes consisting of creation and annihilation operators on a suitable Fock space, which is sometimes referred to as ``second quantization.'' Defining these concepts in a non-local theory is difficult because the kinetic term in the action contains infinitely many derivatives, thereby leading to substantial differences of the Hamiltonian formulation as compared to the local theory which only features second derivatives at the most.

Instead, we shall focus on the Green function method. To that end, the classical field equations take the form
\begin{align}
\left[\mathcal{D} - V(X)\right] \varphi(X) = 0 \, .
\end{align}
We define a causal Green function $\ts{\mathcal{G}}^\bullet(X',X)$ as a solution of
\begin{align}
\left[\mathcal{D} - V(X)\right] \ts{\mathcal{G}}^\bullet(X',X) = -\delta^{(2)}(X'-X) \, ,
\end{align}
and shall refer to these Green functions in the presence of the potential term $V(X)$ as ``interacting Green functions,'' whereas we also define a free causal Green function $\mathcal{G}^\bullet(X'-X)$ as a solution of
\begin{align}
\mathcal{D} \mathcal{G}^\bullet(X'-X) = -\delta^{(2)}(X'-X) \, .
\end{align}
Free Green functions, due to the translational invariance of Minkowski space, depend only on the difference of the arguments $X'-X$. The symbol ``$\bullet$'' denotes the fact that there are several choices for causal Green functions, corresponding to their analytic definition in the complex plane and their corresponding causal properties, see Ch.~\ref{ch:ch2} for more details. In the local case, that is, when $\ell=0$ and hence $\mathcal{D} = \Box-m^2$ we define the separate symbols
\begin{align}
\left[\mathcal{D} - V(X)\right] \ts{G}^\bullet(X',X) &= -\delta^{(2)}(X'-X) \, , \\
\mathcal{D} G^\bullet(X'-X) &= -\delta^{(2)}(X'-X) \, .
\end{align}
To summarize, we denote Green functions with a bold face symbol when they solve the corresponding equation with $V\not=0$, that is, in the presence of the potential term. Conversely, the regular font is reserved for the free Green functions in the case $V=0$. In order to distinguish local and non-local Green functions we reserve the regular font for the local Green functions and employ the calligraphic font for non-local Green functions.

\section{Vacuum fluctuations around a $\delta$-potential}
\label{sec:ch6:vacuum}

Let us now focus on a particular potential that allows us to extract some physical insight. As in the case of quantum scattering discussed in Ch.~\ref{ch:ch5} we shall focus on a $\delta$-shaped potential, which, simple as it may be, can be used to model the plate potentials one encounters in the Casimir effect \cite{Bordag:1992,Munoz-Castaneda:2013yga}.
\begin{align}
V(X) = V(x) = \lambda\delta(x) \, ,
\end{align}
where $\lambda>0$ has dimensions of an inverse length (or a mass) and parametrizes the strength of the repulsive potential. Since $V(X)$ is spacetime-dependent, the effective Lorentz invariance of the theory is spoiled. In this case, however, the potential is static in the $tx$-reference frame, and from now on we shall limit our considerations to that preferred reference frame.

Moreover, due to the staticity of the situation, we may now perform a temporal Fourier decomposition of both the Green functions and the field $\varphi$ itself. Note that due to staticity both the free and interacting Green functions must be a function of the time difference $t'-t$, and for this reason one may perform a temporal Fourier transform in the time coordinate $t'-t$ directly. We write
\begin{align}
\varphi_\omega(x) = \int\limits_{-\infty}^\infty \dd t \, e^{i\omega t} \varphi(t,x) \, , \quad
\varphi(t,x) = \int\limits_{-\infty}^\infty \frac{\dd\omega}{2\pi} e^{-i\omega t} \varphi_\omega(x) \, .
\end{align}
Similarly, the Fourier transform of the operator $\mathcal{D}$ can be performed such that $\Box= \omega^2+\partial_x^2$ and
\begin{align}
\mathcal{D}_\omega = \exp\left[\ell^{2N}(-\partial_x^2-\varpi^2)^N\right](\partial_x^2+\varpi^2) \, , \quad \varpi = \sqrt{\omega^2-m^2} \, .
\end{align}
We denote the temporal Fourier transforms of the interacting Green functions and free Green functions as
\begin{align}
\label{eq:ch6:green-functions-fourier}
\ts{\mathcal{G}}^\bullet_\omega(x',x) = \int\limits_{\infty}^\infty \dd t \, e^{i\omega t} \ts{\mathcal{G}}^\bullet(t,x',x) \, , \quad
\mathcal{G}^\bullet_\omega(x'-x) = \int\limits_{\infty}^\infty \dd t \, e^{i\omega t} \mathcal{G}^\bullet(t,x'-x) \, ,
\end{align}
respectively, and similarly for the local Green functions $\ts{G}_\omega(x',x)$ and $G_\omega(x'-x)$. The physical situation in the presence of the $\delta$-potential is invariant under spatial reflections $x\rightarrow-x$ which implies
\begin{align}
\ts{\mathcal{G}}^\bullet_\omega(x',x) = \ts{\mathcal{G}}^\bullet_\omega(-x',-x) \, , \quad \ts{G}^\bullet_\omega(x',x) = \ts{G}^\bullet_\omega(-x',-x) \, .
\end{align}

\subsection{Free Green functions}
\label{sec:ch6:free-green-functions}
Even though already presented in Ch.~\ref{ch:ch2}, let us briefly recall the exact expressions and definitions for the free Green functions in the temporal Fourier domain, paying special attention to the appearance of the mass term $m$ as well as to the non-local modification $\Delta\mathcal{G}_\omega(x)$. It is useful to define the two auxiliary quantities
\begin{align}
\varpi = \sqrt{\omega^2-m^2} \, , \quad \varkappa = \sqrt{m^2-\omega^2} \, .
\end{align}
Then, the relevant free Green functions take the form
\begin{align}
G^\text{F}_\omega(x) &= \begin{cases} \frac{i}{2\varpi} e^{i\varpi|x|} \quad & \text{for~} |\omega| \ge m \, ; \\ \frac{1}{2\varkappa}e^{-\varkappa|x|} \quad & \text{for~} |\omega| < m \, , 
\end{cases} \\[10pt]
G^\text{R}_\omega(x) &= \begin{cases} \frac{i\text{sgn}(\omega)}{2\varpi} e^{i\text{sgn}(\omega)\varpi|x|} \quad & \text{for~} |\omega| \ge m \, ; \\ \frac{1}{2\varkappa}e^{-\varkappa|x|} \quad & \text{for~} |\omega| < m \, , 
\end{cases} \\[10pt]
G{}^{(1)}_\omega(x) &= \frac{\cos\varpi x}{\varpi}\theta(|\omega|-m) \, .
\end{align}
The non-local causal Green functions (that is, the retarded Green function and the Feynman Green function) pick up a non-local modification term as compared to the local case,
\begin{align}
\mathcal{G}^\text{F,R}_\omega(x) &= G^\text{F,R}_\omega(x) + \Delta\mathcal{G}_\omega(x) \, , \\
\Delta\mathcal{G}_\omega(x) &= \int\limits_{-\infty}^\infty \frac{\dd q}{2\pi} \cos(qx) \frac{1-\alpha(\varpi^2-q^2)}{\varpi^2-q^2} \, , \quad \alpha(z) = e^{-(-z\ell^2)^N} \, .
\end{align}
Since the Hadamard\index{Green function!Hadamard} function is a solution to a homogeneous equation, and the ghost-free non-local operator does not introduce any new poles in the propagator, the free non-local Hadamard function coincides with the free local Hadamard function,
\begin{align}
\mathcal{G}_\omega^{(1)}(x) = G{}^{(1)}_\omega(x) = \frac{\cos\varpi x}{\varpi}\theta(|\omega|-m) \, .
\end{align}
For more details on the free Green functions we refer to Ch.~\ref{ch:ch2} as well as appendix \ref{app:2d-massive-green-functions}.

\subsection{Interacting Green functions}
Let us now find the interacting Green functions, which can be done analytically in the case of a $\delta$-shaped potential. In particular, we will present two independent methods and prove that they yield the same result. Once the interacting Green functions are known we can construct the Hadamard\index{Green function!Hadamard} function and, after a suitable renormalization, extract the vacuum fluctuations around the $\delta$-potential in both the local and non-local case. We shall perform all calculations in temporal Fourier space, and only in the very end perform an inverse temporal Fourier transformation to obtain the vacuum fluctuations in spacetime.

\subsubsection{Hadamard prescription}
\label{sec:ch6:hadamard}
Recall from Ch.~\ref{ch:ch5} that an exact solution for the Fourier mode $\varphi_\omega(x)$ can be given in terms of the free retarded Green function $\mathcal{G}^\text{R}_\omega(x)$ and a free solution $\varphi^0_\omega(x)$ via the Lippmann--Schwinger equation \cite{Lippmann:1950}\index{Lippmann--Schwinger equation},
\begin{align}
\varphi_\omega(x) = \varphi_\omega^0(x) - \int\dd x' \mathcal{G}^\text{R}_\omega(x-x') V(x') \varphi_\omega(x') \, .
\end{align}
In the case of $V(x)=\lambda\delta(x)$ we can rewrite
\begin{align}
\label{eq:ch6:solution-phi}
\varphi_\omega(x) = \varphi_\omega^0(x) - \Lambda_\omega \varphi^0_\omega(0)\mathcal{G}^\text{R}_\omega(x) \, , \quad \Lambda_\omega = \frac{\lambda}{1+\lambda\mathcal{G}^\text{R}_\omega(0)} \, ,
\end{align}
where we have assumed that the condition $1+\lambda\mathcal{G}^\text{R}_\omega(0)\not=0$ is satisfied. Formally one may also employ the advanced Green function for these considerations since the situation is static, and we shall comment on that further below again. We can now construct the interacting Hadamard function $\ts{\mathcal{G}}^{(1)}_\omega(x',x)$, whose real-space expression is
\begin{align}
\ts{\mathcal{G}}^{(1)}(X',X) = \langle\hat{\varphi}(X')\hat{\varphi}(X)+\hat{\varphi}(X)\hat{\varphi}(X')\rangle \, , \quad \ts{\mathcal{G}}^{(1)}(X',X) = \ts{\mathcal{G}}^{(1)}(X,X') \, .
\end{align}
After a temporal Fourier transform one finds for each mode
\begin{align}
\ts{\mathcal{G}}^{(1)}_\omega(x',x) = \langle\hat{\varphi}_\omega(x')\hat{\varphi}_{-\omega}(x)+\hat{\varphi}_{-\omega}(x)\hat{\varphi}_\omega(x')\rangle \, ,
\end{align}
where the symmetry in $X\leftrightarrow X'$ implies that 
\begin{align}
\label{eq:ch6:hadamard-symmetry}
\ts{\mathcal{G}}^{(1)}_{-\omega}(x',x) = \ts{\mathcal{G}}^{(1)}_\omega(x,x') \, .
\end{align}
These are of course formal expressions since we have not defined the non-local vacuum expectation value, and the fields are field operators. However, see the considerations in Ch.~\ref{ch:ch2}, the non-local free Hadamard\index{Green function!Hadamard} function \emph{coincides} with the local free Hadamard function. For that reason, using Eq.~\eqref{eq:ch6:solution-phi}, we can provide a unique prescription for the non-local interacting Hadamard function by requiring
\begin{align}
\mathcal{G}{}^{(1)}_\omega(x',x) = G{}^{(1)}_\omega(x',x) = \langle \hat{\varphi}^0_\omega(x') \hat{\varphi}^0_{-\omega}(x) + \hat{\varphi}^0_{-\omega}(x)\hat{\varphi}^0_\omega(x') \rangle
\end{align}
such that
\begin{align}
\begin{split}
\label{eq:ch6:hadamard-1}
\ts{\mathcal{G}}^{(1)}_\omega(x',x) = G{}^{(1)}_\omega(x'-x) &- \Lambda_\omega\mathcal{G}^\text{R}_\omega(x')G{}^{(1)}_{-\omega}(x) - \Lambda_{-\omega}\mathcal{G}^\text{R}_{-\omega}(x)G{}^{(1)}_\omega(x') \\
& + G{}^{(1)}_\omega(0) \Lambda_\omega\mathcal{G}_\omega(x') \Lambda_{-\omega}\mathcal{G}_{-\omega}(x) \, .
\end{split}
\end{align}
This relation is of central importance for our following considerations since it expresses the non-local interacting Hadamard function in terms of the free, local Hadamard function and additional $\lambda$-terms, which in turn are different in the local and non-local case, respectively.

Let us close by remarking on some properties of \eqref{eq:ch6:hadamard-1}. First, it is proportional to free Hadamard functions, and for that reason it is proportional to $\theta(|\omega|-m)$. Let us also notice that
\begin{align}
\label{eq:ch6:hdm-reflection-omega}
\ts{\mathcal{G}}^{(1)}_{-\omega}(x',x) = \ts{\mathcal{G}}^{(1)}_\omega(x',x) \, .
\end{align}
Together with \eqref{eq:ch6:hadamard-symmetry} this implies
\begin{align}
\label{eq:ch6:hadamard-symmetry-2}
\ts{\mathcal{G}}^{(1)}_{\omega}(x',x) = \ts{\mathcal{G}}^{(1)}_{|\omega|}(x',x) \, .
\end{align}
This property reflects the staticity of the situation: it is also possible to invert the direction of time and work exclusively with advanced Green functions. To that end we can define the advanced Green function as well as the corresponding $\Lambda_\omega$-coefficient
\begin{align}
\label{eq:ch6:adv-1}
\mathcal{G}^\text{A}_\omega(x) = \mathcal{G}^\text{R}_{-\omega}(x) \, , \quad \Lambda^\text{A}_\omega = \Lambda_{-\omega} = \frac{\lambda}{1+\lambda\mathcal{G}^\text{A}_\omega(0)} \, .
\end{align}
We can then define
\begin{align}
\begin{split}
\label{eq:ch6:hdm-result-adv}
\ts{\mathcal{G}}^{(1)}_{\omega}(x',x)_\ind{A} = {G}^{(1)}_{\omega}(x'-x) &-\Lambda^\ind{A}_\omega {\mathcal{G}}^{A}_{\omega}(x') {G}^{(1)}_{-\omega}(x) -\Lambda^\ind{A}_{-\omega}{\mathcal{G}}^{A}_{-\omega}(x) {G}^{(1)}_{\omega}(x')\\
&+{G}^{(1)}_{\omega}(0) \Lambda^\ind{A}_\omega {\mathcal{G}}^{A}_{\omega}(x') \Lambda^\ind{A}_{-\omega}{\mathcal{G}}^{A}_{-\omega}(x)\, ,
\end{split}
\end{align}
but due to Eqs.~\eqref{eq:ch6:adv-1} as well as \eqref{eq:ch6:hdm-reflection-omega} this expression coincides with Eq.~\eqref{eq:ch6:hadamard-1} obtained from the retarded Green function,
\begin{align}
\ts{\mathcal{G}}^{(1)}_{\omega}(x',x) = \ts{\mathcal{G}}^{(1)}_{\omega}(x',x)_\ind{A} \, .
\end{align}
This property is expected but represents a consistency check of our methods.

\subsubsection{Lippmann--Schwinger method for Green functions}
One might wonder whether the previous construction of the interacting, non-local Hadamard\index{Green function!Hadamard} function is unique, and for that reason we will present here an alternative derivation of the causal Green functions, which in turn can be used to extract an equivalent representation of the Hadamard function. Since the following considerations apply for all causal Green functions (that is, for the retarded, advanced, and Feynman Green function) we shall denote the relevant interacting Green function as $\ts{\mathcal{G}}^\text{C}_\omega(x',x)$ and the free Green function as $\mathcal{G}^\text{C}_\omega(x',x)$. Moreover, let us decompose the interacting Green function as
\begin{align}
\label{eq:ch6:a-omega}
\ts{\mathcal{G}}^\text{C}_\omega(x',x) = \mathcal{G}^\text{C}_\omega(x',x) + \mathcal{A}_\omega(x',x) \, ,
\end{align}
where the function $\mathcal{A}_\omega(x',x)$ is an auxiliary quantity satisfying
\begin{align}
\big[ \mathcal{D}_\omega - V(x) \big] \mathcal{A}_\omega(x,x') = V(x)\mathcal{G}^\text{C}_\omega(x-x') \, .
\end{align}
One may verify by inspection that the solution for $\mathcal{A}_\omega(x',x)$ has the form
\begin{align}
\mathcal{A}_\omega(x',x) = - \int\limits_{-\infty}^\infty \dd x'' \ts{\mathcal{G}}^\text{C}_\omega(x',x'')V(x'')\mathcal{G}^\text{C}_\omega(x''-x) \, .
\end{align}
In some sense the above relation can be interpreted as the Lippmann--Schwinger\index{Lippmann--Schwinger equation} representation of the interacting causal Green function. In the simple case $V(x) = \lambda\delta(x)$ the integral can be taken and one finds
\begin{align}
\mathcal{A}_\omega(x',x) = - \lambda\ts{\mathcal{G}}^\text{C}_\omega(x',0)\mathcal{G}^\text{C}_\omega(x) \, .
\end{align}
Inserting this into Eq.~\eqref{eq:ch6:a-omega} yields
\begin{align}
\ts{\mathcal{G}}^\text{C}_\omega(x',x) = \mathcal{G}^\text{C}_\omega(x'-x) - \lambda\ts{\mathcal{G}}^\text{C}_\omega(x',0) \mathcal{G}^\text{C}_\omega(x) \, .
\end{align}
For $x=0$ the above reduces to a consistency relation,
\begin{align}
\ts{\mathcal{G}}^\text{C}_\omega(x',0) = \mathcal{G}^\text{C}_\omega(x') - \lambda\ts{\mathcal{G}}^\text{C}_\omega(x',0) \mathcal{G}^\text{C}_\omega(0) \, ,
\end{align}
which---provided that $1+\lambda\mathcal{G}^\text{C}_\omega(0)\not=0$---amounts to the condition
\begin{align}
\ts{\mathcal{G}}^\text{C}_\omega(x',0) = \frac{\mathcal{G}^\text{C}_\omega(x')}{1+\lambda\mathcal{G}^\text{C}_\omega(0)} \, .
\end{align}
Finally, we arrive at the explicit representation for all causal propagators,
\begin{align}
\label{eq:ch6:causal-propagator}
\ts{\mathcal{G}}^\text{C}_\omega(x',x) = \mathcal{G}^\text{C}_\omega(x'-x) - \lambda\frac{\mathcal{G}^\text{C}_\omega(x')\mathcal{G}^\text{C}_\omega(x)}{1+\lambda\mathcal{G}^\text{C}_\omega(0)} \, .
\end{align}
Using the properties of the free Feynman Green function implies that the interacting Feynman Green function satisfies
\begin{align}
\ts{\mathcal{G}}^\text{F}_\omega(x',x) = \ts{\mathcal{G}}^\text{F}_\omega(x,x') \, , \quad
\ts{\mathcal{G}}^\text{F}_{-\omega}(x',x) = \ts{\mathcal{G}}^\text{F}_\omega(x',x) = \ts{\mathcal{G}}^\text{F}_{|\omega|}(x',x) \, ,
\end{align}
as well as
\begin{align}
\label{eq:ch6:hadamard-theta}
\Im\left[ \ts{\mathcal{G}}^\text{F}_\omega(x',x) \right] = 0 \quad \text{for} \quad |\omega| < m \, .
\end{align}
The retarded propagator, on the other hand, satisfies
\begin{align}
\ts{\mathcal{G}}^\text{R}_\omega(x',x) = \ts{\mathcal{G}}^\text{R}_\omega(x,x') \, , \quad
\ts{\mathcal{G}}^\text{R}_{-\omega}(x',x) = \overline{\ts{\mathcal{G}}^\text{R}_\omega}(x',x) \, ,
\end{align}
where the bar denotes complex conjugation.

\subsubsection{Equivalence of both methods}
It follows from basic definitions, see Sec.~\ref{sec:ch2:green-functions-quantum-theory}, that all free Green functions are related via
\begin{align}
\mathcal{G}^\text{F}(X'-X) = \frac12\left( \mathcal{G}^\text{R}(X'-X) + \mathcal{G}^\text{A}(X'-X) + i \mathcal{G}^{(1)}(X'-X) \right) \, ,
\end{align}
In the static case this also holds true for each Fourier mode,
\begin{align}
\mathcal{G}_\omega^\text{F}(x'-x) = \frac12\left( \mathcal{G}_\omega^\text{R}(x'-x) + \mathcal{G}_\omega^\text{A}(x'-x) + i \mathcal{G}_\omega^{(1)}(x'-x) \right) \, ,
\end{align}
where $\mathcal{G}_\omega^\text{A}(x'-x)$ is the advanced Green function that we can define as the time reversed retarded Green function,
\begin{align}
\mathcal{G}_\omega^\text{A}(x'-x) = \mathcal{G}_{-\omega}^\text{R}(x'-x) \, .
\end{align}
Having the representation \eqref{eq:ch6:causal-propagator} for all causal propagators at our disposal one may also verify that the above relation remains true in the interacting case,
\begin{align}
\label{eq:ch6:master}
\ts{\mathcal{G}}_\omega^\text{F}(x',x) = \frac12\left( \ts{\mathcal{G}}_\omega^\text{R}(x',x) + \ts{\mathcal{G}}_\omega^\text{A}(x',x) + i \ts{\mathcal{G}}_\omega^{(1)}(x',x) \right) \, .
\end{align}
In particular, this implies that the Hadamard\index{Green function!Hadamard} function can be written as
\begin{align}
\label{eq:ch6:relation-feynman-hadamard}
\ts{\mathcal{G}}^{(1)}_\omega(x',x) = 2\Im\left[ \ts{\mathcal{G}}^\text{F}_\omega(x',x) \right] = 2\Im \left[ \mathcal{G}^\text{F}_\omega(x'-x) - \lambda\frac{\mathcal{G}^\text{F}_\omega(x')\mathcal{G}^\text{F}_\omega(x)}{1+\lambda\mathcal{G}^\text{F}_\omega(0)} \right] \, .
\end{align}
Evidently, similar relations to the above hold in the absence of the potential for $\lambda=0$ as well as in the local theory. On the other hand, in Sec.~\ref{sec:ch6:hadamard} we have shown
\begin{align}
\tag{\ref*{eq:ch6:hadamard-1}}
\begin{split}
\ts{\mathcal{G}}^{(1)}_\omega(x',x) = G{}^{(1)}_\omega(x'-x) &- \Lambda_\omega\mathcal{G}^\text{R}_\omega(x')G{}^{(1)}_{-\omega}(x) - \Lambda_{-\omega}\mathcal{G}^\text{R}_{-\omega}(x)G{}^{(1)}_\omega(x') \\
& + G{}^{(1)}_\omega(0) \Lambda_\omega\mathcal{G}_\omega(x') \Lambda_{-\omega}\mathcal{G}_{-\omega}(x) \, .
\end{split}
\end{align}
As different as these two representations may appear, they are in fact equivalent. One can prove this by inserting the direct expressions for the free non-local Green functions as delineated in Sec.~\ref{sec:ch6:free-green-functions}, but the calculations are lengthy and we hence present them in appendix \ref{app:master-eq}.

\subsection{Vacuum polarization}
We can now define the vacuum polarization as follows:
\begin{align}
\begin{split}
\label{eq:ch6:fluctuations}
\langle\varphi^2(x)\rangle_\text{ren} &= \lim\limits_{X'\rightarrow X} \Big[ \langle\hat{\varphi}(X')\hat{\varphi}(X)\rangle_{\lambda\not=0} - \langle\hat{\varphi}(X')\hat{\varphi}(X)\rangle_{\lambda=0} \Big] \\
&= \frac12 \lim\limits_{X'\rightarrow X} \left[ \ts{\mathcal{G}}^{(1)}(X',X) - \mathcal{G}^{(1)}(X',X) \right] = \frac12 \lim\limits_{X'\rightarrow X} \left[ \ts{\mathcal{G}}^{(1)}(X',X) - G^{(1)}(X',X) \right] \, .
\end{split}
\end{align}
In two dimensions this quantity is dimensionless. We subtract the vacuum contribution such that the polarization vanishes in the absence of the potential. Moreover, in the last equality we used the fact that the free Hadamard\index{Green function!Hadamard} functions coincide in the local and non-local case. Employing our exact representation \eqref{eq:ch6:hadamard-1} for the interacting Hadamard function, and determining the real-space expression via \eqref{eq:ch6:green-functions-fourier} in combination with \eqref{eq:ch6:hadamard-symmetry-2} yields
\begin{align}
\begin{split}
\label{eq:phisquared-hadamard}
\langle \varphi^2(x) \rangle_\ind{ren} &= -\int\limits_m^\infty \frac{\dd \omega}{2\pi} \Big[ 2\Lambda_\omega {\mathcal{G}}^\ind{R}_{\omega}(x) {G}^\ind{(1)}_{-\omega}(x) - G^\ind{(1)}_{\omega}(0) \left|\Lambda_\omega {\mathcal{G}}^\ind{R}_{\omega}(x)\right|^2 \Big] \, .
\end{split}
\end{align}
Alternatively, inserting \eqref{eq:ch6:causal-propagator} into \eqref{eq:ch6:fluctuations} as well as making use of the interrelation \eqref{eq:ch6:relation-feynman-hadamard} yields
\begin{align}
\label{eq:phisquared-feynman}
\langle \varphi^2(x) \rangle_\ind{ren} = - \Im \left[ \int\limits_m^\infty \frac{\dd\omega}{2\pi} \lambda\,\frac{\left[{\mathcal{G}}^\ind{F}_\omega(x)\right]^2}{1+\lambda \,{\mathcal{G}}^\ind{F}_\omega(0)} \right] \, .
\end{align}
The integration limits follow directly from \eqref{eq:ch6:hadamard-theta}. As guaranteed by our proof in appendix \eqref{app:master-eq} these two rather difficult expression are indeed equivalent. Moreover, it is clear that in the limiting case of $\lambda=0$ they reduce to zero.

In what follows, we would like to evaluate \eqref{eq:ch6:fluctuations} for the local theory as well as for the non-local theories $\mathrm{GF_1}$ and $\mathrm{GF_2}$. It is convenient to perform a variable transformation from $\omega$ to $\varpi$ such that we can rewrite both \eqref{eq:phisquared-hadamard} and \eqref{eq:phisquared-feynman} as 
\begin{align}
\label{eq:ch6:main-integral}
\langle \varphi^2(x) \rangle_\ind{ren} &= \int \limits_0^\infty \frac{\dd\varpi}{4\pi} \frac{\Phi_{\omega}(x)}{\sqrt{\varpi^2+m^2}} \, ,
 \quad \Phi_{\omega}(x) = \frac{B^2-\cos^2(\varpi x)-2\cos(\varpi x)BC}{1+C^2} \, , 
\end{align}
where we introduced the abbreviations
\begin{align}
\label{eq:ch6:main-integral-aux}
B = 2g_\omega(x) - \sin(\varpi|x|) \, , \quad C = 2g_\omega(0) + 2\varpi/\lambda \, , \quad g_\omega(x) = \varpi\Delta\mathcal{G}_\omega(x) \, .
\end{align}
In the following sections we will evaluate \eqref{eq:ch6:main-integral} using both analytical and numerical tools. The effects of non-locality are captured solely by the dimensionless function $g_\omega(x)$. Unlike the quantum scattering case in Ch.~\ref{ch:ch5}, the results now depend on the precise form of $g_\omega(x)$ everywhere, for all values of $x$. Note that the appearance of the factor $C$ renders $\lambda$ an effective, $\omega$-dependent coupling:
\begin{align}
C = \frac{2\varpi}{\widetilde{\lambda}_\omega} \, , \quad \widetilde{\lambda}_\omega = \frac{\lambda}{1+\lambda\Delta\mathcal{G}_\omega(0)}
\end{align}
As shown previously in Ch.~\ref{ch:ch2}, the function $g_\omega(x)$ vanishes as $x\rightarrow\infty$ and its contribution towards $\phisq$ disappears. The $C$-term, however, is still present and may affect the asymptotic behavior since it is related to effects of non-locality at the location of the potential at $x=0$.

The effective $\omega$-dependent coupling can become quite large if $\lambda\Delta\mathcal{G}_\omega(0)\sim -1$, and one can show that this is possible for some frequency $\omega_\star$ provided the potential strength surpasses a critical value,
\begin{align}
\lambda\ell \ge \lambda_\star\ell = \frac{\Gamma\left(\tfrac14\right)}{\sqrt{2}} \approx 2.56369\dots \, .
\end{align}
As we have shown in Ch.~\ref{ch:ch5}, for such a frequency $\omega_\star$ the $\delta$-barrier becomes completely opaque with the transmission coefficient vanishing identically. Similarly, the coefficient $C$ vanishes for such a frequency, but numerical investigations (see below) reveal that this property does not affect the vacuum polarization significantly. The effective $\omega$-dependence of the coupling $\lambda$, however, does indeed influence the vacuum polarization in regions where $x>\ell$.

\subsubsection{Local theory}
For now, let us establish a baseline and consider purely the local case where $g_\omega(x)\equiv 0$. This case has been studied earlier, see Refs.~\cite{Bordag:1992,Milton:2004vy} and references therein. In that case we have
\begin{align}
B = -\sin(\varpi|x|) \, , \quad C = \frac{2\varpi}{\lambda} \, ,
\end{align}
and the vacuum polarization \eqref{eq:ch6:main-integral} can be written as
\begin{align}
\label{eq:vac-pol-local}
\langle \varphi^2(x) \rangle_\ind{ren}^\ind{loc.} &= \lambda \int\limits_0^\infty \frac{\dd \varpi}{4\pi} \frac{2\varpi\sin(2\varpi|x|) - \lambda\cos(2\varpi x)}{\sqrt{\varpi^2+m^2}(4\varpi^2 + \lambda^2)} \, .
\end{align}
Provided the mass is non-vanishing, $m>0$, the integral converges. For $x=0$ the integrand simplifies, however, and allows for an analytical treatment:
\begin{align}
\langle \varphi^2(0) \rangle_\ind{ren}^\ind{loc.} &= - \int\limits_0^\infty \frac{\dd \varpi}{4\pi} \frac{1}{\sqrt{\varpi^2+\mu^2}} \frac{1}{1 + 4\varpi^2}
= \begin{cases} \displaystyle -\frac{\text{arcosh}\left(\frac{1}{2\mu}\right)}{4\pi\sqrt{1-4\mu^2}} & \text{for~} \mu < \tfrac12 \, , \\[15pt]
\displaystyle -\frac{1}{4\pi} & \text{for~} \mu=\tfrac12 \, , \\[15pt]
\displaystyle -\frac{\text{arccos}\left(\frac{1}{2\mu}\right)}{4\pi\sqrt{4\mu^2-1}} & \text{for~} \mu > \tfrac12 \, .
\end{cases}
\end{align}
where we have defined the dimensionless parameter $\mu := m/\lambda$. Note that $\langle \varphi^2(0) \rangle_\ind{ren}^\ind{loc.}$ is always negative, and asymptotically one has
\begin{alignat}{3}
\langle \varphi^2(0) \rangle_\ind{ren}^\ind{loc.} &\rightarrow -\infty \quad &&\text{for} \quad \mu \rightarrow 0 \, , \\
\langle \varphi^2(0) \rangle_\ind{ren}^\ind{loc.} &\rightarrow 0 \quad &&\text{for} \quad \mu \rightarrow \infty \, .
\end{alignat}
The divergence in the limit $m\rightarrow 0$ is well known in two-dimensional quantum field theory and corresponds to an infrared divergence, which is one way to realize that the mass $m$ plays the role of a regulator in the present context.

\begin{figure}[!htb]%
    \centering
    \includegraphics[width=0.8\textwidth]{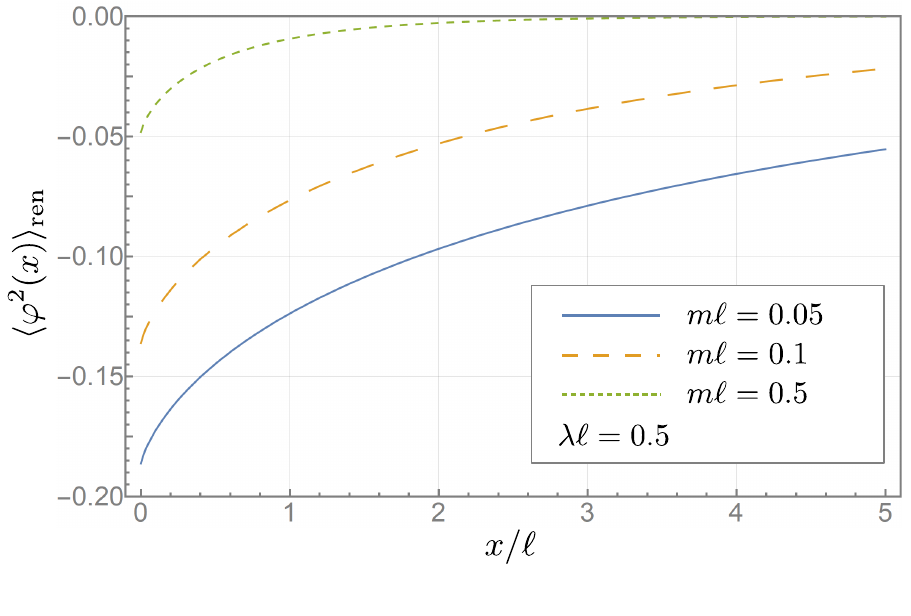}\\[-15pt]
    \caption[Local vacuum polarization.]{The local vacuum polarization $\langle \varphi^2(x) \rangle_\ind{ren}^\ind{loc.}$ as a function of the dimensionless distance $x/\ell$ for a fixed potential parameter ($\lambda\ell=0.5$) and for various values of the dimensionless mass parameter $m\ell$. Because it is reflection symmetric, $\langle \varphi^2(-x) \rangle_\ind{ren}^\ind{loc.} = \langle \varphi^2(x) \rangle_\ind{ren}^\ind{loc.}$, we only plotted it for positive $x$. The numerical evaluations imply that the vacuum polarization is not differentiable at $x=0$. The above is plotted as a function of $x/\ell$ in order to facilitate a future comparison with various \emph{non-local} theories.}
    \label{fig:ch6:phisquared-loc}
\end{figure}

For $x\not=0$ we have to resort to numerical evaluation of $\langle \varphi^2(x) \rangle_\ind{ren}^\ind{loc.}$ which is rather straightforward due to the $\sim 1/\varpi^2$ asymptotics. See Fig.~\ref{fig:ch6:phisquared-loc} for a plot of the local vacuum polarization for various values of the dimensionless coupling $\lambda\ell$. For later convenience we choose to normalize all dimensionful quantities with respect to the scale of non-locality $\ell$. The numerical evaluations imply that $\langle \varphi^2(x) \rangle_\ind{ren}^\ind{loc.}$ is not differentiable at $x=0$, which in turn implies that quantities of the sort $\langle\varphi\partial_x\varphi\rangle$ are not well-defined at the location of the potential at $x=0$, and it will be interesting to track this property in non-local theories.

\subsubsection{Non-local $\mathrm{GF_1}$ theory}
Let us now turn towards the non-local $\mathrm{GF_1}$ theory with the differential operator
\begin{align}
\mathcal{D}_\omega = \exp\left[\ell^2(-\partial_x^2-\varpi^2)\right](\partial_x^2+\varpi^2) \, .
\end{align}
Even though originally evaluated for the case of vanishing mass $m=0$ in Ch.~\ref{ch:ch5}, we can find $\Delta\mathcal{G}_\omega(x)$ analytically also for $m>0$ in a straightforward fashion. For $\omega>m$ we obtain
\begin{align}
\Delta\mathcal{G}_\omega(x) = \frac{1}{2\varpi}\left\{ \sin(\varpi|x|) - \Im \left[ e^{i\varpi |x|} \text{erf}\left( x_+ \right) \right] \right\} \, , \quad x_\pm = \frac{|x|}{2\ell} \pm i\omega\ell \, .
\end{align}
The asymptotics of the error function in the complex plane, for $\Re(z)=\text{const.}$ and $\Im(z)\pm\infty$ are
\begin{align}
\text{erf}(z) \sim -\frac{e^{-z^2}}{\sqrt{\pi} z} \, .
\end{align}
This implies that the coefficients $B$ and $C$ take the form
\begin{align}
B = - \Im \left[ e^{i\varpi |x|} \text{erf}\left( x_+ \right) \right] \, , \quad
C = \frac{2\varpi}{\lambda} - \text{erfi} (\varpi\ell) \, ,
\end{align}
where $\text{erfi}(x)$ is the imaginary error function, whose asymptotics for a real argument $z\rightarrow\infty$ are
\begin{align}
\text{erfi}(z) \sim \frac{e^{z^2}}{\sqrt{\pi} z} \, .
\end{align}
This implies (for finite $\lambda>0$) the following asymptotics in large frequencies for $\varpi\to \infty$:
\begin{align}
B \sim -\frac{1}{\sqrt{\pi}\varpi\ell} e^{\varpi^2\ell^2 - x^2/(4\ell^2)} \, , \quad
C \sim \frac{1}{\sqrt{\pi}\varpi\ell} e^{\varpi^2\ell^2} \, , \quad
\frac{B}{C} \sim -e^{-x^2/(4\ell^2)} \, ,
\end{align}
that is,  while both of these coefficients diverge, the ratio $B/C$ remains finite. This yields
\begin{align}
\Phi_\omega(x) \sim e^{-x^2/(2\ell^2)} - 2\cos(\varpi x) e^{-x^2/(4\ell^2)} \, .
\end{align}
Studying the frequency integral representation of $\langle \varphi^2(x) \rangle_\ind{ren}$ in Eq.~\eqref{eq:ch6:main-integral} we realize that the first term leads to a logarithmic divergence that needs to be manually subtracted in order to obtain a physically meaningful result. This is already the first indication that $\mathrm{GF_1}$ theory is not always well-defined, and we will revisit this issue in Sec.~\ref{sec:ch6:stability}.

Let us parametrize this divergent integral by introducing a UV cutoff $\Omega$ such that
\begin{align}
Z_0=\int\limits_0^\Omega \frac{\dd\varpi}{4\pi} \frac{1}{\sqrt{\varpi^2+m^2}}={1\over 4\pi}\ln\left({\Omega+\sqrt{\Omega^2+m^2}\over m} \right)\, .
\end{align}
In a similar fashion we may define
\begin{align}
Z_1=-\int\limits_0^\infty \frac{\dd\varpi}{4\pi} \frac{2\cos(\varpi x)}{\sqrt{\varpi^2+m^2}}
= -\frac{1}{2\pi} K_0(m|x|) \, ,
\end{align}
where $K_0(x)$ is the modified Bessel function, such that the total expression for $\langle \varphi^2(x) \rangle_\ind{ren}^\mathrm{GF_1}$ can be recast as
\begin{align}
\langle \varphi^2(x) \rangle_\ind{ren}^\mathrm{GF_1}&=e^{-x^2/(2\ell^2)} Z_0+e^{-x^2/(4\ell^2)} Z_1+\Psi(x)\, , \\
 \Psi(x)&=\int \limits_0^\infty \frac{\dd\varpi}{4\pi} \frac{\widetilde{\Phi}_{\omega}(x)}{\sqrt{\varpi^2+m^2}} \, , \quad \widetilde{\Phi}_{\omega}(x) = \Phi_\omega(x) -  e^{-x^2/(2\ell^2)} + 2\cos(\varpi x) e^{-x^2/(4\ell^2)} \, .
\end{align}
Since the term containing $Z_0$ diverges we may subtract it manually and define the remainder as the ``renormalized vacuum polarization'' that we can compare to the expression for $\langle \varphi^2(x) \rangle_\ind{ren}^\text{loc.}$, leaving aside for now the questions about physical motivation for that procedure.

Moreover, as a technical aside, the numerical evaluation of $\Psi(x)$ is greatly simplified by subtracting the convergent subleading asymptotics analytically, since the sub-subleading remainder is a fast-decreasing function of $\varpi$. We will see in the next section on $\mathrm{GF_2}$ theory again that this is a very useful technique for optimizing numerical convergence of integrals.

\begin{figure}[!htb]%
    \centering
    \includegraphics[width=0.8\textwidth]{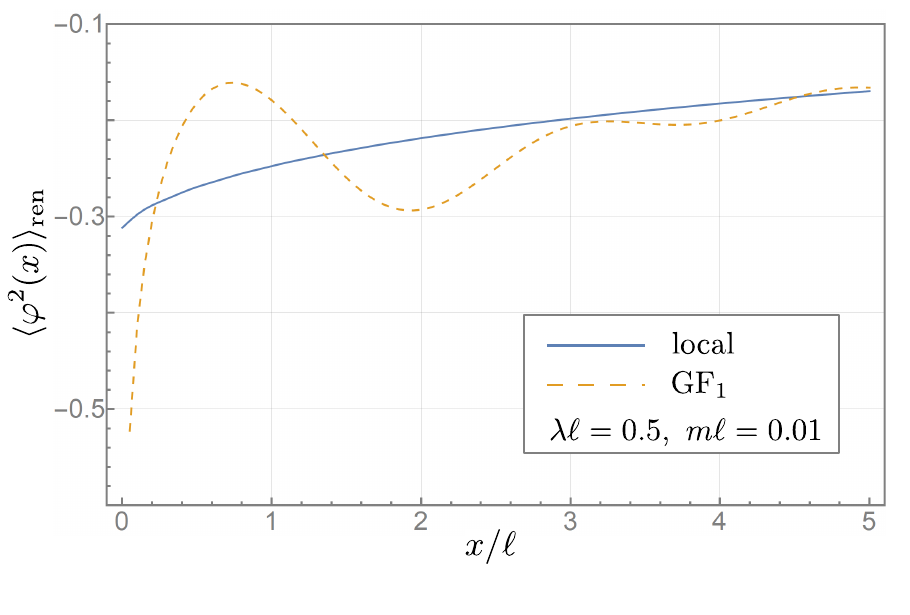}\\[-15pt]
    \caption[Non-local vacuum polarization in $\mathrm{GF_1}$ theory.]{The non-local vacuum polarization in $\mathrm{GF_1}$ theory as compared to the local vacuum polarization plotted as a function of the dimensionless distance $x/\ell$.}
    \label{fig:ch6:phisquared-loc-gf1}
\end{figure}

That being said, see Fig.~\ref{fig:ch6:phisquared-loc-gf1} for a graphical representation of the non-local vacuum polarization in $\mathrm{GF_1}$ theory as compared to the local case. While it undergoes rather strong oscillations in the vicinity of the $\delta$-potential at $x=0$, its asymptotics seem to agree with the local results. This is interesting and gives perhaps some credibility to the subtraction of the divergent $Z_0$-term. It would also be interesting to compare these oscillations to those encountered in linearized gravity around a point particle \cite{Boos:2018bhd}, compare also Sec.~\ref{sec:ch3:friedel}.

On the other hand, the numerical results imply that the renormalized vacuum polarization $\langle \varphi^2(x) \rangle_\ind{ren}^\mathrm{GF_1}$ is still not differentiable at $x=0$, which is unexpected since non-locality is expected to smear sharp features at a characteristic length scale $\ell$. For this reason we have some remaining doubts about our ad hoc $Z_0$-subtraction scheme and do not take the results obtained in the present context very seriously. As has been shown previously \cite{Frolov:2016xhq}, and as we will revisit in Sec.~\ref{sec:ch6:stability}, $\mathrm{GF_1}$ theory is known for instabilities in the time domain, and the present difficulties related to the divergence in the temporal Fourier transform are one manifestation of that fact.

\subsubsection{Non-local $\mathrm{GF_2}$ theory}
\begin{figure}
    \centering
    \includegraphics[width=0.8\textwidth]{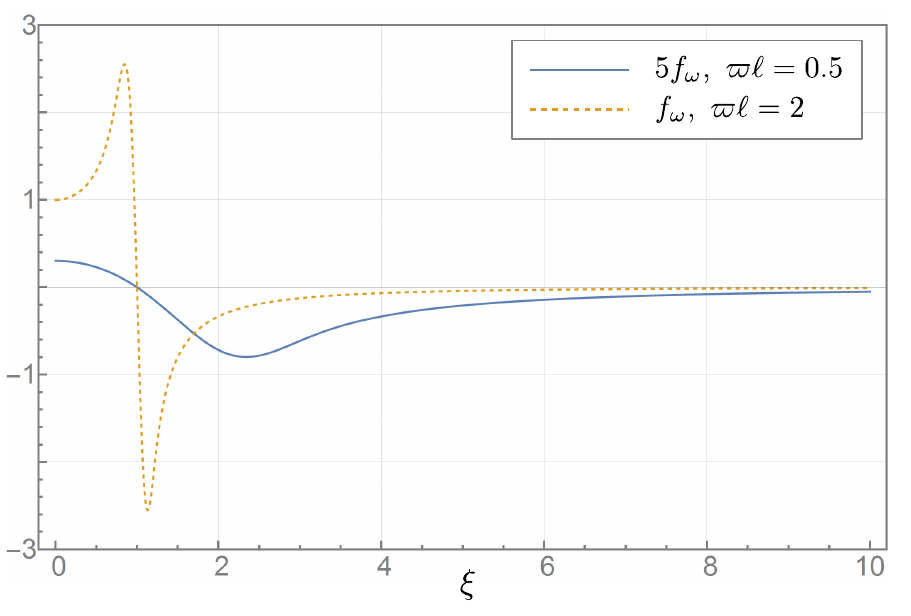}
    \caption[Integrand in the non-local case.]{The shape of the function $f_\omega(\xi)$ which enters the integral \eqref{eq:ch6:int-gf2} depends quite strongly on the value of $\varpi\ell$: For small values it is a smooth function which is numerically small (solid line in the above plot, scaled by a factor of 5 to increase visibility). If $\varpi\ell > \sqrt{2a_2} \approx 1.058\dots$, however, there is a rather strong feature at $\xi\sim 1$ that fundamentally affects the Fourier transform (dashed line). For this reason it is useful to approximate the function $f_\omega$ by two different analytical functions in both cases.}
    \label{fig:ch6:plot-fb-1}
\end{figure}

While it is expected that $\mathrm{GF_2}$ theory is well-behaved both in the spatial and temporal Fourier domain, its analytical treatment is somewhat more involved due to the quartic exponential in the differential operator,
\begin{align}
\mathcal{D}_\omega = \exp\left[\ell^4(\partial_x^2+\varpi^2)^2\right](\partial_x^2+\varpi^2) \, .
\end{align}
That being said, it is possible to determine $\Delta\mathcal{G}_\omega(0)$ analytically,
\begin{align}
\Delta\mathcal{G}_\omega(0) = \frac{\sqrt{2} \varpi^2\ell^3}{6\Gamma\left(\tfrac34\right)}~{}_\ind{2}F_\ind{2}\left[\tfrac34,\tfrac54;\tfrac32,\tfrac74;-(\varpi\ell)^4\right]
-\Gamma\left(\tfrac34\right) \frac{\ell}{\pi} ~{}_\ind{2}F_\ind{2}\left[\tfrac14,\tfrac34;\tfrac12,\tfrac54;-(\varpi\ell)^4\right] \, ,
\end{align}
see appendix \ref{app:deltaG-0-gf2} for more details. Moreover, $\Delta\mathcal{G}_\omega(x)$ for $x\not=0$ needs to be evaluated numerically, which is involved and hence presented in detail in appendix \ref{app:deltaG-x-gf2}. However, it is possible to study the overall behavior of the function $g_\omega(x) = \varpi\Delta\mathcal{G}_\omega(x)$ in a bit more detail. It is given by
\begin{align}
\label{eq:ch6:int-gf2}
g_\omega(x) &= \int\limits_0^\infty\frac{\dd \xi}{\pi} \cos(\xi \tilde{x})f_{\omega}(\xi) \, , \quad
f_{\omega}(\xi) =\frac{1-e^{-(\varpi\ell)^4(1-\xi^2)^2}}{1-\xi^2} \, , \quad \tilde{x} = \varpi x \, .
\end{align}
For small values of $\omega\ell$ the function $f_\omega(\xi)$ is a smooth function with a slowly-decaying tail, whereas for larger values of $\omega\ell$ beyond a critical value there exists a sharp feature. We visualize this behavior in Fig.~\ref{fig:ch6:plot-fb-1}.

For large values of $\varpi\ell$ we are able to extract the following asymptotics (see appendix \ref{app:deltaG-x-gf2-asymptotics} for details) for the dimensionless $g_\omega(x) = \varpi\Delta\mathcal{G}_\omega(x)$:
\begin{align}
g_\omega(0) &= -\frac{1}{4\sqrt{\pi} \varpi^2\ell^2} + \mathcal{O}\left(\varpi^{-6}\right) \, , \\
g_\omega(x) &= \frac{\sin(\varpi |x|)}{2} - \frac{a_2}{3\pi} \left( 2 + e^{-4a_2^2} \right) \frac{x \sin(\varpi x)}{\varpi\ell^2} - \frac{a_2}{2\pi} \left( 3-e^{-4a_2^2} \right)\frac{\cos(\varpi x)}{\varpi^2\ell^2}  + \mathcal{O}\left(\varpi^{-4}\right) \, .
\end{align}
Here $a_2$ is a parameter that appears frequently in our numerical approximation scheme,\footnote{We thank Don Page for providing the approximation $\alpha_2\approx 445/794$, which also admits a suitable mnemonic \cite{Page:2020b}.}
\begin{align}
a_2 \approx 0.5604532115\dots \, .
\end{align}
This implies the following large-$\varpi$ behavior for the coefficients $B$ and $C$ in $\langle \varphi^2(x) \rangle_\ind{ren}^\mathrm{GF_2}$,
\begin{align}
B \sim -\frac{2a_2}{3\pi} \left( 2 + e^{-4a_2^2} \right) \frac{x \sin(\varpi x)}{\varpi\ell^2} \, , \quad
C \sim \frac{2\varpi}{\lambda} - \frac{1}{4\sqrt{\pi} \varpi^2\ell^2} \, .
\end{align}
Observe that unlike in $\mathrm{GF_1}$ theory the non-local contributions are decreasing with larger values of $\varpi$, which was expected from the different non-local form factor in $\mathrm{GF_2}$ theory as compared to $\mathrm{GF_1}$ theory. One obtains the following asymptotic behavior for the integrand,
\begin{align}
\Phi_\omega(x) \sim &-\frac{\lambda^2\cos^2(\varpi x)}{4\varpi^2 + \lambda^2} + \frac{8a_2\lambda}{3\pi\ell^2}\left(1-e^{-4a_2^2}\right) \frac{x\cos(\varpi x)\sin(\varpi x)}{4\varpi^2 + \lambda^2} \, .
\end{align}
Together with the leading factor of $1/\sqrt{\varpi^2+m^2}$ the integral for $\langle \varphi^2(x) \rangle_\ind{ren}^\mathrm{GF_2}$ is convergent, see Eq.~\eqref{eq:ch6:main-integral}, although an analytic evaluation is impossible since not even $\Delta\mathcal{G}_\omega(x)$ is known analytically. We present our numerical methods in great detail in appendix \ref{app:deltaG-gf2}.

\begin{figure}[!htb]%
    \centering
    \subfloat{{\includegraphics[width=0.8\textwidth]{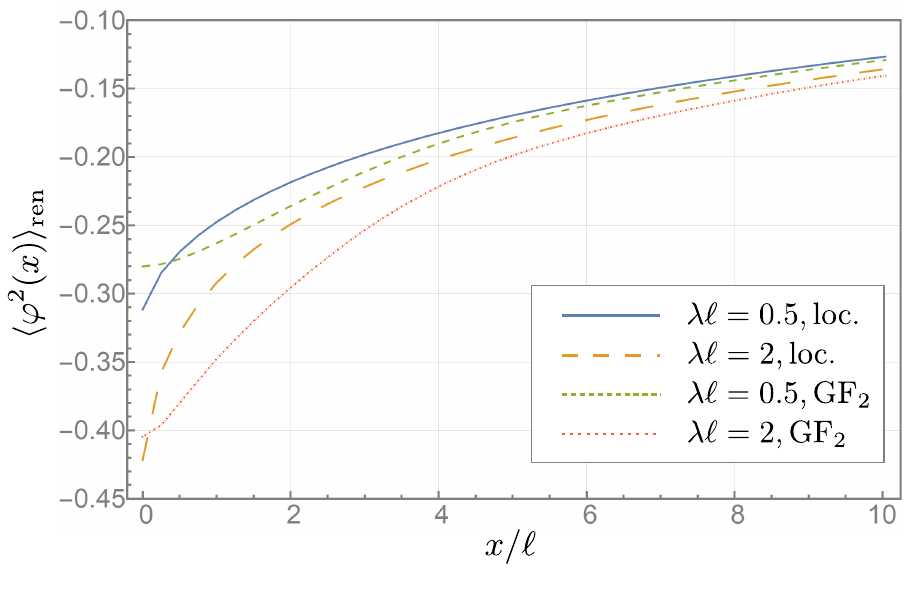} }}\\[-15pt]
    \caption[Non-local vacuum polarization in $\mathrm{GF_2}$ theory.]{The non-local vacuum polarization in $\mathrm{GF_2}$ theory as compared to the local vacuum polarization plotted as a function of the dimensionless distance $x/\ell$ for two different dimensionless couplings.}
    \label{fig:ch6:phisquared-loc-gf2}
\end{figure}

With the numerical methods under control, we can now plot $\langle \varphi^2(x) \rangle_\ind{ren}^\mathrm{GF_2}$ for a rather wide range of $x$ and $\lambda$ and compare it to the local result $\langle \varphi^2(x) \rangle_\ind{ren}^\text{loc.}$, see Fig.~\ref{fig:ch6:phisquared-loc-gf2}. Interestingly, the behavior at $x=0$ is very different from the local case: it is numerically smooth, implying that $\langle\varphi\partial_x\varphi\rangle$ is finite.

\begin{figure}[!htb]%
    \centering
    \subfloat{{\includegraphics[width=0.8\textwidth]{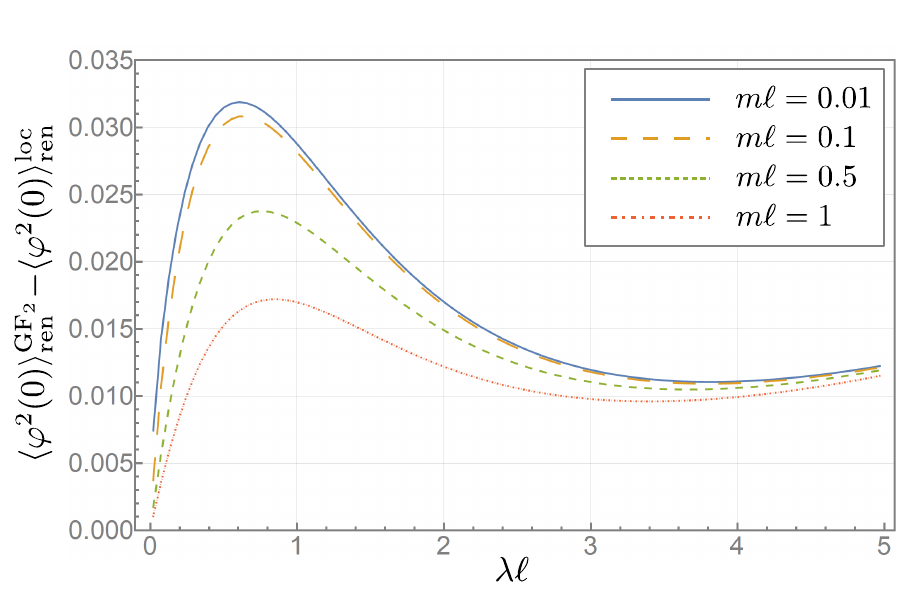} }}\\[-10pt]
    \caption[Comparison of local and non-local vacuum polarization at $x=0$.]{The difference of the vacuum polarization at the location of the potential, $x=0$, as a function of the potential strength $\lambda\ell$. The difference scales with the dimensionless mass parameter $m\ell$, leaving the shape of the curve almost unchanged. In the limiting case $\lambda\rightarrow 0$ the renormalized vacuum polarization vanishes as expected.}
    \label{fig:ch6:phisquared0-difference}
\end{figure}

There are a few features in the non-local vacuum polarization as compared to the local case:
\begin{itemize}
\item[(i)] \textit{Asymptotics:} As expected, the non-local vacuum polarization approaches the local expressions for large values $x\gg\ell$. As this is theoretically built in to the theory of ghost-free non-locality this demonstrates that our numerical methods are reliable.
\item[(ii)] \textit{Smoothing:} At small distance scales $x\sim\ell$ the shape of the vacuum polarization is drastically different in the non-local theory and approaches a constant value with vanishing slope, whereas the slope is non-zero in the local theory. This implies that in the non-local theory expressions $\sim\partial_x\varphi^2$ are well-behaved at the location of the potential.
\item[(iv)] \textit{Overshoot:} For a large range of mass and potential parameters (quite possibly for \emph{all} choices) the numerical value of the vacuum polarization at the location of the potential at $x=0$ is numerically larger in the non-local case than in the local case. We call this ``overshoot'' and attempt to capture this behavior in Fig.~\ref{fig:ch6:phisquared0-difference}.
\item[(iv)] \textit{Crossing:} In intermediate distance scales, $x\sim\ell$, the vacuum polarizations in the local theory and non-local theory cross, which implies that the difference between the local and non-local vacuum polarization can be both positive and negative. In $\mathrm{GF_1}$ theory this feature is even more pronounced, see also Fig.~\ref{fig:ch6:phisquared-loc-gf1}. These crossings, or oscillations, appear to be a generic feature of non-local theories as opposed to local theories, see also \cite{Boos:2018bhd} and Sec.~\ref{sec:ch3:friedel}.
\end{itemize}

\section{Stability properties of non-local QFT}
\label{sec:ch6:stability}
As shown before, in a non-local $\mathrm{GF_N}$ theory the non-local modification term to the free Green functions can be written as
\begin{align}
\Delta\mathcal{G}_\omega(x) &= \int\limits_{-\infty}^\infty \frac{\dd q}{2\pi} \cos(qx) \frac{1-\alpha(\varpi^2-q^2)}{\varpi^2-q^2} \, , \quad \alpha(z) = e^{-(-z\ell^2)^N} \, , \quad N \in \mathbb{N} \, .
\end{align}
Whereas the above integral converges for any value of $N$, applying a Fourier transform to the above integral is not always possible since the large-$\omega$ asymptotics of the integrand are
\begin{align}
\frac{1-\alpha(\varpi^2-q^2)}{\varpi^2-q^2} \sim -\frac{1}{\varpi^2} e^{-(-\varpi\ell^2)^N} \, ,
\end{align}
which diverges for odd $N=2n-1$ and converges for even $N=2n$ with $n=1,2,\dots$. This divergence can be exemplified analytically in the case of $\mathrm{GF_1}$ theory, see Ch.~\ref{ch:ch2} where we present an analytical expression for $\mathcal{G}_\omega(x)$ that asymptotically grows exponentially in $\omega$. The origin of this divergence stems from the Lorentzian signature and, hence, the hyperbolic nature of the $\Box$-operator. For purely spatial problems where $\Box$ is an elliptic operator there is no such problem.

The argument presented above is formal in some sense, since the non-local modification by itself is not a physical observable. However, it has been shown explicitly in \cite{Frolov:2016xhq} that time-dependent sources of radiation lead to divergences in the near-field oscillation amplitudes around that source. In the present context of the previous chapter we have demonstrated explicitly that another observable diverges for $\mathrm{GF_1}$ theories: the vacuum polarization. This constitutes a physical divergence in non-local $\mathrm{GF_1}$ quantum field theory. This is somewhat surprising since our scenario, unlike that presented in Ref.~\cite{Frolov:2016xhq}, is manifestly static and hence time-independent.

For these reasons great care should be administered whenever studying $\mathrm{GF}_{2n-1}$ theories in a Lorentzian setting.

\section{Thermal fluctuations around a $\delta$-potential}
\label{sec:ch6:thermal}
With the methods introduced above it is also possible to study the fluctuations around a $\delta$-potential at finite temperature see \cite{Biswas:2009nx,Biswas:2010yx,Biswas:2012ka,Dijkstra:2019wcz}. Recall that the Hadamard\index{Green function!Hadamard} function is defined as 
\begin{align}
G{}^{(1)}_{\bullet}(\ts{x}',\ts{x}) = \langle \hat{\varphi}(\ts{x}')\hat{\varphi}(\ts{x}) + \hat{\varphi}(\ts{x})\hat{\varphi}(\ts{x}') \rangle_\bullet \, ,
\end{align}
where ``$\bullet$'' denotes the quantum state in which the quantum expectation value is performed. Similarly, in case of a static potential, one may consider the temporal Fourier representation
\begin{align}
G{}^{(1)}_{\omega,\bullet}(x',x)= \langle \hat{\varphi}_\omega(x')\hat{\varphi}_{-\omega}(x) + \hat{\varphi}_{-\omega}(x)\hat{\varphi}_\omega(x') \rangle_\bullet \, ,
\end{align}
where the quantum average is again taken over the state ``$\bullet$.'' In the local theory the thermal Hadamard function\index{Green function!Hadamard (thermal)} is given by
\begin{align}
G{}^{(1)}_{\omega,\beta}(x',x) &= \coth\left( \frac{\beta|\omega|}{2} \right) \frac{\cos[\varpi(x'-x)]}{\varpi}\theta(\omega^2-m^2) \, , \quad \varpi = \sqrt{\omega^2-m^2} \, ,
\end{align}
where we denoted the thermal state by $\beta = 1/(k_\text{B}T)$, $k_\text{B}$ is the Boltzmann constant, and $T$ denotes the temperature. One can rewrite the hyperbolic cotangent in terms of the Bose--Einstein distribution\index{Bose--Einstein distribution},
\begin{align}
\coth\left( \frac{\beta|\omega|}{2} \right) = 1 + 2n_{|\omega|,\beta} \, , \quad
n_{\omega,\beta} = \frac{1}{e^{\beta\omega}-1} \, .
\end{align}
In the limiting case $T\rightarrow 0$, that is, $\beta \rightarrow \infty$, one finds
\begin{align}
G{}^{(1)}_{\omega,\beta\rightarrow\infty}(x',x) = G{}^{(1)}_{\omega}(x',x) \, ,
\end{align}
thereby recovering the vacuum expression employed in our previous considerations in this chapter. We posit that in the non-local theory this thermal Hadamard function\index{Green function!Hadamard (thermal)} remains identical to the local expression,
\begin{align}
\mathcal{G}{}^{(1)}_{\omega,\beta}(x',x) = G{}^{(1)}_{\omega,\beta}(x',x) \, ,
\end{align}
since the considerations that led us to the vacuum relation $\mathcal{G}{}^{(1)}_\omega(x',x) = G{}^{(1)}_\omega(x',x)$ did not depend on the quantum state, and, rather, followed directly from the fact that the Hadamard function is a solution of the homogeneous equation
\begin{align}
\mathcal{D}_\omega \mathcal{G}{}^{(1)}_{\omega,\beta}(x',x) = 0 \, ,
\end{align}
which is insensitive to the non-local form factors considered in this thesis.

Recall now that the interacting Hadamard function is linear in the free Hadamard function, and hence we find in complete analogy to Eq.~\eqref{eq:ch6:hadamard-1} the thermal interacting Hadamard function
\begin{align}
\begin{split}
\ts{\mathcal{G}}^{(1)}_{\omega,\beta}(x',x) = G{}^{(1)}_{\omega,\beta}(x'-x) &- \Lambda_\omega\mathcal{G}^\text{R}_\omega(x')G{}^{(1)}_{-\omega,\beta}(x) - \Lambda_{-\omega}\mathcal{G}^\text{R}_{-\omega}(x)G{}^{(1)}_{\omega,\beta}(x') \\
& + G{}^{(1)}_{\omega,\beta}(0) \Lambda_\omega\mathcal{G}_\omega(x') \Lambda_{-\omega}\mathcal{G}_{-\omega}(x) \, .
\end{split}
\end{align}
We are now ready to define the renormalized thermal fluctuations around a $\delta$-shaped potential in both local and non-local theories, where we again subtract the vacuum contributions in the case $\lambda=0$ but keep the vacuum contributions at $\lambda>0$:
\begin{align}
\begin{split}
\label{eq:ch6:thermal-main-result}
\langle \varphi^2(x)\rangle^{\lambda,\beta}_\ind{ren} &= \int\limits_m^\infty \frac{\dd\omega}{2\pi} \left[\ts{\mathcal{G}}^{(1)}_{\omega,\beta}(x,x) - \frac{1}{\sqrt{\omega^2-m^2}} \right] = \Psi(x,\beta,m,\lambda,\ell) + \Pi(\beta m) \\
\Psi(x,\beta,m,\lambda,\ell) &= \int\limits_0^\infty\frac{\dd\varpi}{4\pi} \frac{\Phi_\omega(x,\beta,m,\lambda,\ell)}{\sqrt{\varpi^2+m^2}} \, , \\
\Phi_\omega(x,\beta,m,\lambda,\ell) &= \frac{B^2-\cos^2(\varpi x) - 2\cos(\varpi x)BC}{1+C^2} D \, , \\
B &= 2\varpi\Delta\mathcal{G}_\omega(x) - \sin(\varpi|x|) \, , \quad
C = 2\varpi/\widetilde{\lambda}_\omega \, \quad D = \coth\left( \frac{\beta\sqrt{\varpi^2 + m^2}}{2} \right) \, , \\
\widetilde{\lambda}_\omega &= \frac{\lambda}{1+\lambda\Delta\mathcal{G}_\omega(0)} \, .
\end{split}
\end{align}
This should be compared with Eqs.~\eqref{eq:ch6:main-integral} and \eqref{eq:ch6:main-integral-aux}, which is readily reproduced in the limiting case of $\beta\rightarrow\infty$ and hence $D=1$.

\begin{figure}[!htb]%
    \centering
    \includegraphics[width=0.8\textwidth]{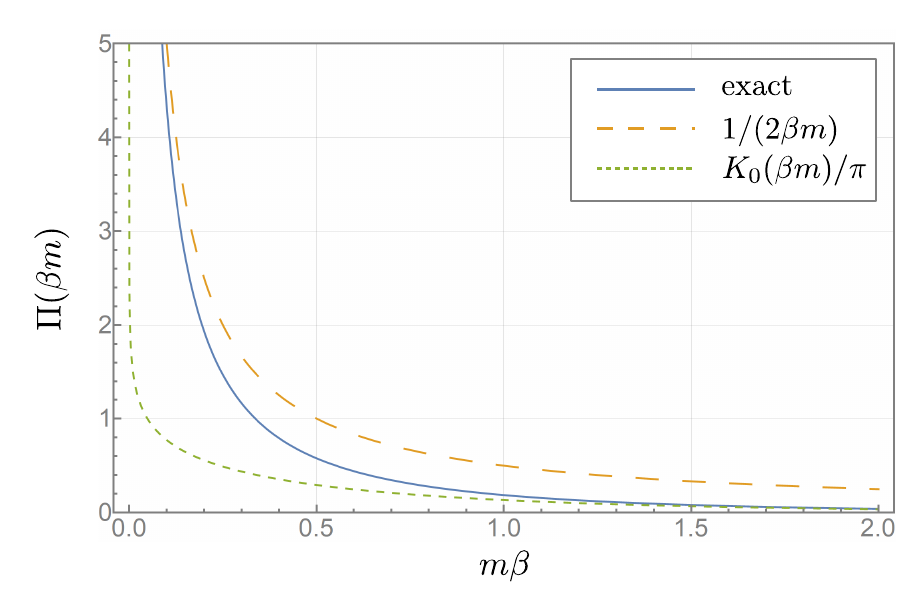}\\[-15pt]
    \caption[Thermal background for both local and non-local case.]{The thermal background $\Pi(\beta m)$ plotted as a function of the dimensionless combination $m\beta$, together with the two analytic approximations for small and large temperatures.}
    \label{fig:ch6:pp-background}
\end{figure}

The expression $\Pi(\beta m)=\langle \varphi^2(x)\rangle^{\lambda=0,\beta}_\ind{ren}$ is a universal, $\lambda$-independent expression that solely describes the thermal fluctuations of the vacuum case. Moreover, due to translation invariance $\Pi$ does not depend on $x$ and it is given by the following expression:
\begin{align}
\label{eq:ch6:pp-background}
\Pi(\beta m) = \int\limits_0^\infty \frac{\dd\varpi}{\pi\sqrt{\varpi^2+m^2}} \frac{1}{\exp(\beta\sqrt{\varpi^2+m^2}) - 1}\, .
\end{align}
We plot this function in Fig.~\ref{fig:ch6:pp-background}, together with its asymptotics
\begin{align}
\Pi(\beta m) \approx \begin{cases} \displaystyle \frac{1}{2\beta m} & \text{ for } \beta m \ll 1 \, , \\[10pt]
\displaystyle \frac{1}{\pi} K_0(\beta m) & \text{ for } \beta m \gg 1 \, . \end{cases}
\end{align}
The $\lambda$-dependent expression is captured by the function $\Psi$, which is identical to the vacuum case except for the linear appearance of the thermal $D$-factor acting as a density function determining how much each frequency $\omega$ contributes to the total fluctuations given the inverse temperature $\beta$. At large frequencies and fixed and finite $\beta$ this density factor regularizes the frequency integral exponentially. In principle we may now evaluate \eqref{eq:ch6:thermal-main-result} for a choice of the parameters $\{x, \beta, m, \lambda, \ell \}$, but before doing so it is sensible to discuss a few physical conditions on these parameters.

First, in a thermal bath of temperature $T$ the number density of massive scalar quanta is highly suppressed if the rest mass $m$ is much larger than $T$. For this reason we demand
\begin{align}
\label{eq:ch6:param-cond-1}
\frac{m}{T} = m\beta < 1 \, .
\end{align}
Second, the temperature may also not be too large, because then the influence of the potential will be almost negligible, whence we also demand
\begin{align}
\label{eq:ch6:param-cond-2}
\frac{\lambda}{T} = \lambda\beta > 1 \, .
\end{align}
With these choices, let us discuss the local case first and then move on to the non-local $\mathrm{GF_2}$ case. One may also be interested in considering the non-local $\mathrm{GF_1}$ case since at finite temperature the Boltzmann distribution for large frequencies regularizes the $\omega$-integration, but due to the doubts expressed in the previous section we shall limit our considerations to $\mathrm{GF_2}$ theory.

\subsection{Local theory}
In the local case one has $\Delta\mathcal{G}_\omega(x)\equiv 0$ and $\ell=0$, which implies that the function $\Psi$ depends on the dimensionless variables $\beta m$, $\lambda/m$, and $xm$. For the convenience of comparing the local results no the non-local results we shall instead use $\ell>0$ as a normalization parameter, as we have done in several instances in this thesis before. The numerical evaluation of $\Psi$ is nearly identical to those required in the vacuum case presented in Sec.~\ref{sec:ch6:vacuum} since the thermal factor $D$ does not exacerbate the convergence properties of the frequency integral \eqref{eq:ch6:thermal-main-result}---if anything, for finite $\beta$ it enhances the convergence properties at large $\omega$.

\begin{figure}[!htb]%
    \centering
    \includegraphics[width=0.8\textwidth]{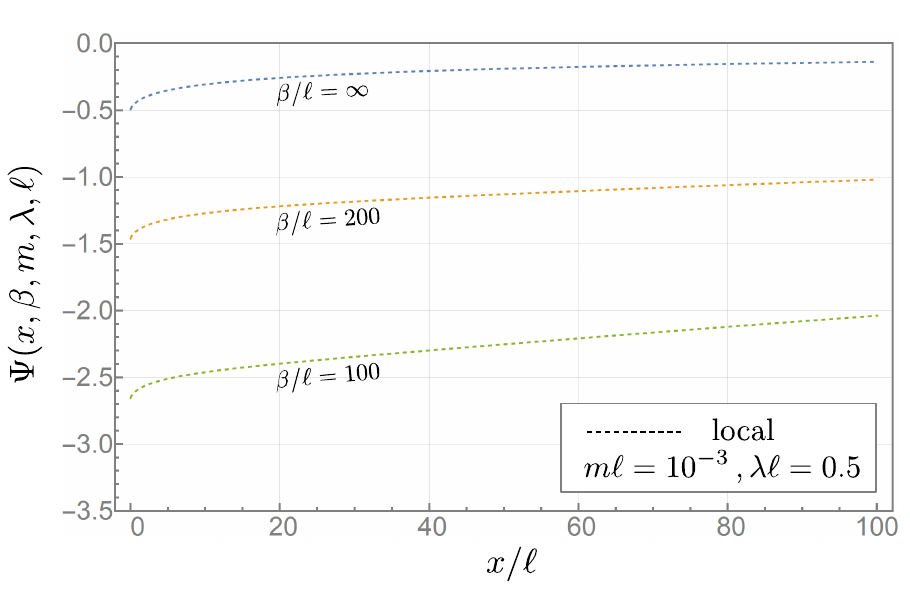}\\[-15pt]
    \caption[Thermal fluctuations in the local theory.]{Local thermal fluctuations, plotted for various values of the inverse temperature and for the vacuum case, subject to conditions \eqref{eq:ch6:param-cond-1} and \eqref{eq:ch6:param-cond-2}, against the dimensionless distance $x/\ell$. The scale of non-locality $\ell$ is employed purely as a reference scale to facilitate the comparison to the non-local theory.}
    \label{fig:ch6:fluctuation-loc}
\end{figure}

Given a physically sensible choice of mass and potential parameters, see Fig.~\ref{fig:ch6:fluctuation-loc} for a plot of the function $\Psi$ for a range of temperatures, where we also included the zero-temperature case as a reference. These thermal fluctuations, much like in the vacuum case, exhibit a sharp peak at the location of the potential at $x=0$, and the primary effect of the thermal state appears to be an effective negative shift of the fluctuations while leaving the overall shape mostly intact.

\subsection{Non-local $\mathrm{GF_2}$ theory}
\begin{figure}[!htb]%
    \centering
    \includegraphics[width=0.8\textwidth]{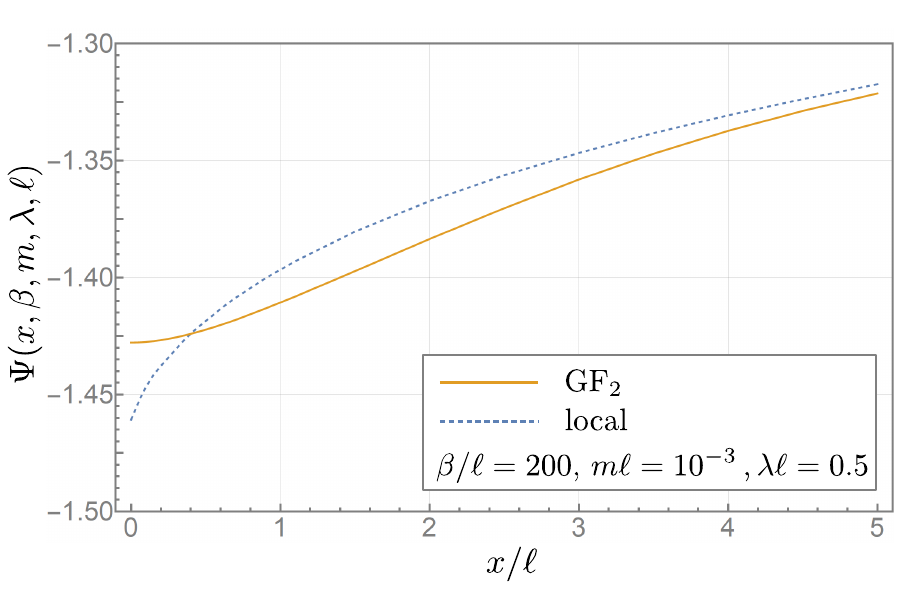}\\[-15pt]
    \caption[Thermal fluctuations in the non-local $\mathrm{GF_2}$ theory.]{Comparison of the local and $\mathrm{GF_2}$ thermal fluctuations for a physically sensible value of temperature ($\beta\ell=200$), plotted as a function of the dimensionless distance $x/\ell$. Mass and potential parameters are chosen in accordance with Eqs.~\eqref{eq:ch6:param-cond-1} and \eqref{eq:ch6:param-cond-2}.}
    \label{fig:ch6:fluctuation-gf2}
\end{figure}

Similarly to the local case, the non-local $\mathrm{GF_2}$ case is affected solely by the appearance of the thermal factor $D$ in the main integral \eqref{eq:ch6:thermal-main-result}, but of course now the non-local modification term $\Delta\mathcal{G}_\omega(x) \not=0$. We have devoted considerable efforts into evaluating this contribution numerically and have detailed our steps in Sec.~\ref{sec:ch6:vacuum} such that we can now focus solely on the effects of the finite temperature; see a plot of the thermal fluctuations for $\mathrm{GF_2}$ theory, compared to the local case, in Fig.~\ref{fig:ch6:fluctuation-gf2}.

The overall result is again quite similar to the local case. While the presence of a thermal bath appears to shift the thermal fluctuations to smaller, more negative values, the overall shape is still similar to the vacuum case of $\mathrm{GF_2}$ theory. Moreover, $\mathrm{GF_2}$ theory removes the sharp peak at $x=0$ and it appears that the quantity $\langle\varphi\partial_x\varphi\rangle$ remains differentiable also in the case of finite temperature in the presence of non-locality. At large distances, $x/\ell\gg 1$, the local and non-local expressions approach each other.

One may wonder what happened to the influence of the $\omega$-dependent coupling $\widetilde{\lambda}_\omega$, but our numerical investigations have shown that even though the factor $\widetilde{\lambda}_\omega$ becomes large at some intermediate frequency $\omega_\star$ the overall effect averages out and does not seem to survive the $\omega$-integration.\footnote{Our numerical plots are taken for a potential barrier $\lambda<\lambda_\star$, where $\lambda>\lambda_\star$ is required for the effective coupling to diverge. Because it did not affect the overall shape of the thermal fluctuations we decided to generate the plots for smaller values of $\lambda$ which are numerically faster to evaluate.}

The decisive property of well-behaved non-local $\mathrm{GF_{2n}}$ theories seems to lie in its short-scale regulatory influence on the thermal fluctuations, while leaving the asymptotic, large-distance behavior numerically unchanged. In this sense these considerations present an explicit UV improvement due to non-locality as modelled by $\mathrm{GF_2}$ theory.

\section{Fluctuation-dissipation theorem}
The fluctuation-dissipation theorem \cite{Nyquist:1928,Callen:1951,Landau:1980a,Landau:1980b} is a powerful concept that relates fluctuations in a physical observable (such as the vacuum polarization) to properties of a susceptibility. For the local theory the fluctuation-dissipation theorem has the form \cite{Landau:1980a,Landau:1980b}
\begin{align}
\langle \varphi^2(x)\rangle^{\lambda,\beta} &= \coth\left( \frac{\beta\omega}{2} \right) 2\Im \left[ \ts{{G}}_\omega^\ind{R}(x,x) \right] \, .
\end{align}
Note that this is a relation between the \emph{unrenormalized} vacuum fluctuations evaluated in Fourier space on the one side, and the interacting Green function on the other side. Similarly, it is possible to find a relation between the interacting, point-split Hadamard function\index{Green function!Hadamard} and the interacting, point-split Green function,
\begin{align}
\ts{\mathcal{G}}^{(1)}_\omega(x',x) = 2 \, \text{sgn}\left(\omega\right) \Im \left[ \ts{\mathcal{G}}^\ind{R}_\omega(x',x) \right] \, .
\end{align}
Note that this is a relation between \emph{non-local Green functions}, which can be checked for the exact solutions presented in these sections. One may then take the limit $x\rightarrow x'$ which yields
\begin{align}
\label{eq:fdt}
\langle \varphi^2(x)\rangle^{\lambda,\beta} &= \coth\left( \frac{\beta\omega}{2} \right) 2\Im \left[ \ts{\mathcal{G}}_\omega^\ind{R}(x,x) \right] \, ,
\end{align}
which we dub the ``non-local fluctuation-dissipation theorem.'' This constitutes a somewhat non-trivial consistency check of the interaction of thermal fluctuations with non-local physics, and while we have only proven the above relation in the somewhat special case of a $\delta$-potential we expect it to hold true in more general cases as well.

\section{Concluding remarks}

In this chapter we have studied the vacuum polarization and thermal fluctuations around a $\delta$-potential in quite some detail. Paying careful attention to DeWitt's asymptotic causality conditions, see Ch.~\ref{ch:ch2} for more details, we were also able to explicitly prove the validity of the fluctuation-dissipation theorem. While the presence of non-locality affects the vacuum polarization and thermal fluctuations at large distances as well, this effect turned out to be numerically very small. It is interesting to note that the necessary subtraction of divergent vacuum contributions in non-local ghost-free quantum theory remains identical to the local case. The most significant impact of non-locality appears in the vicinity of the potential: while in local quantum field theory the fluctuations are finite at the location of the potential, they are not differentiable. Non-locality smoothes the polarization such that quantities like $\langle\varphi\partial_x \varphi\rangle$ are now regular everywhere.

An interesting subject of future study should involve a non-linear interacting, non-local ghost-free quantum theory. With this chapter we have put the notion of non-local Green functions on a stronger foundation, and we hope that these results provide useful in this line of investigations.

%%%%%%%%%%%%%%%%%%%%%%%%%%%%%%%%%%%%%%%%%%%%%%%%%%%%%%%%%%%%%%%%%%%%%%%%%%%%%%%%%%%%%%%%%%%%%%%%%%%
%
% Chapter: 
%
\chapter{Black holes, generalized Polyakov action, and Hawking radiation}
\label{ch:ch7}
\textit{In this final chapter we will investigate implications of non-locality for quantum field theories in curved spacetime. We will introduce a non-local generalization of Polyakov's effective action and show that while it does not affect a black hole's Hawking temperature there is a non-vanishing effect on the black hole's entropy and on the conformal anomaly. This chapter is based on Ref.~\cite{Boos:2019vcz}.}

\section{Introduction}

Black holes are not only a fascinating prediction of Einstein's theory of gravity, they may very well provide us with insights into the quantum nature of gravity. A famous example is the prediction of Hawking radiation \cite{Hawking:1974sw,Bekenstein:1974ax,Page:1976df,Unruh:1976db,Hawking:1982dh,Kiefer:2012} that emerges in semiclassical gravity wherein the black hole background is treated as classical but the matter field on that background is described by quantum field theoretical creation and annihilation operators. It can be shown that the early-time and late-time vacuum states of that quantum field do not coincide: they are related by a Bogoliubov transformation, where the specific coefficients are related to a thermal distribution. As a result, black holes are thought to emit thermal radiation of a characteristic temperature which has hence been dubbed \emph{Hawking temperature}.\index{Hawking temperature}

In the context of the present thesis one might wonder whether and how the presence of non-locality at a small length scale $\ell$ affects that Hawking temperature. Na\"ively, since non-local infinite-derivative gravity was constructed as the effective field theory for a UV-complete description of gravity, one would expect that it should leave some imprints in the semiclassical description of gravity. However, there are some underlying arguments we wish to elucidate first.

As already presented in Ch.~\ref{ch:ch6}, it is indeed possible to define a well-behaved concept of temperature in non-local field theories in terms of the free Hadamard\index{Green function!Hadamard} and free Wightman\index{Green function!Wightman} functions. These functions, viewed as Green functions, are solutions to a homogeneous equation of the form
\begin{align}
\mathcal{D}\mathcal{G}(X',X) = 0 \, ,
\end{align}
where $\mathcal{D}$ is the differential operator of the theory under consideration. As argued in Sec.~\ref{sec:ch1:on-shell-off-shell} and revisited in more detail in the previous chapter, the class of non-local theories studied in this thesis does not yield additional solutions for homogeneous equations. For this reason this class of non-local theories is also referred to as ``ghost-free:'' it does not introduce new propagating modes. This directly implies that the free local theory and free non-local theory have the \emph{identical Hadamard and Wightman functions.} In case there exists an interaction with other fields, however, this statement is no longer valid, as we exemplified in Sec.~\ref{sec:ch6:thermal} of the previous chapter.

In the context of black holes it is of central importance to extend the studies from flat Minkowski spacetime to curved spacetimes. A good approximation for the near-horizon geometry of a Schwarzschild black hole are Rindler coordinates, where the acceleration parameter is directly related to the surface gravity of the black hole. A natural question to ask is how the presence of non-locality would impact the temperature measured by an Unruh--DeWitt detector. There has been some controversy in the recent literature \cite{Nicolini:2009dr,Modesto:2017ycz,Kajuri:2017jmy,Kajuri:2018myh}, and in this thesis we take the following point of view:

Frolov and Zelnikov \cite{Frolov:2011} propose a rather straightforward method for measuring the temperature in a non-local field theory using an Unruh--DeWitt detector. In particular, they parametrize the Unruh--DeWitt detector as a local quantum system following the usual local laws of quantum mechanics, and assume that it is in thermal contact with a non-local quantum field. Assuming that the Hilbert space factorizes they are able to relate the measured temperature to the Wightman functions of the non-local scalar field. If this process takes place in vacuum the Wightman functions are insensitive to non-locality, as argued above, and the presence of non-locality does not affect the measurement of the temperature.

In the case of Hawking radiation, however, it has been shown that the flux as measured by an observer at infinity is described by the retarded Green function \cite{Frolov:1981,Frolov:1981mz}. For this reason it seems plausible to us that the asymptotic flux may depend on the presence of non-locality, albeit the effects might be too small to detect. It is hence our goal to determine the vacuum expectation value of the energy momentum tensor of a non-local scalar field on a given black hole background, and study whether and how these expressions depend on the presence of non-locality.

It should be emphasized that the studies presented in this chapter correspond to a test of non-local physics in the strong field regime in proximity of a black hole. In particular, we will focus on a two-dimensional example because then it is possible to extract the quantum average of the stress energy tensor by means of the Polyakov action\footnote{In what follows, ``Polyakov action'' shall refer to the effective action of a scalar field in two-dimensional curved spacetime and should not be confused with the Polyakov action encountered when describing string worldsheets.}\index{Polyakov action} \cite{Polyakov:1981}, which in turn can be derived after quantizing a conformal scalar field on a given background and performing a functional integral over the conformal anomaly \cite{Luscher:1980fr,Dowker:1993rt,Ribeiro:2018pyo}.

In what follows we will first discuss the relevant details for the local case and establish our notation. In a second step we introduce a modification of the Polyakov action by inserting a non-local form factor and then analyze how this affects the effective action as well as the energy momentum tensor. Then, we will focus on two applications: first we determine the contribution of the non-local terms to the entropy of a black hole in two dimensions, and secondly we use the conformal anomaly to derive the Hawking flux at spatial infinity for a static, two-dimensional black hole \cite{Christensen:1977jc}. In a third step, we focus on a concrete black hole solution from two-dimensional dilaton gravity and develop semi-analytical methods to determine both the non-local entropy corrections as well as the effects on the energy momentum tensor.

Before delving into the calculational details, the main idea can be summarized like this: suppose the total action for a theory involving a classical gravitational field $g{}_{\mu\nu}$ and a quantum scalar field $\hat{\psi}$ takes the form
\begin{align}
S[g{}_{\mu\nu}, \hat{\psi}] = S_\text{g}[g{}_{\mu\nu}] + S_\text{m}[g{}_{\mu\nu}, \hat{\psi}] \, .
\end{align}
The dynamics of the classical gravitational field are given by the corresponding field equations, but the quantum scalar field's dynamics are given by the path integral over all possible configurations. It is possible to rewrite this theory in terms of a so-called \emph{effective action}\index{effective action} for the scalar field $\hat{\psi}$ such that the classical equations of motion yield the same result as the path integral approach. This effective action, substituting the classical solution for $\hat{\psi}$, then takes the form
\begin{align}
W[g{}_{\mu\nu}] = S_\text{g}[g{}_{\mu\nu}] + W_\text{m}[g{}_{\mu\nu}] \, .
\end{align}
The quantum field $\hat{\psi}$ no longer appears and has hence been ``integrated out'' of the theory. One may now treat the resulting effective action $W[g{}_{\mu\nu}]$ as the effective action that describes how the gravitational dynamics are changed due to the presence of a quantum scalar field. In what follows, we will follow this approach.

\section{2D conformal anomaly and the Polyakov action}
From now on we shall work in a two-dimensional spacetime\footnote{In this chapter, all Greek indices are spacetime indices, and we reserve Greek letters from the beginning of the alphabet ($\alpha,\beta,\gamma,\dots$) for repeated indices.} of signature $(-,+)$ with a metric $g{}_{\mu\nu}$, and $\hat{\psi}$ is a conformally invariant quantum field on that background. The classical trace of the field's energy momentum tensor vanishes due to conformal invariance, but the quantum average does not. This is called the conformal anomaly and stems from the fact that the two operations of renormalization and taking the trace do not commute \cite{Duff:1993wm}. One may write
\begin{align}
\label{eq:ch7:anomaly}
\langle \hat{T}^{\mu\nu}\rangle g_{\mu\nu} = 2b R \, ,
\end{align}
where $R$ is the two-dimensional Ricci scalar, and $b=1/(48\pi)$ for a conformal massless real scalar field. Polyakov \cite{Polyakov:1981} demonstrated that this non-vanishing trace of the energy-momentum tensor can be derived from the following effective action at one-loop (which, for this reason, is often referred to as \emph{Polyakov action}\index{Polyakov action}; see also L\"uscher \textit{et al.} \cite{Luscher:1980fr}):
\begin{align}
\label{eq:ch7:polyakov}
W_\text{Pol}[g_{\mu\nu}] = -\frac{b}{2} \int \dd^2 x\, \sqrt{-g} \, R \, \frac{1}{\Box} R \, .
\end{align}
The trace turns out to be
\begin{align}
T=T^{\mu\nu}g_{\mu\nu} = \frac{2}{\sqrt{-g}}{\delta W_\text{Pol}\over \delta g_{\mu\nu}} g_{\mu\nu} =2 bR\, .
\end{align}
At this point it is helpful to notice that the Polyakov action \eqref{eq:ch7:polyakov} can be rewritten using an auxiliary, non-minimally coupled classical scalar field $\varphi$ such that
\begin{align}
\label{eq:ch7:polyakov-aux}
W_\text{Pol}[g{}_{\mu\nu},\varphi]=b \int \dd^2 x\ \sqrt{-g}\left[\frac12 \varphi \,\Box \varphi -R\varphi\right]\, .
\end{align}
The scalar field equation is
\begin{align}
\label{eq:ch7:scalar-field-aux-local}
\Box \varphi = R
\end{align}
and hence, on-shell, imposing this relation, one recovers \eqref{eq:ch7:polyakov}. Let us remark that the action \eqref{eq:ch7:polyakov-aux} is a functional of both the metric $g{}_{\mu\nu}$ \emph{and} the classical field $\varphi$. It reproduces the Polyakov action on-shell, but the action itself is \emph{not} conformally invariant, and the auxiliary scalar field $\varphi$ is \emph{not} to be confused with the original quantum field $\hat{\psi}$ that has physical significance.

Using the effective action, the components of the effective energy-momentum tensor are \cite{Frolov:1981}
\begin{align}
\label{eq:ch7:tmunu-local}
T^{\mu\nu}={2\over \sqrt{-g}} {\delta W_\text{Pol}\over \delta g_{\mu\nu}} = b \Big[\varphi^{;\mu}\varphi^{;\nu}-2 \varphi^{;\mu\nu}
-g^{\mu\nu}\Big( {1\over 2}\varphi^{;\alpha}\varphi_{;\alpha}-2\Box\varphi  \Big)\Big] \, ,
\end{align}
where we understand $\varphi = 1/\Box R$ and denote covariant differentiation by the semicolon, $\varphi{}^{;\mu\nu} = \nabla{}^\nu\nabla{}^\mu\varphi$. One may verify that this energy-momentum tensor is indeed conserved, $T{}^{\mu\alpha}{}_{;\alpha} = 0$. Moreover, the expression $1/\Box$ is to be taken as the appropriate causal Green function in accordance with the physical boundary conditions of the problem under consideration, and we will revisit this point in much detail below. For no incoming fluxes from the past, for example, one would identify $1/\Box$ with the retarded Green function.

Let us now turn our attention towards two-dimensional black holes. As we detail in appendix \ref{app:2d-static}, a two-dimensional static black hole geometry can be written as
\begin{align}
\label{eq:ch7:geometry}
\dd s^2=-f\, \dd t^2+ \frac{\dd r^2}{f} \, ,
\end{align}
where $f = -\xi{}^\alpha\xi{}_\alpha$ and $\ts{\xi} = \partial_t$ is the timelike Killing vector. A solution of Eq.~\eqref{eq:ch7:scalar-field-aux-local} is
\begin{align}
\label{eq:ch7:scalar-field-aux-local-sol}
\varphi=\Phi_0+\chi \, , \quad \Phi_0=-\ln f \, ,
\end{align}
where $\Box\Phi_0=R$ and $\Box\chi=0$, that is, $\chi$ is a homogeneous solution constructed from zero modes\index{zero mode} of the d'Alembert operator $\Box$. In our case we choose $\chi$ such that the resulting energy-momentum tensor is stationary (see more details in appendix \ref{app:state}),
\begin{align}
\chi= w t+k r_*\, , \quad r_*=\int \frac{\dd r}{f} \, ,
\end{align}
where $w$ and $k$ are two constants and $r_\ast$ is a tortoise coordinate. Both of these fields, $\Phi_0$ and $\chi$, contribute to the effective energy-momentum tensor. Substituting \eqref{eq:ch7:scalar-field-aux-local-sol} into \eqref{eq:ch7:tmunu-local} we can define these two contributions as
\begin{align}
T^{\mu\nu} = T_{(\Phi_0)}^{\mu\nu} + T_{(\chi)}^{\mu\nu} \, .
\end{align}
The $\Phi_0$-contributions take the form
\begin{align}
T_{(\Phi_0)}^{\mu\nu} &= b\Big[\Phi_0^{;\mu}\Phi_0^{;\nu}-2\Phi_0^{;\mu\nu}-g^{\mu\nu}\Big( {1\over 2}\Phi_0^{;\alpha}\Phi_0{}_{;\alpha}-2\Box\Phi_0  \Big)\Big] , \\
T_{(\Phi_0)}{}^{\mu}{}_{\nu} &= b \left(
   \begin{array}{cc}
       -2f''+{f'^2\over 2 f} & 0 \\
       0 & -{f'^2\over 2 f} \\
     \end{array}
   \right) \, ,
\end{align}
whereas the $\chi$ zero-mode contributions are
\begin{align}
\label{eq:ch7:tmunu-chi-local}
T_{(\chi)}{}^{\mu}{}_{\nu}= b \left(
   \begin{array}{cc}
       -{k^2+w^2\over 2 f} & wk \\
       -{wk\over f^2} & {k^2+w^2\over 2 f} \\
     \end{array}
   \right) \, .
\end{align}
The two constants $w$ and $k$ encode the quantum state of the field $\chi$: when either $w$ or $k$ vanish, the off-diagonal components of $T_{(\chi)}{}^{\mu}{}_{\nu}$ vanish as well, implying the absence of any fluxes. A different choice where $w=0$ and $k=\kappa$, where here $\kappa$ denotes the surface gravity,
\begin{align}
\kappa = \frac12 f'|_{r=r_g} \, ,
\end{align}
corresponds to the Hartle--Hawking state. Last, $w=\kappa$ and $k=-\kappa$ define the Unruh vacuum state, and for more details we refer to appendix \ref{app:state}.

\section{Ghost-free modification of the Polyakov action}
Let us now, with these considerations in mind, extend the description to non-local field theory. To that end, we begin by considering a non-local modification of the auxiliary Polyakov effective action \eqref{eq:ch7:polyakov-aux} where
\begin{align}
\label{eq:ch7:gf-polyakov-aux}
W_\text{GF}[g{}_{\mu\nu}, \varphi] = \frac{1}{48\pi} \int\dd^2 x \sqrt{-g} \left[ \frac12 \varphi A \varphi - R\varphi \right] \, , \quad A = \Box \, e^{P(\Box)} \, , \quad P(z) = (-\ell^2 z)^N \, .
\end{align}
Here, $\ell>0$ denotes the scale of non-locality, $N=1,2,\dots$ is a positive integer, and for a given $N$ we refer to the above non-local theory as $\mathrm{GF_N}$. Now, however, the scalar field equation takes the form
\begin{align}
\Box e^{P(\Box)} \varphi = R \, ,
\end{align}
where again the scalar curvature $R$ acts as a source for the now non-local field $\varphi$. Note that in the limiting case $\ell\rightarrow 0$ we recover the previous results. Inserting this relation back into \eqref{eq:ch7:gf-polyakov-aux} we obtain the \emph{ghost-free modification of the Polyakov action},
\begin{align}
\label{eq:ch7:gf-polyakov}
W_\text{GF}[g{}_{\mu\nu}] &= -\frac{1}{96\pi}\int\dd^2 x \sqrt{-g} R A^{-1} R 
= -\frac{1}{96\pi}\int\dd^2 x \sqrt{-g} R \frac{e^{-(-\ell^2\Box)^N}}{\Box} R \, .
\end{align}
In the limiting case $\ell\rightarrow 0$ we recover the Polyakov action encountered in local field theory \eqref{eq:ch7:polyakov}. For this reason the above relation constitutes a possible ghost-free modification of the Polyakov action. Our aim is to study how this effective action, as a starting point, affects physical observables.

In the simplest case of $\mathrm{GF_1}$ theory the action can be written as the parametric integral
\begin{align}
\label{eq:ch7:polyakov-gf-decomposition}
W_\text{GF}=W_\text{Pol}+W_{\ell} \, , \quad
W_{\ell}=\int\limits_0^{\ell^2}\dd s \, \widetilde{W}(s)\, , \quad
\widetilde{W}(s)=-\frac{1}{96\pi} \int\dd^2 x \sqrt{-g} \, R\, e^{s\Box} R \, . 
\end{align}
This formulation is convenient because it captures the operator $1/\Box$ as a parametric integral, but in the cases of $N \ge 2$ this method does not apply directly. Again, in the case of $\ell\rightarrow 0$ the integral collapses and we recover the previous result \eqref{eq:ch7:polyakov}.

\subsection{Effective energy-momentum tensor}
\label{sec:ch7:tmunu}
We will now use the ghost-free modification of the Polyakov action \eqref{eq:ch7:gf-polyakov} to derive the trace as well as the tensorial components of the energy-momentum tensor.

\subsubsection{Trace}
For this calculation it is useful to work in the conformal gauge where the two-dimensional metric takes the form
\begin{align}
g{}_{\mu\nu} = e^{2\sigma}\eta{}_{\mu\nu} \, , \quad \sqrt{-g} = e^{2\sigma} \, .
\end{align}
In this representation the d'Alembert operator and the scalar curvature are given by
\begin{align}
\Box = e^{-2\sigma}\overline{\Box} \, , \quad R = -2\Box\sigma \, ,
\end{align}
where $\bar{\Box} = \eta{}^{\alpha\beta}\partial_\alpha\partial_\beta$ is the flat d'Alembertian, and we can recast the ghost-free Polyakov action into a functional of $\sigma$:
\begin{align}
W_\text{GF}[\sigma] &= -\frac{1}{24\pi} \int\dd^2 x \,e^{2\sigma} \sigma  e^{-(-\ell^2\Box)^N}\Box \sigma \, .
\end{align}
Then, the trace can be written as a variational derivative with respect to the conformal factor,
\begin{align}
T = g{}_{\mu\nu}T{}^{\mu\nu} = \frac{2g{}_{\mu\nu}}{\sqrt{-g}} \frac{\delta W_\text{GF}}{\delta g{}_{\mu\nu}} = e^{-2\sigma} \frac{\delta W_\text{GF}}{\delta \sigma} \, .
\end{align}
In order to perform the variational derivative we need to integrate by parts, as is common with most variational procedures, see also the relevant discussion in Sec.~\ref{sec:ch1:action}. While this is straightforward for any differential operator of finite order, the infinite-derivative operator $P(\Box)$ requires special treatment. Variations of such operators have been studied in the literature \cite{Snider:1964,Wilcox:1967}, and one may express them in terms of a parametric integral according to
\begin{align}
\label{ch7:variation}
\delta \big[ e^{\hat{B}} \big] = \int\limits_0^1 \dd \xi e^{(1-\xi)\hat{B}} \big[ \delta\hat{B} \big] e^{\xi\hat{B}} \, ,
\end{align}
where $\hat{B}$ denotes a finite-order, self-adjoint differential operator, in our case $\hat{B} = \Box$. Moreover, in the conformal gauge one has
\begin{align}
\delta(\sqrt{-g}\Box) = \delta\overline{\Box} = 0 \, , \quad \delta\Box = -2\delta\sigma\Box \, .
\end{align}
With these relations, and limiting our considerations to $\mathrm{GF_1}$ theory, the trace now takes the form
\begin{align}
\label{eq:ch7:trace-gf1}
T = \frac{1}{24\pi} e^{\ell^2\Box} R +\frac{\ell^2}{48\pi}\int\limits_0^1\dd\xi \left[ e^{(1-\xi)\ell^2\Box} R \right] \left[ e^{\xi \ell^2 \Box} R \right]  .
\end{align}
In the limiting case of $\ell\rightarrow 0$ we again recover the well known conformal anomaly of the scalar field of Eq.~\eqref{eq:ch7:anomaly}, $T = R/(24\pi)$.

\subsubsection{Tensorial components}
We may also perform a direct variation of \eqref{eq:ch7:gf-polyakov} with respect to the metric $g{}_{\mu\nu}$, yielding the non-local counterpart to Eq.~\eqref{eq:ch7:tmunu-local}. We find
\begin{align}
T^{\mu\nu}=\frac{1}{48\pi}&\Big[
\varphi^{;\mu}(e^{-\ell^2\Box}\varphi)^{;\nu}-{1\over 2}g^{\mu\nu}\varphi^{;\alpha}(e^{-\ell^2\Box}\varphi)_{;\alpha} -2 \varphi^{;\mu\nu} +2g^{\mu\nu}\Box\varphi \Big] \nonumber \\ \label{eq:ch7:tmunu-gf}
-{\ell^2\over 48\pi}&\int\limits_0^1\dd\xi\Big\{
\Big(e^{-(1-\xi)\ell^2\Box}\Box\varphi\Big)^{;\mu} \Big(e^{-\xi \ell^2\Box}\varphi\Big)^{;\nu}\\
&\hspace{35pt}-{1\over 2}g^{\mu\nu}\Big(e^{-(1-\xi)\ell^2\Box}\Box\varphi\Big)^{;\alpha} \Big(e^{-\xi \ell^2\Box}\varphi\Big)_{;\alpha} -{1\over 2}g^{\mu\nu}\Big(e^{-(1-\xi)\ell^2\Box}\Box\varphi\Big) \Big(e^{-\xi \ell^2\Box}\Box\varphi\Big) \Big\} \, . \nonumber
\end{align}
Indeed, inserting $\ell=0$ reproduces Eq.~\eqref{eq:ch7:tmunu-local}. Moreover, taking the trace of \eqref{eq:ch7:tmunu-gf} correctly reproduces \eqref{eq:ch7:trace-gf1}. The scalar field is subject to the on-shell condition
\begin{align}
\label{eq:ch7:phi-eom-gf}
\varphi=A^{-1}R={e^{\ell^2\Box}\over\Box}R \, ,
\end{align}
where the precise form of $\varphi$ depends on the quantum state that is to be chosen, and we will discuss that in the next section. While our considerations were limited to the relatively simple case of $\mathrm{GF_1}$ theory they can in principle be extended to higher $N=2,3,\dots$ by choosing a different $\hat{B}$ and applying the variational chain rule.

\subsection{State dependence}
As in the local case, the solution to Eq.~\eqref{eq:ch7:phi-eom-gf} consists of a homogeneous ``zero mode'' solution $\chi$ and an inhomogeneous solution $\Phi$ such that
\begin{align}
\label{eq:ch7:phi-sol-gf}
\varphi=\Phi+\chi \, .
\end{align}
Since the operator $P(\Box)$ is invertible we can express $\Phi$ in terms of the local solution,
\begin{align}
\label{eq:ch7:phi-sol-gf1-2}
\Phi=e^{\ell^2\Box}\Phi_0 \, , \quad \Phi_0=-\ln f\, ,
\end{align}
For the same reason the homogeneous solution is the same as in the local case, see also Sec.~\ref{sec:ch1:on-shell-off-shell}, which corresponds to a significant difference between ghost-free non-local theories and higher-derivative theories, in the latter of which the last statement is not true. Again, we may insert the solution \eqref{eq:ch7:phi-sol-gf} into \eqref{eq:ch7:tmunu-gf} and obtain two contributions,
\begin{align}
T^{\mu\nu} = T_{(\Phi)}^{\mu\nu} + T_{(\chi)}^{\mu\nu} \, .
\end{align}
The inhomogeneous contribution amounts to replacing $\varphi$ by $\Phi$ in Eq.~\eqref{eq:ch7:tmunu-gf} such that
\begin{align}
T_{(\Phi)}^{\mu\nu}=\frac{1}{48\pi}&\Big[
\Phi^{;\mu}(e^{-\ell^2\Box}\Phi)^{;\nu}-{1\over 2}g^{\mu\nu}\Phi^{;\alpha}(e^{-\ell^2\Box}\Phi)_{;\alpha} -2 \Phi^{;\mu\nu} +2g^{\mu\nu}\Box\Phi \Big] \nonumber \\
-{\ell^2\over 48\pi}&\int\limits_0^1\dd\xi\Big\{
\Big(e^{-(1-\xi)\ell^2\Box}\Box\Phi\Big)^{;\mu} \Big(e^{-\xi \ell^2\Box}\Phi\Big)^{;\nu}\\
&\hspace{35pt}-{1\over 2}g^{\mu\nu}\Big(e^{-(1-\xi)\ell^2\Box}\Box\Phi\Big)^{;\alpha} \Big(e^{-\xi \ell^2\Box}\Phi\Big)_{;\alpha} -{1\over 2}g^{\mu\nu}\Big(e^{-(1-\xi)\ell^2\Box}\Box\Phi\Big) \Big(e^{-\xi \ell^2\Box}\Box\Phi\Big) \Big\} \, . \nonumber
\end{align}
The homogeneous contributions, however, take the form
\begin{align}
T_{(\chi)}^{\mu\nu} = \frac{1}{48\pi} \Big[ \chi^{;\mu}\chi^{;\nu} - \frac12 g^{\mu\nu}\chi^{;\alpha}\chi_{;\alpha}-2 \chi^{;\mu\nu} + \Phi_0^{;\mu}\chi^{;\nu}+\chi^{;\mu}\Phi_0^{;\nu}-g^{\mu\nu}\,\chi_{;\alpha}\Phi_{0}^{\alpha} \Big] \, .
\end{align}
This part of the energy-momentum tensor is of central importance to our considerations. First, imposing the on-shell condition $\Box\chi=0$ one may show that $g{}_{\mu\nu}T_{(\chi)}^{\mu\nu} = 0$. And second, note that  this expression is entirely independent of the non-local scale $\ell$ and hence coincides with the corresponding expression found earlier in Eq.~\eqref{eq:ch7:tmunu-chi-local}.

Recall that the Hawking flux at spatial infinity is proportional to the off-diagonal components of the energy-momentum tensor. Moreover, it is possible to show that these contributions stem entirely from $T_{(\chi)}^{\mu\nu}$. This implies that the Hawking flux at spatial infinity, in the non-local case and provided one ignores backreaction effects onto the metric, perfectly agrees with the local results, confirming similar claims in Ref.~\cite{Kajuri:2018myh}.

The diagonal components of $T_{(\Phi)}^{\mu\nu}$, however, \emph{do} depend on the presence of non-locality and it is conceivable that the backreaction of these terms will affect the parameters of the black hole. We will leave this question for a future study.

\section{Black hole entropy}
Leaving aside the effective energy-momentum tensor for now, it is also of interest to determine the impact of non-locality on the entropy of a black hole in two dimensions. To that end, the representation of the ghost-free modification of the Polyakov action in Eq.~\eqref{eq:ch7:gf-polyakov-aux} proves useful.

After Wald's seminal discovery of the Noether charge technique \cite{Wald:1993nt}, Myers proved that these methods can be applied to non-local theories as well \cite{Myers:1994sg}. Since the total action of our gravitational theory is the sum of the standard, local gravitational action and the effective action given by the Polyakov expression, the entropy is also given by a sum of two contributions. Since this standard Wald entropy is well known for almost all (if not all) black hole solutions, we will focus on the contributions to the black hole entropy generated by the presence of non-locality.

Following Ref.~\cite{Myers:1994sg} and making use of the representation \eqref{eq:ch7:gf-polyakov-aux} it is possible to show that the ghost-free, non-local contribution to the entropy is given by\footnote{We will use the letter ``S'' for black hole entropy in this section and hope that it does not lead to possible confusion with the action which, unfortunately, traditionally is denoted by the same letter.}
\begin{align}
S_\text{GF} = \frac{1}{12} \varphi \big|_{r=r_g} \, .
\end{align}
Since the scalar field $\varphi$ is subject to the field equations \eqref{eq:ch7:phi-eom-gf}, and in what follows we will restrict our attention again to $\mathrm{GF_1}$ theory, we obtain the on-shell relation
\begin{align}
\label{eq:ch7:bh-entropy-gf1}
S_\text{GF} = \frac{1}{12} \frac{e^{\ell^2\Box}}{\Box} R \Big|_{r=r_g} \, .
\end{align}
It is clear that in the limit $\ell\rightarrow 0$ this reduces to
\begin{align}
S_\text{Pol} = \frac{1}{12} \frac{1}{\Box} R \Big|_{r=r_g} \, ,
\end{align}
which is the standard contribution to the black hole entropy due to the scalar hair $\hat{\psi}$. At this point it is important to emphasize that both the function $\varphi$ as well as the propagator $\exp(\ell^2\Box)/\Box$ depend on the quantum state. For the Hartle--Hawking state the scalar field $\varphi$ is finite on the bifurcation horizon (which is a point in two-dimensional spacetime; for more information on bifurcation horizons we refer to the original work by Boyer \cite{Boyer:1969}) and is given by
\begin{align}
\varphi=\Phi+r_*+c \, ,
\end{align}
where $c$  is a constant and $\Phi$ is given in Eq.~\eqref{eq:ch7:phi-sol-gf1-2}. In the local case one has instead
\begin{align}
\varphi=\Phi+r_*+c \, , \quad \Phi_0 = -\ln f \, ,
\end{align}
where the constant $c$ is the same in both cases since the zero modes coincide. The absolute value of this constant can be fixed, see Myers \cite{Myers:1994sg}. In our case we are interested in the difference of the local case and the non-local case,
\begin{align}
\Delta S = S_\text{GF} - S_\text{Pol} = \frac{\Phi+\ln f}{12} \, ,
\end{align}
which does not depend on the state.

\section{Hawking flux}
While studying the effective energy-momentum tensor in Sec.~\ref{sec:ch7:tmunu} we already saw that the off-diagonal components do not depend on the presence of non-locality. In this section we would like to re-derive this result by employing a method developed by Christensen and Fulling \cite{Christensen:1977jc} that employs a general representation of a stationary energy-momentum tensor in two dimensions.

Let us adopt the coordinates $\{t,r\}$ wherein the stationary energy-momentum tensor conservation yields
\begin{align}
\label{eq:ch7:tmunu-conservation}
\partial_r T{}^r{}_t = 0 \, , \quad \partial_r( f T{}^r{}_r ) = \frac12 f' T{}^\alpha{}_\alpha \, .
\end{align}
The Hawking flux at spatial infinity is then given by
\begin{align}
\frac{\dd E}{\dd t} = \frac12 \int\limits_{r_g}^\infty \dd r f'(r) T{}^\alpha{}_\alpha(r) \, .
\end{align}
Here, $r=r_g$ is the coordinate location of the black hole horizon where $g^{rr} = (\nabla r)^2 = f(r_g) = 0$. Integrating by parts, the conservation law \eqref{eq:ch7:tmunu-conservation} implies that the contributions to the Hawking radiation stem from the product $f T{}^r{}_r$ evaluated at the horizon and at spatial infinity. At the horizon one has $f(r_g)=0$, and for that reason the $T{}^r{}_r$-component must be singular if there is to be any non-zero contribution from the horizon. Conversely, if $T{}^r{}_r$ is a regular function at the horizon there will not be any contribution to the Hawking flux as measured at infinity.

Since the ghost-free modification of the Polyakov effective action can be written as the sum of a local piece for $\ell=0$ and a non-local piece containing $\ell$, see Eq.~\eqref{eq:ch7:polyakov-gf-decomposition}, the effective energy momentum trace $T$ and its tensor component $T{}^r{}_r$ also consist of two pieces,
\begin{align}
\label{eq:ch7:delta-t-tmunu}
T = T_\text{(Pol)} + \int\limits_0^s \dd s \, \widetilde{T} \, , \quad
T{}^r{}_r = T_\text{(Pol)}{}^r{}_r  + \int\limits_0^s \dd\tilde{s} \, \widetilde{T}{}^r{}_r \, , \quad
\widetilde{T}^{\mu\nu}=\frac{2}{\sqrt{-g}} \frac{\delta \widetilde{W}(s)}{\delta g_{\mu\nu}} \, .
\end{align}
In what follows let us delineate a method for determining the $\widetilde{T}{}^{\mu\nu}$ tensor components for a given metric of the form \eqref{eq:ch7:geometry}. First, the general variation of $\widetilde{W}(s)$ takes the form
\begin{align}
\delta\widetilde{W}(s) = \int \dd^2x \sqrt{-g} \, \widetilde{T}^{\mu\nu} \delta g_{\mu\nu} \, .
\end{align}
It is important to first perform the variation and only then substitute a given metric, in our case the one specified in Eq.~\eqref{eq:ch7:geometry}. Restricting ourselves to spatial and static variations,
\begin{align}
\delta(\dd s^2) = -\left[ \dd t^2 + \frac{\dd r^2}{f^2} \right] \delta f = \delta g{}_{\mu\nu} \dd x{}^\mu \dd x{}^\nu \, ,
\end{align}
we obtain
\begin{align}
2f\frac{\delta \widetilde{W}(s)}{\delta f} = \widetilde{T}{}^r{}_r -  \widetilde{T}{}^t{}_t \, .
\end{align}
We can combine these results and express $\widetilde{T}$ as well as $\widetilde{T}{}^r{}_r$ purely in terms of non-local infinite-derivative operators as well as the curvature scalar,
\begin{align}
\widetilde{T} &= \frac{1}{48\pi} \Big[ 2f e^{s\Box} R'' + 2f'\left( e^{s\Box} R \right)' + R e^{s\Box} R + sf \int\limits_0^1 \dd \xi \left( e^{(1-\xi)s\Box}R \right)' \left( e^{\xi s \Box} R \right)' \Big] \, , \\
\widetilde{T}^r{}_r &= \frac{1}{96\pi}\Big[ 4f \left(e^{s\Box}R \right)'' + 2f'\left( e^{s\Box R}\right)' + R e^{s\Box}R + s \partial_r \int\limits_0^1 \dd \xi f \left( e^{(1-\xi)s\Box}R \right) \left( e^{\xi s\Box}R \right)' \Big] \, ,
\end{align}
where we abbreviated $(\dots)'=\partial_r(\dots)$. Recall that these expressions are parametric functions of $s$ that enter the definition of the full contributions in Eq.~\eqref{eq:ch7:delta-t-tmunu}.

We can now consider these expressions in two spatial regimes: at the horizon, where $f(r_g)=0$, and at spatial infinity, where $f=1$ and due to asymptotic flatness the scalar curvature vanishes, $R=0$. In the latter case the above contributions vanish. At the horizon, however, both $\widetilde{T}$ and $\widetilde{T}^r{}_r$ are non-zero, albeit well-behaved. This regularity at the horizon, as explained already above, implies
\begin{align}
\lim\limits_{r\rightarrow r_g} f \widetilde{T}^r_r = 0 \, .
\end{align}
This constitutes our proof that the presence of non-locality does not affect the flux of Hawking radiation at infinity. We will discuss this (and the previous discussions) in more detail in a concrete example in the following section.

\section{Example: two-dimensional dilaton black hole}
For the sake of concreteness we would like to demonstrate the developed tools and techniques in a ``real life'' example of a static black hole encountered in two-dimensional dilaton gravity whose gravitational action takes the form
\begin{align}
S[g{}_{\mu\nu}] = \frac12 \int \dd^2x \sqrt{-g}\, e^{-2\phi}[R+4(\nabla\phi)^2+4\lambda^2] \, .
\end{align}
In the above, $R$ is the scalar curvature, $\phi$ is the dilaton field, and $\lambda$ is a constant. The action arises in string theory \cite{Fradkin:1985ys,Callan:1985ia} and it admits black hole solutions \cite{Witten:1991,Mandal:1991,McGuigan:1991qp,Frolov:1992xx}. There are also solutions featuring additional matter fields in the framework of what has been called the CGHS model \cite{Callan:1992rs}; see also Refs.~\cite{Grumiller:2002nm,Fabbri:2005}.

\subsection{Black hole metric}
For our purposes we will study a black hole solution given by
\begin{align}
\label{eq:ch7:dilaton-metric-1}
\dd s^2 = f \dd t^2 + \frac{\dd r^2}{f} \, , \quad f = 1-\frac{M}{\lambda}e^{-2\lambda r} \, , \quad \phi = -\lambda r \, , \quad r_g = \frac{1}{2\lambda} \ln \left( \frac{M}{\lambda} \right) \, ,
\end{align}
where $M$ is the mass parameter of the black hole, and the horizon is located at $f(r_g) = 0$. Before delving into the detailed study of effective energy-momentum tensors let us reduce the number of physical parameters by singling out $\lambda$ as a physical scale. To that end, we introduce the new, dimensionless coordinates $\{\tau, x\}$ and define a ``physical'' metric $\dd\bar{s}^2$ such that
\begin{align}
\label{eq:ch7:dilaton-metric-2}
\tau = 2\lambda t \, , \quad x = 2\lambda(r-r_g) \, , \quad \dd\bar{s}^2 = \frac{1}{4\lambda^2} \, \dd s^2 \, , \quad \dd s^2= -f \, \dd\tau^2 + \frac{\dd x^2}{f} \, , \quad f = 1-e^{-x} \, .
\end{align}
In the following we will perform all calculations in the rescaled ``physical'' metric and restore the correct dimensionality whenever required. One example is the surface gravity: for the dimensionful metric it takes the value $\bar{\kappa} = \lambda$ whereas in the dimensionless case it is simply $\kappa=1/2$. As the dimensionless curvature scalar is of remarkably simple form,
\begin{align}
R = e^{-x} \, ,
\end{align}
it is convenient to introduce the Ricci curvature as a radius variable. Expressed in the coordinates $\{\tau, R\}$ the metric \eqref{eq:ch7:dilaton-metric-2} takes the form
\begin{align}
\label{eq:ch7:dilaton-metric-3}
\dd s^2 = -(1-R) \dd\tau^2 + \frac{\dd R^2}{R^2 (1-R)} \, .
\end{align}
The horizon is located at $R=1$ and spatial infinity corresponds to $R=0$. The static d'Alembert operator, expressed in these coordinates, takes the form
\begin{align}
\Box = R \, \partial_R \left [ (1-R) R\, \partial_R \right] \, .
\end{align}

\subsection{Analytical considerations}
We are interested in evaluating the non-local corrections to the trace anomaly as well as its impact on the black hole entropy, and for the simplest case of $\mathrm{GF_1}$ theory in both cases the following expression appears (we defined $s = \ell^2$ for convenience):
\begin{align}
\label{eq:ch7:f-def}
F(s,R) = e^{s\Box} R \, , \quad (\partial_s - \Box) F(s,R) = 0 \, , \quad F(0,R) = 1 \, .
\end{align}
Because the effective action can be written as a sum of the Polyakov contribution and an integral over non-local modification directly involving the function $F(s,R)$, see Eq.~\eqref{eq:ch7:polyakov-gf-decomposition}, the corrections to the trace anomaly and black hole entropy can also be expressed in terms of parametric $s$-integrals over this function $F(s,R)$. Finding this function is possible in terms of a spectral representation method. The calculations are a bit involved and for this reason we refer to appendix \ref{app:spectral}.

Using these methods, the function $F(s,R)$ can be represented as the following integral:
\begin{align}
\label{eq:ch7:f-integral}
F(s,R) = \int\limits_0^\infty \dd p \, \rho_p \, e^{-p^2 s} \Psi_p(R) ,
\end{align}
where we defined the following auxiliary quantities:
\begin{align}
\Psi_p(R) &= \sqrt{\frac{2}{p\tanh(\pi p)}} \Big[ \Re(c_p)\Im(Z_p(R)) - \Im(c_p)\Re(Z_p(R)) \Big] \, , \nonumber \\
Z_p(R) &= R^{ip} {}_2 F_1\left( ip, ip+1; 2ip+1; R \right) \, , \quad c_p = - \frac{4^{ip}\Gamma\left(ip+\tfrac12\right)}{\sqrt{\pi}\Gamma(ip)} \, , \quad p \in \mathbb{R} \, , \label{eq:ch7:f-sol} \\
\rho_p &= \frac{\sqrt{p\sinh(2\pi p)}}{\sqrt{2}\pi^{3/2}(1+p^2)} \Re \left[ \frac{i+p}{4^{ip}} \Gamma(ip)\Gamma\left(\tfrac12-ip\right) {}_3 F{}_2(1+ip,1+ip,ip;~ 2+ip, 1+2ip;~ 1 ) \right] \, . \nonumber
\end{align}
As we prove in appendix \eqref{app:spectral}, the mode functions $\Psi_p(R)$ are real-valued and orthonormal, correspond to plane waves at spatial infinity ($R=0$) and are finite at the horizon ($R=1$). The density measure $\rho_p$ is chosen such that $F(0,R)=R$. While we have been unable to evaluate the integral \eqref{eq:ch7:f-integral} analytically, it is quite well behaved due to the exponential factor $e^{-p^2s}$ and numerical evaluations for a wide range of values $s$ can be easily performed.

\subsection{Quasilocal approximation}
While the previous analytic results are promising we also developed a quasilocal approach that utilizes the series expansion of the exponential operator. This is useful because it allows us to systematically compare the two approaches for calculating $F(s,R)$. To that end we express $F(s,R)$ as follows:
\begin{align}
\begin{split}
F(s,R) &= e^{s\Box} R = \sum\limits_{n=0}^\infty \frac{s^n \Box^n}{n!} R \\
&\approx \sum\limits_{n=0}^N \frac{s^n \Box^n}{n!} = R + s \partial_r f \partial_r R + \frac12 s^2 (\partial_r f \partial_r)^2 R + \dots + \mathcal{O}\left( s^{N+1} \right) \, .
\end{split}
\end{align}
Note that the boundary condition $F(0,R) = R$ is trivially satisfied. It seems reasonable to expect that for small parameters $sR\ll 1$ the above approximation scheme converges sufficiently fast, and we will show that this is indeed correct. We dub this approximation method ``quasilocal'' because each order of the expansion contains more derivatives, becoming fully non-local in the limit $N\rightarrow\infty$.

\subsection{Non-local corrections to the trace anomaly}
With the an explicit representation of $F(s,R)$ available we can insert \eqref{eq:ch7:f-sol} into Eq.~\eqref{eq:ch7:trace-gf1} for the for the trace anomaly and we obtain
\begin{align}
T &= T{}_\text{Pol} + \Delta T \, , \quad
T{}_\text{Pol} = \frac{1}{24\pi} R \, , \\
\Delta T &= \frac{1}{24\pi} \left[ F(s,R) - R \right] + \frac{s}{48\pi}\int\limits_0^1\dd\xi F[(1-\xi)s,R] F[\xi s,R] ,
\end{align}
Since the integral \eqref{eq:ch7:f-integral} converges reliably, we are able to permute the integration procedures. First integrating over $\xi$ and \emph{then} performing the numerical integration over $p$ allows us to keep the number of numerical steps to 1.

Utilizing instead the quasilocal approximation we may write
\begin{align}
\begin{split}
T = \sum\limits_{n=0}^\infty \frac{s^n}{n!} T_n(R) \, , \quad
T_n(R) = \frac{1}{48\pi} \left[ 2\Box^n R + \sum\limits_{p=0}^{n-1} (\Box^p R)(\Box^{n-p-1}R) \right] \, .
\end{split}
\end{align}
The terms $T_n$ with $n\ge 1$ parametrize the quasilocal corrections to the trace anomaly, and for $n=0$ one recovers $T_0 = T_\text{Pol}(R) = R/(24\pi)$ which is the Polyakov result in the absence of non-locality.

\begin{figure*}[!hbt]
    \centering
    \subfloat{{ \includegraphics[width=0.8\textwidth]{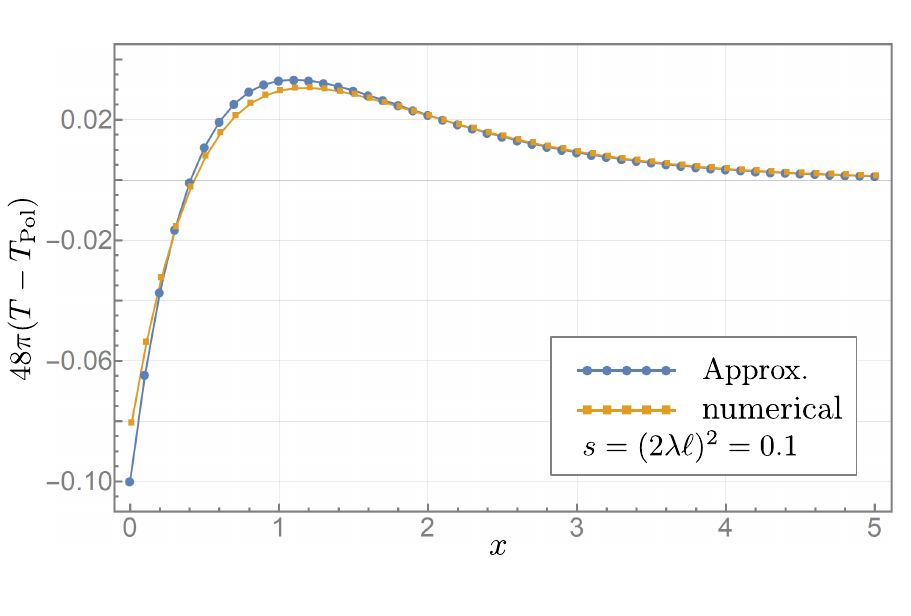} }} \\
    \subfloat{{ \includegraphics[width=0.8\textwidth]{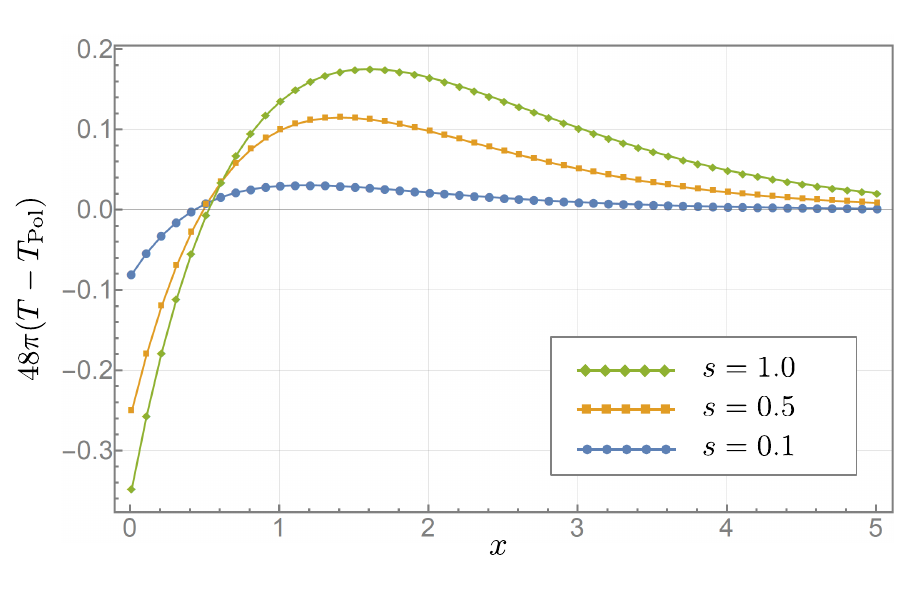} }}
    \caption[Non-local contributions to the trace anomaly.]{Top: We plot the non-local contributions to the trace anomaly as a function of the dimensionless distance $x$ for $s=(2\lambda \ell)^2 = 0.1$, where $1/(2\lambda)$ is a characteristic length scale of the background black hole, and $\ell>0$ denotes the scale of non-locality. The two different calculational methods are labelled as ``numerical'' and ``Approx.,'' and they agree well. (The approximation is performed only to linear level in $s$.) Bottom: Having gained some trust in the numerical methods, we plot the non-local contributions to the trace anomaly for another few values of non-locality $s$.}
    \label{fig:ch7:trace}
\end{figure*}

With these two methods available we may plot the non-local corrections $\Delta T = T - T_\text{Pol}$ for each method, see Fig.~\ref{fig:ch7:trace}. These results imply that the trace anomaly is reduced at the black hole horizon, and then picks up the majority of the modifications at $x\sim 1$, before rapidly decreasing to zero as $x\rightarrow\infty$. For small values of non-locality, $s=(2\lambda\ell)^2<1$, the quasilocal expressions agree remarkably well with our numerical method, which instils some confidence into the numerical methods. For values $s \gtrsim 1$ we need to rely solely on our numerical methods, and Fig.~\ref{fig:ch7:trace} suggests that while the contributions have the overall similar shape for larger values of non-locality their magnitude increases. Interestingly, there appears to be a critical distance $x_0 \sim \tfrac12$ where all non-local modifications of the trace anomaly vanish identically.

\subsection{Non-local corrections to black hole entropy}
Let us now move on to the modifications of the black hole entropy due to the presence of non-locality. The total entropy can be written as
\begin{align}
S_\text{GF} &= S_\text{Pol} + \Delta S \, ,
\end{align}
where $\Delta S$ captures the impact of non-locality, and $S_\text{Pol}$ is the standard Polyakov contribution which takes the form \cite{Myers:1994sg}
\begin{align}
S_\text{Pol} = -\frac16\phi = -\frac16 \lambda r_g = -\frac{1}{12}\ln\left(\frac{M}{\lambda}\right) \, .
\end{align}
Recall that $\phi$ is the classical dilaton field and has nothing to do with the auxiliary scalar fields employed earlier in the context of the local Polyakov action. Utilizing Eqs.~\eqref{eq:ch7:bh-entropy-gf1} and \eqref{eq:ch7:f-def} we can write the non-local entropy modification as
\begin{align}
\Delta S &= \frac{1}{12} \frac{e^{\ell^2\Box}-1}{\Box} R \Big|_{r=r_g} = \frac{1}{12} \int\limits_0^s \dd \tilde{s} F(\tilde{s}, 1) = \frac{1}{12} \int\limits_0^\infty \dd p \, \rho_p \Psi_p(1) \frac{1-e^{-sp^2}}{p^2} ,
\end{align}
where in the last step we performed the integration of $\tilde{s}$. Evaluating $F(s,R)$ at the horizon is well-defined since great care was taken to define the modes $\Psi_p(R)$ to be regular everywhere, see also the details presented in appendix \ref{app:spectral}. Note that in the limiting case $\ell\rightarrow 0$ we have $s = (2\lambda\ell)^2\rightarrow 0$ and hence $\Delta S = 0$, as it must.

\begin{figure}[!htb]%
    \centering
    \includegraphics[width=0.8\textwidth]{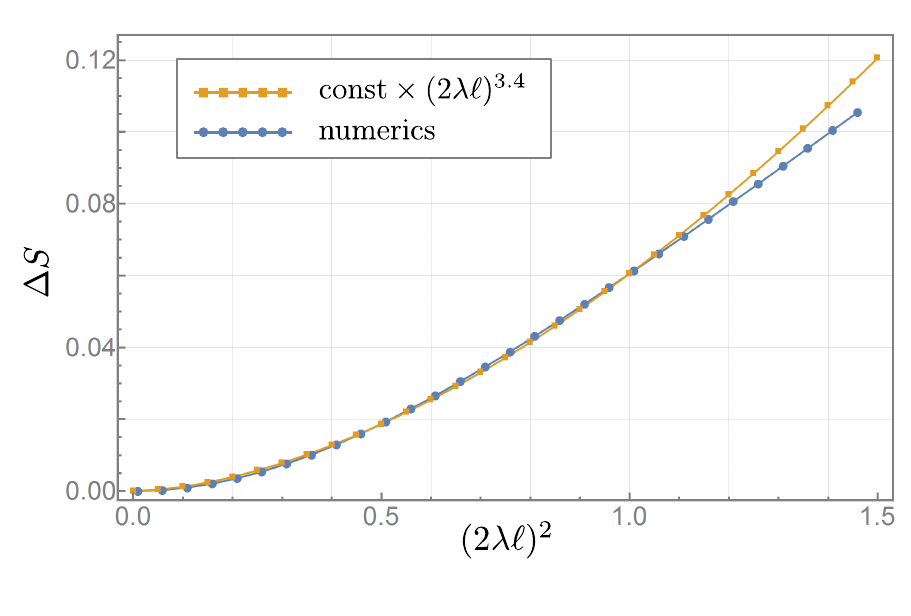}\\[-15pt]
    \caption[Non-local corrections to the black hole entropy.]{Non-local corrections to the black hole entropy, $\Delta S$, plotted as a function of the dimensionless non-locality $s = (2\lambda \ell)^2$. For small values numerical investigations indicate that it is possible to capture the essence in a power law.}
    \label{fig:ch7:deltaS}
\end{figure}

Due to the complicated dependence of the final expression on $p$ we are not able to find an analytic expression for the entropy corrections, but one may show that the integration still converges reliably. This is not obvious since the integrand is no longer proportional to $e^{-s p^2}$ but follows from the properties of both $\Psi_p$ and $\rho_p$ as functions of $p$. We plot the entropy corrections in Fig.~\eqref{fig:ch7:deltaS} as a function of dimensionless non-locality $(2\lambda\ell)^2$.

Our results indicate that the entropy corrections are increasing with non-locality $\ell$, and numerical investigations suggest that for smaller values of $\ell$ they can even be captured by a power law,
\begin{align}
\Delta S[\ell < (2\lambda)^{-1}] \sim \text{const} \times s^{3.4} \, .
\end{align}
For $\ell=0$ the corrections vanish, as they must. For larger values of non-locality the above power law no longer applies, but we were not able to find a closed form expression.

This change of entropy due to the presence of non-locality implies that there exists a non-trivial backreaction of non-locality onto the black hole geometry. As we have shown above, if these backreaction effects are ignored the flux of Hawking radiation at spatial infinity remains unchanged. These results seem to imply that if backreaction effects are taken into account the parameters of the black hole would change, and hence there would be an impact of non-locality on the asymptotic flux of Hawking radiation.

\section{Concluding remarks}
In this chapter we have analysed the impact of non-locality on quantum field theory in curved spacetime. In order to address this question we defined a non-local ghost-free modification of the well-known Polyakov action by replacing $\Box \rightarrow \Box \exp[(-\ell^2\Box)^N]$,
\begin{align}
W_\text{Pol}[g{}_{\mu\nu}] &= -\frac{1}{96\pi}\int\dd^2 x \sqrt{-g} R \frac{1}{\Box} R \, , \quad
W_\text{GF}[g{}_{\mu\nu}] = -\frac{1}{96\pi}\int\dd^2 x \sqrt{-g} R \frac{e^{-(-\ell^2\Box)^N}}{\Box} R \, .
\end{align}
While the former action can be derived from first principles, see Ref.~\cite{Duff:1993wm} and references therein, the latter is merely postulated. However, we were able to construct it by generalizing the local derivation of the Polyakov action with the aid of a non-minimally coupled auxiliary field $\varphi$. An important difference that we could not resolve lies in the fact that form factors $\exp(-\ell^2\Box)$ manifestly break conformal invariance, and for that reason it might be worthwhile to study whether it is possible to introduce non-locality in a conformally invariant fashion. It is well known that the d'Alembert operator transforms as a conformal density in two dimension \cite{Dabrowski:2008kx}, and for that reason it seems possible that another scalar field $\chi$ that itself transforms as a density of the opposite weight might ameliorate that fact via a dilaton-type interaction of the form $\exp(-\ell^2\Box\chi)$.

We have demonstrated that while the presence of non-locality, in the given framework, does not affect the flux of Hawking radiation at spatial infinity, it does affect the black hole entropy and the trace anomaly. For that reason it is conceivable that non-locality---once backreaction effects are taken into account--will affect black hole parameters, such as its mass, surface gravity, and Hawking temperature. We will leave these considerations for a future study.

%%%%%%%%%%%%%%%%%%%%%%%%%%%%%%%%%%%%%%%%%%%%%%%%%%%%%%%%%%%%%%%%%%%%%%%%%%%%%%%%%%%%%%%%%%%%%%%%%%%
%
% Chapter: Summary of results
%
\chapter{Conclusions}
\label{ch:ch8}

In this thesis we explored the effects of non-locality as mediated by infinite-derivative form factors $f(\Box)$ in both gravity and quantum theory and demonstrated its implications in several explicit examples. The cornerstone of our considerations is the notion of non-local Green functions whose properties we studied in some detail. The non-locality studied in this thesis is of a special kind: it preserves Lorentz invariance and is ghost-free. While it is convenient to maintain Lorentz invariance, the ghost-free criterion has rather far reaching consequences. Namely, it singles out those non-local theories whose propagator does not pick up new poles.

We have shown in this thesis that this very property has interesting consequences: first, since there are no new poles in the propagator, homogeneous solutions in the local and non-local theories coincide. And second, the property $f(0) = 1$ guarantees that at far distances away from the sources, one typically recovers the local solutions. Let us now briefly summarize our key findings:

\section{Summary of key results}
\begin{itemize}
\item Asymptotic causality is determined by analytical properties of local Green functions. Ghost-free non-local modifications decrease in timelike and spatial directions on characteristic scale $r/\ell\gg 1$, consistent with DeWitt's principle of asymptotic causality.
\item Static Green functions of non-local ghost-free theories asymptotically approach those of local theories, and are manifestly regular in the coincidence limit. They can be determined analytically in many cases.
\item In weak-field non-local gravity, static Green functions can be used to construct static and stationary solutions that asymptotically agree with General Relativity. But \emph{unlike} in General Relativity, the solutions are manifestly regular at the location of $\delta$-shaped matter sources.
\item Based on these stationary solutions one can generate the metric of ultrarelativistic objects in a suitable Penrose limit, which are regular at the location of the $\delta$-shaped matter sources.
\item The scattering coefficients of a non-local quantum field are susceptible to the presence of non-locality. In particular, there exists a critical frequency that is completely reflected.
\item The vacuum polarization and thermal fluctuations around a $\delta$-shaped potential in non-local quantum field theory are smoothed by the presence of non-locality.
\item The thermal fluctuations around the $\delta$-potential obey the fluctuation-dissipation theorem.
\item One can define a non-local ghost-free modification of the Polyakov effective action. The presence of non-locality enters the trace anomaly and the expectation value of the energy-momentum tensor components. In two dimensions this does not influence the asymptotic flux of Hawking radiation but can modify a black hole's entropy.
\end{itemize}

\section{Open problems}
There are several interesting open problems in the field of non-local ghost-free physics. Here we would like to mention a few of them:

\begin{itemize}
\item Since most of this thesis has been devoted to the study of linear non-local equations, which can be solved exactly via non-local Green functions, a natural step consists of advancing beyond the linear regime. This could perhaps be achieved perturbatively, employing our developed non-local Green functions.
\item Second, it would be worthwhile to construct a self-consistent non-local variational problem and thereby place the non-local action on a firmer footing.
\item Last, it would be very interesting to find more exact solutions of the full, non-linear field equations of ghost-free gravity. Then it would be possible to finally address the fate of gravitational singularities as well as the role of the event horizon, with possible ramifications in many areas of theoretical and mathematical physics.
\end{itemize}

The area of non-local ghost-free physics remains a fascinating field of study, with many unexpected results that challenge the way we think about space and time. We enjoyed our journey and eagerly anticipate new insights, waiting to be discovered at the non-local horizon.

%%%%%%%%%%%%%%%%%%%%%%%%%%%%%%%%%%%%%%%%%%%%%%%%%%%%%%%%%%%%%%%%%%%%%%%%%%%%%%%%%%%%%%%%%%%%%%%%%%%
%
% Appendix
%
%\part*{Appendix}
%\addcontentsline{toc}{part}{Appendix}

\appendix
\chapter{Calculational details}
\label{app:calc}

\section{Retarded Green function two dimensions}
\label{app:2d-retarded}
Here we would like to show
\begin{align}
G^\text{R}(t'-t,x'-x) &= \int\limits_{-\infty}^\infty \frac{\dd k}{2\pi} e^{ik(x'-x)} G^\text{R}_k(t'-t) = \frac12 \theta\left[ (t-t')^2 - (x-x')^2 \right] \quad \text{if~} t' > t \, , \\
G^\text{R}_k(t'-t) &= \frac{\sin k(t'-t)}{k} \, .
\end{align}
Abbreviating $t'-t=\Delta t>0$ and $x'-x=\Delta x$ we begin by rewriting the $k$-integral, 
\begin{align}
G^\text{R}(t'-t,x'-x) &= \int\limits_{-\infty}^\infty \frac{\dd k}{2\pi} e^{ik\Delta x} \frac{\sin k\Delta t}{k}
= \int\limits_{-\infty}^\infty \frac{\dd k}{4\pi i k} e^{ik\Delta x} \left( e^{ik\Delta t} - e^{-ik\Delta t} \right) \\
&= \int\limits_{-\infty}^\infty \frac{\dd k}{4\pi i k} \left( e^{ik(\Delta x + \Delta t)} - e^{ik(\Delta x - \Delta t)} \right) = \int\limits_{-\infty}^\infty \frac{\dd k}{4\pi i k} e^{ik(\Delta x + \Delta t)} - \int\limits_{-\infty}^\infty \frac{\dd k}{4\pi i k} e^{ik(\Delta x - \Delta t)} \, . \nonumber 
\end{align}
Defining now $v = \Delta x + \Delta t$ and $u = \Delta x - \Delta t$ we may rewrite this as
\begin{align}
G^\text{R}(t'-t,x'-x) &= \int\limits_{-\infty}^\infty \frac{\dd k}{4\pi i k} e^{ikv} - \int\limits_{-\infty}^\infty \frac{\dd k}{4\pi i k} e^{iku} \\
&= \frac12 \int\limits_{-\infty}^v \dd \tilde{v} \int\limits_{-\infty}^\infty \frac{\dd k}{2\pi} e^{ik\tilde{v}} - \frac12 \int\limits_{-\infty}^u \dd \tilde{u} \int\limits_{-\infty}^\infty \frac{\dd k}{2\pi} e^{ik\tilde{u}}  = \frac12 \int\limits_{-\infty}^v \dd \tilde{v} \delta(\tilde{v}) - \frac12 \int\limits_{-\infty}^u \dd \tilde{u} \delta(\tilde{u}) \\
&= \frac12 \left[ \theta(v) - \theta(u) \right] = \frac12\left[ \theta(\Delta x + \Delta t) - \theta(\Delta x - \Delta t) \right] \, .
\end{align}
The integration over $\tilde{u}$ and $\tilde{v}$ has to be performed with a suitable regulator $k \rightarrow k \pm i\alpha$ and $\alpha>0$, after which the limit $\alpha=0$ is to be taken. Since by assumption $\Delta t > 0$ we may use the identity
\begin{align}
\theta(\Delta x + \Delta t) + \theta(\Delta x - \Delta t) = \theta\left[ (\Delta t)^2 - (\Delta x)^2 \right] \, ,
\end{align}
which is non-zero only for timelike directions.

\section{Two-dimensional massive Green functions}
\label{app:2d-massive-green-functions}
We are interested in the temporal Fourier transform
\begin{align}
\label{eq:app:spatial-fourier}
G_\omega(x'-x) = \int\limits_{-\infty}^\infty \frac{\dd k}{2\pi} \, e^{ik(x'-x)} G_{\omega,k} \, ,
\end{align}
where $G{}_{\omega,k}$ denote the Fourier coefficients of the free Green function for a massive scalar field in two dimensions,
\begin{align}
G_{\omega,k} = \frac{-1}{\omega^2-k^2-m^2} = \frac{1}{k^2+m^2-\omega^2} \, .
\end{align}
For convenience let us define
\begin{align}
\varpi = \sqrt{\omega^2-m^2} \, , \quad \varkappa = \sqrt{m^2-\omega^2} \, .
\end{align}
It is our goal to evaluate the one-dimensional temporal Fourier transform \eqref{eq:app:spatial-fourier}. For $\omega^2-m^2<0$ one can integrate directly,
\begin{align}
G_\omega(x'-x) = \frac{1}{2\varkappa}e^{-\varkappa|x'-x|} \, .
\end{align}
Let us now focus on the case $\omega^2-m^2>0$. Then, the integrand has simple poles at $k_\pm = \pm\varpi$. Regularizing these poles via a suitable $i\epsilon$-prescription gives rise to the retarded, advanced, Feynman, and anti-Feynman Green functions. The prescriptions are as follows:
\begin{align}
G^\text{R}_{\omega,k} &= \frac{-1}{(\omega+i\epsilon)^2-k^2-m^2} \, , \\
G^\text{A}_{\omega,k} &= \frac{-1}{(\omega-i\epsilon)^2-k^2-m^2} \, , \\
G^\text{F}_{\omega,k} &= \frac{-1}{\omega^2-k^2-m^2+i\epsilon} \, , \\
G^{\overline{\text{F}}}_{\omega,k} &= \frac{-1}{\omega^2-k^2-m^2-i\epsilon} \, .
\end{align}
Due to the different pole structure for each regularization we need to evaluate Eq.~\eqref{eq:app:spatial-fourier} separately for each case. Moreover, since the integrand in Eq.~\eqref{eq:app:spatial-fourier} vanishes for $k\rightarrow\pm\infty$, depending on the sign of $x$, we can evaluate these integrals using contour integration. In what follows we will detail the required calculations; we visualize the chosen contours in Fig.~\ref{fig:app:g-fourier-2d-contours}.

\subsection{Inhomogeneous Green functions}

\subsubsection{Retarded Green function}
The poles of $G^\text{R}_{\omega,k}$ in the complex $k$-plane are located at
\begin{align}
k_\pm = \pm\sqrt{(\omega+i\epsilon)^2-m^2} \approx \pm \left[ \varpi + \frac{i\epsilon\omega}{\varpi} \right] \hat{=}~ \pm \left[ \varpi + i\epsilon\text{sgn}(\omega) \right] \, .
\end{align}
In the last step we have rescaled $\epsilon$ by $\varpi>0$. Note that the sign of $\omega$ appears, which is crucial for the following configurations. Let us now evaluate \eqref{eq:app:spatial-fourier} using contour integration. There are four different cases we need to treat separately:
\begin{itemize}
\item[(i)] $\omega>0$ and $x>0$\\[-1.4\baselineskip]
\item[(ii)] $\omega>0$ and $x<0$\\[-1.4\baselineskip]
\item[(iii)] $\omega<0$ and $x>0$\\[-1.4\baselineskip]
\item[(iv)] $\omega<0$ and $x<0$
\end{itemize}
In cases (i) and (iii) we may close the contour in the upper half-plane in a counter-clockwise curve $\mathscr{C}_+$, whereas in the cases (ii) and (iv) we need to close it in the lower half-plane in a clockwise curve $\mathscr{C}_-$. Note that for clockwise (mathematically negative) orientations the residue theorem of contour integration picks up an additional minus sign. Moreover, the sign of $\omega$ shifts the poles above or below the $\text{Re}\,k$-axis; see Fig.~\ref{fig:app:retarded-1} for $\omega>0$ and Fig.~\ref{fig:app:retarded-2} for $\omega<0$. We obtain
\begin{align}
G^\text{R}_\omega(x) &= \int\limits_{-\infty}^\infty \frac{\dd k}{2\pi} \frac{e^{ikx}}{(k-k_+)(k-k_-)} \\
\overset{(i)}{=}&~ (+2\pi i) \, \text{Res}\left[ \frac{e^{ikx}}{2\pi(k-k_+)(k-k_-)},~ k=k_+ \right] = +\frac{i}{2\varpi} e^{i\varpi x} \\
\overset{(ii)}{=}&~ (-2\pi i) \, \text{Res}\left[ \frac{e^{ikx}}{2\pi(k-k_+)(k-k_-)},~ k=k_- \right] = +\frac{i}{2\varpi} e^{-i\varpi x} \\
\overset{(iii)}{=}&~ (+2\pi i) \, \text{Res}\left[ \frac{e^{ikx}}{2\pi(k-k_+)(k-k_-)},~ k=k_- \right] = -\frac{i}{2\varpi} e^{-i\varpi x} \\
\overset{(iv)}{=}&~ (-2\pi i) \, \text{Res}\left[ \frac{e^{ikx}}{2\pi(k-k_+)(k-k_-)},~ k=k_+ \right] = -\frac{i}{2\varpi} e^{+i\varpi x} \, .
\end{align}
This may be summarized as
\begin{align}
G{}^\text{R}_\omega(x) = \frac{i\text{sgn}(\omega)}{2\varpi} \exp\big[ i\varpi\text{sgn}(\omega)|x| \big] \, .
\end{align}
In Sec.~\ref{sec:ch2:physical-interpretation} of the main body of this thesis we prove that this function is indeed the retarded Green function. In the massless limit $m\rightarrow 0$ one has $\varpi\rightarrow|\omega|$ such that
\begin{align}
\lim\limits_{m\rightarrow 0}G{}^\text{R}_\omega(x) = \frac{i}{2\omega} e^{i\omega|x|} \, .
\end{align}

\subsubsection{Advanced Green function}
The poles of $G^\text{A}_{\omega,k}$ in the complex $k$-plane are located at
\begin{align}
k_\pm = \pm\sqrt{(\omega-i\epsilon)^2-m^2} \approx \pm \left[ \varpi - \frac{i\epsilon\omega}{\varpi} \right] \hat{=}~ \pm \left[ \varpi - i\epsilon\text{sgn}(\omega) \right] \, .
\end{align}
Again, there are four different cases we need to treat separately:
\begin{itemize}
\item[(i)] $\omega>0$ and $x>0$\\[-1.4\baselineskip]
\item[(ii)] $\omega>0$ and $x<0$\\[-1.4\baselineskip]
\item[(iii)] $\omega<0$ and $x>0$\\[-1.4\baselineskip]
\item[(iv)] $\omega<0$ and $x<0$
\end{itemize}
See Fig.~\ref{fig:app:advanced-1} for $\omega>0$ and Fig.~\ref{fig:app:advanced-2} for $\omega<0$ for the choice of contours. We obtain
\begin{align}
G^\text{A}_\omega(x) &= \int\limits_{-\infty}^\infty \frac{\dd k}{2\pi} \frac{e^{ikx}}{(k-k_+)(k-k_-)} \\
\overset{(i)}{=}&~ (+2\pi i) \, \text{Res}\left[ \frac{e^{ikx}}{2\pi(k-k_+)(k-k_-)},~ k=k_- \right] = -\frac{i}{2\varpi} e^{-i\varpi x} \\
\overset{(ii)}{=}&~ (-2\pi i) \, \text{Res}\left[ \frac{e^{ikx}}{2\pi(k-k_+)(k-k_-)},~ k=k_+ \right] = -\frac{i}{2\varpi} e^{+i\varpi x} \\
\overset{(iii)}{=}&~ (+2\pi i) \, \text{Res}\left[ \frac{e^{ikx}}{2\pi(k-k_+)(k-k_-)},~ k=k_+ \right] = +\frac{i}{2\varpi} e^{+i\varpi x} \\
\overset{(iv)}{=}&~ (-2\pi i) \, \text{Res}\left[ \frac{e^{ikx}}{2\pi(k-k_+)(k-k_+)},~ k=k_- \right] = +\frac{i}{2\varpi} e^{-i\varpi x} \, .
\end{align}
This may be summarized as
\begin{align}
G{}^\text{A}_\omega(x) = -\frac{i\text{sgn}(\omega)}{2\varpi} \exp\big[ -i\varpi\text{sgn}(\omega)|x| \big] \, .
\end{align}
In the massless limit $m\rightarrow 0$ one has $\varpi\rightarrow|\omega|$ such that
\begin{align}
\lim\limits_{m\rightarrow 0}G{}^\text{A}_\omega(x) = -\frac{i}{2\omega} e^{-i\omega|x|} \, .
\end{align}

\subsubsection{Feynman Green function}
The poles of $G^\text{F}_{\omega,k}$ in the complex $k$-plane are located at
\begin{align}
k_\pm = \pm\sqrt{\varpi^2+i\epsilon} \approx \pm \left[ \varpi + \frac{i\epsilon}{2\varpi} \right] \hat{=}~ \pm \left[ \varpi + i\epsilon \right] \, ,
\end{align}
where in the last equality we rescaled $\epsilon$ by the positive factor $2\varpi$. Here, unlike in the retarded and advanced case, the sign of $\omega$ does \emph{not} enter. For this reason there are only two cases to treat:
\begin{itemize}
\item[(i)] $x>0$\\[-1.4\baselineskip]
\item[(ii)] $x<0$
\end{itemize}
See Fig.~\ref{fig:app:feynman} for the choice of contours. We obtain
\begin{align}
G^\text{F}_\omega(x) &= \int\limits_{-\infty}^\infty \frac{\dd k}{2\pi} \frac{e^{ikx}}{(k-k_+)(k-k_-)} \\
\overset{(i)}{=}&~ (+2\pi i) \, \text{Res}\left[ \frac{e^{ikx}}{2\pi(k-k_+)(k-k_-)},~ k=k_+ \right] = +\frac{i}{2\varpi} e^{+i\varpi x} \\
\overset{(ii)}{=}&~ (-2\pi i) \, \text{Res}\left[ \frac{e^{ikx}}{2\pi(k-k_+)(k-k_-)},~ k=k_- \right] = +\frac{i}{2\varpi} e^{-i\varpi x} \, .
\end{align}
This may be summarized as
\begin{align}
G{}^\text{F}_\omega(x) = \frac{i}{2\varpi} \exp\big[ i\varpi|x| \big] \, .
\end{align}
In the massless limit $m\rightarrow 0$ one has $\varpi\rightarrow|\omega|$ such that
\begin{align}
\lim\limits_{m\rightarrow 0}G{}^\text{F}_\omega(x) = \frac{i}{2|\omega|} e^{i|\omega x|} \, .
\end{align}

\subsubsection{Anti-Feynman Green function}
\begin{figure*}[!htb]%
    \centering
    \subfloat[Retarded Green function, $\omega>0$.]{{\includegraphics[width=0.4\textwidth]{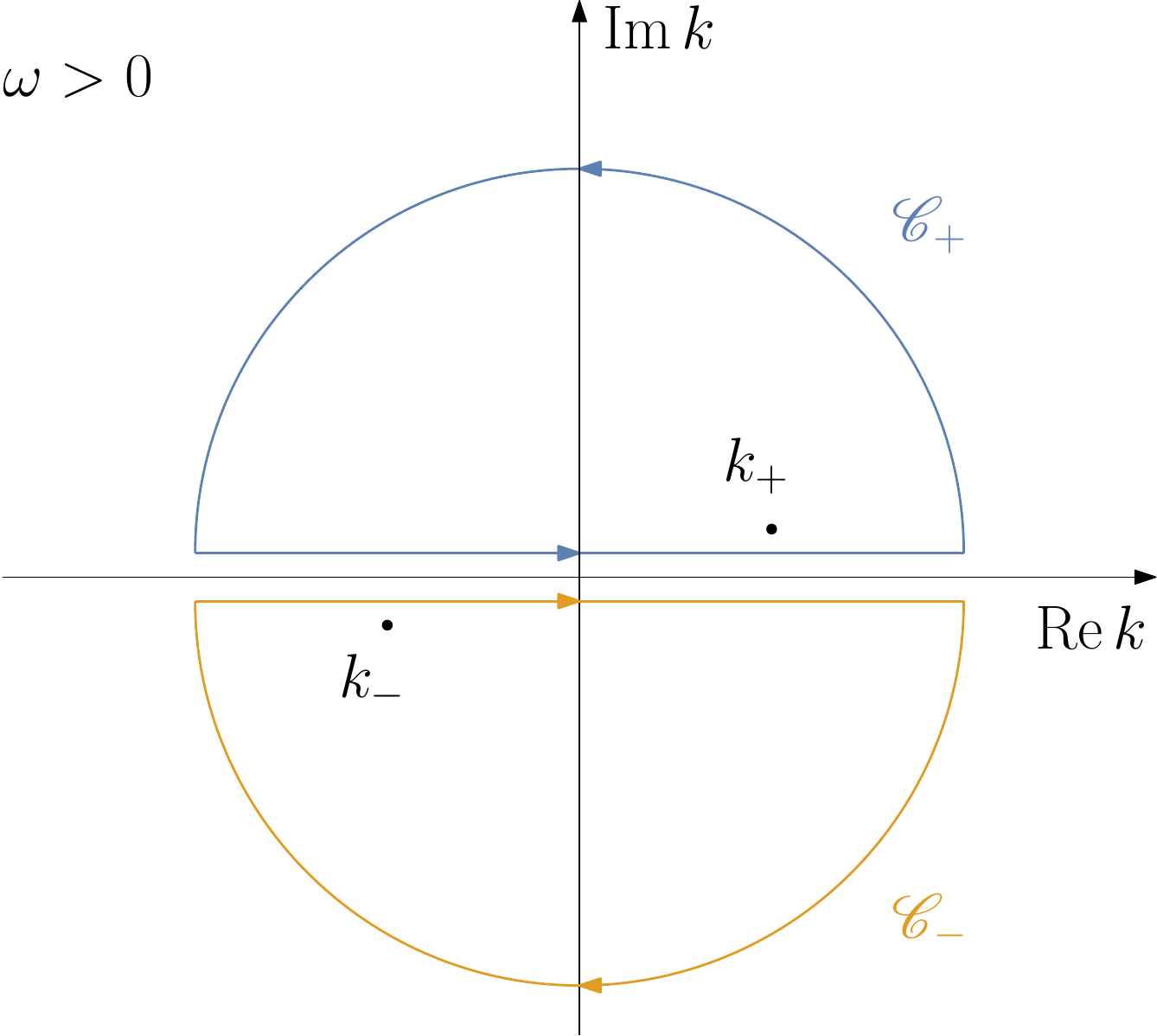} \label{fig:app:retarded-1} }}%
    \qquad
    \subfloat[Retarded Green function, $\omega<0$.]{{\includegraphics[width=0.4\textwidth]{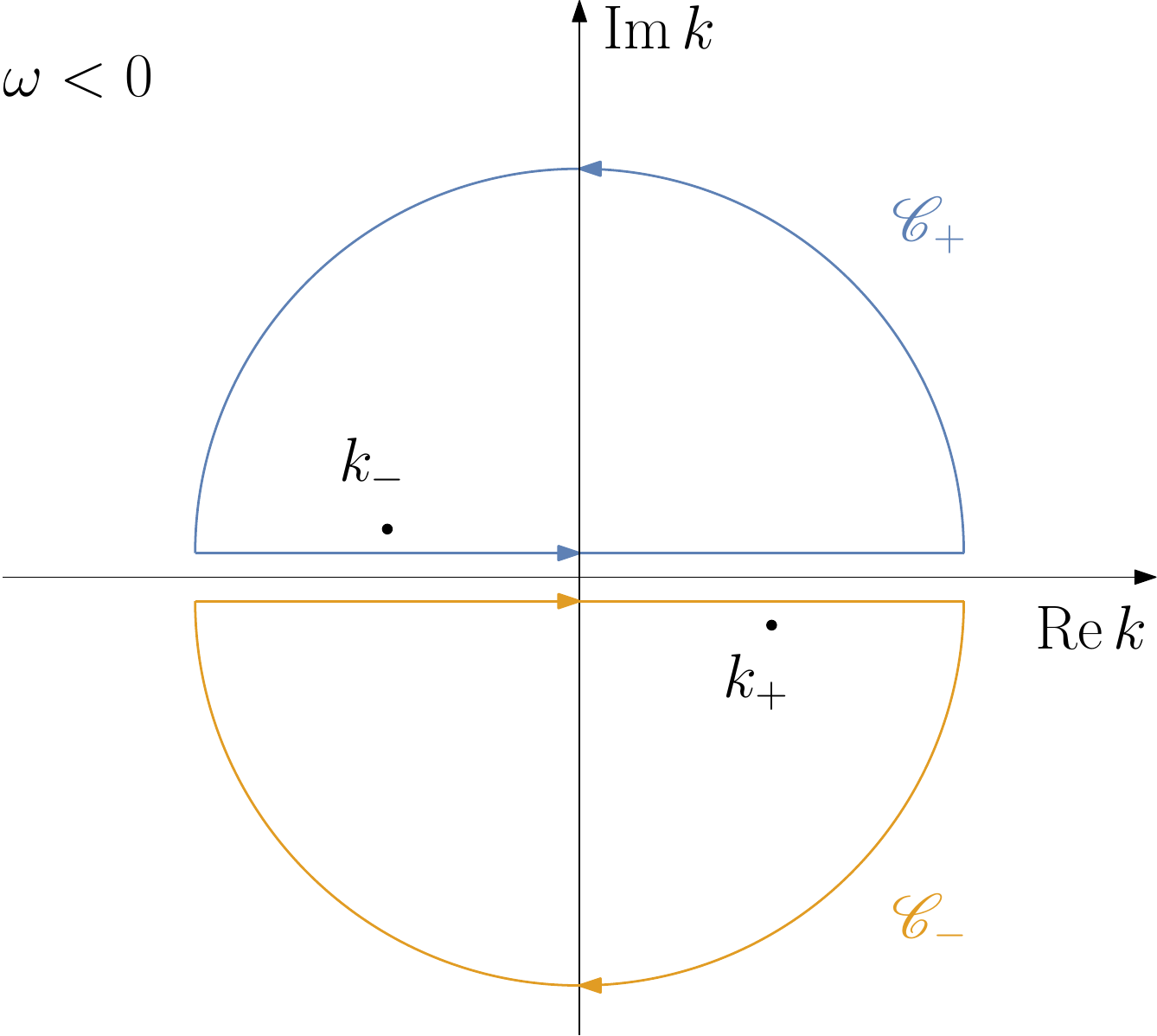} \label{fig:app:retarded-2} }}

    \subfloat[Advanced Green function, $\omega>0$.]{{\includegraphics[width=0.4\textwidth]{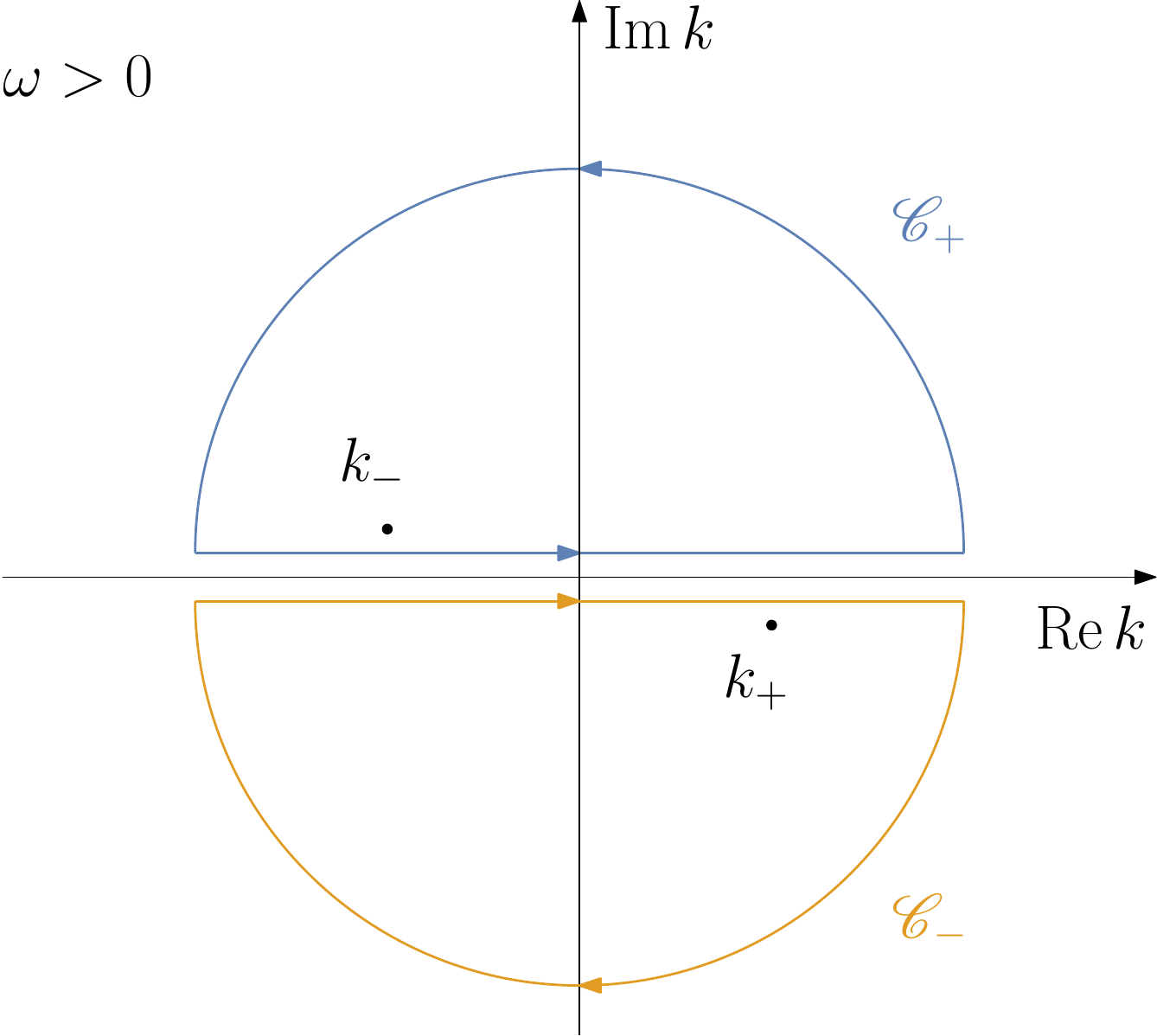} \label{fig:app:advanced-1} }}%
    \qquad
    \subfloat[Advanced Green function, $\omega<0$.]{{\includegraphics[width=0.4\textwidth]{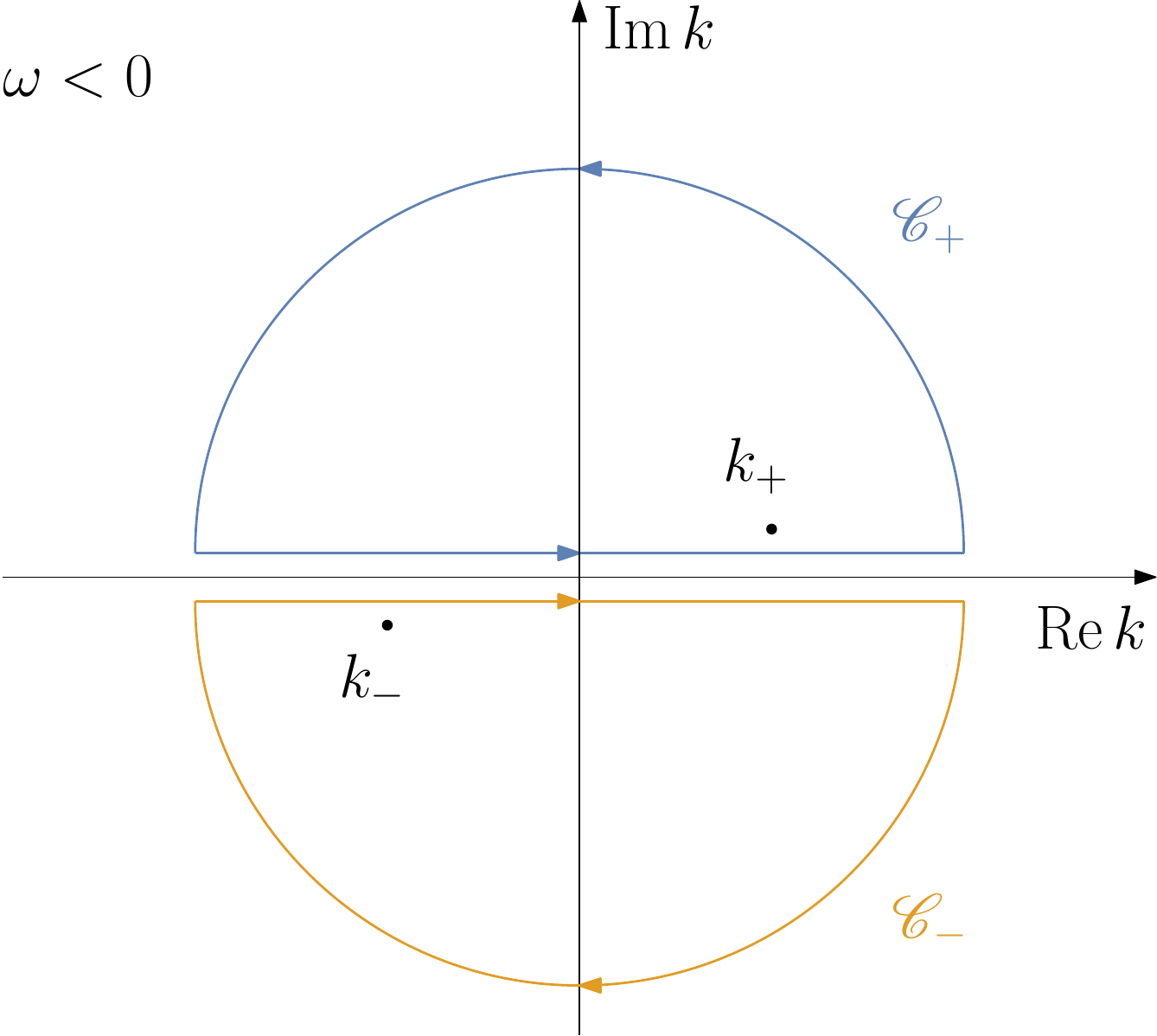} \label{fig:app:advanced-2} }}
    
    \subfloat[Feynman Green function.]{{\includegraphics[width=0.4\textwidth]{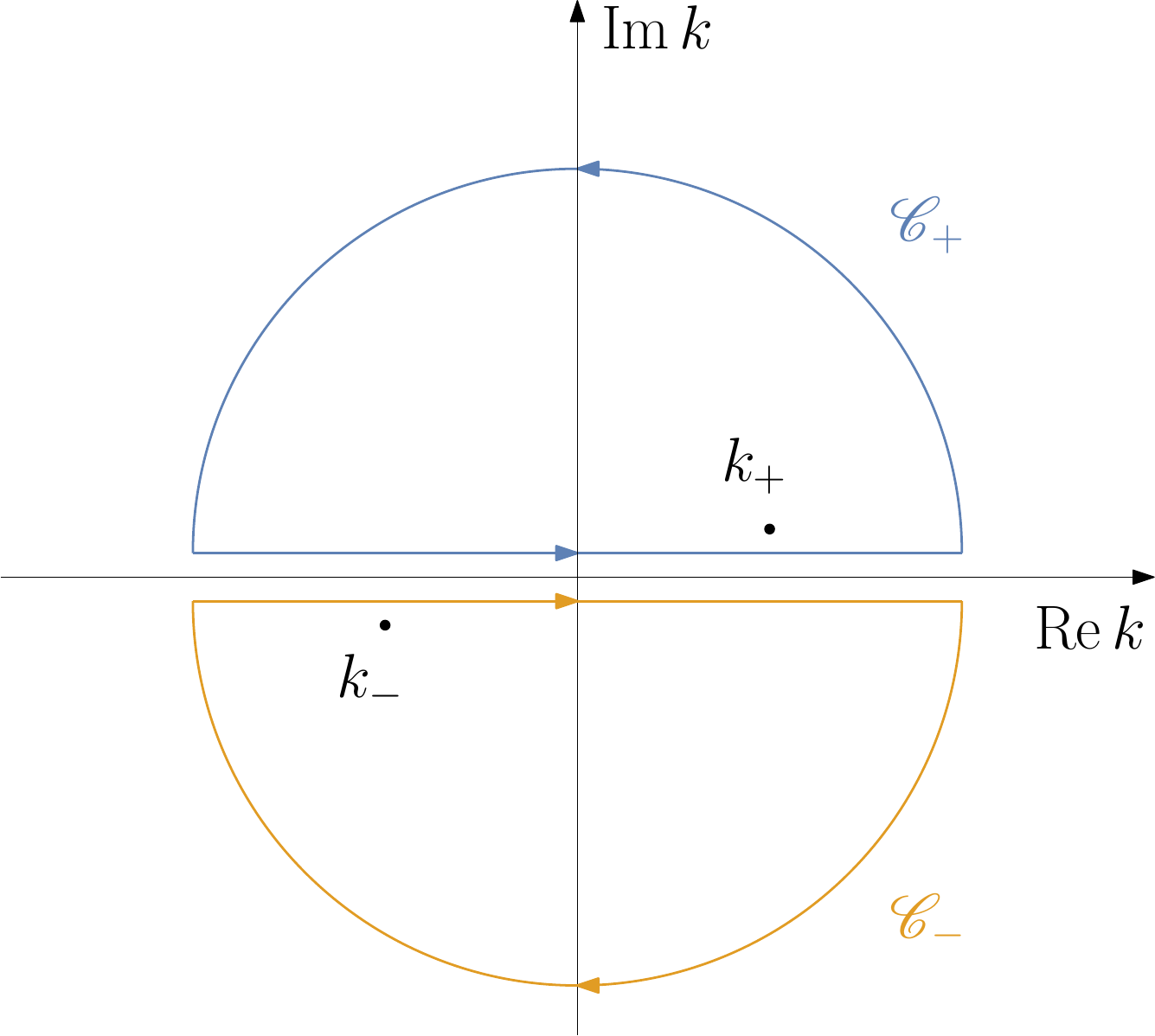} \label{fig:app:feynman} }}%
    \qquad
    \subfloat[Anti-Feynman Green function.]{{\includegraphics[width=0.4\textwidth]{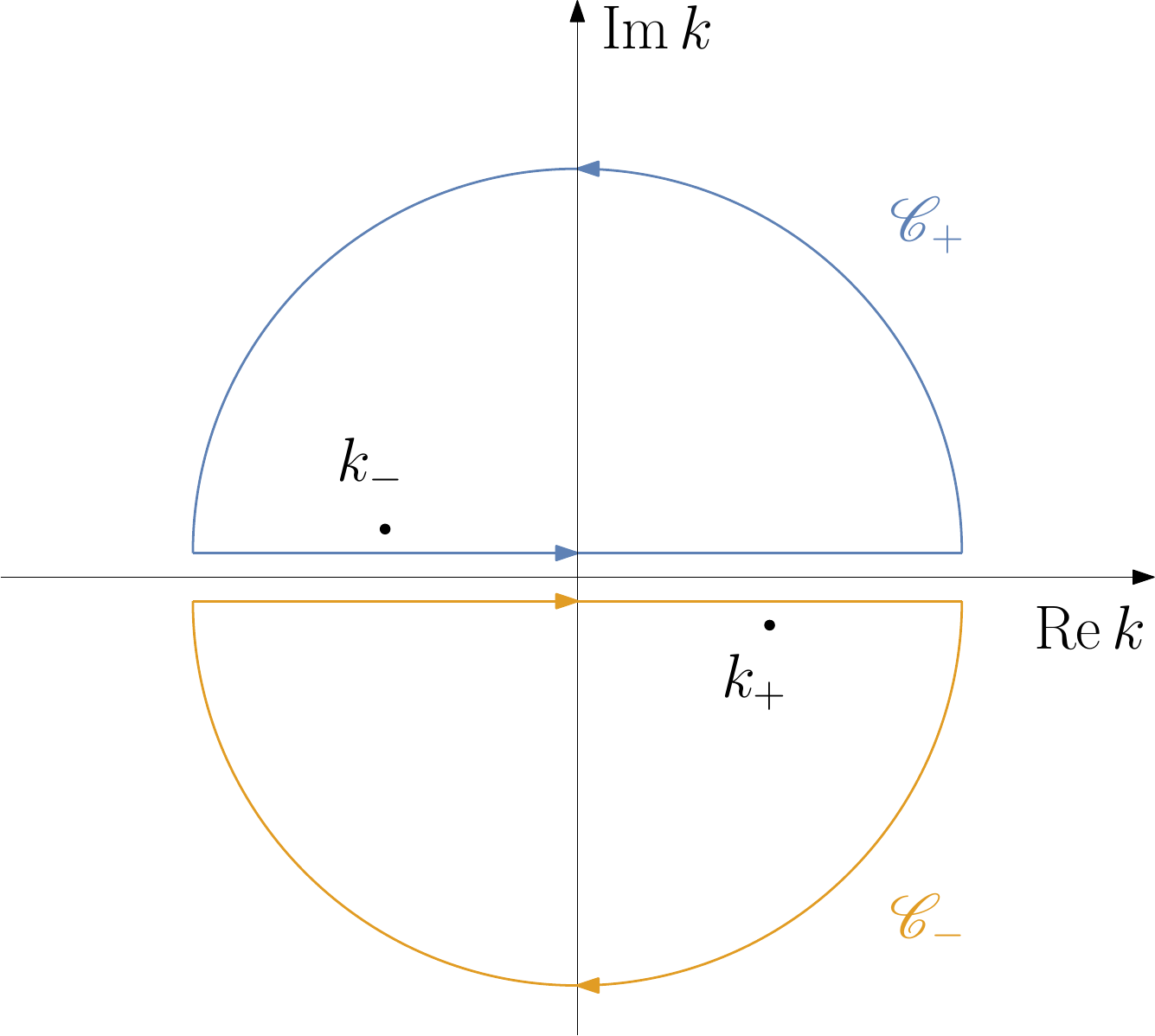} \label{fig:app:anti-feynman} }}

    \caption[Integration contours for local, causal Green functions.]{Integration contours for the temporal Fourier transform of the local, causal Green functions. Note that under a time reversal transformation $\omega\rightarrow-\omega$ the location of the poles are reflected along the real axis.}
    \label{fig:app:g-fourier-2d-contours}
\end{figure*}

The time-reverse version of the Feynman Green function is sometimes called anti-Feynman Green function, and we mention it here for completeness. The poles of $G^{\bar{\text{F}}}_{\omega,k}$ in the complex $k$-plane are located at
\begin{align}
k_\pm = \pm\sqrt{\varpi^2-i\epsilon} \approx \pm \left[ \varpi - \frac{i\epsilon}{2\varpi} \right] \hat{=}~ \pm \left[ \varpi - i\epsilon \right] \, ,
\end{align}
where again there are only two cases to treat:
\begin{itemize}
\item[(i)] $x>0$\\[-1.4\baselineskip]
\item[(ii)] $x<0$
\end{itemize}
See Fig.~\ref{fig:app:anti-feynman} for the choice of contours. We obtain
\begin{align}
G{}^{\bar{\text{F}}}_\omega(x) &= \int\limits_{-\infty}^\infty \frac{\dd k}{2\pi} \frac{e^{ikx}}{(k-k_+)(k-k_-)} \\
\overset{(i)}{=}&~ (+2\pi i) \, \text{Res}\left[ \frac{e^{ikx}}{2\pi(k-k_+)(k-k_-)},~ k=k_- \right] = -\frac{i}{2\varpi} e^{-i\varpi x} \\
\overset{(ii)}{=}&~ (-2\pi i) \, \text{Res}\left[ \frac{e^{ikx}}{2\pi(k-k_+)(k-k_-)},~ k=k_+ \right] = -\frac{i}{2\varpi} e^{+i\varpi x} \, .
\end{align}
This may be summarized as
\begin{align}
G{}^{\bar{\text{F}}}_\omega(x) = -\frac{i}{2\varpi} \exp\big[ -i\varpi|x| \big] \, .
\end{align}
In the massless limit $m\rightarrow 0$ one has $\varpi\rightarrow|\omega|$ such that
\begin{align}
\lim\limits_{m\rightarrow 0}G{}^{\bar{\text{F}}}_\omega(x) = -\frac{i}{2|\omega|} e^{-i|\omega x|} \, .
\end{align}

\subsection{Homogeneous Green functions}
\label{app:green-functions-homogeneous}
\begin{figure*}[!htb]%
    \centering
    \includegraphics[width=0.75\textwidth]{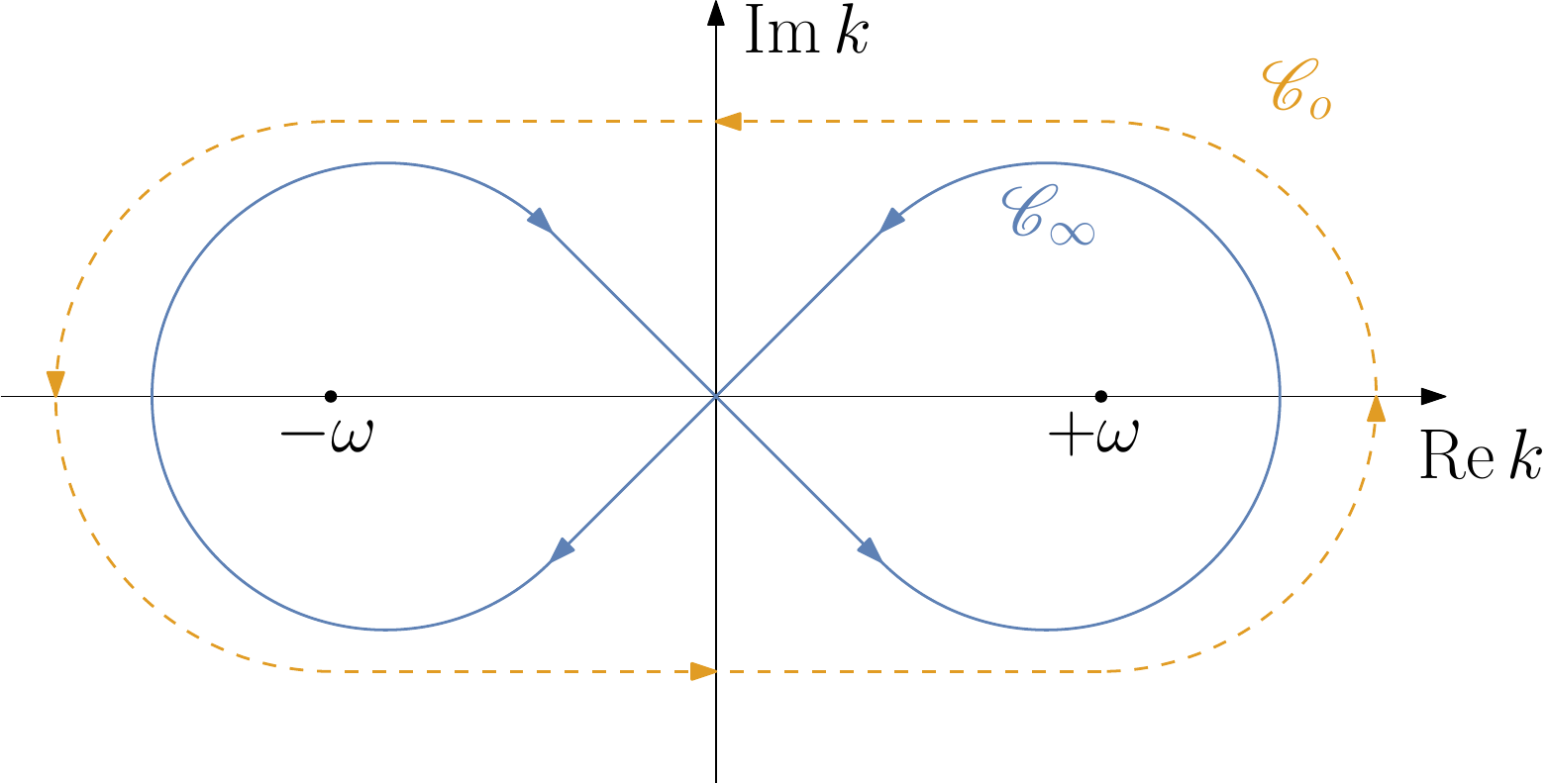} 
    \caption[Closed integration contours for homogeneous Green functions.]{Closed integration contours for the temporal Fourier transform of the homogeneous Green functions. While $\mathscr{C}_\infty$ gives rise to the Hadamard function, $\mathscr{C}_o$ gives rise to its complement, which is sometimes called Pauli--Jordan function.}
    \label{fig:app:homogenenous}
\end{figure*}

The homogeneous Green functions in Fourier space are solutions to the equation
\begin{align}
\left( \partial_x^2 + \omega^2 \right) G{}^\bullet_\omega(x) = 0 \, ,
\end{align}
where $\bullet=(1),+,-$ for the Hadamard and Wightman functions, respectively.\index{Green function!Hadamard}\index{Green function!Wightman} They are given by
\begin{align}
G^{(1)}_\omega(x) &= \frac{\cos\big(\varpi x\big)}{\varpi} \theta(\omega^2-m^2) \, , \\
G^+_\omega(x) &= \frac{\cos\big(\varpi x\big)}{\varpi} \theta(\omega-m) \, , \\
G^-_\omega(x) &= \frac{\cos\big(\varpi x\big)}{\varpi} \theta(-\omega-m) \, ,
\end{align}
where the Hadamard function can be conveniently written as a summation of the two positive and negative energy Wightman functions. Moreover, there is also another solution, called the Pauli--Jordan function\index{Green function!Pauli--Jordan} $\tilde{G}_\omega(x)$, which is the other linearly independent solution
\begin{align}
\tilde{G}_\omega(x) &= \frac{\sin\big(\varpi x\big)}{\varpi} \theta(\omega^2-m^2) \, .
\end{align}
We will not use this function in the present thesis and just mention it here for completeness. The homogeneous Green functions can also be derived from contour integration,
\begin{align}
G^{(1)}_\omega(x) = (-i) \oint\limits_{\mathscr{C}_\infty} \frac{\dd k}{2\pi} \frac{e^{i\omega x}}{k^2-\omega^2} \, , \quad
\tilde{G}_\omega(x) = - \oint\limits_{\mathscr{C}_o} \frac{\dd k}{2\pi} \frac{e^{i\omega x}}{k^2-\omega^2} \, .
\end{align}
Unlike the contours for the inhomogeneous functions, these contours are finite-sized. The functions automatically evaluate to zero for $\omega<m$ because then the integrand is a regular function. See Fig.~\ref{fig:app:homogenenous} for the integration contours, and see DeWitt \cite{DeWitt:1965} (in particular, pp.~29--32) for beautiful explanations and depictions of the contours. Caveat: DeWitt uses a different notation for the Green functions compared to the one employed in this thesis.

\section{Proof of Eq.~(\ref*{eq:ch6:master})}
\label{app:master-eq}
Let us prove the central relation between the temporal Fourier components of the Feynman Green function on the one side, and the retarded, advanced and Hadamard Green function on the other side,
\begin{align}
\tag{\ref*{eq:ch6:master}}
\ts{\mathcal{G}}{}^\ind{F}_\omega(x',x) = \frac12 \left( \ts{\mathcal{G}}{}^\ind{R}_\omega(x',x) + \ts{\mathcal{G}}{}^\ind{A}_\omega(x',x) + i \, \ts{\mathcal{G}}{}^\ind{(1)}_\omega(x',x) \right) \, .
\end{align}
Because $\ts{\mathcal{G}}{}^\ind{F}_\omega(x',x) = \ts{\mathcal{G}}{}^\ind{F}_{-\omega}(x',x)$ we can restrict the proof to the case of $\omega > 0$. This is just a reflection of the staticity of the interacting Green function under consideration, since $\omega\rightarrow-\omega$ corresponds to a time reversal transformation.

Let us begin by expressing each free Green function in terms of the local expression and its possible non-local modification (except for the Hadamard function, which is the same in the free case for both the local and non-local theory). We have
\begin{align}
\begin{split}
\label{eq:app:main}
\mathcal{G}{}^\ind{R}_\omega(x) &= \mathcal{G}{}^\ind{F}_\omega(x) = G{}^\ind{F}_\omega(x) + \Delta\mathcal{G}{}_\omega(x) \, , \\
\mathcal{G}{}^\ind{A}_\omega(x) &= \mathcal{G}{}^\ind{R}_{-\omega}(x) = \overline{G{}^\ind{F}_\omega}(x) + \Delta\mathcal{G}{}_\omega(x) \, , \\
\mathcal{G}{}^\ind{(1)}_\omega(x) &= G{}^{(1)}_\omega(x) = -i\left[ G{}^\ind{F}_\omega(x) - \overline{G{}^\ind{F}_\omega}(x) \right] \, .
\end{split}
\end{align}
Note that $\Delta\mathcal{G}{}_\omega(x) \in \mathbb{R}$ as well as $\Delta\mathcal{G}{}_{-\omega}(x) = \Delta\mathcal{G}{}_\omega(x)$. Let us also define (for $\omega>0$)
\begin{align}
\Lambda{}_\omega := \frac{\lambda}{1 + \lambda G^\ind{F}_\omega(0) + \lambda \Delta\mathcal{G}^\ind{F}_\omega(0)} \, ,
\qquad \Lambda_\omega^\ind{R} = \Lambda_\omega \, ,
\qquad \Lambda_{-\omega}^\ind{R} = \overline{\Lambda_\omega} \, ,
\end{align}
where the bar denotes complex conjugation. In a second step we may insert these relations into the interacting Green functions to obtain
\begin{align}
\label{eq:app:aux-1}
\ts{\mathcal{G}}{}^\ind{F}_\omega(x,x') &= G{}^\ind{F}_\omega(x-x') + \Delta \mathcal{G}{}_\omega(x-x') - \Lambda_\omega \left[G^\ind{F}_\omega(x) + \Delta\mathcal{G}_\omega(x)\right]\left[G^\ind{F}_\omega(x') + \Delta\mathcal{G}_\omega(x')\right] \, , \\
\label{eq:app:aux-2}
\ts{\mathcal{G}}{}^\ind{R}_\omega(x,x') &= G{}^\ind{F}_\omega(x-x') + \Delta \mathcal{G}{}_\omega(x-x') - \Lambda_\omega \left[G^\ind{F}_\omega(x) + \Delta\mathcal{G}_\omega(x)\right]\left[G^\ind{F}_\omega(x') + \Delta\mathcal{G}_\omega(x')\right] \, , \\
\label{eq:app:aux-3}
\ts{\mathcal{G}}{}^\ind{A}_\omega(x,x') &= \overline{G{}^\ind{F}_\omega}(x-x') + \Delta \mathcal{G}{}_\omega(x-x') - \overline{\Lambda_\omega}\left[\overline{G^\ind{F}_\omega}(x) + \Delta\mathcal{G}_\omega(x)\right]\left[\overline{G^\ind{F}_\omega}(x') + \Delta\mathcal{G}_\omega(x') \right] \, , \\
i\ts{\mathcal{G}}{}^\ind{(1)}_\omega(x,x') &= G{}^\ind{F}_\omega(x-x') - \overline{G{}^\ind{F}_\omega}(x-x') \nonumber \\
&\hspace{12pt} - \Lambda_\omega \left[G^\ind{F}_\omega(x) + \Delta\mathcal{G}_\omega(x)\right] \left[ G{}^\ind{F}_\omega(x') - \overline{G{}^\ind{F}_\omega}(x') \right] -\overline{\Lambda_\omega} \left[\overline{G^\ind{F}_\omega}(x') + \Delta\mathcal{G}_\omega(x')\right] \left[ G{}^\ind{F}_\omega(x) - \overline{G{}^\ind{F}_\omega}(x) \right]  \nonumber \\
\label{eq:app:aux-4}
&\hspace{12pt} + \Lambda_\omega \left[G^\ind{F}_\omega(x) + \Delta\mathcal{G}_\omega(x)\right] \overline{\Lambda_\omega} \left[\overline{G^\ind{F}_\omega}(x') + \Delta\mathcal{G}_\omega(x')\right] \left[ G{}^\ind{F}_\omega(0) - \overline{G{}^\ind{F}_\omega}(0) \right] \, .
\end{align}
In the last line we can recast the term proportional to $\Lambda_\omega\overline{\Lambda_\omega}$ as follows:
\begin{align}
\label{eq:app:lambda}
\Lambda_\omega\overline{\Lambda_\omega}\left[ G{}^\ind{F}_\omega(0) - \overline{G{}^\ind{F}_\omega}(0) \right] = \overline{\Lambda_\omega} - \Lambda_\omega \, .
\end{align}
Inserting these expressions into \eqref{eq:ch6:master} and comparing the terms independent of $\Lambda_\omega$ as well as linear terms in $\Lambda_\omega$ then yields the identity. In that step, Eq.~\eqref{eq:app:lambda} is crucial since it relates expressions quadratic in $\Lambda_\omega$ into a linear combination thereof.
Utilizing the fact that $\ts{\mathcal{G}}^\ind{R}_\omega(x,x') + \ts{\mathcal{G}}^\ind{A}_\omega(x,x')$ is real-valued, taking the imaginary part of \eqref{eq:ch6:master} yields
\begin{align}
\ts{\mathcal{G}}^\ind{(1)}_\omega(x,x') = 2\Im \left[ \ts{\mathcal{G}}{}^\ind{F}_\omega(x,x') \right] \, .
\end{align}
Using the expressions \eqref{eq:app:aux-1}--\eqref{eq:app:aux-4} this last equality can be verified explicitly.
Note that the above considerations hold true for any choice of form factor since the only assumptions made  lie in the properties of the local propagators as well as the real-valuedness of $\Delta\mathcal{G}_\omega(x,x')$.

\section{$\Delta\mathcal{G}_\omega$ in $\mathrm{GF_2}$ theory}
\label{app:deltaG-gf2}
This section is devoted to both analytical and numerical considerations of the non-local modification term $\Delta\mathcal{G}_\omega(x)$ in $\mathrm{GF_2}$ theory. It will be useful to consider the dimensionless non-local modification $g_\omega(x) = \varpi\Delta\mathcal{G}_\omega(x)$ given by the integral
\begin{align}
\label{eq:app:gf2-main-integral}
g_\omega(x) &= \int\limits_0^\infty \frac{\dd \xi}{\pi} \cos(\xi \tilde{x}) f_b(\xi) \, , \quad f_b(\xi) := \frac{1-e^{-b^2(1-\xi^2)^2}}{1-\xi^2} \, , \qquad \tilde{x} = \varpi x \, , \qquad b = (\varpi\ell)^2 \, .
\end{align}
For notational convenience in this section we use the dimensionless parameter $b$ and denote the integrand as  $f_b(\xi)$, whereas in the main body of the text we denoted it as $f_\omega(\xi)$. They are the same object.

\subsection{Analytical evaluation of $\Delta\mathcal{G}_\omega(0)$}
\label{app:deltaG-0-gf2}
The function $f_b(\xi)$ can be written as a double integral,
\begin{align}
f_b(\xi) = {1\over 2b\sqrt{\pi}}\int\limits_{-\infty}^{\infty}\dd y\, e^{-{y^2\over 4b^2}} ~{1-\cos[(1-\xi^2)y]\over 1-\xi^2}={1\over 2b\sqrt{\pi}}\int\limits_{-\infty}^{\infty}\dd y\, e^{-{y^2\over 4b^2}}\int\limits_0^y \dd z \sin[(1-\xi^2)z] \, .
\end{align}
Inserting this back into \eqref{eq:app:gf2-main-integral} gives
\begin{align}
\pi g_{\omega}(\tilde{x}) &= \int\limits_0^\infty \dd \xi\, \cos(\xi \tilde{x}) f_b(\xi)={1\over 2b\sqrt{\pi}}\int\limits_{-\infty}^{\infty}\dd y\, e^{-{y^2\over 4b^2}}\int\limits_0^y \dd z P(z,\tilde{x}) \, , \\
P(z,\tilde{x})&=\int\limits_0^\infty \dd\xi\,\cos(\xi \tilde{x})\sin[(1-\xi^2)z]=-{\sqrt{\pi}\over 2\sqrt{z}}\cos\left({\tilde{x}^2\over 4z}+z+{\pi\over 4}\right) \, .
\end{align}
At $\tilde{x}=0$ we can proceed further,
\begin{align}
\int\limits_0^y \dd z P(z,0)= -{\pi\over 2}\left[C\left(\sqrt{\tfrac{2y}{\pi}}\right)-S\left(\sqrt{\tfrac{2y}{\pi}}\right)\right] \, ,
\end{align}
where $C$ and $S$ are the Fresnel integrals \cite{Olver:2010}. Then the $y$-integral can be performed analytically as well and we obtain the final result,
\begin{align}
\pi g_\omega(0)={\sqrt{b}}
\left[{\sqrt{2}\pi b\over 6\Gamma\left(\tfrac34\right)}~{}_\ind{2}F_\ind{2}\left(\tfrac34,\tfrac54;\tfrac32,\tfrac74;-b^2\right)-\Gamma\left(\tfrac34\right)~{}_\ind{2}F_\ind{2}\left(\tfrac14,\tfrac34;\tfrac12,\tfrac54;-b^2\right)\right] \, .
\end{align}
The asymptotics are given by
\begin{align}
\label{app:eq:gf2-g0-asymptotics}
g_\omega(0) \sim \begin{cases}
\displaystyle -{\Gamma\left(\tfrac34\right)\over{\pi}} \sqrt{b} + \mathcal{O}\left(b^{3/2}\right) &\text{for}~ b \ll 1 \, , \\[15pt]
\displaystyle -{1\over 4\sqrt{\pi}\, b} + \mathcal{O}\left(b^{-3}\right) \quad &\text{for~} b \gg 1 \, .
\end{cases}
\end{align}
For a plot of this function we refer to Fig.~\ref{fig:app:plot-n}.

\subsection{Numerical evaluation of $\Delta\mathcal{G}_\omega(x)$}
\label{app:deltaG-x-gf2}
For $\tilde{x}\not=0$ an analytic evaluation, while desirable, is not possible to the best of our knowledge. Performing a direct numerical calculation, while in principle possible, is not the most ideal solution because the integrand is oscillatory and the shape of the function $f_b(\xi)$ changes drastically depending on the value of $b=(\varpi\ell)^2$. For this reason it makes sense to approximate the function $f_b(\xi)$ by an approximative function which can be integrated analytically, and then integrate the remainder numerically.

The function $f_b(\xi)$ has a local maximum at $\xi=0$ and a minimum at $\xi=\xi_+$. If $b$ is large enough, there is another local maximum at $\xi=\xi_-$. The locations are given by
\begin{align}
\xi_\pm = \sqrt{1 \pm \frac{2a_2}{b}} \approx 1 \pm \frac{a_2}{b} \, , \quad \, a_2 = \frac{1}{2\sqrt{2}} \sqrt{ -1 - 2W_{-1}\left( -\frac{1}{2\sqrt{e}} \right) } = 0.5604532115\dots \, ,
\end{align}
where $W_k(x)$ denotes the Lambert W function. The maximum $\xi_-$ only appears if $b>2a_2$, and for this reason it makes sense to develop two approximation schemes for $b<b_0$ and $b>b_0$ where we fix $b_0=3a_2$. Specifically, for $b<b_0$ we can approximate
\begin{align}
b < b_0 : \quad g &\approx -\frac{e^{-|\tilde{x}|}}{2}  + \int\limits_0^{\xi_\infty} \frac{\dd \xi}{\pi} \cos(\xi\tilde{x}) \left[ \frac{1-e^{-b^2(1-\xi^2)^2}}{1-\xi^2} + \frac{1}{1+\xi^2} \right] + E^<_{b_0, \xi_\infty}(\tilde{x}) \, , \\
E^<_{b_0, \xi_\infty}(\tilde{x}) &= \int\limits_{\xi_\infty}^\infty \frac{\dd \xi}{\pi} \cos(\xi\tilde{x}) \left[ \frac{1-e^{-b^2(1-\xi^2)^2}}{1-\xi^2} + \frac{1}{1+\xi^2} \right] \, ,
\end{align}
where $E^<_{b_0, \xi_\infty}(\tilde{x})$ denotes the error of this approximation. For $b>b_0$, on the other hand, we need to capture the maximum at $\xi_-$, and the following approximation works well:
\begin{align}
\label{eq:app:gf2-approx}
f_b(\xi) \approx f^\approx_b(\xi) = \begin{cases}
\displaystyle f_b^1(\xi) = \frac{1-c_1 e^{-b^2(1-\xi)}}{1-\xi^2} & \displaystyle \text{for} ~ \xi \le \xi_- \, , \\[15pt]
\displaystyle f_b^2(\xi) = m\xi + n & \displaystyle \text{for} ~ \xi_- < \xi < \xi_+ \, , \\[15pt]
\displaystyle f_b^3(\xi) = \frac{1-c_3 e^{-b^2(\xi-1)}}{1-\xi^2} & \displaystyle \text{for} ~ \xi > \xi_+ \, .
\end{cases}
\end{align}
The parameters $c_1$, $c_3$, $m$, and $n$ are chosen such that the jump at $\xi=\xi_\pm$ is of order $\mathcal{O}(b^{-2})$, and a suitable choice is
\begin{align}
c_1 &= \exp\left[ -4a_2^2 - b^2\left( 1 - \sqrt{1+\frac{2a_2}{b}} \right) \right] \, , \quad
c_3 = \exp\left[ -4a_2^2 + b^2\left( 1 - \sqrt{1-\frac{2a_2}{b}} \right) \right] \, , \\
m &= \frac12 \left( 1-e^{-4a_2^2} \right) \left( \frac12 - \frac{b^2}{a_2^2} \right) \, , \quad
n = \frac12 \left( 1-e^{-4a_2^2} \right) \left( \frac{b^2}{a_2^2} - 1 \right) \, .
\end{align}
The benefit of the approximation \eqref{eq:app:gf2-approx} is not that it captures the function $f_b(\xi)$ as precisely as possible. Rather, it allows for an analytical integration. For $\tilde{x}=0$ (which will be a useful test of our numerics, since we also have the analytic solution available) one finds for the indefinite integral
\begin{align}
\begin{split}
\int \frac{\dd \xi}{\pi} f^{1,3}_b(\xi) &= \frac{1}{2\pi}\ln\left(\frac{1+\xi}{1-\xi}\right) + \frac{c_{1,3}}{2\pi} \left\{ \text{Ei}[\mp b^2(1-\xi)] - e^{\mp 2b^2} \text{Ei}[\pm b^2(1+\xi)] \right\} \, ,
\end{split}
\end{align}
whereas for $\tilde{x}\not=0$ we find
\begin{align}
\begin{split}
\int \frac{\dd \xi}{\pi} \cos(\xi \tilde{x}) f^{1,3}_b(\xi) &= \frac{\cos{\tilde{x}}}{2\pi} \left\{ \text{Ci}[\tilde{x}(1+\xi)] - \text{Ci}[\tilde{x}(1-\xi)] \right\} + \frac{\sin{\tilde{x}}}{2\pi}\left\{ \text{Si}[\tilde{x}(1+\xi)] - \text{Si}[\tilde{x}(1-\xi)] \right\} \\
&\hspace{12pt}+\frac{c_{1,3}}{2\pi} \Re \left\{ e^{\pm i\tilde{x}}\text{Ei}[\mp(b^2+i\tilde{x})(1-\xi)] - e^{\mp 2b^2} e^{-i\tilde{x}} \text{Ei}[\pm(b^2+i\tilde{x})(1+\xi)] \right\} \, .
\end{split}
\end{align}
Here, $\text{Si}(x)$, $\text{Ci}(x)$, and $\text{Ei}(x)$ denote the sine integral, cosine integral, and exponential integral, respectively \cite{Olver:2010}:
\begin{align}
\text{Si}(x) &:= \int\limits_0^x \! \dd t \, \frac{\sin t}{t} \, , \quad \text{Ci}(x) := \gamma +\ln x + \int\limits_0^x \! \dd t \, \frac{\cos t-1}{t} \, , \\
\text{Ei}(x) &:= \gamma + \ln x + \int\limits_{-x}^0 \!\dd t \frac{1-e^{-t}}{t} \, .
\end{align}
We can collect these results and write the numerical integration as the sum of an analytical, approximative expression on the one hand, and a numerical contribution of the remainder on the other hand:
\begin{align}
b>b_0: \quad g &\approx \int\limits_0^\infty \frac{\dd \xi}{\pi} \cos(\xi\tilde{x}) f^\approx_b(\xi) + \int\limits_0^{\xi_\infty} \frac{\dd\xi}{\pi} \cos(\xi\tilde{x}) \left[ \frac{1-e^{-b^2(1-\xi^2)^2}}{1-\xi^2} - f^\approx_b(\xi) \right] + E^>_{b_0, \xi_\infty}(\tilde{x}) \, , \\
E^>_{b_0, \xi_\infty}(\tilde{x}) &= \int\limits_{\xi_\infty}^\infty \frac{\dd\xi}{\pi} \cos(\xi\tilde{x}) \left[ \frac{1-e^{-b^2(1-\xi^2)^2}}{1-\xi^2} - f^\approx_b(\xi) \right] \, ,
\end{align}
where again $E^>_{b_0, \xi_\infty}(\tilde{x})$ denotes the error of the approximation. See Fig.~\ref{fig:app:gf2-numerics-2} for a graphical visualization of the numerical integration scheme, which works quite well in practice. When performing the analytical calculations there is a small subtlety regarding branch cuts in the cosine integral $\text{Ci}(x)$ and the exponential integral $\text{Ei}(x)$, and for that reason we would like to give the full expressions for the definite integrals. They are rather lengthy and take the form ($\tilde{x}=0$)
\begin{align}
\begin{split}
\int\limits_0^\infty \frac{\dd \xi}{\pi} f_b^\approx(\xi) &= \int\limits_0^{\xi_-} \frac{\dd \xi}{\pi} f_b^1(\xi) + \int\limits_{\xi_-}^{\xi_+} \frac{\dd \xi}{\pi} f_b^2(\xi) + \int\limits_{\xi_+}^\infty \frac{\dd \xi}{\pi}  f_b^3(\xi) \\
&=\hspace{10pt}\frac{1}{\pi}\Big\{ \frac{m}{2}(\xi_+^2-\xi_-^2) + n(\xi_+-\xi_-) \Big\} -\frac{1}{2\pi}\left\{ \ln\left(\frac{\xi_+-1}{\xi_++1}\right) - \ln\left(\frac{1-\xi_-}{1+\xi_-}\right) \right\} \\
&\hspace{11pt}+\frac{c_1}{\pi}\Big\{ \text{Ei}\left[-b^2(1-\xi_-)\right] - e^{-2b^2}\text{Ei}\left[b^2(1+\xi_-)\right] - \text{Ei}\left(-b^2\right) + e^{-2b^2}\text{Ei}\left( b^2\right) \Big\} \\
&\hspace{11pt}-\frac{c_3}{\pi}\Big\{ \text{Ei}\left[b^2(1-\xi_+)\right] - e^{2b^2}\text{Ei}\left[-b^2(1+\xi_+)\right] \Big\} \, ,
\end{split}
\end{align}
as well as ($\tilde{x}\not=0$)
\begin{align}
\begin{split}
\int\limits_0^\infty \frac{\dd \xi}{\pi} \cos(\xi\tilde{x}) f_b^\approx(\xi) &= \int\limits_0^{\xi_-} \frac{\dd \xi}{\pi} \cos(\xi\tilde{x}) f_b^1(\xi) + \int\limits_{\xi_-}^{\xi_+} \frac{\dd \xi}{\pi} \cos(\xi\tilde{x}) f_b^2(\xi) + \int\limits_{\xi_+}^\infty \frac{\dd \xi}{\pi} \cos(\xi\tilde{x}) f_b^3(\xi) \\
&=\hspace{10pt} \frac{\cos\tilde{x}}{2\pi}\Big\{ \text{Ci}\left[\tilde{x}\left(1+\xi_-\right)\right] - \text{Ci}\left[\tilde{x}\left(1-\xi_-\right)\right] - \text{Ci}\left[\tilde{x}\left(1+\xi_+\right)\right] + \text{Ci}\left[\tilde{x}\left(\xi_+-1\right)\right] \Big\} \\
&\hspace{11pt}+\frac{\sin\tilde{x}}{2\pi}\Big\{ \pi + \text{Si}\left[\tilde{x}\left(1+\xi_-\right)\right]-\text{Si}\left[\tilde{x}\left(1-\xi_-\right)\right] - \text{Si}\left[\tilde{x}\left(1+\xi_+\right)\right] + \text{Si}\left[\tilde{x}\left(1-\xi_+\right)\right] \Big\} \\
&\hspace{11pt}+\frac{1}{\pi\tilde{x}^2}\Big\{ m\left[\cos\left(\tilde{x}\xi_+\right)-\cos\left(\tilde{x}\xi_-\right)\right] + (m\xi_+ + n)\tilde{x}\sin\left(\tilde{x}\xi_+\right) - (m\xi_- + n)\tilde{x}\sin\left(\tilde{x}\xi_-\right) \Big\} \\
&\hspace{11pt}+\frac{c_1}{2\pi}\Re\Big\{ e^{i\tilde{x}}\text{Ei}\left[(-b^2-i\tilde{x})(1-\xi_-)\right] - e^{-i\tilde{x}-2b^2}\text{Ei}\left[(b^2+i\tilde{x})(1+\xi_-)\right] \Big\} \\
&\hspace{11pt}-\frac{c_1}{2\pi}\Re\Big\{ e^{i\tilde{x}}\text{Ei}\left(-b^2-i\tilde{x}\right) - e^{-i\tilde{x}-2b^2}\text{Ei}\left(b^2+i\tilde{x}\right) \Big\} \\
&\hspace{11pt}-\frac{c_3}{2\pi}\Re\Big\{ e^{-i\tilde{x}}\widetilde{\text{Ei}}\left[(b^2+i\tilde{x})(1-\xi_+)\right] - e^{-i\tilde{x}+2b^2}\widetilde{\text{Ei}}\left[-(b^2+i\tilde{x})(1+\xi_+)\right] \Big\} \, ,
\end{split}
\end{align}
where we defined
\begin{align}
\widetilde{\text{Ei}}(z) := \begin{cases} \text{Ei}(z) \quad &\text{for~} \Re(z) \ge 0 \, , \\
\text{Ei}(z) + i\pi &\text{for~} \Re(z) < 0 \, , \end{cases}
\end{align}
which implements the branch cut of the exponential integral for arguments with negative real part.

\subsection{Asymptotics of $\Delta\mathcal{G}_\omega(x)$}
\label{app:deltaG-x-gf2-asymptotics}
\begin{figure}[!htb]

    \centering
    \subfloat{{\includegraphics[width=0.47\textwidth]{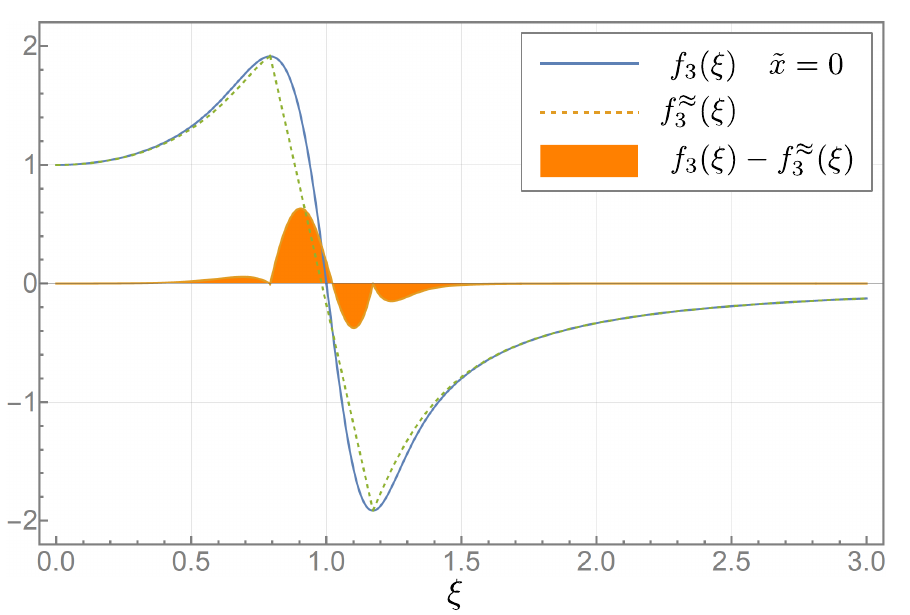} }}%
    \subfloat{{\includegraphics[width=0.47\textwidth]{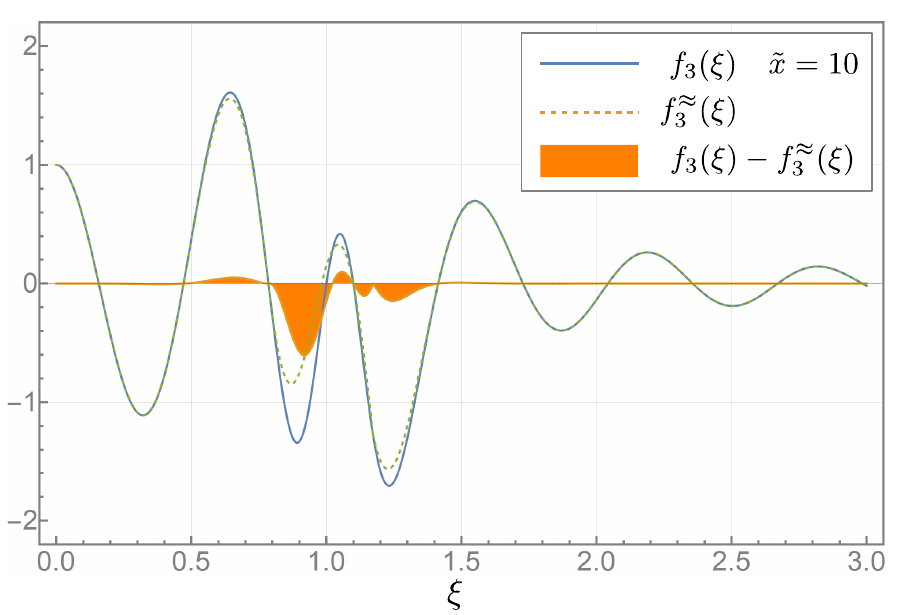} }}
    \caption[Numerical integration scheme for $\Delta\mathcal{G}_\omega(x)$ in $\mathrm{GF_2}$ theory.]{Numerical integration scheme: the subtraction of the approximative function $f^\approx_b(\xi)$ (dashed line) from the exact expression (solid line) improves the falloff behavior of the integrand drastically, enabling us to perform numerical integration over a finite range of $\xi$. The contributions to the numerical integrals are visualized as the shaded area under the curves, implying that the range of contributions to the numerical integral is finite for $\tilde{x}=0$ as well as for $\tilde{x}\not=0$.}
    \label{fig:app:gf2-numerics-2}

    \centering
    \subfloat{{\includegraphics[width=0.47\textwidth]{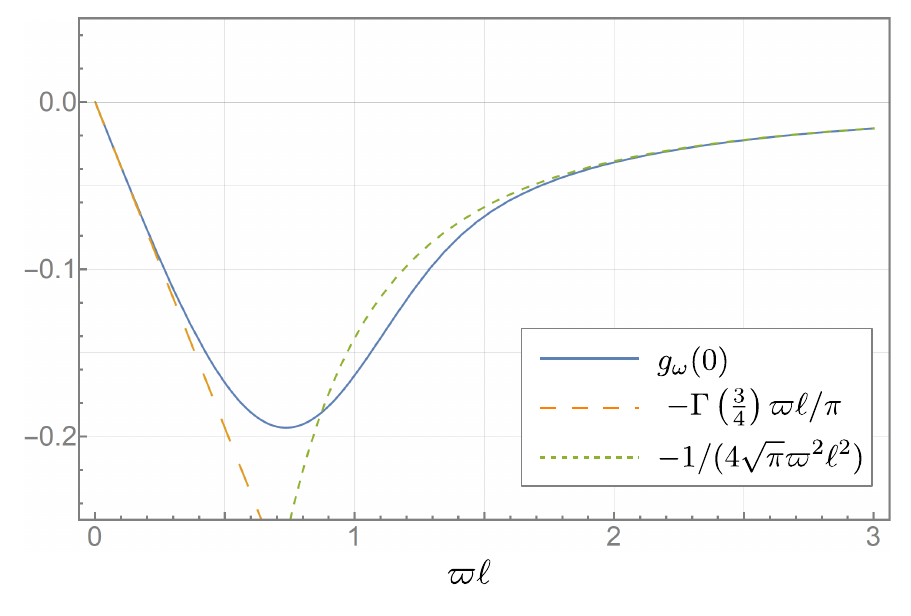} }}%
    \subfloat{{\includegraphics[width=0.47\textwidth]{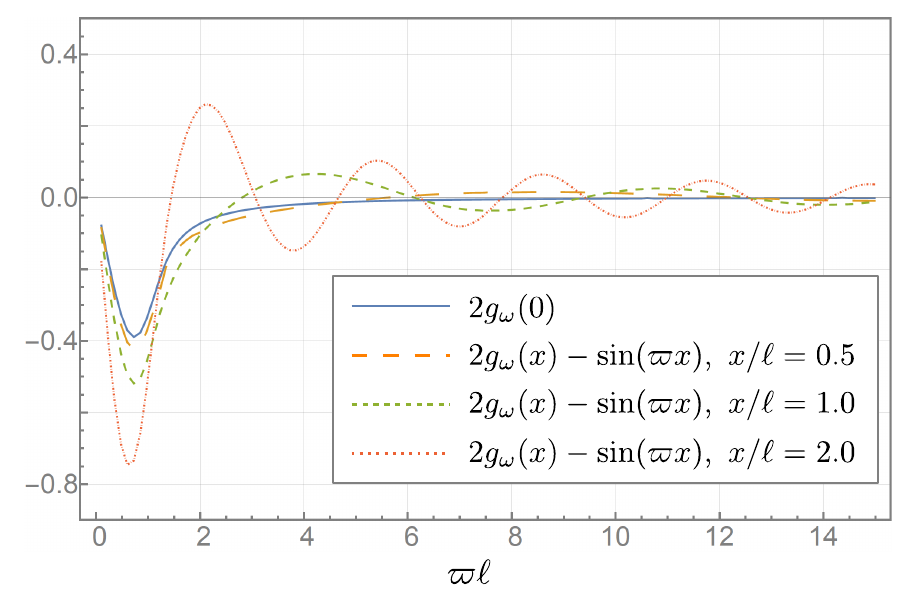} }}
    \caption[Non-local modification $g_\omega(x)$ for $\mathrm{GF_2}$ theory.]{The non-local modification term $g_\omega(x)$ of $\mathrm{GF_2}$ theory for various values of $x$. Left: Analytic result for $x=0$, including the asymptotics. Right: Numerical results for various values of the dimensionless distance $x/\ell$, where we subtracted the leading-order oscillating terms $\sin(\varpi x)$. The remainder is a decreasing function of $\varpi\ell$.}
    \label{fig:app:plot-n}

\end{figure}

The approximation \eqref{eq:app:gf2-approx} works well for for $b\gg 1$ such that we may use it to extract the large-$\varpi$ asymptotics of $g_\omega(x)$ using this method. For large $b$ and fixed $\tilde{x} = \varpi x$ we find
\begin{align}
g_\omega(x) &\approx \frac{\sin\tilde{x}}{2} - \frac{a}{2\pi b} \left( 3-e^{-4a_2^2} \right)\cos(\tilde{x}) - \frac{a}{3\pi b} \left( 2 + e^{-4a_2^2} \right) \tilde{x} \sin(\tilde{x}) + \mathcal{O}\left(b^{-2}\right) \, ,
\end{align}
which implies that for large $\varpi$ the function $g_\omega(x)$ behaves like an oscillatory term of amplitude $\tfrac12$. To obtain the above asymptotics we employed
\begin{align}
\begin{split}
\text{Si}(x\rightarrow 0) &\approx x \, , \quad
\text{Si}(x\pm\epsilon) \approx \text{Si}(x) \pm \frac{\sin x}{x}\epsilon \, , \quad
\text{Ci}(x\pm\epsilon) \approx \text{Ci}(x) + \frac{\cos x}{x}\epsilon \, , \\
\text{Si}(x\rightarrow\infty) &\approx \frac{\pi}{2} - \frac{\cos x}{x} \, , \quad
\text{Ei}(x\rightarrow\pm\infty) \approx \pm\frac{e^{\pm x}}{x} \, .
\end{split}
\end{align}
See Fig.~\ref{fig:app:plot-n} for a graphic confirmation of these asymptotics, which are rather accurate.

\subsection{Remarks on $\langle\varphi^2(x)\rangle_\text{ren}$ in $\mathrm{GF_{2n}}$ theories for larger $n$}
\label{app:gf2n}
It is possible to consider the large-$\varpi$ asymptotics of $g_\omega(0)$ for any $\mathrm{GF_{2n}}$ theory by using the approximation method of ``steepest descent.'' Let us elaborate and begin again with
\begin{align}
g_\omega(0) &= \int\limits_{-\infty}^\infty \frac{\dd \xi}{2\pi} f_\beta(\xi) = \frac{1}{2\pi} g(\beta) \, , \quad f_\beta(\xi) = \frac{1-e^{-\beta(1-\xi^2)^{2n}}}{1-\xi^2} \, , \quad \beta = (\varpi\ell)^{4n} \, .
\end{align}
Let us differentiate the integral with respect to $\beta$ to obtain the simpler expression
\begin{align}
\partial_\beta g(\beta) = \int\limits_{-\infty}^\infty \dd\xi (1-\xi^2)^{2n-1} e^{-\beta(1-\xi^2)^{2n}} \approx h_- + h_+ \, ,
\end{align}
which we approximated by the largest contributions around the maxima at $\xi_\pm = \pm 1$ which appear at large values of $\beta$. We introduce new variables $y_\pm := \xi \mp 1$ and find
\begin{align}
h_\pm &= \int\limits_{-\infty}^\infty \dd y_\pm (-y_\pm^2 \mp 2y_\pm)^{2n-1} \exp\left[ -\beta(-y_\pm^2 \mp 2y_\pm)^{2n} \right] \\
&= \int\limits_{-\infty}^\infty \dd z_\pm (1+z_\pm) (2z_\pm)^{2n-1} e^{-\beta(2z_\pm)^{2n}} \\
&\approx \int\limits_{-\infty}^\infty \dd z_\pm z_\pm (2z_\pm)^{2n-1} e^{-\beta(2z_\pm)^{2n}} = \frac{2^{2n-1}}{\alpha^{\frac{2n+1}{2n}}}\frac{\Gamma\left(\frac{2n+1}{2n}\right)}{n} \, , \quad \alpha := 2^{2n}\beta = [2(\varpi\ell)^2]^{2n} \, ,
\end{align}
where $z_\pm := \mp y_\pm - \frac12 y_\pm^2$. Now we may readily integrate over $\beta$ to arrive at the original expression,
\begin{align}
g_\omega(0) \approx \frac{1}{4\pi n} \Gamma\left(\frac{2n+1}{2n}\right)\left( c - \frac{2n}{\varpi^2\ell^2} \right) \, .
\end{align}
The constant $c$ is undetermined, but we can fix it by comparing the above results with the previous exact results for $\mathrm{GF_2}$ theory, which, in this case, corresponds to $n=1$. Inserting $n=1$ into the above one has in the limit $\varpi\ell \gg 1$
\begin{align}
g_\omega(0) \approx \frac{1}{4\pi}\Gamma\left(\frac32\right)\left(c - \frac{2}{\varpi^2\ell^2}\right) = \frac{1}{8\sqrt{\pi}}\left(c - \frac{2}{\varpi^2\ell^2}\right) \, .
\end{align}
Setting $c=0$ readily reproduces the asymptotics already encountered in \eqref{app:eq:gf2-g0-asymptotics}, which instils some hope into this approximate scheme for higher values of $n$.

\chapter{Two-dimensional ghost-free modification of the Polyakov action}
\label{app:2d}

\section{General relations for static geometries}
\label{app:2d-static}
We discuss here the geometric properties of an asymptotically flat two-dimensional metric $g{}_{\mu\nu}$ that admits a timelike Killing vector $\ts{\xi}$ such that $\xi{}_{(\mu;\nu)} = 0$. Denoting the negative norm of the Killing vector as $f = -\ts{\xi}\cdot\ts{\xi}$, and we assume that $\ts{\xi}$ is normalized such that $f=1$ at spatial infinity (which we assume to be flat). If $g{}_{\mu\nu}$ describes a black hole, then $f=0$ at the event horizon. In the following section we will focus on the region where $f\ge 0$. Let us now also define the 1-form $\xi = \xi{}_\alpha \dd x{}^\alpha$. Denoting the coderivative as $\delta=\star\dd\star$, the trace of the Killing equation for $\ts{\xi}$ can be written as $\delta\xi=0$ where ``$\star$'' denotes the Hodge dual and ``$\dd$'' is the exterior derivative\index{exterior derivative}. Then
\begin{align}
\dd\star\xi = 0 \quad \Rightarrow \quad \star\xi = \eta = \dd r \, ,
\end{align}
where we introduced a new scalar function $r$. Recall that in two dimensions the dual of a 1-form is again a 1-form. Observe that another two-dimensional identity is
\begin{align}
\xi \wedge \dd \xi = 0 \quad \Rightarrow \quad \xi = -\beta \dd t \, .
\end{align}
Here, $t$ and $\beta$ are scalar functions and we chose the negative sign for later convenience. The Killing vector $\ts{\xi}$ alone hence allows us to define the coordinates $\{t,r\}$ such that in these coordinates
\begin{align}
g{}^{tr} = g{}^{\alpha\beta} t{}_{,\alpha} r{}_{,\beta} = - \frac{f}{\beta^2} \, , \quad
g{}^{rr} = g{}^{\alpha\beta} r{}_{,\alpha} r{}_{,\beta} = (\nabla r)^2 = (\star\xi, \star\xi) = f \, .
\end{align}
Hence the metric can be written as
\begin{align}
\dd s^2 = -\frac{\beta^2}{f} \dd t^2 + \frac{\dd r^2}{f} \, .
\end{align}
We may now use the relation $\xi{}^\alpha\nabla{}_\alpha \ts{\xi}\cdot\ts{\xi} = 0$ to show that $f=f(r)$. The relation $\xi{}^{(t;r)}=0$ for the above metric gives the additional relation
\begin{align}
\beta'f - f'\beta = 0 \quad \Rightarrow \quad \beta = \beta_0(t) f \, ,
\end{align}
where the primes denote differentiation with respect to the radial coordinate $r$. By rescaling the time coordinate one may set $\beta_0(t) = 1$, whence the metric takes the final form
\begin{align}
\dd s^2 = -f \dd t^2 + \frac{\dd r^2}{f} = e^\sigma\left( -\dd t^2 + \dd r_\ast^2 \right) \, ,
\end{align}
where $\sigma = \tfrac12\ln f$ and $r_\ast$ denotes a tortoise coordinate.

We can use the above representation of the metric to find the zero modes\index{zero mode} of the d'Alembert operator. Note that there is the general identity
\begin{align}
\Box\sigma = -R{}_{\alpha\beta} \frac{\xi{}^\alpha\xi{}^\beta}{\ts{\xi}\cdot\ts{\xi}} \, .
\end{align}
Inserting the two-dimensional identity $R{}_{\mu\nu} = \tfrac12 R g{}_{\mu\nu}$ this yields
\begin{align}
\Box\sigma = -\frac12 R \, .
\end{align}
This implies that the equation $\Box\varphi = R$ is solved by
\begin{align}
\varphi = -2\sigma + \chi = -\ln f + \chi \, , \quad \Box\chi = 0 \, .
\end{align}
As it turns out, we can proceed to show that the functions $t$ and $r_\ast$ are both zero mode\index{zero mode} solutions:
\begin{align}
\Box t &= \delta (\dd t) = -\delta\left(\frac{\xi}{f}\right)=-\star \dd \star\left(\frac{\xi}{f}\right) =
- \star \left( \dd\frac{\star\xi}{f}\right) = -\frac{\delta\xi}{f} + \frac{1}{f^2}\star(\star\xi\wedge \dd f) = 0 \, , \\
\Box r_\ast &= \delta\left(\frac{\dd r}{f}\right) = \star \dd\left( \frac{\star \dd r}{f} \right) = \star \dd\left( \frac{\xi}{f} \right) = \star \dd^2 t = 0 \, ,
\end{align}
where we used $\delta\xi=0$ and that both $\star\xi$ and $\dd f$ are proportional to $\dd r$. Hence a general solution will be a superposition of $t$ and $r_\ast$ with constant coefficients. We discard the constant solution since it is irrelevant in the present context. Due to the linearity of the problem we may also consider the two functions
\begin{align}
u = t - r_\ast \, , \quad v = t + r_\ast \, ,
\end{align}
which are the retarded and advanced time coordinates. For an eternal black hole the retarded (or outgoing) coordinate $u$ is regular at the past horizon, and the advanced (or ingoing) coordinate $v$ is regular at the future horizon.

Let us conclude this section by deriving useful relations for the mass function and the surface gravity of static two-dimensional black holes, largely based on Ref.~\cite{Frolov:1992xx}. To that end, assume that there exists a conserved energy-momentum tensor $T{}^{\mu\nu}$. Defining the Killing current $J{}_\mu = T{}_{\mu\alpha}\xi^\alpha$ we can introduce the 1-form $J = J{}_\alpha \dd x{}^\alpha$, and one has
\begin{align}
\dd \star J = \star\delta J = 0 \, .
\end{align}
This implies that $\star J$ is closed and hence locally exact,
\begin{align}
\dd m = -\star J \, .
\end{align}
Here, $m$ is the mass function, and for a stationary energy-momentum tensor one has
\begin{align}
m = -\int\dd r T{}^t{}_t \, .
\end{align}
This expression is useful for calculating the influence of an energy density onto the black hole mass that is sourced by a test field propagating on the black hole background. One can also prove the following relation for the black hole's surface gravity \cite{Frolov:1992xx},
\begin{align}
\kappa = \frac12 \int\limits_\Sigma R \xi{}^\alpha \dd \Sigma{}_\alpha \, ,
\end{align}
where $\Sigma$ is a line between the horizon and spatial infinity, and $\dd\Sigma{}_\mu$ is the surface element of that line. In the $\{t,r\}$ coordinates we can rewrite this as
\begin{align}
\kappa = \frac12 \int\limits_{r_g}^\infty \dd r R = \frac12 f\big|_{r=r_g} \, .
\end{align}

\section{Energy-momentum tensor}
\label{app:state}
The zero mode solutions described in the previous section play an important role in the tensorial structure of the effective energy-momentum tensor. In this section we will demonstrate what choices for $\chi$ correspond to which quantum state. For convenience we define
\begin{align}
t{}^\mu{}_\nu = \frac{1}{b} T{}^\mu{}_\nu \, ,
\end{align}
where $b$ is the model-dependent prefactor of the conformal anomaly.

\subsection{Boulware vacuum}
Let us set $\varphi=-\ln f$, which corresponds to $\chi = 0$. The calculations give
\begin{align}
t{}^\mu{}_\nu = \mbox{diag} \left(\frac{f'^2}{2 f}-2f'',-\frac{f'^2}{2 f}\right) \, , \quad t{}^\alpha{}_\alpha = -2f'' = 2R \, .
\end{align}
This vanishes at $\mathscr{I}^+$ and $\mathscr{I}^-$ and is singular at both the future and past horizon, and it therefore reproduces the quantum average of the energy-momentum tensor in the Boulware vacuum state.

\subsection{Hartle--Hawking vacuum}
Let us set $\varphi=-\ln f+k r_\ast$, which corresponds to $\chi = k r_\ast$. Then one has
\begin{align}
t{}^\mu{}_\nu =\mbox{diag}\left(\frac{f'^2-k^2}{2 f}-2f'',-\frac{f'^2-k^2}{2 f}\right) \, .
\end{align}
For a general value of $k$ this energy-momentum tensor diverges at the horizons where $f=0$. However, one may choose $k$ to be related to the surface gravity $\kappa$ to render it finite, $k=f'|_{r_g}=2\kappa$. Then at infinity
\begin{align}
t{}^\mu{}_\nu \sim \mbox{diag} (-2\kappa^2 ,2\kappa^2) \, .
\end{align}
The corresponding state in this case is the Hartle--Hawking vacuum.

\subsection{Unruh vacuum}
Putting $\varphi=-\ln f+\kappa u$, which corresponds to $\chi = \kappa u$, in $\{t,r\}$ coordinates one has
\begin{align}
t{}_\mu{}^\nu = \begin{pmatrix} -2f''+\frac{f'-2\kappa^2}{2f} & -\kappa^2 \\
\frac{\kappa^2}{f^2} & -\frac{f'-2\kappa^2}{2f} \end{pmatrix} \, .
\end{align}
Let us denote by $u_{\mu}=(-f,1)$ a null covector which is regular at infinity. Then, at large distances $r$, one can show $t{}_\mu{}^\nu \sim \kappa^2 u_\mu u^\nu$. For this reason the corresponding energy-momentum tensor describes an outgoing ($u_r>0$) flux of null radiation at $\mathscr{I}^+$.

Finally, let us demonstrate that this energy-momentum tensor is regular at the future horizon, and in this context it is useful to work with an ingoing null coordinate $v = t + r_\star$ that is regular at the future horizon. One may calculate in $\{v, r\}$-coordinates
\begin{align}
t{}_{\mu\nu} = \begin{pmatrix} 2f f''-\frac12 f'^2+\kappa^2 & -4f''+\frac{f'^2-4\kappa^2}{2f} \\ -4f''+\frac{f'^2-4\kappa^2}{2f} & 2\frac{f''}{f}-\frac{f'^2-4\kappa^2}{f^2} \end{pmatrix} \, .
\end{align}
Near the horizon, $r=r_g$, one may approximate
\begin{align}
f(r) = 2\kappa (r-r_g) + \frac12 f_2 (r-r_g)^2 + \frac16 f_3 (r-r_g)^3 + \mathcal{O}[(r-r_g)^4] \, .
\end{align}
The energy-momentum tensor therefore behaves as
\begin{align}
t{}_{\mu\nu} \sim \begin{pmatrix} -\kappa^2 & -f_2 \\ -f_2 & \frac{f_3}{\kappa} \end{pmatrix} + \mathcal{O}(r-r_g) \, ,
\end{align}
which is manifestly regular, and for this reason this choice of $\chi$ leads to the correct boundary conditions consistent with the Unruh vacuum state. One may also show that the negative energy flux through the horizon, $t_{vv}|_{r_g}=-\kappa^2$ coincides with the negative of the outgoing energy flux at $\mathscr{I}^+$, $t_{uu}|_{\mathscr{I}^+}=\kappa^2$.

\section{Spectral representation of $F(s,R)$}
\label{app:spectral}
This section is devoted to finding an integral representation for the operator
\begin{align}
\tag{\ref*{eq:ch7:f-def}}
F(s,R) = e^{s\Box} R \, , \quad (\partial_s - \Box) F(s,R) = 0 \, , \quad F(0,R) = 1 \, .
\end{align}
In the coordinates employed in this context the d'Alembert operator takes the form
\begin{align}
\Box = R \, \partial_R \left[ (1-R) R \, \partial_R \right] \, .
\end{align}
As a first step, let us consider the following eigenvalue problem:
\begin{align}
\label{eq:app7:eq1}
\Box \Psi(R) = \lambda \Psi(R) \, ,
\end{align}
where $\Psi$ is real-valued and finite at the horizon ($R=1$) and at spatial infinity ($R=0$), where $R = e^{-x}$ in terms of the dimensionless distance coordinate $x$. Inserting $R=1$ and $R=0$ into \eqref{eq:app7:eq1} it is straightforward to show that 
\begin{align}
\Psi(R\rightarrow 1) \sim a_{-1}\ln(1-R) + a_0 \, , \quad \Psi(R\rightarrow 0)\sim a_+ R^{\sqrt{\lambda}} + a_- R^{-\sqrt{\lambda}} \, . \label{eq:app7:eq2}
\end{align}
We can now prove that only $\lambda<0$ allows for finite solutions both at the horizon and at infinity, and we shall prove it by contradiction. We begin by defining a new radial coordinate
\begin{align}
R_\ast = -\ln\left(\frac{R}{1-R}\right) \, .
\end{align}
The eigenvalue equation \eqref{eq:app7:eq1} then takes the simple form
\begin{align}
\frac{\dd^2 \Psi}{\dd R_\ast^2} = \lambda (1-R)\Psi \, ,
\end{align}
where $R_\ast$ increases monotonically from $-\infty$ (at the horizon) to $+\infty$ (at spatial infinity), like a tortoise coordinate. Because we assume that $\Psi$ is finite at the horizon, Eq.~\eqref{eq:app7:eq2} implies that $a_{-1} = 0$, which in turn implies to leading order
\begin{align}
\lim \limits_{R_\ast \rightarrow -\infty}\Psi = a_0 \, , \quad \lim\limits_{R_\ast \rightarrow -\infty} \frac{\dd\Psi}{\dd R_*} = 0 \, .
\end{align}
Note that while $a_0$ depends on the normalization of $\Psi$ we can always chose it to be positive. Integrating the eigenvalue relation \eqref{eq:app7:eq1} we may now write
\begin{align}
\frac{\dd \Psi}{\dd R_*}=\lambda \int\limits_{-\infty}^{R_*}(1-R)\Psi \dd R_*=\lambda \int\limits_0^R \frac{\dd R}{R}\Psi > 0 \, ,
\end{align}
which implies that $\dd \Psi/\dd R_*$ is a growing function in $R_\ast$. This means that $\Psi$ cannot be bounded at infinity, indicating that $\lambda$ cannot be positive. For this reason $\lambda<0$ and we will from now on parametrize $\lambda = -p^2$ for $p \in \mathbb{R}$. Equation~\eqref{eq:app7:eq1} then takes the form
\begin{align}
\label{eq:app7:eq3}
\Box \Psi_p(R) = -p^2 \Psi_p(R) \, ,
\end{align}
where we label the eigenfunctions with the parameter $p$ for later convenience. Note that for $p\in\mathbb{R}$ the asymptotics at infinity ($R=0$) are finite, $R^{\pm ip} = e^{\pm ipx}$, which in turn implies that the eigenvalue problem \eqref{eq:app7:eq3} has a \emph{continuous spectrum.} For this reason a solution of Eq.~\eqref{eq:ch7:f-def} can be written in the form
\begin{align}
\tilde{F}(s,R) = \int \dd p \, \rho_p \, e^{-p^2 s} \Psi_p(R) \, ,
\end{align}
where $\rho_p$ is the spectral density that we will determine later, and the factor $e^{-p^2 s}$ solves Eq.~\eqref{eq:app7:eq3}. Let us now focus on the eigenfunctions $\Psi_p(R)$, as they can be found analytically in this case. To that end, the complex-valued function
\begin{align}
Z_p(R) &= R^{ip} {}_2 F_1\left( ip, ip+1; 2ip+1; R \right) \, , \quad p \in \mathbb{R} \, , 
\end{align}
solves Eq.~\eqref{eq:app7:eq3}, and one has $Z_{-p}(R) = \overline{Z_p}(R)$. Using this property one may construct real-valued solutions such as $\Re[{Z}_p(R)]=1/2[Z_p(R)+Z_{-p}(R)]$ and $\Im[{Z}_p(R)]=(1/2i)[Z_p(R)-Z_{-p}(R)]$, and for that reason for each value $p\in\mathbb{R}$ there exist \emph{two} real-valued solutions. As we will demonstrate now, demanding the finiteness of the mode function $\Psi_p$ at the horizon lifts this degeneracy. Expanding $Z_p(R)$ at the horizon ($R=1$) gives
\begin{align}
Z_p(R) &\approx b_p + c_p \log(1-R) + \mathcal{O}\left(1-R\right) \, , \\
b_p &= - \frac{4^{ip}\Gamma\left(ip+\tfrac12\right)}{p\sqrt{\pi}\Gamma(ip)}\big[ -i + 2p\gamma + 2p\psi(ip) \big] \, , \\
c_p &= - \frac{4^{ip}\Gamma\left(ip+\tfrac12\right)}{\sqrt{\pi}\Gamma(ip)} \, ,
\end{align}
where $\Psi(ip)$ denotes the digamma function \cite{Olver:2010}. In order to cancel the divergent $c_p$-term one can construct the unique, real-valued expression
\begin{align}
\Psi_p(R) = f_p \Big\{ \Re(c_p)\Im[Z_p(R)] - \Im(c_p)\Re[Z_p(R)] \Big\} \, .
\end{align}
The degeneracy is lifted, and the above solution, for a given $p \ge 0$ is a real-valued solution of \eqref{eq:app7:eq3} that is also finite at the horizon and behaves like a plane wave at spatial infinity. The coefficient $f_p$ is a normalization factor we will discuss below.

\subsection{Orthogonality and normalization of eigenfunctions}
Having derived a set of physically well-behaved eigenfunctions $\Psi_p(R)$ we may now study their properties in more detail. The Wronskian\index{Wronskian} of these functions takes the form
\begin{align}
W[\Psi_p, \Psi_q] = R(1-R)\Big[ \Psi_p(R) \overset{\leftrightarrow}{\partial}_R \Psi_q(R) \Big] \, ,
\end{align}
where $f \overset{\leftrightarrow}{\partial}_R g = f \partial_R g - g \partial_R f$. Because the solutions $\Psi_p$ are finite both at the horizon ($R=1$) and at infinity ($R=0$) the Wronskian vanishes there, too, and we may write
\begin{align}
0 = \int\limits_0^1 \dd R \, \partial_R W[\Psi_p, \Psi_q]  = (q^2-p^2) \langle \Psi_p, \Psi_q\rangle \, , \quad
\langle \Psi_p, \Psi_q\rangle = \int\limits_0^1 \frac{\dd R}{R} \Psi_p(R) \Psi_q(R) \, .
\end{align}
This shows that eigenfunctions with different eigenvalues $p \not= q$ are indeed orthogonal, and with a proper normalization $f_p$ (see below) these functions are even orthonormal,
\begin{align}
\label{eq:app7:eq4}
\int\limits_0^1 \frac{\dd R}{R} \Psi_p(R) \Psi_q(R) = \delta(p-q) \, .
\end{align}
Now that this is only true for the eigenfunctions $\Psi_p$ that are finite at the horizon; the above considerations do \emph{not} hold for the complex-valued functions $Z_p$ since they typically diverge at the horizon.

In order to find the correct normalization $f_p$ we follow Refs.~\cite{Landau:1965,Baz:1969}. The asymptotics $Z_p(R\rightarrow 0) \approx R^{ip} = e^{ipx}$ imply
\begin{align}
\Psi_p(R\rightarrow0) \approx -f_p \Big[ \Re(c_p)\sin(px) + \Im(c_p) \cos(px) \Big] \, .
\end{align}
Let us consider now two functions $\Psi_p$ and $\Psi_k$ at spatial infinity, $R\approx 0$, where $k$ is very close to $p$. This is always possible since $p$ and $k$ belong to the continuous spectrum. Then we find
\begin{align}
\Psi_p \Psi_{k\approx p} \sim \frac12 |c_p|^2 f_p^2 \cos[(p-k)x] + \text{oscillating terms} \, ,
\end{align}
where oscillating terms are sine and cosine functions that depend on the sum $(p+k)\approx 2p \gg 0$. Performing a similar calculation for plane waves $\varphi_p(R)$ instead one finds $\varphi_p\varphi_{k\approx p} \sim 1/(2\pi)$. Since our eigenfunctions $\Psi_p$ asymptotically behave like plane waves it is useful to demand the same coefficient, and that implies
\begin{align}
f_p = \sqrt{\frac{2}{\pi|c_p|^2}} = \sqrt{\frac{2}{p\tanh(\pi p)}} \, .
\end{align}
Note that there is an additional factor of $\sqrt{2}$ that appears due to the normalization of cosines (as opposed to that of exponentials),
\begin{align}
\int\limits_{-\infty}^\infty \dd x \cos(px)\cos(kx) = \pi\delta(p+k) + \pi\delta(p-k) \, .
\end{align}
We can now evaluate $\Psi_p$ at the horizon explicitly, $R=1$, and find the manifestly finite value
\begin{align}
\Psi_p(1) = \sqrt{2p\coth(\pi p)} \, .
\end{align}
With all these considerations completed, we arrive at a regular, real-valued integral representation
\begin{align}
\label{eq:app7:eq5}
F(s,R) = \int\limits_0^\infty \dd p \, \rho_p \, e^{-p^2 s} \Psi_p(R) ,
\end{align}
and this representation will be at the center of our following considerations.

\subsection{Spectral measure}
We still need to determine the form of the spectral measure $\rho_p$, and it is related to the property $F(0,R) = R$. Inserting this relation into \eqref{eq:app7:eq5} yields
\begin{align}
1 = \int\limits_0^\infty \dd p \, \rho_p \, \frac{\Psi_p(R)}{R} \, ,
\end{align}
where we may now multiply on both sides with another eigenfunction $\Psi_q$ and integrate over $R$ in order to exploit the orthonormality relation \eqref{eq:app7:eq4},
\begin{align}
\begin{split}
\int\limits_0^1 \dd R \, \Psi_q(R) &= \int\limits_0^1 \dd R \int\limits_0^\infty \dd p \, \rho_p \, \Psi_q(R) \frac{\Psi_p(R)}{R} = \int\limits_0^\infty \dd p \rho_p \int\limits_0^1 \frac{\dd R}{R} \Psi_p(R) \Psi_q(R) \\
&= \int\limits_0^\infty \dd p \rho_p \delta(p-q) = \rho_q \, .
\end{split}
\end{align}
Fortunately, the integral in the very first expression on the left can be evaluated analytically:
\begin{align}
\begin{split}
d_p &= \int\limits_0^1 \dd R ~ Z_p(R) = \frac{1}{1+ip} {}_3 F{}_2(1+ip,1+ip,ip;~ 2+ip, 1+2ip;~ 1 ) \, , \\
\rho_p &= f_p \left[ \Re(c_p)\Im(d_p) - \Im(c_p) \Re(d_p) \right] \\
&= \frac{\sqrt{p\sinh(2\pi p)}}{\sqrt{2}\pi^{3/2}(1+p^2)} \Re \left[ \frac{i+p}{4^{ip}} \Gamma(ip)\Gamma\left(\tfrac12-ip\right) {}_3 F{}_2(1+ip,1+ip,ip;~ 2+ip, 1+2ip;~ 1 ) \right] \, .
\end{split}
\end{align}
Numerically one may verify that
\begin{align}
\int\limits_0^\infty \dd p \, \rho_p^2 = \frac 12 \, .
\end{align}

%%%%%%%%%%%%%%%%%%%%%%%%%%%%%%%%%%%%%%%%%%%%%%%%%%%%%%%%%%%%%%%%%%%%%%%%%%%%%%%%%%%%%%%%%%%%%%%%%%%
%
% Bibliography
%
%\chapter*{Bibliography}
%\label{ch:bibliography}
\thispagestyle{chapter_new}
%\addcontentsline{toc}{chapter}{Bibliography}
%\markright{Bibliography}

%\begingroup

%\endgroup

%%%%%%%%%%%%%%%%%%%%%%%%%%%%%%%%%%%%%%%%%%%%%%%%%%%%%%%%%%%%%%%%%%%%%%%%%%%%%%%%%%%%%%%%%%%%%%%%%%%
%
% Index
%
%\thispagestyle{chapter_new}
%\chapter{}
%\label{ch:index}
%\markright{Index}
\printindex
%\addcontentsline{toc}{chapter}{Index}
  
\end{document}